\documentclass[reqno]{amsart}
\usepackage{amssymb}
\usepackage{amsfonts}

\usepackage{mathrsfs}
\usepackage{caption}
\usepackage{mathtools}
\usepackage{enumerate}
\usepackage{pifont}
\usepackage[numbers,sort&compress]{natbib}
\usepackage{multirow}
\makeatletter
\@addtoreset{equation}{section}
\makeatother

\DeclareMathOperator{\re}{Re}
\DeclareMathOperator{\supp}{supp}
\DeclareMathOperator{\diag}{diag}

\newtheorem{theorem}{Theorem}[section]

\newtheorem{corollary}[theorem]{Corollary}
\newtheorem{proposition}[theorem]{Proposition}
\newtheorem{lemma}[theorem]{Lemma}

\newtheorem{remark}[theorem]{Remark}

\usepackage{hyperref}
\allowdisplaybreaks[3]

\begin{document}




\title[2D Behaviors of a 3D Bose Gas in G-P Regime]{The Second Order 2D Behaviors of a 3D Bose Gas in the Gross-Pitaevskii Regime}
\date{\today}

\author[X. Chen]{Xuwen Chen}
\address{Department of Mathematics, University of Rochester, Rochester, NY 14627, USA}

\email{xuwenmath@gmail.com}

\author[J. Wu]{Jiahao Wu}
\address{School of Mathematical Sciences, Peking University, Beijing, 100871, China}
\email{wjh00043@gmail.com}

\author[Z. Zhang]{Zhifei Zhang}
\address{School of Mathematical Sciences, Peking University, Beijing, 100871, China}

\email{zfzhang@math.pku.edu.cn}

\begin{abstract}
 We consider a system of $N$ bosons interacting in a three-dimensional box endowed with periodic boundary condition that is strongly confined in one direction such that the normalized thickness of the box $d\ll1$. We assume particles to interact through a repulsive, radially symmetric and short-range interaction potential with scattering length scale $a\ll d$. We present a comprehensive study of such system in the Gross-Pitaevskii regime, up to the second order ground state energy, starting from proving optimal Bose-Einstein condensation results which were not previously available. The fine interplay between the parameters $N$, $a$ and $d$ generates three regions. Our result in one region on the one hand, is compatible with the classical three-dimensional Lee-Huang-Yang formula. On the other hand, it reveals a new mechanism exhibiting how the second order correction compensates and modifies the first order energy, which was previously thought of as containing a jump, and thus explains how a three-dimensional Bose gas system smoothly transits into two-dimensional system. Delving into the analysis of this new mechanism exclusive to the second order, we discover a dimensional coupling correlation effect, deeply buried away from the expected 3D and quasi-2D renormalizations, and calculate a new second order correction to the ground state energy. This mechanism proves mathematically the effect of confinement mode coupling also called confinement-induced resonances in theoretical physics.
\end{abstract}

\maketitle
\tableofcontents

\section{Introduction}
\par Understanding how the geometry of domain effects the physics laws is a longlasting and  extensive project in both physics and mathematics. In the emergence, presence and description of the Bose-Einstein condensation, the effect of geometry has been an intriguing and significant subject among physicists for decades. A Bose–Einstein condensate (BEC) was first predicted by 1924 and is now also known as the 5th state of matters. At this state, a large amount of quantum particles occupy the same quantum state, such that microscopic quantum mechanical phenomena become macroscopic. This new state of matter can be used to explore fundamental questions in quantum mechanics, such as the superfluidity, quantized vortices, interference and decoherence. For example, superfluidity is one of the peculiar phenomena highly correlates with BEC. Due to the macroscopic occupation of the same state, system flows as a collective unit and no viscosity arises in the superfluid. Another example is the existence of quantum quantized vortices, which is believed to be responsible for superfluid phase transitions. For their achievement in producing the first gaseous condensation, Cornell, Wieman, and Ketterle received the 2001 Nobel Prize in Physics. The method used to cool and trap atoms with lasers is also awarded the 1997 Nobel Prize in Physics. After the initial qualitative studies, the concurrent research asks for more quantitative analysis. It is thus a natural problem to investigate the macroscopic effect of geometry on such quantum state of matter, with higher accuracy to reveal more inner mechanisms.

\par Next to the 3D experiments on BEC, another straightforward yet non-trivial geometrical domain would be the 2D plane, on which more than one version of electromagnetism could exist. But it is known that, theoretically speaking, in true 2D, a condensate can exist only at temperature $T=0$. In fact, since we are living in a 3D space, 2D domains can only be realized via thin planar 3D objects. On the other hand, since the first quantum encrypted video conference in 2017, it is desirable to realize BEC on micro-chips in the field of Quantum Computing and Quantum Communication. We can expect that, due to their sizes, these chips are to be modeled as, away from stacking, thin planar objects that are 3D regions but effectively 2D. In such a setting, the number of particles $N$ would not be so large and the thickness of the domain $d$ would be very small, while highly accurate description of the system is very much needed. Thus a second order ground state energy approximation is a natural next step for more accurate applications and devices. However, the second order prediction of the Lee-Huang-Yang formula for such a 3D-to-2D problem has not yet been given or verified in physics or mathematics.. Therefore, it is reasonable to investigate the systems of bosons confined in a thin planar trap, or in other words, its motion is strongly confined in one direction such that the system is so-called quasi-2D, different from the true 2D problem. Mathematically, we refer to this problem as a dimensional reduction problem. One can also consider the dimensional reduction problem from 3D to 1D. Physical theories of Bose-Einstein condensation trapped in quasi-2D were discussed extensively, for example, in \cite{BECinquasi2D}. Throughout these years, corresponding Bose-Einstein condensation has been established experimentally in various shapes of quasi-2D trap. For example, \cite{Vortexpairsquasi2d} realized Bose–Einstein condensates of $\sideset{^{23}}{}{\mathop{\mathrm{Na}}}$ gas in a pancake-shaped optical dipole trap, \cite{ATOMICBEC} demonstrated Bose–Einstein condensates of $\sideset{^{87}}{}{\mathop{\mathrm{Rb}}}$ atoms loaded into a pair of twisted-bilayer optical lattices, and \cite{MOLECULARBEC} reported the observation of Bose–Einstein condensates of $\sideset{}{_2}{\mathop{\mathrm{Cs}}}$ molecules in a two-dimensional, flat-bottomed trap.


\par On the other hand, if we only look at the leading order energy expansion, there seems to be a jump between the transition of different parameter regions. The Gross-Pitaevskii energy functional with a 3D coupling constant $g=ad^{-1}$ (that is to say in a 3D region) is given by
\begin{equation*}
  \mathcal{E}_{2D}^{GP}[\varphi]=\int_{\mathbb{T}^2}\vert\nabla\varphi\vert^2+4\pi Nad^{-1}\vert\varphi\vert^2
\end{equation*}
while the Gross-Pitaevskii energy functional in a quasi-2D region is equipped with a quasi-2D coupling constant
\begin{equation*}
  g=\Big\vert\ln(Nd^2)-\frac{d}{a}\Big\vert^{-1}.
\end{equation*}
Many efforts have been undertaken by physicists to explain this jump, such as \cite[Appendix B]{202405}, but so far satisfactory answers have not been reached. A related physical phenomenon is the Confinement-induced resonances (CIRs), which arise when particle scattering is set in the strongly anisotropic harmonic traps, that spatially confines the motion of the particles in one or two directions. Our study encounters a similar structure as the thickness $d$ decreasing and the system entering a quasi-2D parameter region. There, different from the resemblance to 3D for not so small $d$, the 2D correlation becomes noticeably appreciatable and will eventually contribute to the first order energy expansion. This effect explains how the leading order smoothly transits from a 3D coupling constant $g=ad^{-1}$ to a quasi-2D coupling constant. Moreover, a \textquotedblleft dimensional coupling \textquotedblright correlation arises.

\par The mathematics of Bose gas and its condensation has been investigated for almost a century and saw many progresses in recent years. Many important and interesting rigorous studies were collected and produced by Lieb, Seiringer, Solovej and Yngvason in \cite{lieb2005mathematics}. In mathematics, the dimensional reduction problem for a particle system was first studied in \cite{1dBEC} by Lieb, Seiringer and Yngvason, who gave a comprehensive answer about the first order ground state energy of Bose gas in highly elongated traps that can be considered as quasi-1D. In \cite{2005Bosons}, Schnee and Yngvason studied the quasi-2D ground state behavior of bosons in planar traps up to the first order. Recently, for the pure 3D Bose gas, Boccato, Brennecke, Cenatiempo and Schlein provided a mathematically rigorous proof of the ground state energy up to second order in \cite{2018Bogoliubov} in the Gross-Pitaevksii regime, and has greatly motivated the finer study of this subject. The recent progress includes, for example, the second order energy approximation in the thermodynamic limit \cite{thermo1,thermo2}. We follow the lead of these great works to the dimensional reduction problem in the Gross-Pitaevskii regime and offer a comprehensive study of the 3D to 2D problem. During the course, the geometry has led us to find more complete 3D formulae, and discover new mechanisms. On the one hand, we have found that, the second order correction compensates and modifies the first order energy, which was previously thought of as containing a jump, and thus explains how a three-dimensional Bose gas system smoothly transits into two-dimensional system. Moreover, after long computations ($\sim $100 pages) carrying out the sort of expected 3D and quasi-2D renormalizations (with needed modifications here), we unearth a correlation which contains energy at the order of the 3D and quasi-2D correlations. We call this the dimensional coupling correlation effect. These 3D, quasi-2D and dimensional coupling correlations jointly contribute to the energy of the first and second order. This is the 1st time effects of CIR shows up in the proof of mathematics.\footnote{Here, the resonance should be understood as a very long time effect in physics as we are studying the ground state.}

\par At a glance, the subject we are discussing here seems to be very specialized. It is actually the intersection of flourishing and deep research areas, like the rigorous analysis of (classical and quantum)
many-body systems which can also be further divided into the static, dynamic, $\delta $-potentials, Coulomb potentials, ... cases, (See, for example, \cite{CXWHolmer2,CXWHolmer3,CXW4,lewinNamSerfatySolovej2015bogoliubov,SSL,S2015,ESY1,ESY2,ESY3,GM1,GM2}) and the general area of finding the second order energy of a PDE/system. (See, for example, \cite{LWP2012,LW2023}). On the other hand, an overall physics review of CIRs can be found in \cite{202402}. The theoretical studies of CIRs in quasi-1D or quasi-2D boson systems were given, for example, in \cite{202404} and \cite{202401} respectively, while recent experimental results can be found in, for example, \cite{202403}. The study of CIRs to fermions is also of great interests among physicists, one can learn more information in, for instance, \cite{202405}. However, there has not been any comprehensive mathematical study so far, and this is the first time this physical phenomenon arise in the rigorous proof. Moreover, this dimensional coupling correlation effect we discovered has similar structure with the confined induced mode-coupling problem which arises in the study of dimensional crossover Anderson localization and in the study of the mode separation, also known as Migdal momentum shell renormalization method (The mode separation can be found, for example, \cite{Chaikin_Lubensky_1995}, and for the renormalization effect in the dimensional crossover, one can see, for example, \cite{PhysRevB.77.064205}). Our work may also provide some mathematical insights to the known to very difficult but highly-valued dimension crossover problem in Anderson localization studied in, for instance \cite{PhysRevB.77.064205}.

\par In this work, we consider a system of $N$ spinless bosons trapped in the 3D torus  $\Lambda_d=[-\frac{1}{2},\frac{1}{2}]^2\times[-\frac{d}{2},\frac{d}{2}]\subset\mathbb{R}^3$ with periodic boundary conditions. The particle motion is strongly confined in one direction in the sense that $d\to 0$. For $i=1,\dots,N$, $\mathbf{x}_i\in\Lambda_d$ describes the position of the i-th particle. Also for some $\mathbf{x}\in\mathbb{R}^3$, we may adopt the notation $\mathbf{x}=(x,z)$ with $x\in\mathbb{R}^2$ and $z\in\mathbb{R}$. The wave function of the system should be in the Hilbert space $L_s^2(\Lambda_d^N)$ consisting of functions that are symmetric with permutations of N particle, which is appropriate for describing the system of bosons. The N-body Hamilton operator is given by
\begin{align}\label{Hamiltonian}
 H_{N,a,d}=\sum_{j=1}^{N}-\Delta_{\mathbf{x}_j}+\sum_{1\leq i<j\leq N}v_{a}(\mathbf{x}_i-\mathbf{x}_j)
\end{align}
acting on the Hilbert space $L^2_s(\Lambda_d^N)$ with
\begin{equation}\label{interaction potential}
  v_a(\mathbf{x})=\frac{1}{a^2}v\left(\frac{\mathbf{x}}{a}\right).
\end{equation}
We may write $H_N=H_{N,a,d}$ for short. The main subject of study is the ground state energy of $H_N$, which we denote by
\begin{equation}\label{ground state energy}
  E_{N,a,d}=\inf_{\substack{\psi\in L^2_s(\Lambda_d^N)\\
\Vert\psi\Vert_2=1}}\langle H_N\psi,\psi\rangle.
\end{equation}
We may also put $E_N=E_{N,a,d}$ for short.
\par We require the interaction potential $v$ to be non-negative, radially-symmetric and compactly supported in a 3D ball $B_{R_0}$. Moreover, we assume $v$ has scattering length $\mathfrak{a}_0\geq0$. Hereafter we will always assume a function to be with these three properties whenever it is referred to as the \textit{interaction potential}.
\par The scattering length of an interaction potential $v$ is a vastly studied subject, here we follow the definition in \cite[Appendix C]{lieb2005mathematics} and define it through the following zero-energy scattering equation
\begin{equation}\label{zero-energy scattering equation}
  \left\{\begin{aligned}
   &-\Delta_{\mathbf{x}} f(\mathbf{x})+\frac{1}{2}v(\mathbf{x})f(\mathbf{x})=0,\quad \mathbf{x}\in\mathbb{R}^3.\\
   &\lim_{\vert \mathbf{x}\vert\to\infty}f(\mathbf{x})=1.
  \end{aligned}\right.
\end{equation}
There exists a non-negative constant $\mathfrak{a}_0$, which is known as the scattering length of $v$, such that for $\vert \mathbf{x}\vert>R_0$ we have
\begin{equation}\label{definition of scattering length}
  f(\mathbf{x})=1-\frac{\mathfrak{a}_0}{\vert \mathbf{x}\vert}.
\end{equation}
On the other hand, the scattering length $\mathfrak{a}_0$ can also be recovered by
\begin{equation}\label{recovery of scattering length via integral}
  \int_{\mathbb{R}^3}v(\mathbf{x})f(\mathbf{x})d\mathbf{x}=8\pi\mathfrak{a}_0.
\end{equation}
By scaling, the scattering length of $v_a$ is $a\mathfrak{a}_0$. The particle number $N$ should be large enough and the scattering parameter $a$ should be small enough in the sense that $N\to\infty,\,a\to0$. Since $v_a$ is supported on $B_{aR_0}$, then it is also reasonable to put $\frac{a}{d}\to0$ so that the support of $v_a$ is contained in the box $\Lambda_d$.

\par As we are considering $d\to0$, there is in fact an intrinsic 2D effect hiding in (\ref{Hamiltonian}). Apart from the 3D interaction potential $v$, it is also useful to consider a certain corresponding 2D interaction potential $u$, which is non-negative, radially-symmetric in 2D and compactly supported in a 2D ball $\mathcal{B}_{\rho_0}$. Similar to (\ref{zero-energy scattering equation}), we define the 2D scattering length of $u$ through
\begin{equation}\label{zero-energy scattering equation 2D}
  \left\{\begin{aligned}
   &-\Delta_{x} f_{\mathrm{2D}}(x)+\frac{1}{2}u(x)f_{\mathrm{2D}}(x)=0,\quad x\in\mathbb{R}^2.\\
   &\lim_{\vert x\vert\to\infty}\frac{f_{\mathrm{2D}}(x)}{\ln\vert x\vert}=1.
  \end{aligned}\right.
\end{equation}
Here we must impose the boundary condition $f_{\mathrm{2D}}(x)/\ln\vert x\vert\to1$ due to the $\ln\vert x\vert$ order divergence of the fundamental solution of the Laplacian in 2D. There exists a non-negative constant $\mathfrak{a}_u$, which is referred to as the 2D scattering length of $u$, such that for all $\vert x\vert>\rho_0$
\begin{equation}\label{definition of scattering length 2D}
  f_{\mathrm{2D}}(x)=\ln\frac{\vert x\vert}{\mathfrak{a}_u},
\end{equation}
while in 2D case,
\begin{equation}\label{recovery of scattering length via integral 2D}
  \int_{\mathbb{R}^2}u(x)f_{\mathrm{2D}}(x)dx=4\pi,
\end{equation}
and does not recover the scattering length $\mathfrak{a}_u$ as (\ref{recovery of scattering length via integral}).


Inspired by \cite{2005Bosons}, the coupling constant\footnote{One can consider a wider range of confining 3D trap potentials of bosonic systems (See for example \cite{2005Bosons}). The confining potential is given by $L^{-2}V_{ext}(L^{-1}x)+d^{-2}V_{ext}^{\perp}(d^{-1}z)$. Let $s(z)$ and $e^{\perp}$ be the ground state wave function and the ground state energy of operator $-\Delta_z+V_{ext}^\perp(z)$ respectively, then the effective 2D scattering length is given by $a_{\mathrm{2D}}=d\exp\Big({-\frac{d}{2a\mathfrak{a}_{0}}}\big(\int s(z)^4dz\big)^{-1}\Big)$ and $g=\vert\ln(\bar{\rho}a_{\mathrm{2D}}^2)\vert^{-1}$ ($\bar{\rho}$ is the mean density) is the coupling constant of 2D Gross-Pitaevskii functional


\begin{equation*}
  \mathcal{E}^{\mathrm{GP}}_{\mathrm{2D}}[\varphi]
  =\int_{\mathbb{R}^2}\vert\nabla_x\varphi(x)\vert^2+L^{-2}V_{ext}(L^{-1}x)\vert\varphi(x)\vert^2+4\pi
  Ng\vert\varphi(x)\vert^4dx.
\end{equation*}
In our setting in the Gross-Pitaevskii regime, we consider box trap potential such that $L=1$, $\bar{\rho}=N/L^2=N$, $\int s(z)^4dz=1$ and $e^\perp=0$ and thus $g$ becomes (\ref{coupling constant g}).}  $g$ under our setting (a 3D to 2D problem) is given by
\begin{equation}\label{coupling constant g}
  g=\vert\ln(Na_{\mathrm{2D}}^2)\vert^{-1},
\end{equation}
with $a_{\mathrm{2D}}$ the effective 2D scattering length given by
\begin{equation}\label{a_2D}
  a_{\mathrm{2D}}=d\exp\Big({-\frac{d}{2a\mathfrak{a}_{0}}}\Big).
\end{equation}
Plugging (\ref{a_2D}) into (\ref{coupling constant g}) we can rewrite $g$ as
\begin{equation}\label{rewrite coupling constant g}
  g=\Big\vert\ln(Nd^2)-\frac{d}{a\mathfrak{a}_0}\Big\vert^{-1}.
\end{equation}
This definition of $g$ leaves a seemingly jump in the first order energy, as we will discuss further near (\ref{leading order more precise}) and (\ref{leading order more precise II}) and in Remark \ref{remark1}, that this previously thought of jump actually contains a hidden smooth transition mechanism if we go to the second order. The term $\frac{d}{a\mathfrak{a}_0}$ in (\ref{rewrite coupling constant g}) indicates the 3D effect since we retrieve the classical 3D coupling constant if we simply let $d=1$, while the term $\ln(Nd^2)$ represent the 2D effect of a 3D system due to the smallness of $d$. These two effects compete with each other, or in other words, determine the physical behavior (3D or 2D) of the system of Bose gas. Their relationship therefore prompts a partition of parameter region. We divide the parameters into three regions due to different correlations between them.
\begin{equation}\label{Region I II III}
   \left\{\begin{aligned}
&\text{Region I: $\frac{d}{a}\gg\vert\ln(Nd^2)\vert$, $Nd^2\gg 1$}\\
&\text{Region II: $\frac{d}{a}\gg\vert\ln(Nd^2)\vert$, $Nd^2\lesssim 1$}\\
&\text{Region III: $\frac{d}{a}\sim\vert\ln(Nd^2)\vert$}
\end{aligned}\right.
\end{equation}
  We refer to the region where $\frac{d}{a}\gg\vert\ln(Nd^2)\vert$ and $Nd^2\gg 1$ as \textbf{Region I} or a 3D region, the region where $\frac{d}{a}\gg\vert\ln(Nd^2)\vert$ and $Nd^2\lesssim 1$ as \textbf{Region II} or an intermediate region, and the region where $\frac{d}{a}\sim\vert\ln(Nd^2)\vert$ as \textbf{Region III} or a quasi-2D region. When we say \textbf{Gross-Pitaevskii condition} or \textbf{Gross-Pitaevskii regime} or \textbf{Gross-Pitaevskii limit}, apart from the requirements that $N$ tends to infinity and $a$, $d$, and $\frac{a}{d}$ tends to $0$, we additionally require that $\frac{Na}{d}=1$ in Region I and II, and $Ng=\mathfrak{a}_0$ in Region III. In other words, we normalize $Ng\sim\mathfrak{a}_0$ in the Gross-Pitaevskii regime. Here in this paper, we are mainly interested in the relative size of the normalized thickness of the box $d$ with respect to $N$, thus we always set $\frac{d}{a}\sim N$. Therefore in Region III we always consider $\frac{d}{a}\sim\vert\ln(Nd^2)\vert$. One could investigate a larger Region $\mathrm{III}^\prime$ where $\frac{d}{a}\lesssim\vert\ln(Nd^2)\vert$. In fact, our result even go a bit further beyond the barrier $\frac{d}{a}\sim\vert\ln(Nd^2)\vert$ (See Theorem \ref{core III}). For technical reason, we further divide Region II into two sub-regions in the Gross-Pitaevskii regime
\begin{equation*}\label{Region IIt1 and IIt2}
  \left\{\begin{aligned}
&\text{Region $\textrm{II}_{\mathrm{I}}$: $e^{-CN^{t_1}}\lesssim d\lesssim N^{-\frac{1}{2}}$ for some fix $t_1\in (0,1)$}\\
&\text{Region $\textrm{II}_{\mathrm{III}}$: $d\lesssim e^{-CN^{t_2}}$ for some fix $t_2\in (0,1)$ and $N=\frac{d}{a}\gg\vert\ln(Nd^2)\vert$}
\end{aligned}\right.
\end{equation*}
Here we denote $C$ as some universal constant. We will choose $t_1=\frac{1}{72}$ and $t_2$ to be any fixed number less than $t_1$. Notice that Regions $\textrm{II}_{\mathrm{I}}$ and $\textrm{II}_{\mathrm{III}}$ intersect if $t_1>t_2$.

\par In this work, we are mainly interested in the relative size of the normalized thickness of the box $d$ with respect to $N$. Moreover, $\ln d^{-1}$ appears to be a second order correction to the leading order ground state energy (see Theorem \ref{core}). Thus we demonstrate the partition of parameter regions with respect to $\frac{\ln\ln d^{-1}}{\ln N}$, at the scale of energy correction per particle in the logarithmic sense, in the following Figure \ref{Region} since we prefer a bounded graph here (The $\frac{\ln d^{-1}}{ N}$ scale will result in a much magnified size of Region III and a much reduced size of Region I. Either way, they mean the same thing).
\begin{figure}[h]
\centering
\includegraphics[width=1\textwidth]{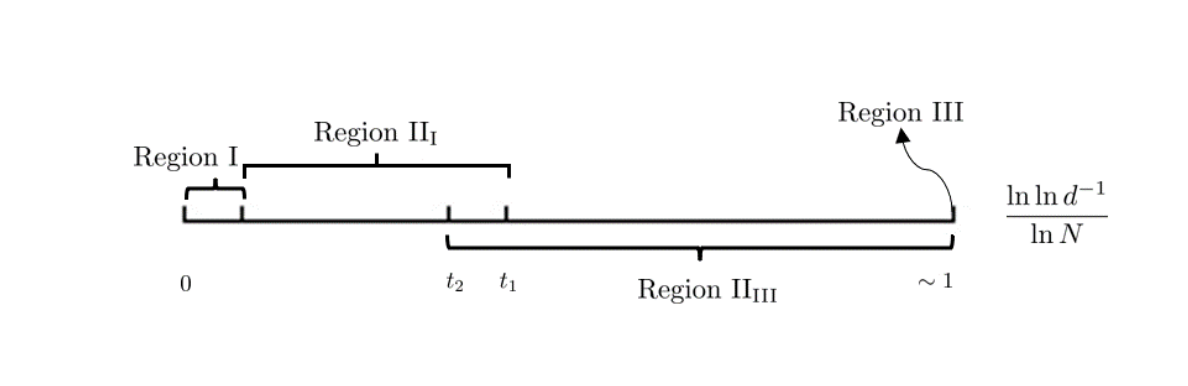}
\captionsetup{font=small}
\caption{Partition of parameter regions (not up to scale nor proportional to how often each case happens)}
\label{Region}
\end{figure}

\par In Region I, $d$ decays slower than $N^{-\frac{1}{2}}$, and it is reasonable to presume 3D behavior dominates the system through constraint. On the other hand, in Region III, $d$ decays exponentially such that the system is more constrained in a 2D space, thus we expect the 2D effect outweighs the 3D effect. The intermediate Regions $\mathrm{II}_{\mathrm{I}}$ and $\mathrm{II}_{\mathrm{III}}$ are transition regions, we will prove how the 3D system smoothly transits into 2D, despite a seemingly jump in the first order energy. In fact, all transitions between regions are smooth. Moreover, we uncover a dimensional coupling effect that contributes to the second order ground state energy.

\par Notice that here the three-dimensional density $\rho_{\mathrm{3D}}=N/d$ since we have normalized $L$ to $1$, then the case $Nd^2\gg 1$ in Region I corresponds to the condition $\rho_{\mathrm{3D}}^{-\frac{1}{3}}\gg d$ for the 3D mean interparticle distance, which coheres with the physics terminology of 3D region in the 3D-to-2D problem. On the other hand, even in the quasi-2D region under the G-P condition, we still have $N\gtrsim\vert\ln(Nd^2)\vert$ i.e. $d^2\gtrsim N^{-1}e^{-N}$ , which implies that under our configurations, the problem can never be directly regarded as a true 2D problem.

\par A mathematically rigorous analysis of the ground state energy $E_N$ (\ref{ground state energy}) was first presented by Schnee and Yngvason in \cite{2005Bosons}. It is proved that, to the leading order in the Gross-Pitaevskii regime
\begin{equation}\label{leading order}
  E_N=4\pi N^2g\big(1+o(1)\big).
\end{equation}
Although it was not explicitly shown in \cite{2005Bosons}, we can still follow their calculations to obtain a numerically more precise result
\begin{equation}\label{leading order more precise}
  E_N-4\pi N^2a\mathfrak{a}_0d^{-1}=\left\{\begin{aligned}
&O(N^{\frac{15}{17}}d^{\frac{2}{17}}),\quad \text{In Region I}\\
&O\big(N^{\frac{8}{9}}+\big\vert\ln(Nd^2)\big\vert\big),\quad \text{In Region $\textrm{II}_{\mathrm{I}}$}
\end{aligned}
\right.
\end{equation}
and
\begin{equation}\label{leading order more precise II}
  E_N-4\pi N^2g=\left\{\begin{aligned}
&O(N^{1-s}),\quad\text{In Region $\textrm{II}_{\mathrm{III}}$}\\
&O(Na^{\frac{1}{9}}d^{-\frac{1}{9}}),\quad \text{In Region III}
\end{aligned}
\right.
\end{equation}
Here $s\in (0,1)$ depending only on $t_2$ and may tend to $0$ when $t_2$ tends to $0$ and tend to $\frac{1}{9}$ when $t_2$ tends to $1$. Notice that we have normalized $Ng\sim \mathfrak{a}_0$.

\par For the leading order in the Gross-Pitaevskii regime, \cite{2005Bosons} actually showed its result in two parts in (\ref{leading order more precise}) and (\ref{leading order more precise II}) above. But as the interplay of parameters cross the border from Region $\mathrm{I}$ to Region $\mathrm{III}$, the $\frac{N^2a}{d}$ leading order in (\ref{leading order more precise}) becomes larger than $N$ and thus (\ref{leading order more precise}) is no longer legitimate in Region $\mathrm{III}$, which leaves a jump between different parameter regions. Our result in Theorem \ref{core} will explain how the leading order shown in (\ref{leading order more precise}) smoothly transits into (\ref{leading order more precise II}).

\par In this paper, we calculate, for all three regions, beyond the first order ground state energy approximation, to provide more accurate digits in realistic systems.

\subsection{Main Theorems}

We first establish, in the following Theorem \ref{core}, a higher order approximation of the ground state energy in both Regions I and $\mathrm{II}_{\mathrm{I}}$.

\begin{theorem}[For Regions I \& $\mathrm{II}_{\mathrm{I}}$]\label{core}
  Let $v$ be a smooth, non-negative, radially-symmetric and compactly supported function. Then in Regions $\mathrm{I}$ or $\mathrm{II}_{\mathrm{I}}$ in the Gross-Pitaevskii regime, that is in the limit $N\to\infty$, $a,\,d,\,\frac{a}{d}\to0$ while $\frac{Na}{d}=1$ and $d\gtrsim e^{-CN^{t_1}}$ with $t_1=\frac{1}{72}$ and $C$ some universal constant, the ground state energy $E_N$ of $H_N$ given in (\ref{Hamiltonian}) has the form
       \begin{align}\label{thm E_N I}
          E_N=4\pi(N-1)\frac{Na}{d}\mathfrak{a}_0+\mathfrak{e}_d+E_{Bog}^{(d)}+
               O(d^{\frac{1}{4}}\ln d^{-1}+N^{-\frac{1}{8}+t_1}),
       \end{align}
        where
       \begin{equation}\label{thm e_d I}
          \mathfrak{e}_d=2\mathfrak{a}_0^2d^2
          -\lim_{M\to\infty}\sum_{\substack{p\in\mathbb{Z}^3\backslash\{0\}\\
          \vert p_1\vert,\vert p_2\vert,\vert p_3\vert\leq
          M}}\frac{4\mathfrak{a}_0^2\cos(d\vert\mathcal{M}_dp\vert)}
          {\vert\mathcal{M}_dp\vert^2},
       \end{equation}
         and
        \begin{equation}\label{thm E_bod^d I}
         E_{Bog}^{(d)}=\frac{1}{2}\sum_{p\in2\pi\mathbb{Z}^3\backslash\{0\}}
          e_p^{(d)},
        \end{equation}
        where
         \begin{equation}\label{thm e_bog^d I}
         e_p^{(d)}=-\vert\mathcal{M}_dp\vert^2-8\pi\mathfrak{a}_0+
          \sqrt{\vert\mathcal{M}_dp\vert^4+16\pi\mathfrak{a}_0\vert\mathcal{M}_dp\vert^2}
          +\frac{(8\pi\mathfrak{a}_0)^2}{2\vert\mathcal{M}_dp\vert^2}.
         \end{equation}
          Here we let $\mathcal{M}_{d}=\diag{(1,1,\frac{1}{d})}$, so that for all 3D vectors $p=(p_1,p_2,p_3)$
         \begin{equation}\label{thm M_dp}
           \vert\mathcal{M}_dp\vert^2=p_1^2+p_2^2+\frac{p_3^2}{d^2}.
         \end{equation}
          Moreover, $\mathfrak{e}_d$ and $E_{Bog}^{(d)}$ are exactly of the order
         $\ln d^{-1}$ and $1$ respectively, and we can write explicitly
          \begin{equation}\label{thm order of e_d I}
           \mathfrak{e}_{d}=-8\pi\mathfrak{a}_0^2\ln d^{-1}+O(1).
          \end{equation}
\end{theorem}

\begin{remark}\label{remark1}
\
  \begin{enumerate}[$(1)$]
    \item If we formally take $d=1$ in (\ref{thm E_N I}) in Theorem \ref{core}, we immediately recover the pure 3D ground state energy approximation up to second order. In the thermodynamic limit, it is well-known as the Lee-Huang-Yang formula established through a series of pioneering works \cite{LHY1,LHY2,LHY3}. In the Gross-Pitaevskii regime, this formula was mathematically rigorous proved in \cite{2018Bogoliubov}. In fact, even though our result in Theorem \ref{core} seems to hold true under the requirement that $d\to0$, we can modify the calculation details in our proof such that our argument is universally legal in the region where $e^{-CN^{t_1}}\lesssim d\lesssim1$. In fact, we can make use of the cut-off parameter $\nu$ in Lemma \ref{commutator of H_21,H_4with B'} to gain a new estimate instead  of (\ref{E_21^B'}), and we will demand the parameter $\textit{l}=N^{-\alpha}$ for some $0<\alpha<1$ in Section \ref{Proof of the Main Theorem for Region I}, then use the algorithm in \cite{hainzlSchleinTriay2022bogoliubov} to complete it. Due to this observation, we know that our result is in fact compatible with the pure 3D result, or in other words, the system transits smoothly from the case $d\sim1$ to $d\ll1$.
    \item The second order term $E^{(d)}_{Bog}$ given in (\ref{thm E_bod^d I}) is asymptotically independent of $d$, i.e. it is in fact of order $1$\footnote{We thank Arnaud Triay for pointing out this interesting fact.}. Actually, it can be rewritten as
        \begin{equation}\label{2025}
           E_{Bog}^{(d)}=\frac{1}{2}\sum_{\substack{p\in2\pi\mathbb{Z}^3\backslash\{0\}\\
           p_3=0}}
          e_p^{(d)}+O(d^2).
        \end{equation}
        For a proof to (\ref{2025}) one can see (\ref{WTH2}) and (\ref{est of e_bog}). Here we preserve the form (\ref{thm E_bod^d I}) to compare our result with the $3D$ result given in \cite[Theorem 1.1]{2018Bogoliubov}.
    \item The constant $\mathfrak{e}_d$, which was thought of a correction to the scattering length $\mathfrak{a}_0$ due to the finiteness of the box $\Lambda_d$ in \cite{2018Bogoliubov}, reveals a new mechanism exhibiting how the second order correction compensates and modifies the first order erengy. From its logarithmic dependence on $d$ shown in (\ref{thm order of e_d I}), we notice that $\mathfrak{e}_d$ acts as an intermediate correction between the leading order $\frac{N^2a}{d}$ and the order $1$ remainder. If we take a modified coupling constant $\tilde{g}$ to be
        \begin{equation}\label{modified g}
          \tilde{g}=\big\vert\ln\big(N\tilde{a}_{\mathrm{2D}}^2\big)\big\vert^{-1}
        \end{equation}
        with the modified effective 2D scattering length $\tilde{a}_{\mathrm{2D}}$ being
        defined by
        \begin{equation}\label{modified a_2D}
          \tilde{a}_{\mathrm{2D}}=N^{-\frac{1}{2}}d\exp\Big(-\frac{d}{2a\mathfrak{a}_0}\Big)
        \end{equation}
        then as a corollary of our result in Theorem \ref{core}, we have
        \begin{equation}\label{modified result I}
          E_N=4\pi N^2\tilde{g}+O(1).
        \end{equation}
        The equivalence of different definitions of effective 2D scattering length was discussed in \cite[(1.19)]{2005Bosons}, but only for the leading order. As discussed around (\ref{leading order more precise}) and (\ref{leading order more precise II}), for the leading order in the Gross-Pitaevskii regime, \cite{2005Bosons} actually left a jump between different parameter regions. Our calculations explain how the leading order shown in (\ref{leading order more precise}) smoothly transits into (\ref{leading order more precise II}) due to the correction of $\mathfrak{e}_d$.
        \item In Theorem \ref{core}, we can consider a more general anisotropic 3D torus $\Lambda_{(d_1,d_2,d_3)}$ given by
            \begin{equation*}
              \Lambda_{(d_1,d_2,d_3)}=[-\frac{d_1}{2},\frac{d_1}{2}]\times
              [-\frac{d_2}{2},\frac{d_2}{2}]\times[-\frac{d_3}{2},\frac{d_3}{2}]
            \end{equation*}
            with parameters $d_i$ satisfy $d_1\geq d_2\geq d_3$ and some other suitable conditions. We can go through all the calculations in the proof of Theorem \ref{core} carefully and reach a more general ground state energy approximation for Hamiltonian defined on the anisotropic 3D torus $\Lambda_{(d_1,d_2,d_3)}$, which is a small but useful extension of the 3D Lee-Huang-Yang formula, as a perfect cube does not exist in reality.
  \end{enumerate}
\end{remark}

\par For the result in Regions $\mathrm{II}_{\mathrm{III}}$ and III, we also obtain the second order energy, but its format is relatively less explicit compared to Theorem \ref{core} due to a dimensional coupling effect. We reduce the approximation of the ground state energy up to second order to several one-particle scattering equations, and this dimensional coupling effect follows from these equations. An energy $\mathcal{I}_{N,a,d}$ related to these equations provides the second order result. To clarify the definition of $\mathcal{I}_{N,a,d}$, we first provide these scattering equations. We first consider the following ground state problem of a three-dimensional, one-particle scattering equation equipped with Neumann boundary condition.
\begin{equation}\label{introduction asymptotic energy pde on the ball}
  \left\{\begin{aligned}
  &(-\Delta_{\mathbf{x}}+\frac{1}{2}v)f_\textit{l}=\lambda_\textit{l}f_\textit{l},\quad \vert \mathbf{x}\vert\leq \frac{d}{a}\textit{l},\\
  &\left.\frac{\partial f_\textit{l}}{\partial \mathbf{n}}\right\vert_{\vert \mathbf{x}\vert=\frac{d}{a}\textit{l}}=0,\quad \left.f_{\textit{l}}\right\vert_{\vert \mathbf{x}\vert=\frac{d}{a}\textit{l}}=1.
  \end{aligned}\right.
\end{equation}
for some $\textit{l}\in (0,\frac{1}{2})$. In Theorem \ref{core III}, we choose
\begin{equation}\label{intro choose l}
  \textit{l}=\frac{1}{4}.
\end{equation}
Equation (\ref{introduction asymptotic energy pde on the ball}) can be interpreted as an asymptotic equation to the behavior of a single one boson among a large bosonic system interacting in three-dimensional space. We then let for $\mathbf{x}\in\Lambda_d$
\begin{equation}\label{intro define eta}
  \eta(\mathbf{x})=-\frac{1}{\sqrt{d}}\Big(1-f_{\textit{l}}\big(\mathbf{x}/a\big)\Big).
\end{equation}
Here we make a constant extension to $f_{\textit{l}}$ due to the Neumann boundary condition. For  any $p\in2\pi\mathbb{Z}^3$ we denote $\eta_p$ as the Fourier coefficients of $\eta$ on the torus $\Lambda_d$. More precisely
\begin{equation}\label{intro define eta_p}
  \eta_p=\frac{1}{\sqrt{d}}\int_{\Lambda_d}\eta(\mathbf{x})e^{-p^{T}\mathcal{M}_d\mathbf{x}}
d\mathbf{x}.
\end{equation}
Recall that $\mathcal{M}_{d}=\diag{(1,1,\frac{1}{d})}$. We also let
\begin{equation}\label{intro define W_p}
  W(\mathbf{x})=\frac{\lambda_{\textit{l}}}{a^2\sqrt{d}}f_{\textit{l}}\big(\mathbf{x}/a\big)
\chi_{d\textit{l}}(\mathbf{x}).
\end{equation}
Here $\chi_{d\textit{l}}(\mathbf{x})$ is the characteristic function of the closed 3D ball $\overline{B}_{d\textit{l}}$. We also denote $W_p$ as the Fourier coefficients of $W$ on the torus $\Lambda_d$.

\par When we enter Region $\mathrm{II}_{\mathrm{III}}$ or even Region III, where $d$ decays exponentially with respect to some power of $N$, the two-dimensional interaction effect dominates. We then consider another two-dimensional, one-particle scattering equation equipped with Neumann boundary condition. Here, one version of the induced 2D interaction potential $u$ is
\begin{equation}\label{intro define u}
  u(x)=\frac{2(d\textit{l})^3}{\sqrt{d}}\int_{-1}^{1}W\left(d\textit{l}\cdot\mathbf{x}\right)dz
\end{equation}
or in other words
\begin{equation}\label{intro define u_dl}
  u_{d\textit{l}}(x)=\frac{1}{(d\textit{l})^2}u\left(\frac{x}{d\textit{l}}\right)
=\frac{2}{\sqrt{d}}\int_{-d\textit{l}}^{d\textit{l}}
W(\mathbf{x})dz.
\end{equation}
With $u$ introduced, we consider the following equation
\begin{equation}\label{introduction asymptotic energy pde on the ball 2D}
  \left\{\begin{aligned}
  &(-\Delta_x+\frac{1}{2}u)g_\textit{h}=\mu_\textit{h}g_\textit{h},\quad \vert x\vert\leq \frac{h}{d\textit{l}},\\
  &\left.\frac{\partial g_\textit{h}}{\partial \mathbf{n}}\right\vert_{\vert x\vert=\frac{h}{d\textit{l}}}=0,\quad \left.g_h\right\vert_{\vert x\vert=\frac{h}{d\textit{l}}}=1
  \end{aligned}\right.
\end{equation}
for some $h\in(0,\frac{1}{2})$. In Theorem \ref{core III}, we choose
\begin{equation}\label{intro choose h}
  h=N^{-\frac{13}{2}}.
\end{equation}
Equation (\ref{introduction asymptotic energy pde on the ball 2D}) can be regarded as a two-dimensional approximation of a single boson inside a large bosonic system, which interacts in a two-dimensional space. In our setting, when $d$ is small enough compared to $N$, this approximation dominates the first order energy. We then similarly let for $\mathbf{x}=(x,z)\in\Lambda_d$
\begin{equation}\label{intro define xi}
  \xi(\mathbf{x})=-\frac{1}{\sqrt{d}}\Big(1-g_{h}\big(x/(d\textit{l})\big)\Big).
\end{equation}
Here we also make a constant extension to $g_{h}$ due to the Neumann boundary condition. We also define for $\mathbf{x}=(x,z)\in\Lambda_d$
\begin{equation}\label{intro define Y}
  Y(\mathbf{x})=\frac{\mu_h}{(d\textit{l})^2\sqrt{d}}{g}_h\big(x/(d\textit{l})\big)
\chi^{\mathrm{2D}}_h(x)+\Big(W(\mathbf{x})
-\frac{1}{2\sqrt{d}}u_{d\textit{l}}(x)\Big){g}_h\big(x/(d\textit{l})\big),
\end{equation}
where $\chi_{h}^{\mathrm{2D}}(x)$ is the characteristic function of the 2D closed ball $\mathcal{B}_{h}$. We also denote for $p\in2\pi\mathbb{Z}^3$, $\xi_p$ and $Y_p$ as the Fourier coefficients of $\xi$ and $Y$ on the torus $\Lambda_d$ respectively.

\par To derive a second order asymptotic formula to the ground state energy (\ref{ground state energy}) in Regions $\mathrm{II}_{\mathrm{III}}$ or $\mathrm{III}$, where the 2D effect dominates, we must take a dimensional coupling effect into account. We define
\begin{align}
  \mathfrak{D}(\mathbf{x})&=\Big(\frac{1}{2}v_a(\mathbf{x})-
\sqrt{d}W(\mathbf{x})\Big)\xi(\mathbf{x})\label{intro define d(x)}\\
k(\mathbf{x})&=\sqrt{d}\eta(\mathbf{x})\xi(\mathbf{x})\label{intro define k}\\
q(\mathbf{x})&=-2\sqrt{d}\nabla_{\mathbf{x}}\eta(\mathbf{x})
\cdot\nabla_{\mathbf{x}}\xi(\mathbf{x})-\sqrt{d}
\eta(\mathbf{x})\Delta_{\mathbf{x}}\xi(\mathbf{x})\label{intro define q}
\end{align}
From (\ref{introduction asymptotic energy pde on the ball}), we know that $k(\mathbf{x})$ satisfies the following dimensional coupling scattering equation on $\Lambda_d$
\begin{equation}\label{intro dimensional coupling scattering equation}
  -\Delta_{\mathbf{x}}k(\mathbf{x})+\frac{1}{2}v_a(\mathbf{x})k(\mathbf{x})
+\mathfrak{D}(\mathbf{x})=q(\mathbf{x}).
\end{equation}
We may denote respectively $\mathfrak{D}_p$, $k_p$ and $q_{p}$ as the Fourier coefficients of $\mathfrak{D}$, $k$ and $q$ on the torus $\Lambda_d$.

\par Recall that the induced 2D interaction potential $u_{d\textit{l}}$ given in (\ref{intro define u_dl}) is constructed by taking the average value of $2\sqrt{d}W(\mathbf{x})$ in the $z$ direction. Thus $\mathfrak{D}$ defined in (\ref{intro define d(x)}) measures the difference between the original interaction potential $v_a$ and the induced potential $2\sqrt{d}W$. Their discrepancy suggests a modification to the second order energy, since simply replacing the original interaction potential $v_a$ with the induced 2D interaction potential $2\sqrt{d}W$ is no longer enough for the calculation of energy of higher order. To characterize $\mathfrak{D}$, we introduce the dimensional coupling equation (\ref{intro dimensional coupling scattering equation}). By the definition (\ref{intro define k}) of $k$, where a 3D approximation $\eta$ and a 2D approximation $\xi$ entangle with each other, equation (\ref{intro dimensional coupling scattering equation}) can be construed as the characterization of a single boson inside a large bosonic system interacting in 3D space, where its movement is strongly limited in one direction. The 2D approximation and this dimensional coupling structure are absorbed by the 3D effect and will contribute to the second or lower order energy when $d$ is relatively large. When $d$ enters a especially thin region, Regions $\mathrm{II}_{\mathrm{III}}$ or $\mathrm{III}$ to be precise, their scales will be large enough to compete with the classical leading order generated by a 3D approximation. Next to the 3D renormalization, the quasi-2D renormalization is expected, which is the 2D approximation characterized by the 2D equation (\ref{introduction asymptotic energy pde on the ball 2D}), and we will prove that it modifies the leading order shown in Theorem \ref{core}. But even after these 3D and quasi-2D renormalizations, there is still an energy contribution of at least order $O(N^2a^2d^{-2})$ left hidden in the excitation Hamiltonian, where we discover that the dimensional coupling correlation structure is responsible for this energy contribution. The dimensional coupling structure is the residue characterized by the dimensional coupling scattering equation (\ref{intro dimensional coupling scattering equation}), and the energy driven by it will become one of the main components of the second order energy (The problem, as we will explain, takes about 100 pages of calculation to see it).

\par These three equations (\ref{introduction asymptotic energy pde on the ball}), (\ref{introduction asymptotic energy pde on the ball 2D}) and (\ref{intro dimensional coupling scattering equation}) are crucial to concluding both Theorems \ref{core} and \ref{core III}. We need to obtain more properties of them so that we can construct unitary operators to extract energy contributions to the leading or second order using these three equations. Therefore, a more thorough analysis on them will be carried out in Section \ref{Scattering Equations with Neumann condition}. We want to remark that although it seems straightforward to use directly the dimensional coupling scattering equation (\ref{intro dimensional coupling scattering equation}) to characterize the one-particle behavior rather than going through the labyrinthine process starting from a 3D scattering equation then a 2D version and finally the dimensional coupling scattering equation (See the end of this section or Section \ref{Excitation Hamiltonians} for a more comprehensive lay-out), it is in fact difficult to compute mathematically the corresponding energy of equation (\ref{intro dimensional coupling scattering equation}) due to the entanglement of two different dimensions and we still need equations (\ref{introduction asymptotic energy pde on the ball}) and (\ref{introduction asymptotic energy pde on the ball 2D}) to attain an explicit formula of the leading order energy.

\par With all the above preparations, we can present the result of higher order approximation of ground state energy in both Regions $\mathrm{II}_{\mathrm{III}}$ and III.
\begin{theorem}[For Regions III \& $\mathrm{II}_{\mathrm{III}}$]\label{core III}
  Let $v$ be a smooth, non-negative, radially-symmetric and compactly supported function. Then in Regions $\mathrm{II}_{\mathrm{III}}$ or $\mathrm{III}$ in Gross-Pitaevskii regime, that is in the limit $N\to\infty$, $a,\,d,\,\frac{a}{d}\to0$ while $Ng=\mathfrak{a}_0$, $\frac{a}{d}\sim N^{-1}$ and $d\lesssim e^{-CN^{t_2}}$ with $0<t_2<t_1=\frac{1}{72}$ fixed and $C$ some universal constant. Then the ground state energy $E_N$ of $H_N$ given in (\ref{Hamiltonian}) has the form

  \begin{align}\label{thm E_N III}
   E_N=4\pi(N-1)Ng+\mathcal{I}_{N,a,d}+O\Big\{\Big(N\Big(\frac{a}{d}\Big)^{\frac{9}{8}}
+\Big(\frac{a}{d}\Big)^{\frac{1}{8}}\ln N\Big)\Big\}.
  \end{align}
Here the second order term $\mathcal{I}_{N,a,d}$ (or $\mathcal{I}_N$ for short) is given by
\begin{align}\label{thm I_N,a,d III}
\mathcal{I}_{N}=&(N-1)N\big(\mathcal{C}_N-4\pi g\big)
+\frac{1}{2}\sum_{p\in2\pi\mathbb{Z}^3\backslash\{0\}}\Big\{
-\vert\mathcal{M}_dp\vert^2-2N\mathcal{C}_N\nonumber\\
&+\sqrt{\vert\mathcal{M}_dp\vert^4+4N\mathcal{C}_N\vert\mathcal{M}_dp\vert^2
+4N^2\big(\mathcal{C}_N^2-(q_p+Y_p)\big)}\Big\}.
\end{align}
with
\begin{align}\label{thm C_N III}
  \mathcal{C}_N=\Big(W_0+\sum_{p\neq0}W_p\eta_p+\sum_{p\neq0}\big(W_p+Y_p+\mathfrak{D}_p\big)\xi_p
+\sum_{p\neq0}\big(2Y_p+\mathfrak{D}_p+q_p\big)k_p\Big).
\end{align}
The coefficients arising in (\ref{thm C_N III}) are defined around equations (\ref{introduction asymptotic energy pde on the ball}), (\ref{introduction asymptotic energy pde on the ball 2D}) and (\ref{intro dimensional coupling scattering equation}) with
\begin{equation}\label{thm choose l,h}
 \textit{l}=\frac{1}{4},\quad h=N^{-\frac{13}{2}}.
\end{equation}
Notice in (\ref{thm C_N III}), 3 different types of correlation energy are addressed. The $\sum_{p\neq0}W_p\eta_p$ part is the contribution of the 3D correlation structure, the $\sum_{p\neq0}\big(W_p+Y_p+\mathfrak{D}_p\big)\xi_p$ part comes from the so-called quasi-2D correlation structure, and the $\sum_{p\neq0}\big(2Y_p+\mathfrak{D}_p+q_p\big)k_p$ part reveals the effect of the dimensional coupling correlation structure. Moreover, it can be bounded that
\begin{equation}\label{thm bound I_N}
  \mathcal{I}_{N}=O\Big(N\sqrt{\frac{a}{d}}+\ln N\Big)\ll N.
\end{equation}
Furthermore, the above results still hold true when we improve $\frac{a}{d}\sim N^{-1}$ to
\begin{equation}\label{thm condition III}
  N\Big(\frac{a}{d}\Big)^{\frac{19}{18}-r}\to0
\end{equation}
for some $r\in (0,\frac{1}{18})$ (not necessarily fixed).
\end{theorem}

\begin{remark}
  \
\begin{enumerate}[$(1)$]
  \item Since Regions $\mathrm{II}_{\mathrm{I}}$ and $\mathrm{II}_{\mathrm{III}}$ intersect with each other, we know indeed by comparing Theorem \ref{core} and Theorem \ref{core III} that in the overlapping region where $e^{-CN^{t_1}}\lesssim d\lesssim e^{-CN^{t_2}}$, we have
\begin{equation*}
  4\pi(N-1)Ng+\mathcal{I}_{N,a,d}
=4\pi(N-1)\frac{Na}{d}\mathfrak{a}_0+\mathfrak{e}_d+E_{Bog}^{(d)}+o(1).
\end{equation*}
This observation provides a more concrete representation of $\mathcal{I}_{N,a,d}$ in part of Region $\mathrm{II}$, and shows in a way that the results in Theorem \ref{core} and Theorem \ref{core III} are compatible and transitive.

  \item Even though we do not obtain a more explicit formula of $\mathcal{I}_{N,a,d}$, we can use the estimate (\ref{thm bound I_N}) to improve $t_1$ from $\frac{1}{72}$ to $\frac{1}{4}-$ such that Theorem \ref{core} applies to a wider range of parameters. In fact, as a direct corollary of Theorems \ref{core} and \ref{core III}, we have in Region $\mathrm{II}_{\mathrm{I}}^\prime$
\begin{equation}\label{corollary III}
  E_N=4\pi N^2a\mathfrak{a}_0d^{-1}+O(N^{\frac{1}{2}}).
\end{equation}
Here Region $\mathrm{II}_{\mathrm{I}}^\prime$ is the refinement of Region $\mathrm{II}_{\mathrm{I}}$ where we improve $t_1$ from $\frac{1}{72}$ to $\frac{1}{4}-$. We use (\ref{corollary III}) instead of (\ref{leading order more precise}) in Section \ref{proof II1} to reach a finer result that Theorem \ref{core} also holds in Region $\mathrm{II}_{\mathrm{I}}^\prime$.
  \item In this paper, we mainly focus on the relative scale of the thickness $d$ that confines one direction of the system (See Figure \ref{Region}), while we still conjecture that the barrier (\ref{thm condition III}) can be removed such that Theorem \ref{core III} holds for a larger Region $\mathrm{III}^\prime$. The main difficulty that hinders us from acquiring a more general result is that the estimates to the coefficients involved are not optimal. Especially the estimates concerning $q$, $Y$ and $\mathfrak{D}$. It is reasonable to believe once we reach the sharp estimates, the barrier (\ref{thm condition III}) can be removed.
\end{enumerate}
\end{remark}

\subsection{Outline of the Proof}
\par Now we sketch the ideas in the proof of Theorems \ref{core} and \ref{core III}. The truncated Fock space $F_{N,d}$ constructed over the orthogonal complement $L^2_{\perp}(\Lambda_{d})$ of the condensate wave function $\phi_0^{(d)}(\mathbf{x})=\frac{1}{\sqrt{d}}$, and the formalism of second quantization that writes $H_N$ defined in (\ref{Hamiltonian}) in the form of creation and annihilation operators $a_p^*$ and $a_p$
\begin{equation}\label{intro H_N}
  H_N=\sum_{p}\vert \mathcal{M}_{d}p\vert^2 a_p^*a_p+\frac{1}{2\sqrt{d}}\sum_{p,q,r}v_{r}^{(a,d)}a_{p+r}^*a_q^*a_pa_{q+r},
\end{equation}
are two basic tools for our computations. We launch calculations around this Fock space formalism.

\par
\begin{figure}[h]
\centering
\includegraphics[width=1\textwidth]{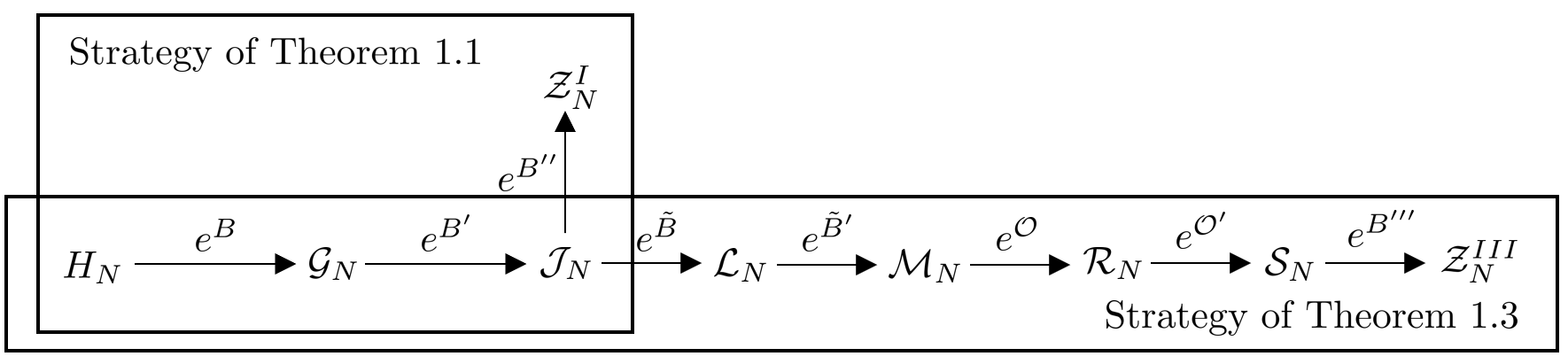}
\captionsetup{font=small}
\caption{Strategy of Proof}
\label{layout}
\end{figure}

 To obtain the second order formula, we would need to prove an optimal BEC result for this system first. Different from the pure 3D and pure 2D works in which the needed result is from \cite{boccatoBrenCena2020optimal,CaraciCenaSchlein2021} respectively, this result is not available. At the same time, to apply the Bogoliubov transformation to \textquotedblleft diagonalize\textquotedblright $H_N$, we are going to conjugate $H_N$ with several suitable unitary operators, such that we can at a long last correctly generate the correlation structures that contribute to the ground state energy. These operations, which are more well-known as renormalizations, allow us to derive Propositions \ref{Optimal BEC} and \ref{Optimal BEC III} of optimal BEC for all regions, which have not been proved before. Hence we start from here. We need to apply several renormalizations. Our strategy of renormalization is demonstrated in Table \ref{TAble} and Figure \ref{layout}, we give a brief outline of how we arrange these renormalizations and use them to reach the optimal BEC and hence the main Theorem \ref{core} and \ref{core III}. We leave the more thorough analysis of each excitation Hamiltonians in Section \ref{Excitation Hamiltonians}.

\begin{table}[!ht]
  \centering
  \caption{Classification of Renormalizations}\label{TAble}
  \begin{tabular}{|c|c|c|}
    \hline
    \multirow{2}{*}{3D}& Quadratic renormalization& $e^B$ \\ \cline{2-3}
    \multirow{2}{*}{}& Cubic renormalization& $e^{B^{\prime}}$ \\ \hline
    \multirow{2}{*}{Quasi-2D}& Quadratic renormalization& $e^{\tilde{B}}$ \\ \cline{2-3}
    \multirow{2}{*}{}& Cubic renormalization& $e^{{\tilde{B}}^{\prime}}$ \\ \hline
    \multirow{2}{*}{Dimensional coupling}& Quadratic renormalization& $e^{\mathcal{O}}$ \\ \cline{2-3}
    \multirow{2}{*}{}& Cubic renormalization& $e^{{\mathcal{O}}^{\prime}}$ \\ \hline
    \multirow{2}{*}{Bogoliubov transformation}& For Theorem \ref{core}& $e^{B^{\prime\prime}}$ \\ \cline{2-3}
    \multirow{2}{*}{}& For Theorem \ref{core III}& $e^{B^{\prime\prime\prime}}$ \\ \hline
  \end{tabular}
\end{table}

\par With the presence of Bose-Einstein condensation, it is intuitive to test the Hamilton operator $H_N$ with factorized wave function $(\varphi_0^{(d)})^{\otimes N}$. Using the second quantized form of $H_N$ (\ref{split H_N}), it is easy to compute
\begin{equation}\label{factorized state intro}
  \big\langle H_N(\varphi_0^{(d)})^{\otimes N},(\varphi_0^{(d)})^{\otimes N}\big\rangle=
\frac{a}{2d}N(N-1)\widehat{v}(0).
\end{equation}
By the observation that $f<1$ on the support of $v$ due to its subharmonicity, we know that
\begin{equation}\label{inequality between b_0 and a_0 intro}
  \widehat{v}(0)=\int_{\mathbb{R}^3}v(\mathbf{x})d\mathbf{x}
>\int_{\mathbb{R}^3}v(\mathbf{x})f(\mathbf{x})d\mathbf{x}=8\pi\mathfrak{a}_0.
\end{equation}
Comparing (\ref{factorized state intro}) with (\ref{leading order}), we find that (\ref{factorized state intro}) is always bigger than the true ground state energy of $H_N$, and thus $(\varphi_0^{(d)})^{\otimes N}$ is not a good approximation to the ground state of $H_N$. The reason that causes such difference is the inter-particle correlation structure. In Regions I and II in the Gross-Pitaevskii regime, (\ref{factorized state intro}) provides the correct leading $N$ order of ground state energy since we require $\frac{Na}{d}=1$,
while the expectation of $H_N$ on the factorized state (\ref{factorized state intro}) still does not recover the accurate leading order term and an error of order $N$ is left. This order $N$ error will be compensated by a 3D correlation structure of the Hamilton operator, which is now implicitly in the form of (\ref{Hamiltonian}) or (\ref{intro H_N}) but will surface explicitly and correctly during renormalizations.

\par On the other hand, since $\frac{a}{d}> N^{-1}$ in Region III in the Gross-Pitaevskii regime, the correlation structure here is way more special and even contains stronger energy than the one in Region I. Here in the region that $d$ decays acutely fast, the main inter-particle correlation structure is not only determined by the 3D effect, but also a quasi-2D effect. 3D and quasi-2D correlation structures together correct the leading order energy to (\ref{leading order}). Although via a direct observation, taking 3D and quasi-2D correlation structures into account seems enough for the computation, while in fact we have to moreover look into a dimensional coupling effect to compute the ground state energy up to second order, and it takes a large amount of computations (all the way to Section \ref{6}) to discover this dimensional coupling correlation structure. The analysis will thereafter be more intricate.
\par In the pure 2D problem studied in \cite{caraciCenaSchlein2022excitation}, the expectation of Hamiltonian on factorized states is also of the order larger than $N$, and hence the pure 2D correlation structure there affects the first order ground state energy similarly to the quasi-2D correlation structure here. Notice that in \cite{caraciCenaSchlein2022excitation}, an additional quartic renormalization was applied to handle the quartic term $\mathcal{V}_N$ (or in our word, $H_4$) of the Hamiltonian. Since in the pure 2D case, the intrinsic obstacle is that the quartic part $\mathcal{V}_N$ of the excitation Hamiltonian is not negligible on uncorrelated states, and it is essential for the derivation of upper bounds to control the quartic operator from above. That is a speciality of the pure 2D problem, and we do not observe that here (See, for example, \cite[(5.15)]{caraciCenaSchlein2022excitation} and (\ref{bound on H_4 on Z}) of this paper). In our 3D-to-2D setting, there is no such problem on controlling the quartic term, since, at the end of the day, it is still set in 3D space despite the confinement in one direction. The problem exclusive to 3D-to-2D setting arises here is that after the expected 3D and quasi-2D renormalizations, the remaining quadratic and cubic terms still contain energy that contributes to the second order approximation. Therefore, we furthermore need the additional quadratic and cubic dimensional coupling renormalizations to extract the energy unforeseen explicitly in the remaining quadratic and cubic terms.

\par The grand scheme of proving Theorem \ref{core} is sort of intermediate between the ones in \cite{2018Bogoliubov,hainzlSchleinTriay2022bogoliubov}, where a classical pure 3D setting was studied, but it is not a simple generalization of the pure 3D problem. Due to the extra d-dependence, more subtle estimates are needed in the proof of Theorem \ref{core}.

 \par We start by conjugating the Hamilton operator $H_N$ with two unitary operator respectively, the 3D quadratic transformation $e^{B}$ and the 3D cubic transformation $e^{B^\prime}$ with
\begin{align*}
  B&=\frac{1}{2}\sum_{p\neq0}\eta_p(a_p^*a_{-p}^*a_0a_0-h.c.),\\
  B^{\prime}&=\sum_{p,q,p+q\neq 0}\eta_p\chi_{\vert\mathcal{M}_dq\vert\leq\kappa}
  (a_{p+q}^*a_{-p}^*a_qa_0-h.c.).
\end{align*}
Here $\kappa$ is a cut-off parameter and $\eta_p$ are defined through the 3D scattering equation with Neumann boundary condition (\ref{introduction asymptotic energy pde on the ball}). Notice that here the choice of $\eta_p$ is inspired by \cite{2018Bogoliubov}, but the skew-symmetric operators $B$ and $B^{\prime}$ are defined using classical creation and annihilation operators $a_p^*$ and $a_p$ (not in the form of modified version presented in Section \ref{Modified Creation and Annihilation Operators}), which resemble the ones given in \cite{hainzlSchleinTriay2022bogoliubov} (That is, $B$ and $B^{\prime}$ are defined like in \cite{hainzlSchleinTriay2022bogoliubov} but with $\eta_p$ like in \cite{2018Bogoliubov}). This kind of definition helps us simplify the computations, and at the same time, can lead to a concrete and explicit expression of energy approximation since the equation (\ref{introduction asymptotic energy pde on the ball}) has been thoroughly analyzed. We write the excitation Hamilton operator
\begin{equation*}
  \mathcal{G}_N=e^{-B}H_Ne^{B},\quad\mathcal{J}_N=e^{-B^\prime}\mathcal{G}_Ne^{B^\prime}.
\end{equation*}

\par In Regions I and $\mathrm{II}_{\mathrm{I}}$, the above renormalizations actually model the correct correlation structure driven by the 3D effect, and the excitation Hamiltonian $\mathcal{J}_N$ can be approximated by the sum of a quadratic Hamiltonian and the quartic non-zero momentum sum of potential operator $H_4$ (the restriction of the potential energy to the orthogonal complement of the condensate wave function $\phi_0^{(d)}$). This allows us to apply the generalized Bogoliubov transformation denoted by $e^{B^{\prime\prime}}$ with
 \begin{equation*}
  B^{\prime\prime}=B(\tau)=\frac{1}{2}\sum_{p\neq0}\tau_p(b_p^*b_{-p}^*-h.c.).
\end{equation*}
Here $b_p^*$ and $b_p$ are so-called modified creation and annihilation operators used to simplify the computation of Bogoliubov transformation. Then we reach a diagonalized excitation Hamiltonian
\begin{equation*}
  \mathcal{Z}^{I}_N=e^{-B^{\prime\prime}}\mathcal{J}_Ne^{B^{\prime\prime}}.
\end{equation*}
This excitation Hamiltonian can be approximated by the sum of a quadratic diagonalized Hamiltonian and the quartic non-zero momentum sum of potential operator $H_4$ up to errors that can be ignored on low energy state. Observe that $H_4$ is on one hand non-negative, and on the other hand generates lower order energy on the low energy eigenspaces of the diagonalized part of $\mathcal{Z}^{I}_N$. Then Theorem \ref{core} follows by comparing the eigenvalues of $H_N$ with the diagonalized part of $\mathcal{Z}^{I}_N$ using min-max principle.

\par In Regions $\mathrm{II}_{\mathrm{III}}$ and III where $d$ decays acutely, simply extracting a 3D correlation structure is far from enough to reach a precise second order approximation to the ground state energy. In Region III, even the leading order approximation can not be correctly recovered due to the fact that $\frac{Na}{d}> 1$. We continue to conjugate $\mathcal{J}_N$ with two unitary operators respectively, the quasi-2D quadratic transformation $e^{\tilde{B}}$ and the quasi-2D cubic transformation $e^{\tilde{B}^\prime}$ with
\begin{align*}
  \tilde{B}&=\frac{1}{2}\sum_{p\neq0}\xi_p(a_p^*a_{-p}^*a_0a_0-h.c.),\\
  \tilde{B}^{\prime}&=\sum_{p,q,p+q\neq 0}\xi_p(a_{p+q}^*a_{-p}^*a_qa_0-h.c.),
\end{align*}
and we let
\begin{equation*}
  \mathcal{L}_N=e^{-\tilde{B}}\mathcal{J}_Ne^{\tilde{B}},\quad \mathcal{M}_N=e^{-\tilde{B}^\prime}\mathcal{L}_Ne^{\tilde{B}^{\prime}}.
\end{equation*}
$\xi_p$ are defined through a 2D scattering equation with Neumann condition given in (\ref{introduction asymptotic energy pde on the ball 2D}). The choice of $\xi_p$ is inspired by \cite{caraciCenaSchlein2022excitation}, while the 2D scattering equation is now induced by the 3D scattering equation since we are working on a 3D-to-2D problem. The form of $\tilde{B}$ and $\tilde{B}^{\prime}$ is still similar to one given in \cite{hainzlSchleinTriay2022bogoliubov}, and it makes the computation shorter. These quasi-2D renormalizations extract the correlation structure driven by the 2D effect that contributes to the first and second order terms of energy. Nevertheless, in the regime where $d$ decays acutely, neither simply viewing the system as 3D nor 2D is a good approximation.

\par The real problem now surfaces, after more than 100 pages of computation. The cubic quasi-2D correlation remainder $H_3^{\prime\prime}$ in $\mathcal{M}_N$ still contains energy of at least order $O(N^2a^2d^{-2})$, which can not yet be considered as a lower energy contribution compared with the expected second order energy. On the other hand, to apply the Bogoliubov transformation, we have to control the cubic term $H_3^{\prime\prime}$, or extract the energy contribution hidden in it and transform it into a controllable cubic remainder $H_3^{\prime\prime\prime}$. Here we discover a 3D-to-2D dimensional coupling correlation structure still contributes to the second order ground state energy. This new-found structure is the main feature that distinguish the 3D-to-2D problem considered in this paper with either the pure 3D problem or the pure 2D problem. To reveal the energy contribution of this correlation structure, we conjugate $\mathcal{M}_N$ with another two unitary operators, the dimensional coupling quadratic transformation $e^{\mathcal{O}}$ and the dimensional coupling cubic transformation $e^{\mathcal{O}^\prime}$ with
\begin{align*}
  \mathcal{O}&=\frac{1}{2}\sum_{p\neq0}k_p(a_p^*a_{-p}^*a_0a_0-h.c.),\\
  \mathcal{O}^{\prime}&=\sum_{p,q,p+q\neq 0}k_p(a_{p+q}^*a_{-p}^*a_qa_0-h.c.),
\end{align*}
and we let
\begin{equation*}
  \mathcal{R}_N=e^{-\mathcal{O}}\mathcal{M}_Ne^{\mathcal{O}},\quad \mathcal{S}_N=e^{-\mathcal{O}^\prime}\mathcal{R}_Ne^{\mathcal{O}^{\prime}}.
\end{equation*}
$k_p$ are defined through a coupling scattering equation given in (\ref{intro dimensional coupling scattering equation}). Now the excitation Hamiltonian $\mathcal{S}_N$ can be approximated by the sum of a quadratic Hamiltonian and the quartic non-zero momentum sum of potential operator $H_4$, we then apply another generalized Bogoliubov transformation $e^{B^{\prime\prime\prime}}$ with
\begin{equation*}
  B^{\prime\prime\prime}=B(\tilde{\tau})=\frac{1}{2}\sum_{p\neq0}
\tilde{\tau}_p(b_p^*b_{-p}^*-h.c.).
\end{equation*}
 We write the diagonalized Hamilton operator as follows
\begin{equation*}
  \mathcal{Z}^{III}_N=e^{-B^{\prime\prime\prime}}\mathcal{S}_Ne^{B^{\prime\prime\prime}}.
\end{equation*}
Finally, similar to the concluding part of Theorem \ref{core}, we finish Theorem \ref{core III} by comparing the eigenvalues of $H_N$ with the diagonalized part of $\mathcal{Z}^{III}_N$ using min-max principle.

\par The plan of this paper goes as follows. In Section \ref{Fock Space Formalism}, for the sake of completeness, we present the formalism of truncated Fock space $F_{N,d}$ and define the classical and modified creation and annihilation operators. In Section \ref{Scattering Equations with Neumann condition}, we collect important estimates about three scattering equations given above. In Section \ref{Excitation Hamiltonians}, we present the result concerning excitation Hamiltonians shown in Figure \ref{layout} in turn. For detailed analysis of each excitation Hamiltonians, we leave the computation of $\mathcal{G}_N$ to Section \ref{2}, $\mathcal{J}_N$ to Section \ref{3}. The mathematically rigorous analysis of $\mathcal{Z}_N^I$ is carried out in Section \ref{4}. We also analyze $\mathcal{L}_N$ and $\mathcal{M}_N$ in Section \ref{5}, $\mathcal{R}_N$ and $\mathcal{S}_N$ in Section \ref{6}, and moreover $\mathcal{Z}_N^{III}$ in Section \ref{7}. In Sections \ref{Proof of the Main Theorem for Region I} and \ref{Proof of the Main Theorem for Region III}, we use the result given in previous sections to prove optimal Bose-Einstein condensation results and apply min-max principle to conclude Theorems \ref{core} and \ref{core III} respectively.

\section{Fock Space Formalism}\label{Fock Space Formalism}
\par The Fock space, first introduced by  V. A. Fock in \cite{fock1932konfigurationsraum}, has went through years of development and been widely used in the theory of quantum mechanics. The Standard quantum mechanical Fock space over $L_s^2(\Lambda_d)$ is given by
\begin{equation*}
  \mathcal{F}=\bigoplus_{n=0}^{\infty}L_s^2(\Lambda_d^N).
\end{equation*}
For the sake of integrity, we present in this section, the truncated Fock Space $F_{N,d}$ subjected to the 3D anisotropic torus $\Lambda_{d}=[-\frac{1}{2},\frac{1}{2}]^2\times[-\frac{d}{2},\frac{d}{2}]$, and the operator actions defined over it. The idea is inspired by \cite{2018Bogoliubov}, where a pure 3D case was considered.

\subsection{Truncated Fock Space, Creation and Annihilation Operators}
\
\par Let $\{\phi_p^{(d)}\}$ be an orthonormal basis on $L^2(\Lambda_{d})$ given by
\begin{equation}\label{orthonormal basis on Lambda_d}
  \phi_p^{(d)}(\mathbf{x})=\frac{1}{\sqrt{d}}e^{ip^T\mathcal{M}_{d}\mathbf{x}},\quad p\in 2\pi\mathbb{Z}^3,\quad
  \mathcal{M}_{d}=\diag{(1,1,\frac{1}{d})}.
\end{equation}
For non-negative integers $n,m$, and $\psi\in L^2(\Lambda_{d}^n)$, $\varphi\in L^2(\Lambda_{d}^m)$, the tensor product and the symmetric tensor product between $\psi$ and $\varphi$ can be defined respectively as
\begin{equation*}\label{tensor product}
\begin{aligned}
  \psi\otimes\varphi(\mathbf{x}_1,\dots,\mathbf{x}_{n+m})=&
\psi(\mathbf{x}_1,\dots,\mathbf{x}_n)\varphi(\mathbf{x}_{n+1},\dots,\mathbf{x}_{n+m}),\\
  \psi\otimes_{s}\varphi(\mathbf{x}_1,\dots,\mathbf{x}_{n+m})=& \frac{1}{\sqrt{n!m!(n+m)!}}\\
&\times\sum_{\sigma\in S_{n+m}}\psi(\mathbf{x}_{\sigma(1)},\dots,\mathbf{x}_{\sigma(n)})
  \varphi(\mathbf{x}_{\sigma(n+1)},\dots,\mathbf{x}_{\sigma(n+m)}).
\end{aligned}
\end{equation*}
Under these definitions, it is easy to verify $\psi\otimes\varphi\in L^2(\Lambda_{d}^{n+m})$ and $\psi\otimes_{s}\varphi\in L^2_s(\Lambda_{d}^{n+m})$. Both of the two operations are associative and the latter is even commutative.

\par Let $L^2_{\perp}(\Lambda_{d})\coloneqq (span\{\phi_0^{(d)}\})^{\perp}$ be the orthogonal complement of the condensate wave function $\phi_0^{(d)}$ in $L^2(\Lambda_d)$. Now the truncated Fock space over $L^2_\perp(\Lambda_d)$ can be defined by
\begin{equation*}
  F_{N,d}=\bigoplus_{n=0}^N L^2_{\perp}(\Lambda_{d})^{\otimes_s n}.
\end{equation*}
We endow this vector space with usual Hilbert inner product. As observed in \cite{lewinNamSerfatySolovej2015bogoliubov}, we have the fact that $L^2(\Lambda_{d}^N)=L^2(\Lambda_{d})^{\otimes N}$, $L^2_s(\Lambda_{d}^N)=L^2(\Lambda_{d})^{\otimes_s N}$ for all positive integers $N$. Then for every $N$-particle wave function $\psi\in L^2_s(\Lambda_{d}^N)$, there exists a unique family of $\alpha^{(n)}\in L^2_{\perp}(\Lambda_{d})^{\otimes_s n}$ such that
\begin{equation}\label{represtentation of phi}
  \psi=\sum_{n=0}^{N}\alpha^{(n)}\otimes_s(\phi_0^{(d)})^{\otimes(N-n)}.
\end{equation}
The representation (\ref{represtentation of phi}) of $\psi$ allows us to define an operator $U_{N,d}:L^2_s(\Lambda_{d}^N)\to F_{N,d}$ as follows:
\begin{equation}\label{define U_N,d}
  U_{N,d}\psi=(\alpha^{(0)},\dots,\alpha^{(N)}).
\end{equation}
Furthermore, $U_{N,d}$ is an unitary operator in the sense that
\begin{equation}\label{isometry}
  \Vert\psi\Vert^2=\sum_{n=0}^{N}\Vert\alpha^{(n)}\Vert^2.
\end{equation}
We may sometimes omit the $d$ subscript and simply denote it by $U_N$ in what follows.

\par For all non-negative integers $n$ and $p\in 2\pi\mathbb{Z}^3$, we define the creation operator $a_p^*:L^2_s(\Lambda_{d}^n)\to L^2_s(\Lambda_{d}^{n+1})$ and the annihilation operator $a_p:L^2_s(\Lambda_{d}^n)\to L^2_s(\Lambda_{d}^{n-1})$ as follows:
\begin{equation}\label{define a_p & a_p*}
\left.\begin{aligned}
  (a_p^*\psi)(\mathbf{x}_1,\dots,\mathbf{x}_{n+1}) &= \frac{1}{\sqrt{n+1}}\sum_{i=1}^{n+1}\phi_p^{(d)}(\mathbf{x}_i)
\psi(\mathbf{x}_1,\dots,\hat{\mathbf{x}_i},\dots,\mathbf{x}_{n+1}) \\
  (a_p\psi)(\mathbf{x}_1,\dots,\mathbf{x}_{n-1}) &= \sqrt{n}\int_{\Lambda_{d}}\overline{\phi_p^{(d)}}(\mathbf{x})
\psi(\mathbf{x},\mathbf{x}_1,\dots,\mathbf{x}_{n-1})d\mathbf{x}, \quad n\geq 1
\end{aligned}\right.
\end{equation}
We additionally define for $n=0$ that $a_p\coloneqq 0$, $L^2_s(\Lambda_{d}^0)\coloneqq \mathbb{C}$ and $L^2_s(\Lambda_{d}^{-1})\coloneqq \{0\}$. $a_p$ and $a_p^*$ defined in (\ref{define a_p & a_p*}) are in fact $n$ and $d$ dependent, but we have omitted it to avoid long equations. For all $p\in 2\pi\mathbb{Z}^3$, $a_p^*$ is in fact the adjoint of $a_p$ provided their domains of definition are matched. By a direct calculation we attain the canonical commutation relations
\begin{equation}\label{canonical commutation relations}
  [a_q,a_p^*]=\delta_{p,q},\quad[a_q,a_p]=[a_q^*,a_p^*]=0,\quad p,q\in 2\pi\mathbb{Z}^3,
\end{equation}
and their operator norms can be bounded by
\begin{equation}\label{norm of a_p and a_p*}
  \Vert a_p^*f\Vert\leq\sqrt{n+1}\Vert f\Vert,\quad \Vert a_pf\Vert\leq\sqrt{n}\Vert f\Vert.
\end{equation}
We omit the subscripts of norms for the sake of brevity. With the creation and annihilation operators being defined, it is of great use to define the number of excited particles operator for any positive integer $n$ on $L_s^2(\Lambda_d^n)$ as
\begin{equation}\label{define N_+}
  \mathcal{N}_+^L=\sum_{p\neq0}a_p^*a_p.
\end{equation}
We omit the $n$ dependence here. Observe that for any function $\alpha^{(n)}\in L^2_{\perp}(\Lambda_{d})^{\otimes_s n}$
\begin{equation}\label{define of N_+}
  \mathcal{N}_+^L\alpha^{(n)}=n\alpha^{(n)}.
\end{equation}

\par From (\ref{isometry}), the truncated Fock space $F_{N,d}$ is isometric to $L^2_s(\Lambda_d^N)$ through the unitary operator $U_{N,d}$. Hence it is natural, and will be very useful in our further analysis to have a truncated Fock space $F_{N,d}$ version of creation and annihilation operators defined in (\ref{define a_p & a_p*}). We first adopt some notations $A_p$, $A_p^*$ and $\mathcal{N}_+^F$, which are operators defined on the truncated Fock space $F_{N,d}$:
\begin{equation}\label{A_p,A_p*}
  A_p^*=\begin{pmatrix}
          0 &   &   &   \\
          a_p^*&  0 &   &   \\
            & \ddots& \ddots &   \\
            &   & a_p^* & 0
        \end{pmatrix}_{(N+1)\times(N+1)}
  A_p=\begin{pmatrix}
        0 & a_p &   &   \\
          & 0 & \ddots &   \\
          &   & \ddots & a_p \\
          &   &   & 0
      \end{pmatrix}_{(N+1)\times(N+1)}
\end{equation}
and
\begin{equation}\label{N_+ on F_N,d}
  \mathcal{N}_+^F=\begin{pmatrix}
        \mathcal{N}_+^L &   &   &   \\
          & \mathcal{N}_+^L &   &   \\
          &   & \ddots &   \\
          &   &   & \mathcal{N}_+^L
      \end{pmatrix}_{(N+1)\times(N+1)}=\begin{pmatrix}
         0 &   &   &   \\
           & 1 &   &   \\
           &   &\ddots &   \\
           &   &   & N
       \end{pmatrix}
\end{equation}
Let $Q$ be the orthogonal projection from $L^2(\Lambda_{d})$ to $L^2_{\perp}(\Lambda_{d})$, then $U_{N,d}$ and its inverse can be described explicitly in terms of creation and annihilation operators by
\begin{equation}\label{U_N & U_N*}
\left.\begin{aligned}
 U_{N,d}\psi &= \bigoplus_{n=0}^N Q^{\otimes n}\left[\frac{a_0^{N-n}\psi}{\sqrt{(N-n)!}}\right],\\
  U_{N,d}^*(\alpha^{(0)},\dots,\alpha^{(N)}) &= \sum_{n=0}^{N}\frac{a_0^{*(N-n)}(\alpha^{(n)})}{\sqrt{(N-n)!}}.
\end{aligned}\right.
\end{equation}
Using (\ref{U_N & U_N*}) and the canonical commutation relations (\ref{canonical commutation relations}) we find the following useful formulae for $p,q\neq 0$:
\begin{equation}\label{conjugate relation}
\left.\begin{aligned}
U_{N,d}a_0^*a_0U_{N,d}^*=N-\mathcal{N}^F_+,\quad & U_{N,d}a_p^*a_qU_{N,d}^*=A_p^*A_q,\\
  U_{N,d}a_0^*a_pU_{N,d}^*=\sqrt{N-\mathcal{N}_+^F}A_p,\quad &U_{N,d}a_p^*a_0U_{N,d}^*=A_p^*\sqrt{N-\mathcal{N}_+^F}.
\end{aligned}\right.
\end{equation}
With (\ref{conjugate relation}) and definitions (\ref{A_p,A_p*}) and (\ref{N_+ on F_N,d}) we immediately know that
\begin{equation*}
  \mathcal{N}_+^F=U_{N,d}\mathcal{N}_+^LU_{N,d}^*
\end{equation*}
which implies that $\mathcal{N}_+^F$ is in fact a $F_{N,d}$ version of the number of excited particles operator. We will denote both of $\mathcal{N}_+^L$ and $\mathcal{N}_+^F$ by $\mathcal{N}_+$ for simplicity.
\par On the other hand, we can recursively derive a general commutation relation via canonical commutation relations (\ref{canonical commutation relations}) that for $p\in 2\pi\mathbb{Z}^3$ and any non-negative integers $l,k$
\begin{equation}\label{general commutation relations}
  [a_p^l,a_p^{*k}]=\left\{
\begin{aligned}
   & \sum_{j=1}^{l}c^{(l,k)}_{(l-j,k-j)}a_p^{*(k-j)}a_p^{(l-j)},\quad k\geq l,\\
   & \sum_{j=1}^{k}c^{(l,k)}_{(l-j,k-j)}a_p^{*(k-j)}a_p^{(l-j)},\quad k\leq l,
\end{aligned}
\right.
\end{equation}
for some constants $c^{(l,k)}_{(l-j,k-j)}$. In particular, for any non-negative integers $l,k$ and a suitable $j$ we have $c^{(l,k)}_{(l-j,k-j)}=c^{(k,l)}_{(k-j,l-j)}$. Moreover, $c^{(l,k)}_{(l-1,k-1)}=kl$ and for $k\geq l$, $c^{(l,k)}_{(0,k-l)}=\frac{k!}{(k-l)!}$. With (\ref{general commutation relations}), the operator representations (\ref{U_N & U_N*}), and the fact that $a_0\alpha^{(n)}=0$ for any $\alpha^{(n)}\in L^2_{\perp}(\Lambda_{d})^{\otimes_s n}$, it is also useful to compute for $p\in 2\pi\mathbb{Z}^3\backslash\{0\}$,
\begin{eqnarray}
  &U_{N,d}a_0^*U_{N-1,d}^*=\sqrt{N-\mathcal{N}_+}I_{0},\quad & U_{N-1,d}a_0U_{N,d}^*=I_0^*\sqrt{N-\mathcal{N}_+},\nonumber\\
  &U_{N,d}a_p^*U_{N-1,d}^*=A_p^*I_0,\quad &U_{N-1,d}a_pU_{N,d}^*=I^*_0A_p.\label{general conjugate relation}
\end{eqnarray}
Here we treat $\mathcal{N}_+$, $A_p$ and $A_p^*$ as operators on $F_{N,d}$, and $I_0$ maps $F_{N-1,d}$ to $F_{N,d}$ which is given by
\begin{equation}\label{E_0}
  I_0=\begin{pmatrix}
        Id_{N} \\
        \mathbf{0}
      \end{pmatrix}_{(N+1)\times N}
\end{equation}
Calculations above show that the corresponding version of $a_p$ and $a_p^*$ on $F_{N,d}$ should be $I^*_0A_p$ and $A_p^*I_0$. But they are far from adequate since they do not hold our truncated Fock space invariant. A version of $a_p$ or $a_p^*$ that actually acts on $F_{N,d}$, or in other words, preserving the number of particle $N$ will be of equal importance in our further analysis. For this reason, we define the modified creation and annihilation operators.

\subsection{Modified Creation and Annihilation Operators}\label{Modified Creation and Annihilation Operators}
\

\par Inspired by \cite{2018Bogoliubov,hainzlSchleinTriay2022bogoliubov}, for $p\in 2\pi\mathbb{Z}^3\backslash\{0\}$, and fixed particle number $N$, the modified creation and annihilation operators, acting on $L_s^2(\Lambda_d^N)$, are defined as follows
\begin{equation}\label{define b_p,b_p^* fock}
  b_p^*=a_p^*\frac{a_0}{\sqrt{N}},\quad
  b_p=\frac{a_0^*}{\sqrt{N}}a_p.
\end{equation}
Then it is easy to verify, on $L_s^2(\Lambda_d^N)$, for $p\neq0$:
\begin{equation}\label{b_p^*b_q fock}
  a_p^*a_p=b^*_pb_p+\frac{1}{N}a_p^*\mathcal{N}_+a_p,\quad
  \frac{1}{N}a_p^*a_{-p}^*a_0a_0=b_p^*b_{-p}.
\end{equation}
and their commutators are given by
\begin{equation}\label{commutator b fock}
  [b_p,b_q^*]=\delta_{p,q}-\varepsilon_{p,q},\quad[b_p,b_q]=[b_p^*,b_q^*]=0,
\end{equation}
where
\begin{equation}\label{epsilonpq}
  \varepsilon_{p,q}=\delta_{p,q}N^{-1}\mathcal{N}_++N^{-1}a^*_qa_p.
\end{equation}
Moreover, with conjugation formulae (\ref{conjugate relation}) we can compute directly
\begin{equation}\label{conjugate relation for b}
  U_{N,d}b_p^*U_{N,d}^*=A_p^*\sqrt{1-\frac{\mathcal{N}_+^{\mathcal{F}}}{N}},\quad U_{N,d}b_pU_{N,d}^*=\sqrt{1-\frac{\mathcal{N}_+^{\mathcal{F}}}{N}}A_p.
\end{equation}
Comparing (\ref{conjugate relation for b}) with (\ref{general conjugate relation}), it is intuitive to notice that the modified version of creation and annihilation operators, $b_p^*$ and $b_p$ preserve the number of particles, that is to say, both $U_Nb_pU_N^*$ and $U_Nb_p^*U_N^*$ act on the Fock space $F_{N,d}$. With this observation, we define the generalized Bogoliubov transformation, which is a unitary operator $e^{B(\tau)}$ with $B(\tau)$ having the form
\begin{equation}\label{define B(tau) fock}
  B(\tau)=\frac{1}{2}\sum_{p\neq0}\tau_p(b_p^*b_{-p}^*-h.c.)
\end{equation}
with coefficients $\tau_p$ satisfying $\tau_p=\tau_{-p}=\overline{\tau_p}$. Using (\ref{commutator b fock}), we can calculate
\begin{equation}\label{rrrrrr}
  [b_p^*,B(\tau)]=\tau_pb_{-p}+r_p
\end{equation}
where
\begin{equation}\label{rrrrrrrrr}
  r_p=-\frac{1}{2}\sum_{q\neq0}\tau_q(\varepsilon_{-q,p}b_q+b_{-q}\varepsilon_{q,p}).
\end{equation}
The action of the generalized Bogoliubov transformation can be calculated explicitly using Taylor's formula and the (\ref{rrrrrr}):
\begin{equation}\label{action of Bog trans fock}
\begin{aligned}
  e^{-B(\tau)}b_p^*e^{B(\tau)}&=\cosh\tau_p b_p^*+\sinh\tau_p b_{-p}+d_p^*\\
 e^{-B(\tau)}b_pe^{B(\tau)}&=\cosh\tau_p b_p+\sinh\tau_p b_{-p}^*+d_p
 \end{aligned}
\end{equation}
where we have (we let $t_0=1$ in the following summation):
\begin{equation}\label{dddddddddd}
  d_p^*=\sum_{n=1}^{\infty}\int_{0}^{t_0}dt_1\cdots\int_{0}^{t_{n-1}}dt_n\tau_p^{n-1}
  e^{-t_nB(\tau)}\big(\chi_0(n;2)r_p+\chi_{0}(n+1;2)r_{-p}^*\big)e^{t_nB(\tau)}.
\end{equation}
Here $\chi_{0}(n;2)$ is the only Dirichlet character modulo $2$, satisfies $\chi_{0}(n;2)=1$ at odd integers, and $\chi_{0}(n;2)=0$ at even integers.

\par In this paper, we mainly work on transformations constructed based on the standard creation and annihilation operators since they are more intuitive and more straightforward to calculate, while the generalized Bogoliubov transformation defined above via modified creation and annihilation operators will be used in the last step of energy renormalization since the elegant formula (\ref{action of Bog trans fock}) helps us to conclude the explicit expression of energy $E_N$ defined in (\ref{ground state energy}) up to small errors. Next, we are going to write $H_N$ defined in (\ref{Hamiltonian}) in the form of creation and annihilation operators using the formalism of second quantization.

\subsection{Formalism of Second Quantization}
\
\par So long as $\frac{a}{d}$ is small enough (namely $\frac{a}{d}<\frac{1}{2R_0}$) so that $\supp v_a\subset\Lambda_{d}$, which ensures $v_a$ given in (\ref{interaction potential}) would be a periodic function on $\Lambda_{d}$, using Fourier series, we can write $H_N$ in the form of creation and annihilation operators:
\begin{equation}\label{H_N}
  H_N=\sum_{p}\vert \mathcal{M}_{d}p\vert^2 a_p^*a_p+\frac{1}{2\sqrt{d}}\sum_{p,q,r}v_{r}^{(a,d)}a_{p+r}^*a_q^*a_pa_{q+r},
\end{equation}
where
\begin{equation}\label{v_p^a,d}
  v_{p}^{(a,d)}=\int_{\Lambda_d}v_a(\mathbf{x})
\overline{\phi_p^{(d)}}(\mathbf{x})d\mathbf{x}=\frac{a}{\sqrt{d}}
  \widehat{v}\left(\frac{a\mathcal{M}_dp}{2\pi}\right),\quad p\in2\pi\mathbb{Z}^3.
\end{equation}
The notation $\widehat{v}$ is given by
\begin{equation*}
  \widehat{v}(\mathbf{\xi})=\int_{\mathbb{R}^3}v(\mathbf{x})e^{-2\pi i\mathbf{x}\cdot\mathbf{\xi}}d\mathbf{x},\quad \mathbf{\xi}\in\mathbb{R}^3.
\end{equation*}
Notice that $v_p^{(a,d)}=v_{-p}^{(a,d)}=\overline{v_p^{(a,d)}}$. Equation (\ref{H_N}) is referred to the second quantized form of $H_N$. Using $a_0^*a_0=N-\mathcal{N}_+$ we have
\begin{equation}\label{a_0*a_0*a_0a_0}
  a_0^*a_0^*a_0a_0=a_0^*(a_0a_0^*-1)a_0=(N-\mathcal{N}_+)^2-(N-\mathcal{N}_+).
\end{equation}
This observation allows us to split $H_N$ into several self-adjoint operators
\begin{equation}\label{split H_N}
  H_N=H_{01}+H_{02}+H_{21}+H_{22}+H_{23}+H_3+H_4,
\end{equation}
where
\begin{align*}
H_{01}=&\frac{1}{2\sqrt{d}}v_0^{(a,d)}N(N-1),\quad
H_{02}=-\frac{1}{2\sqrt{d}}v_0^{(a,d)}\mathcal{N}_+(\mathcal{N}_+-1),\\
H_{21}=&{\sum_{p\neq0}}\vert \mathcal{M}_{d}p\vert^2 a_p^*a_p,\quad\quad
H_{22}=\frac{1}{\sqrt{d}}(N-\mathcal{N}_+){\sum_{p\neq0}}v_p^{(a,d)}a_p^*a_p,\\
H_{23}=&\frac{1}{2\sqrt{d}}{\sum_{p\neq0}}v_p^{(a,d)}(a_{p}^*a_{-p}^*a_0a_0+h.c.),\\
H_3=&\frac{1}{\sqrt{d}}{\sum_{p,r,p+r\neq0}}v_r^{(a,d)}(a_{p+r}^*a_{-r}^*a_pa_0+h.c.),\\
H_4=&\frac{1}{2\sqrt{d}}{\sum_{p,q,p+r,q+r\neq0}}v_r^{(a,d)}a_{p+r}^*a_q^*a_pa_{q+r}.
\end{align*}
Operators that have similar structures to $H_{23}$ (their coefficients may differ) are often referred to quadratic terms, and those having similar structures to $H_3$ are often referred to cubic terms.

\par To end this section we point out that the kinetic operator $H_{21}$ and the non-zero momentum sum of potential operator $H_4$ both play special roles in the course of renormalization. For an N-particle wave function $\psi\in L^2_s(\Lambda_{d}^N)$ with (\ref{define U_N,d}) also holds, we calculate directly using (\ref{conjugate relation})
\begin{align}
  \langle H_{21}\psi,\psi\rangle
&=\langle U_NH_{21}U_N^*U_N\psi,U_N\psi\rangle
=\sum_{n=0}^{N}\langle H_{21}\alpha^{(n)},\alpha^{(n)}\rangle\nonumber\\
&=\sum_{n=1}^{N}\int_{\Lambda_d^n}
\sum_{i=1}^{n}\vert\nabla_{\mathbf{x}_i}\alpha^{(n)}\vert^2
=\sum_{n=1}^{N}n\int_{\Lambda_d^n}
\vert\nabla_{\mathbf{x}_1}\alpha^{(n)}\vert^2\label{H_21psi,psi fock}
\end{align}
and
\begin{align}
  &\langle H_4\psi,\psi\rangle=\langle U_NH_4U_N^*U_N\psi,U_N\psi\rangle
  =\sum_{n=0}^{N}\langle H_4\alpha^{(n)},\alpha^{(n)}\rangle\nonumber\\
  =&\sum_{n=2}^{N}\Big\langle \frac{1}{2\sqrt{d}}{\sum_{p,q,r}}v_r^{(a,d)}
  a_{p+r}^*a_q^*a_pa_{q+r}\alpha^{(n)},\alpha^{(n)}\Big\rangle\nonumber\\
  =&\sum_{n=2}^{N}\int_{\Lambda_d^n}\sum_{i<j}^{n}
v_a(\mathbf{x}_i-\mathbf{x}_j)\vert\alpha^{(n)}\vert^2
  =\frac{1}{2}\sum_{n=2}^{N}n(n-1)
\int_{\Lambda_d^n}v_a(\mathbf{x}_1-\mathbf{x}_2)\vert\alpha^{(n)}\vert^2
  \label{H_4psi,psi fock}
\end{align}
Equation (\ref{H_21psi,psi fock}) and (\ref{H_4psi,psi fock}) will be used repeatedly in the up-coming calculations.
\section{Properties of Scattering Equations}\label{Scattering Equations with Neumann condition}
\par This is a preparation section devoted to choose suitable coefficients that will be used in the up-coming renormalization procedure. We will analyze three one-particle scattering equations, which are priorly introduced as (\ref{introduction asymptotic energy pde on the ball}), (\ref{introduction asymptotic energy pde on the ball 2D}) and (\ref{intro dimensional coupling scattering equation}). The first equation (\ref{introduction asymptotic energy pde on the ball}) is a 3D asymptotic equation of a single boson inside a large bosonic system interacting in three-dimensional space, while the second one (\ref{introduction asymptotic energy pde on the ball 2D}) is the corresponding 2D version of the equation of asymptotic behaviors. Most of the results of the first two equations shown here have already been collected or proven in \cite{2018Bogoliubov,caraciCenaSchlein2022excitation} and we just go a bit further. We call the third equation (\ref{intro dimensional coupling scattering equation}) a dimensional coupling scattering equation. It encodes a special correlation structure that is generated by the superposition of the 3D and 2D effects. This structure is one of the main driving force when $d$ decays fast enough, and therefore it can be considered as a unique feature in the 3D-to-2D problem that has never been studied before.
\subsection{3D Scattering Equation}\label{3 3D  scattering equation}
\
\par We first consider the following ground state energy equation with Neumann boundary condition for some parameter $\textit{l}\in(0,\frac{1}{2})$
\begin{equation}\label{asymptotic energy pde on the ball}
  \left\{\begin{aligned}
  &(-\Delta_{\mathbf{x}}+\frac{1}{2}v)f_\textit{l}=\lambda_\textit{l}f_\textit{l},\quad \vert \mathbf{x}\vert\leq \frac{d}{a}\textit{l},\\
  &\left.\frac{\partial f_\textit{l}}{\partial \mathbf{n}}\right\vert_{\vert \mathbf{x}\vert=\frac{d}{a}\textit{l}}=0,\quad \left.f_\textit{l}\right\vert_{\vert \mathbf{x}\vert=\frac{d}{a}\textit{l}}=1.
  \end{aligned}\right.
\end{equation}
Notice that we omit the $a$ and $d$ dependence in the notations of $f_\textit{l}$ and $\lambda_\textit{l}$. Equation (\ref{asymptotic energy pde on the ball}) has been thoroughly analyzed, and one can consult \cite{LSY2006Derivation,2018Bogoliubov,boccatoBrenCena2020optimal} for details. We define $w_{\textit{l}}=1-f_\textit{l}$, then we can make constant extensions to both $f_\textit{l}$ and $w_\textit{l}$ outside of the 3D closed ball $\overline{B}_{\frac{d\textit{l}}{a}}$ such that $f_\textit{l}\in H^2_{loc}(\mathbb{R}^3)$ and $w_\textit{l}\in H^2(\mathbb{R}^3)$. By scaling we let
\begin{equation}\label{scaling f_l &w_l}
  \widetilde{f}_\textit{l}(\mathbf{x})=f_\textit{l}\Big(\frac{\mathbf{x}}{a}\Big),\quad \widetilde{w}_\textit{l}(\mathbf{x})=w_\textit{l}\Big(\frac{\mathbf{x}}{a}\Big).
\end{equation}
Regarding $\widetilde{w}_\textit{l}$ as a periodic function on the torus $\Lambda_d$, we observe that it satisfies the equation
\begin{equation}\label{asymptotic energy pde on the torus}
  \Big(-\Delta_{\mathbf{x}}+\frac{1}{2a^2}v\big(\frac{\mathbf{x}}{a}\big)\Big)
 \widetilde{w}_\textit{l}(\mathbf{x})
  =\frac{1}{2a^2}v\big(\frac{\mathbf{x}}{a}\big)
-\frac{\lambda_\textit{l}}{a^2}\big(1-\widetilde{w}_\textit{l}(\mathbf{x})\big)
  \chi_{d\textit{l}}(\mathbf{x}),\quad \mathbf{x}\in\Lambda_d.
\end{equation}
Here $\chi_{d\textit{l}}$ is the characteristic function of the closed 3D ball $\underline{B}_{d\textit{l}}$, and we choose suitable $\textit{l}\in(0,\frac{1}{2})$ so that $\overline{B}_{d\textit{l}}\subset\Lambda_d$. Standard elliptic equation theory grants the uniqueness of the solution to equation (\ref{asymptotic energy pde on the torus}). By Fourier transform, (\ref{asymptotic energy pde on the torus}) is equivalent to its discrete version
\begin{equation}\label{discrete asymptotic energy pde on the torus}
\begin{aligned}
\left\vert\mathcal{M}_dp\right\vert^2\widetilde{w}_{\textit{l},p}+\frac{1}{2\sqrt{d}}
  \sum_{q\in2\pi\mathbb{Z}^3}v_{p-q}^{(a,d)}\widetilde{w}_{\textit{l},q}
=\frac{1}{2}
  v_p^{(a,d)}+\frac{\lambda_\textit{l}}{a^2}
  \widetilde{w}_{\textit{l},p}-\frac{\lambda_\textit{l}}{a^2\sqrt{d}}\widehat{\chi_{d\textit{l}}}
  \left(\frac{\mathcal{M}_dp}{2\pi}\right),
\end{aligned}
\end{equation}
where $p$ is an arbitrary 3D vector in $2\pi\mathbb{Z}^3$ and the Fourier coefficients are given by
\begin{equation*}
  \widetilde{w}_{\textit{l},p}=\int_{\Lambda_d}\widetilde{w}_{\textit{l}}(\mathbf{x})
  \overline{\phi_p^{(d)}}(\mathbf{x})d\mathbf{x},\quad
\frac{1}{\sqrt{d}}\widehat{\chi_{d\textit{l}}}
  \left(\frac{\mathcal{M}_dp}{2\pi}\right)
=\int_{\Lambda_d}\chi_{d\textit{l}}(\mathbf{x})
  \overline{\phi_p^{(d)}}(\mathbf{x})d\mathbf{x}.
\end{equation*}
The required properties of $f_\textit{l}$ and $w_\textit{l}$ are collected in the next lemma.
\begin{lemma}\label{fundamental est of v,w,lambda}
Let $v$ be a smooth interaction potential with scattering length $\mathfrak{a}_0$. Recall that an interaction potential should be a radially-symmetric, compactly supported and non-negative function. Let $f_\textit{l}$, $\lambda_\textit{l}$, $w_\textit{l}$ and $\widetilde{w}_{\textit{l},p}$ be defined as above. Then for parameter $\textit{l}\in (0,\frac{1}{2})$ satisfying $\frac{a}{d\textit{l}}<C$ for a small constant C independent of $a$, $d$ and $\textit{l}$, there exist some constants, also denoted as C, independent of $a$, $d$ and $\textit{l}$ such that following estimates hold true for $\frac{a}{d\textit{l}}$ small enough.
\begin{enumerate}[$(1)$]
  \item The asymptotic estimate of ground state energy $\lambda_\textit{l}$ is
  \begin{equation}\label{est of lambda_l}
          \left\vert\lambda_\textit{l}-\frac{3\mathfrak{a}_0}{\textit{l}^3}\frac{a^3}{d^3}
          \left(1+\frac{9}{5}\frac{\mathfrak{a}_0}{\textit{l}}\frac{a}{d}\right)\right\vert\leq \frac{C\mathfrak{a}_0^3}{\textit{l}^5}\frac{a^5}{d^5}.
  \end{equation}
  \item $f_\textit{l}$ is radially symmetric and smooth away from the boundary of $B_{\frac{d\textit{l}}{a}}$ and there is a certain constant $0<c<1$ independent of $a$, $d$ and $\textit{l}$ such that
  \begin{equation}\label{est of f_l}
    0<c\leq f_\textit{l}(\mathbf{x})\leq1.
  \end{equation}
  Moreover, for any integer $0\leq k\leq3$
  \begin{equation}\label{est of w_l and grad w_l}
        \vert D_{\mathbf{x}}^kw_\textit{l}(\mathbf{x})\vert\leq\frac{C}{1+\vert \mathbf{x}\vert^{k+1}}.
  \end{equation}
  \item We have
  \begin{equation}\label{est of int vf_l}
          \left\vert\int_{\mathbb{R}^3}v(\mathbf{x})f_\textit{l}(\mathbf{x})d\mathbf{x}
          -8\pi\mathfrak{a}_0\left(1+\frac{3}{2}
          \frac{\mathfrak{a}_0}{\textit{l}}\frac{a}{d}\right)\right\vert\leq
          \frac{C\mathfrak{a}_0^3}{\textit{l}^2}\frac{a^2}{d^2},
  \end{equation}
  and
        \begin{equation}\label{est of int w_l}
          \left\vert\frac{1}{\textit{l}^2}\frac{a^2}{d^2}\int_{\mathbb{R}^3}
          w_\textit{l}(\mathbf{x})d\mathbf{x}-\frac{2}{5}\pi\mathfrak{a}_0\right\vert\leq
          \frac{C\mathfrak{a}_0^2}{\textit{l}}\frac{a}{d}.
        \end{equation}
  \item For all $p\in2\pi\mathbb{Z}^3\backslash\{0\}$
  \begin{equation}\label{est of w_l,p}
          \frac{1}{\sqrt{d}}\vert \widetilde{w}_{\textit{l},p}\vert\leq\frac{Ca}{d}\frac{1}{\vert\mathcal{M}_dp\vert^2}.
  \end{equation}
\end{enumerate}
\end{lemma}
\begin{remark}\label{remark 3d scat eqn}
The construction of $w_\textit{l}$ can not ensure it to be smooth on the boundary of $B_{\frac{d\textit{l}}{a}}$. But we still use the notation $D_{\mathbf{x}}^kw_\textit{l}$ to represent the $k$-th derivative of $w_\textit{l}$ away from the boundary of $B_{\frac{d\textit{l}}{a}}$. Moreover, since $w_\textit{l}$ is supported on $B_{\frac{d\textit{l}}{a}}$, the integral concerning $D^k_{\mathbf{x}}w_\textit{l}$ always means integrating inside of $B_{\frac{d\textit{l}}{a}}$ unless otherwise specified.
\end{remark}
\noindent\emph{Proof.} After a slightly modification of parameters, statements $(1)$ and $(3)$ follow from \cite[Lemma 3.1]{2018Bogoliubov}, statement $(4)$ follows from \cite[Lemma 4.1]{boccatoBrenCena2020optimal}, and statement $(2)$ follows from \cite[Lemma A.1]{LSY2006Derivation} except for the $k=3$ case in (\ref{est of w_l and grad w_l}). For the $k=3$ case in (\ref{est of w_l and grad w_l}), we can just follow the idea of the proof of \cite[Lemma A.1]{LSY2006Derivation} to deduce that, outside the support of $v$ (which we assume it to be $B_{R_0}$)
\begin{equation*}
  \vert D_{\mathbf{x}}^3w_\textit{l}(\mathbf{x})\vert\leq \frac{C}{\vert \mathbf{x}\vert^4},\quad R_0<\vert \mathbf{x}\vert\leq\frac{d\textit{l}}{a}.
\end{equation*}
On the other hand, inside the 3D ball $B_{R_0}$, we just need to use a standard elliptic regularity estimate (See for example \cite[P.340 Theorem 5]{2010Partial}) together with (\ref{est of lambda_l}) and (\ref{est of f_l}) to get, for some integer $m$ large enough
\begin{equation*}
  \vert D_{\mathbf{x}}^3w_\textit{l}(\mathbf{x})\vert\leq C\Vert w_\textit{l}\Vert_{H^m(B_{R_0})}\leq C\Vert w_\textit{l}\Vert_{L^2(B_{R_0})}\leq C.
\end{equation*}
\begin{flushright}
  {$\Box$}
\end{flushright}

\par With Lemma \ref{fundamental est of v,w,lambda}, we thereafter define for all $p\in2\pi\mathbb{Z}^3$
\begin{equation}\label{eta_p}
  \eta_p= -\frac{1}{\sqrt{d}}\widetilde{w}_{\textit{l},p}.
\end{equation}
From (\ref{discrete asymptotic energy pde on the torus}), it is easy to deduce
\begin{equation}\label{eqn of eta_p}
\begin{aligned}
  \left\vert\mathcal{M}_dp\right\vert^2\eta_p+\frac{1}{2\sqrt{d}}
  \sum_{q\in2\pi\mathbb{Z}^3}v_{p-q}^{(a,d)}\eta_q
=-\frac{1}{2\sqrt{d}}
  v_p^{(a,d)}+\frac{\lambda_\textit{l}}{a^2}
 \eta_p+\frac{\lambda_\textit{l}}{a^2d}\widehat{\chi_{d\textit{l}}}
  \left(\frac{\mathcal{M}_dp}{2\pi}\right).
\end{aligned}
\end{equation}
Since $w_\textit{l}$ is real-valued and radially symmetric, we have $\eta_p=\eta_{-p}=\overline{\eta_p}$. Moreover, we let $\eta\in L^2(\Lambda_d)$ be the function with Fourier coefficients $\eta_p$ and $\eta_\perp$ be its orthogonal projection onto $L^2_\perp(\Lambda_d)$. Then with (\ref{est of w_l and grad w_l}) we can deduce
\begin{equation}\label{est of eta and eta_perp}
\begin{aligned}
  \Vert\eta_\perp\Vert_2^2\leq\Vert\eta\Vert_2^2&=\frac{1}{d}\int_{\vert \mathbf{x}\vert\leq d\textit{l}}
  \left\vert\widetilde{w}_{\textit{l}}(\mathbf{x})\right\vert^2d\mathbf{x}=\frac{a^3}{d}
  \int_{\vert \mathbf{y}\vert\leq\frac{d}{a}\textit{l}}\vert w_{\textit{l}}(\mathbf{y})\vert^2d\mathbf{y}\\
&\leq
  \frac{a^3}{d}\int_{\vert \mathbf{y}\vert\leq\frac{d}{a}\textit{l}}\frac{C}{\vert \mathbf{y}\vert^2} d\mathbf{y}
  =Ca^2\textit{l}.
\end{aligned}
\end{equation}
Similarly we have
\begin{align}
  \Vert\nabla_{\mathbf{x}}\eta_\perp\Vert_2^2=\Vert\nabla_{\mathbf{x}}\eta\Vert_2^2
=&\frac{1}{d}\int_{\vert \mathbf{x}\vert\leq d\textit{l}}
  \left\vert\nabla_{\mathbf{x}}\widetilde{w}_{\textit{l}}(\mathbf{x})\right\vert^2d\mathbf{x}
=\frac{a}{d}\int_{\vert \mathbf{y}\vert\leq\frac{d}{a}\textit{l}}\vert\nabla_{\mathbf{y}} w_{\textit{l}}(\mathbf{y})\vert^2d\mathbf{y}\nonumber\\
\leq&\frac{Ca}{d}\left(\int_{\vert \mathbf{y}\vert\leq1}d\mathbf{y}+
\int_{1<\vert \mathbf{y}\vert\leq\frac{d}{a}\textit{l}}\frac{d\mathbf{y}}{\vert \mathbf{y}\vert^4}\right)
\leq\frac{Ca}{d},
\label{est of grad eta}
\end{align}
as well as
\begin{equation}\label{est of D^2 eta & D^3 eta}
  \Vert D^2_{\mathbf{x}}\eta_\perp\Vert^2_2=\Vert D^2_{\mathbf{x}}\eta\Vert^2_2\leq\frac{C}{ad},\quad
  \Vert D^3_{\mathbf{x}}\eta_\perp\Vert^2_2=\Vert D^3_{\mathbf{x}}\eta\Vert^2_2\leq\frac{C}{a^3d}.
\end{equation}
We can bound $\eta_p$ for all $p\in2\pi\mathbb{Z}^3$ in the same way
\begin{equation}\label{est of eta_0}
  \vert\eta_p\vert\leq\frac{1}{d}\int_{\vert \mathbf{x}\vert\leq d\textit{l}}\widetilde{w}_{\textit{l}}
  (\mathbf{x})d\mathbf{x}=\frac{a^3}{d}\int_{\vert \mathbf{y}\vert\leq\frac{d}{a}\textit{l}}w_{\textit{l}}(\mathbf{y})d\mathbf{y}
  \leq\frac{a^3}{d}\int_{\vert \mathbf{y}\vert\leq\frac{d}{a}\textit{l}}\frac{C}{\vert \mathbf{y}\vert} d\mathbf{y}
  \leq Cad\textit{l}^2.
\end{equation}
With (\ref{est of eta_0}) and the fact that $0\leq w_{\textit{l}}\leq1$, we can also bound the $\eta_{\perp}$ point-wisely by
\begin{equation}\label{est of max eta_perp}
  \Vert\eta_\perp\Vert_{\infty}\leq \frac{1}{\sqrt{d}}(1+Cad\textit{l}^2)\leq\frac{C}{\sqrt{d}}
\end{equation}
under the assumptions that $a$ and $d$ tend to $0$ and $\textit{l}\in(0,\frac{1}{2})$. In addition, since $\widetilde{f}_{\textit{l}}=1-\widetilde{w}_\textit{l}$, we deduce via Plancherel's equality
\begin{equation}\label{int vf_l}
  \int_{\mathbb{R}^3}v(\mathbf{x})f_{\textit{l}}(\mathbf{x})d\mathbf{x}
=\frac{\sqrt{d}}{a}\sum_{p}v_p^{(a,d)}\eta_p
  +\frac{\sqrt{d}}{a}v_0^{(a,d)}.
\end{equation}
Combining (\ref{int vf_l}) with (\ref{est of int vf_l}) we find
\begin{equation}\label{有用的屎}
  \frac{\sqrt{d}}{a}\sum_{p}v_p^{(a,d)}\eta_p
  +\frac{\sqrt{d}}{a}v_0^{(a,d)}=8\pi\mathfrak{a}_0\left(1+\frac{3}{2}
          \frac{\mathfrak{a}_0}{\textit{l}}\frac{a}{d}\right)+O\left(
          \frac{\mathfrak{a}_0^3}{\textit{l}^2}\frac{a^2}{d^2}\right)
\end{equation}

\par For further usage, we may let, for $p\in2\pi\mathbb{Z}^3$,
\begin{equation}\label{define W_p}
  W_p=\frac{\lambda_{\textit{l}}}{a^2d}\left(\widehat{\chi_{d\textit{l}}}\left(
  \frac{\mathcal{M}_dp}{2\pi}\right)+d\eta_p\right).
\end{equation}
It is also easy to verify that $W_p=W_{-p}=\overline{W_p}$. With $W_p$ defined, we can rewrite equation (\ref{eqn of eta_p}) into
\begin{equation}\label{eqn of eta_p rewrt}
  \left\vert\mathcal{M}_dp\right\vert^2\eta_p+\frac{1}{2\sqrt{d}}
  \sum_{q\in2\pi\mathbb{Z}^3}v_{p-q}^{(a,d)}\eta_q+\frac{1}{2\sqrt{d}} v_p^{(a,d)}=W_p.
\end{equation}
Let $W=\sum W_p\phi_p^{(d)}\in L^2_s(\Lambda_d)$ be the function with Fourier coefficients $W_p$, then
\begin{equation}\label{define W(x)}
  W(\mathbf{x})=\frac{\lambda_{\textit{l}}}{a^2\sqrt{d}}
(\chi_{d\textit{l}}(\mathbf{x})-\widetilde{w}_{\textit{l}}
  (\mathbf{x})).
\end{equation}
From (\ref{est of lambda_l}) and (\ref{est of f_l}) we know that $W$ is supported and smooth on the 3D ball $B_{d\textit{l}}$, and
\begin{equation}\label{L^infty W}
  0<W(\mathbf{x})\leq\frac{C}{\sqrt{d}}\cdot\frac{a}{(d\textit{l})^3}.
\end{equation}
Using Lemma \ref{fundamental est of v,w,lambda}, (\ref{est of eta and eta_perp}) and (\ref{est of eta_0}) we can estimate under the assumptions that $a,d,\frac{a}{d}\to0$ and $\frac{a}{d\textit{l}}<C$
\begin{equation}\label{L1&L2 norm of W 3dscatt}
 \Vert W\Vert_2\leq Cad^{-2}\textit{l}^{-\frac{3}{2}},\quad
  \Vert W\Vert_1\leq Cad^{-\frac{1}{2}},
\end{equation}
and
\begin{equation}\label{sum_pW_peta_p 3dscatt}
  \vert W_p\vert\leq\frac{Ca}{d},\quad\left\vert\sum_{p\neq0}W_p\eta_p\right\vert\leq Ca^2d^{-2}\textit{l}^{-1}.
\end{equation}
It is also useful to recall that the Fourier transform of the 3D radial symmertic function $\chi_{d\textit{l}}$ is given by
\begin{equation}\label{fourier chi_dl}
  \widehat{\chi_{d\textit{l}}}\left(
\frac{\mathcal{M}_dp}{2\pi}\right)=\frac{4\pi}{\vert\mathcal{M}_dp\vert^2}
\left(\frac{\sin(d\textit{l}\vert\mathcal{M}_dp\vert)}{\vert\mathcal{M}_dp\vert}
-d\textit{l}\cos(d\textit{l}\vert\mathcal{M}_dp\vert)\right)
\end{equation}
Formula (\ref{fourier chi_dl}) together with (\ref{est of w_l,p}) tell us respectively that for $p\neq0$
\begin{equation}\label{moron2}
  d\vert\eta_p\vert\leq Ca\vert\mathcal{M}_dp\vert^{-2},
\quad\left\vert\widehat{\chi_{d\textit{l}}}\left(
\frac{\mathcal{M}_dp}{2\pi}\right)\right\vert\leq
C(d\textit{l})\vert\mathcal{M}_dp\vert^{-2}.
\end{equation}
We combine (\ref{moron2}) with (\ref{define W_p}) and (\ref{est of lambda_l}) to find
\begin{equation}\label{moron3}
  \vert W_p\vert\leq C\frac{a}{d(d\textit{l})^2}\vert\mathcal{M}_dp\vert^{-2}.
\end{equation}
Moreover, using (\ref{eqn of eta_p rewrt}), we can prove the following useful $\ell^1$ estimate of $\{\eta_p\}$.
\begin{lemma}\label{l1 lemma}
Let $\{\eta_p\}$ be defined in (\ref{eta_p}). Assume that $a,d$ and $\frac{a}{d}$ tend to $0$ and $\frac{d}{a}>\frac{C}{\textit{l}}$ for some universal constant $C$. Then we have, for some universal constant, also denoted by $C$,
\begin{equation}\label{l1 norm of eta_p lemma}
  \sum_{p\neq0}\vert\eta_p\vert\leq C\left(1+\frac{a}{d}\ln(a^{-1})\right).
\end{equation}
\end{lemma}
\noindent
\emph{Proof.} For $p\neq0$, dividing (\ref{eqn of eta_p rewrt}) by $\vert\mathcal{M}_dp\vert^2$ we get
\begin{equation}\label{eta_p for p neq 0}
      \eta_p=\left\{-\frac{1}{2\sqrt{d}}\left(
      \sum_{q}v_{p-q}^{(a,d)}\eta_q+
      v_p^{(a,d)}\right)
      +W_p\right\}\vert\mathcal{M}_dp\vert^{-2}.
\end{equation}
Dividing (\ref{eta_p for p neq 0}) into three terms, we first show that
    \begin{equation}\label{晦气1}
      \frac{1}{\sqrt{d}}\sum_{p\neq0}\vert v_p^{(a,d)}\vert\cdot
      \vert\mathcal{M}_dp\vert^{-2}\leq C\left(1+\frac{a}{d}\ln(a^{-1})\right).
    \end{equation}
    Separating high and low momenta at $\epsilon d^{-1}$ for some $\epsilon>1$ to be determined, we obtain for the low momentum part
    \begin{equation}\label{晦气2}
      \frac{1}{\sqrt{d}}\sum_{0<\vert\mathcal{M}_dp\vert<\epsilon d^{-1}}
      \vert v_p^{(a,d)}\vert\cdot\vert\mathcal{M}_dp\vert^{-2}
      \leq C\frac{a}{d}\sum_{0<\vert\mathcal{M}_dp\vert<\epsilon d^{-1}}
      \vert\mathcal{M}_dp\vert^{-2}.
    \end{equation}
Here we were using (\ref{v_p^a,d}) to bound $\vert v_p^{(a,d)}\vert$. We can control the right-hand side of (\ref{晦气2}) using Riemann integrals. Recall that $p\in 2\pi\mathbb{Z}^3\backslash\{0\}$ and $a,d$ and $\frac{a}{d}$ are so small that we can assume all of them are less than $1$.
\begin{equation}\label{WTH1}
  \begin{aligned}
\sum_{\substack{0<\vert\mathcal{M}_dp\vert<\epsilon d^{-1}\\p_1p_2p_3\neq0}}
\vert\mathcal{M}_dp\vert^{-2}
&\leq\frac{1}{8\pi^3}\int_{\vert\mathcal{M}_d\mathbf{y}\vert<2\epsilon d^{-1}}
\frac{1}{\vert\mathcal{M}_d\mathbf{y}\vert^2}dy_1dy_2dy_3=C\epsilon,\\
\sum_{\substack{0<\vert\mathcal{M}_dp\vert<\epsilon d^{-1}\\p_3=0,\,p_1p_2\neq0}}
\vert\mathcal{M}_dp\vert^{-2}
&\leq\frac{1}{2\pi^2}+\frac{1}
{4\pi^2}\int_{\frac{1}{2}<\vert y\vert<2\epsilon d^{-1}}
\frac{1}{\vert y\vert^2}dy_1dy_2=C(1+\ln(\epsilon d^{-1})),\\
\sum_{\substack{0<\vert\mathcal{M}_dp\vert<\epsilon d^{-1}\\p_1=0,\,p_2p_3\neq0\\or\,p_2=0,\,p_1p_3\neq0}}
\vert\mathcal{M}_dp\vert^{-2}
&\leq  2\left(\frac{d^2}{\pi^2(1+d^2)}
+\frac{1}{4\pi^2}\int_{\frac{1}{2d}<\vert(y_2,\frac{y_3}{d})\vert<2\epsilon d^{-1}}
\frac{dy_2dy_3}{\vert(y_2,\frac{y_3}{d})\vert^2}\right)\\
&=C(d^2+d\ln\epsilon),\\
\sum_{\substack{0<\vert\mathcal{M}_dp\vert<\epsilon d^{-1}\\p_2,\,p_3=0,\,p_1\neq0\\or
\,p_1,\,p_3=0,\,p_2\neq0}}\vert\mathcal{M}_dp\vert^{-2}
&\leq2\left
(\frac{1}{2\pi^2}+\frac{1}{\pi}\int_{1}^{2\epsilon d^{-1}}\frac{1}{y_1^2}dy_1\right)\leq C,\\
\sum_{\substack{0<\vert\mathcal{M}_dp\vert<\epsilon d^{-1}\\p_1,\,p_2=0,\,p_3\neq0}}
\vert\mathcal{M}_dp\vert^{-2}
&\leq \frac{d^2}{2\pi^2}+\frac{d^2}{\pi}
\int_{1}^{2\epsilon}\frac{1}{y_1^2}dy_1\leq Cd^2.
\end{aligned}
\end{equation}
With estimates above we can conclude that
    \begin{equation}\label{很晦气1}
      \sum_{0<\vert\mathcal{M}_dp\vert<\epsilon d^{-1}}
      \vert\mathcal{M}_dp\vert^{-2}\leq C(\epsilon+\ln(\epsilon d^{-1})).
    \end{equation}
    Plugging (\ref{很晦气1}) into (\ref{晦气2}) we derive
    \begin{equation}\label{晦气3}
      \frac{1}{\sqrt{d}}\sum_{0<\vert\mathcal{M}_dp\vert<\epsilon d^{-1}}
      \vert v_p^{(a,d)}\vert\cdot\vert\mathcal{M}_dp\vert^{-2}
      \leq Cad^{-1}(\epsilon+\ln(\epsilon d^{-1})).
    \end{equation}
    For the high momentum part we can bound
    \begin{align}
      \frac{1}{\sqrt{d}}\sum_{\vert\mathcal{M}_dp\vert\geq \epsilon d^{-1}}
      \vert v_p^{(a,d)}\vert\cdot\vert\mathcal{M}_dp\vert^{-2}
&\leq\frac{1}{\sqrt{d}}\left(\sum_p\vert v_p^{(a,d)}\vert^2\right)^{\frac{1}{2}}
      \left(\sum_{\vert\mathcal{M}_dp\vert\geq
      \epsilon d^{-1}}\vert\mathcal{M}_dp\vert^{-4}\right)^{\frac{1}{2}}\nonumber\\
      &=Ca^{-\frac{1}{2}}d^{-\frac{1}{2}}\left(\sum_{\vert\mathcal{M}_dp\vert\geq
      \epsilon d^{-1}}\vert\mathcal{M}_dp\vert^{-4}\right)^{\frac{1}{2}}\label{晦气4}
    \end{align}
    where we have used the fact that $\Vert v_a\Vert_2=a^{-\frac{1}{2}}\Vert v\Vert_2$. We claim that
    \begin{equation}\label{很晦气2}
      \sum_{\vert\mathcal{M}_dp\vert\geq \epsilon d^{-1}}\vert\mathcal{M}_dp\vert^{-4}
      \leq C\epsilon^{-1}d^2.
    \end{equation}
    Hence
    \begin{equation}\label{晦气5}
      \frac{1}{\sqrt{d}}\sum_{\vert\mathcal{M}_dp\vert\geq a^{-1}}
      \vert v_p^{(a,d)}\vert\cdot\vert\mathcal{M}_dp\vert^{-2}\leq
      Ca^{-\frac{1}{2}}d^{\frac{1}{2}}\epsilon^{-\frac{1}{2}}.
    \end{equation}
(\ref{很晦气2}) is derived by a similar argument using Riemann integrals
\begin{equation}\label{WTH2}
  \begin{aligned}
\sum_{\substack{\vert\mathcal{M}_dp\vert\geq\epsilon d^{-1}\\p_1p_2p_3\neq0}}
\vert\mathcal{M}_dp\vert^{-4}
&\leq\frac{1}{8\pi^3}\int_{\vert\mathcal{M}_d\mathbf{y}\vert>\frac{1}{2}\epsilon d^{-1}}
\frac{1}{\vert\mathcal{M}_d\mathbf{y}\vert^4}dy_1dy_2dy_3=C\epsilon^{-1}d^2,\\
\sum_{\substack{\vert\mathcal{M}_dp\vert\geq\epsilon d^{-1}\\p_3=0,\,p_1p_2\neq0}}
\vert\mathcal{M}_dp\vert^{-4}
&\leq\frac{1}{4\pi^2}\int_{\vert y\vert>\frac{1}{2}\epsilon d^{-1}}
\frac{1}{\vert y\vert^4}dy_1dy_2=C\epsilon^{-2}d^2,\\
\sum_{\substack{\vert\mathcal{M}_dp\vert\geq\epsilon d^{-1}\\p_1=0,\,p_2p_3\neq0\\or\,p_2=0,\,p_1p_3\neq0}}
\vert\mathcal{M}_dp\vert^{-4}
&\leq  \frac{1}{2\pi^2}\int_{\vert(y_2,\frac{y_3}{d})\vert>\frac{1}{2}\epsilon d^{-1}}
\frac{dy_2dy_3}{\vert(y_2,\frac{y_3}{d})\vert^4}=C\epsilon^{-2}d^3,\\
\sum_{\substack{\vert\mathcal{M}_dp\vert\geq\epsilon d^{-1}\\p_2,\,p_3=0,\,p_1\neq0\\or
\,p_1,\,p_3=0,\,p_2\neq0}}\vert\mathcal{M}_dp\vert^{-4}
&\leq\frac{2}{\pi}\int_{\frac{1}{2}\epsilon d^{-1}}^{\infty}
\frac{1}{y_1^4}dy_1\leq C\epsilon^{-3}d^3,\\
\sum_{\substack{\vert\mathcal{M}_dp\vert\geq\epsilon d^{-1}\\p_1,\,p_2=0,\,p_3\neq0}}
\vert\mathcal{M}_dp\vert^{-4}
&\leq \frac{d^4}{\pi}
\int_{\frac{1}{2}\epsilon}^{\infty}\frac{1}{y_1^4}dy_1\leq C\epsilon^{-3}d^4.
\end{aligned}
\end{equation}
Taking $\epsilon=\frac{d}{a}>1$, (\ref{晦气5}) together with (\ref{晦气3}) give (\ref{晦气1}). We can bound the remaining two terms similarly by taking $\epsilon=\frac{d}{a}$ and $\epsilon=\textit{l}^{-1}$ respectively
\begin{align}
\frac{1}{\sqrt{d}}\sum_{p\neq0}\left\vert\sum_q v_{p-q}^{(a,d)}\eta_q\right\vert\cdot
\vert\mathcal{M}_dp\vert^{-2}
&\leq C\left(1+\frac{a}{d}\ln(a^{-1})\right)\label{晦气-1}\\
\sum_{p\neq0}\vert W_p\vert\cdot
\vert\mathcal{M}_dp\vert^{-2}&\leq C\left(\frac{a}{d\textit{l}}
+\frac{a}{d}\ln[(d\textit{l})^{-1}]\right)\label{晦气-2}
\end{align}
as long as we notice that
\begin{align*}
 \frac{1}{\sqrt{d}}\left\vert\sum_q v_{p-q}^{(a,d)}\eta_q\right\vert
\leq C\frac{a}{d},&\quad\Vert v_a\eta\Vert_2\leq Ca^{-\frac{1}{2}}d^{-\frac{1}{2}},\\
\vert W_p\vert\leq C\frac{a}{d},&\quad\Vert W\Vert_2\leq Cad^{-2}\textit{l}^{-\frac{3}{2}}
\ll Ca^{-\frac{1}{2}}d^{-\frac{1}{2}}.
\end{align*}

\begin{flushright}
  \{$\Box$\}
\end{flushright}
\par Notice in (\ref{晦气-2}), we have derived a bound of $\sum\vert W_p\vert\vert\mathcal{M}_dp\vert^{-2}$. This is crucial to the estimate of error terms in further proofs, and we will state it as a lemma below.


\begin{lemma}\label{W_p lemma}
Assume that $a,d$ and $\frac{a}{d}$ tend to $0$ and $\frac{d}{a}>\frac{C}{\textit{l}}$ for some universal constant $C$. Then we have, for some universal constant, also denoted by $C$,
\begin{equation}\label{sum W_p M_dp^-2}
  \sum_{p\neq0}\vert W_p\vert\cdot
\vert\mathcal{M}_dp\vert^{-2}\leq C\left(\frac{a}{d\textit{l}}
+\frac{a}{d}\ln[(d\textit{l})^{-1}]\right)
\end{equation}
\end{lemma}


\subsection{Induced 2D Scattering Equation}\label{Induced 2D Scattering Equation}
\
\par Let $\Lambda_{\mathrm{2D}}=[-\frac{1}{2},\frac{1}{2}]^2$ denote a 2D torus. Due to $d\ll1$, a 2D effect may come into play especially in Region $\mathrm{III}$ where $d$ is especially small (decaying exponentially in Region III in the Gross-Pitaevskii limit), and the system is dominated by two large scale directions. Here we define an induced 2D interaction potential $u$, and there follows the corresponding 2D scattering equation. For $x\in\Lambda_{\mathrm{2D}}$ and noticing $\mathbf{x}=(x,z)$, the scaled version $u_{d\textit{l}}$ of $u$ is defined by
\begin{equation}\label{define u_dl}
  u_{d\textit{l}}(x)=\frac{1}{(d\textit{l})^2}u\left(\frac{x}{d\textit{l}}\right)
=\frac{2}{\sqrt{d}}\int_{-d\textit{l}}^{d\textit{l}}
W(\mathbf{x})dz.
\end{equation}
Notice that for $\bar{p}\in2\pi\mathbb{Z}^2$ (such that $p=(\bar{p},p_3)\in2\pi\mathbb{Z}^3$)
\begin{equation*}
  \int_{\Lambda_{\mathrm{2D}}}u_{d\textit{l}}(x)e^{-i\bar{p}\cdot x}dx=2W_{(\bar{p},0)}.
\end{equation*}
Then we can write
\begin{equation}\label{define u}
  u(x)=\frac{2(d\textit{l})^3}{\sqrt{d}}\int_{-1}^{1}W\left(d\textit{l}\cdot\mathbf{x}\right)dz.
\end{equation}
From (\ref{L^infty W}) we know that $u$ is supported and smooth in the 2D ball $\mathcal{B}_{1}$, and
\begin{equation}\label{L^infty u}
  0<u(x)\leq\frac{Ca}{d}.
\end{equation}
 The 2D scattering length of $u$ is given by, according to \cite{LiebYng2000,lieb2005mathematics}
\begin{equation}\label{define a_u}
  \mathfrak{a}_u=e^{-\frac{2\pi}{\mathfrak{E}_u}},
\end{equation}
where $\mathfrak{E}_u$ is the ground state energy of the energy functional
\begin{equation}\label{define 2d energy functional}
  \mathcal{E}_u[\phi]=\int_{\mathcal{B}_1}\left\{\vert\nabla_x\phi\vert^2
+\frac{1}{2}u\vert\phi\vert^2\right\}dx
\end{equation}
with the boundary condition $\phi=1$ for $\vert x\vert=1$. The minimizer $\phi_0$ of $\mathcal{E}_u$ satisfies the localized 2D one-particle zero energy scattering equation
\begin{equation*}
  \left\{\begin{aligned}
   &-\Delta_x \phi_0(x)+\frac{1}{2}u(x)\phi_0(x)=0,\quad x\in\mathcal{B}_1\subset\mathbb{R}^2.\\
   &\left.\phi_0(x)\right\vert_{\vert x\vert=1}=1.
  \end{aligned}\right.
\end{equation*}
From \cite[Lemma 4.1]{2005Bosons}, since $u$ is supported and smooth in the 2D ball $\mathcal{B}_{1}$, we have for some universal constant $C$
\begin{equation}\label{E_u}
  \frac{1}{2}I_u\leq\mathfrak{E}_u\leq\frac{1}{2}I_u+C\Vert u\Vert_{L^\infty}I_u,
\end{equation}
where
\begin{equation}\label{define I_u}
  \frac{8\pi a\mathfrak{a}_0}{d}-\frac{Ca^2}{d^2\textit{l}}
\leq I_u=\int_{\mathbb{R}^2}u(x)dx\leq \frac{8\pi a\mathfrak{a}_0}{d}+\frac{Ca^2}{d^2\textit{l}}
\end{equation}
with the help of (\ref{est of lambda_l}) and (\ref{est of int w_l}). Combining (\ref{L^infty u}), (\ref{define a_u}), (\ref{E_u}) and (\ref{define I_u}) we arrive at
\begin{equation}\label{est of a_u}
  \mathfrak{a}_u=\gamma_ue^{-\frac{d}{2a\mathfrak{a}_0}},
\end{equation}
where
\begin{equation}\label{define gamma_u}
  \exp(-C\textit{l}^{-1})\leq\gamma_u=\exp\left\{2\pi\left
(\frac{\mathfrak{E}_ud-4\pi a\mathfrak{a}_0}{4\pi a\mathfrak{a}_0\mathfrak{E}_u}\right)\right\}\leq\exp(C\textit{l}^{-1})
\end{equation}
for some universal constant $C$.

\par Similar to (\ref{asymptotic energy pde on the ball}), we consider the following 2D ground state energy equation with Neumann boundary condition induced by the 3D scattering equation (\ref{asymptotic energy pde on the ball})
\begin{equation}\label{asymptotic energy pde on the ball 2D}
  \left\{\begin{aligned}
  &(-\Delta_x+\frac{1}{2}u)g_\textit{h}=\mu_\textit{h}g_\textit{h},\quad \vert x\vert\leq \frac{h}{d\textit{l}},\\
  &\left.\frac{\partial g_\textit{h}}{\partial \mathbf{n}}\right\vert_{\vert x\vert=\frac{h}{d\textit{l}}}=0,\quad \left.g_h\right\vert_{\vert x\vert=\frac{h}{d\textit{l}}}=1.
  \end{aligned}\right.
\end{equation}
Here $\textit{l}$ is given in (\ref{asymptotic energy pde on the ball}) and $h\in(0,\frac{1}{2})$ is another parameter which will be later chosen so that $\frac{h}{d\textit{l}}$ is large enough. For detailed analysis of (\ref{asymptotic energy pde on the ball 2D}) one can consult \cite{CaraciCenaSchlein2021,caraciCenaSchlein2022excitation}. But we shall notice that the induced 2D interaction potential $u$ here depends on parameters $a$, $d$ and $\textit{l}$. We define $z_h=1-g_h$, and also make constant extensions to both $g_h$ and $z_h$ outside of the 2D ball $\mathcal{B}_{\frac{h}{d\textit{l}}}$ so that $g_h\in H^2_{loc}(\mathbb{R}^2)$ and $z_h\in H^2(\mathbb{R}^2)$. A scaling gives
\begin{equation}\label{scaling g_h & z_h}
  \widetilde{g}_h(x)=g_h\Big(\frac{x}{d\textit{l}}\Big),\quad
\widetilde{z}_h(x)=z_h\Big(\frac{x}{d\textit{l}}\Big).
\end{equation}
Since $\Lambda_{\mathrm{2D}}=[-\frac{1}{2},\frac{1}{2}]^2$ is a 2D torus, $\widetilde{z}_h$ can be regarded as a periodic function on $\Lambda_{\mathrm{2D}}$. Then we can write (\ref{asymptotic energy pde on the ball 2D}) in the form
\begin{equation}\label{asymptotic energy pde on the torus 2D}
  \Big(-\Delta_x+\frac{1}{2(d\textit{l})^2}u\big(\frac{x}{d\textit{l}}\big)\Big)\widetilde{z}_h(x)
=\frac{1}{2(d\textit{l})^2}u\big(\frac{x}{d\textit{l}}\big)-\frac{\mu_h}{(d\textit{l})^2}
\big(1-\widetilde{z}_h(x)\big)\chi^{\mathrm{2D}}_h(x),\quad x\in\Lambda_{\mathrm{2D}}
\end{equation}
with $\chi^{\mathrm{2D}}_h$ being the characteristic function of the closed 2D ball $\overline{\mathcal{B}}_h\subset\Lambda_{\mathrm{2D}}$. Via the Fourier transform, we also have the discrete version of (\ref{asymptotic energy pde on the torus 2D}) written as
\begin{equation}\label{discrete asymptotic energy pde on the torus 2D}
  \begin{aligned}
\vert\bar{p}\vert^2\widetilde{z}_{h,\bar{p}}+\sum_{\bar{q}\in2\pi\mathbb{Z}^2}
W_{(\bar{p}-\bar{q},0)}\widetilde{z}_{h,\bar{q}}
=W_{(\bar{p},0)}+\frac{\mu_h}{(d\textit{l})^2}\widetilde{z}_{h,\bar{p}}-
\frac{\mu_h}{(d\textit{l})^2}\widehat{\chi^{\mathrm{2D}}_h}\left(\frac{\bar{p}}{2\pi}\right)
\end{aligned}
\end{equation}
with $\bar{p}\in2\pi\mathbb{Z}^2$ and
\begin{equation*}
  \widetilde{z}_{h,\bar{p}}=\int_{\Lambda_{\mathrm{2D}}}\widetilde{z}_h(x)e^{-i\bar{p}\cdot x}dx,\quad
\widehat{\chi^{\mathrm{2D}}_h}\left(\frac{\bar{p}}{2\pi}\right)=
\int_{\Lambda_{\mathrm{2D}}}\chi^{\mathrm{2D}}_h(x)e^{-i\bar{p}\cdot x}dx.
\end{equation*}
We collect the needed properties of $g_h$ and $z_h$ in the next lemma. We first denote
\begin{equation}\label{define m}
  m=\ln\big(h(d\textit{l})^{-1}\mathfrak{a}_u^{-1}\big).
\end{equation}
We know that $m$ will tend to infinity, since we would like to have $\frac{d\textit{l}}{a}$ and $\frac{h}{d\textit{l}}$ large enough, and we have the representation (\ref{est of a_u}).
\begin{lemma}\label{fundamental est of g,z,mu}
Recall that $u\in L^2(\mathbb{R}^2)$ defined in (\ref{define u}) is a 2D interaction potential smooth in the 2D ball $\mathcal{B}_1$ with scattering length $\mathfrak{a}_u$ given in (\ref{est of a_u}). Let $g_h$, $\mu_h$, $z_h$, $\widetilde{z}_{h,\bar{p}}$ and $m$ be defined above. Under the same setting of Lemma \ref{fundamental est of v,w,lambda}, and assume further for some universal large constant $C$ (independent of $a$, $d$, $\textit{l}$ and $h$), $\frac{h}{d\textit{l}}>C$, then there exist some universal constants, also denoted by $C$, such that following estimates hold true for all $\frac{h}{d\textit{l}}$ large enough.
\begin{enumerate}[$(1)$]
  \item The asymptotic estimate of ground state energy $\mu_h$ is
  \begin{equation}\label{est of mu_h}
    \left\vert\mu_h-\frac{2(d\textit{l})^2}{h^2m}
\left(1+\frac{3}{4m}\right)\right\vert\leq \frac{C(d\textit{l})^2}{h^2m^3}.
  \end{equation}
  \item $g_h$ is radially symmetric and smooth away from the boundaries of $\mathcal{B}_1$ and $\mathcal{B}_{\frac{h}{d\textit{l}}}$ and
  \begin{equation}\label{est of g_h}
    0\leq g_h(x)\leq 1.
  \end{equation}
  Moreover,
  \begin{equation}\label{est of z_h}
    \vert z_h(x)\vert\leq\left\{
    \begin{aligned}
    &1,\quad \vert x\vert\leq1\\
    &\frac{C}{m}\ln\big(h(d\textit{l})^{-1}\vert x\vert^{-1}\big),\quad
    1\leq\vert x\vert\leq\frac{h}{d\textit{l}}
    \end{aligned}\right.
  \end{equation}
  and for any integer $1\leq k\leq4$
  \begin{equation}\label{est of grad z_h}
    \vert D^k_xz_h(x)\vert\leq\left\{
    \begin{aligned}
    &\frac{C}{m}\frac{1}{1+\vert x\vert},\quad \text{if $k=1$ and $\vert
    x\vert\leq\frac{h}{d\textit{l}}$}\\
    &\frac{C}{m}\frac{1}{\vert x\vert^k},\quad \text{if $2\leq k\leq 4$ and $1\leq\vert x
    \vert\leq\frac{h}{d\textit{l}}$}
    \end{aligned}\right.
  \end{equation}
and
  \begin{equation}\label{est of D^kz_h inside B_1}
    \Vert D^k_xz_h\Vert_{L^2(\mathcal{B}_1)}\leq\left\{
\begin{aligned}
&\frac{C}{m},\quad\text{if $1\leq k\leq 3$.}\\
&\frac{C}{m}\left(\frac{d\textit{l}}{a}\right)^{\frac{1}{2}},\quad
\text{if $k=4$}
\end{aligned}\right.
  \end{equation}
  \item We have
  \begin{equation}\label{est of int ug_h}
    \left\vert\int_{\mathbb{R}^2}u(x)g_h(x)-\frac{4\pi}{m}\left(1+\frac{1}{2m}
\right)\right\vert\leq \frac{C}{m^3}.
  \end{equation}
  \item For all $\bar{p}\in 2\pi\mathbb{Z}^2\backslash\{0\}$
  \begin{equation}\label{est of z_h,p}
    \vert\widetilde{z}_{h,\bar{p}}\vert\leq\frac{C}{m}\frac{1}{\vert\bar{p}\vert^2}.
  \end{equation}
\end{enumerate}
\end{lemma}
\begin{remark}\label{remark 2d scat eqn}
Similar to Remark \ref{remark 3d scat eqn}, the notation $D^k_xz_h$ always means the $k$-th derivative of $z_h$ away from the boundaries of $\mathcal{B}_1$ and $\mathcal{B}_{\frac{h}{d\textit{l}}}$, and the integral of it always means integrating inside of $\mathcal{B}_{\frac{h}{d\textit{l}}}$ unless otherwise specified. On the other hand, we notice that we can not reach a satisfactory point-wise estimate on $D^kz_h$ inside of the 2D ball $\mathcal{B}_1$ as what we have in (\ref{est of w_l and grad w_l}) because the parameters-dependence of $u$ makes it hard to bound $\vert D^k_xz_h\vert$ by $m^{-1}$ inside of $\mathcal{B}_1$ when $k\geq2$.
\end{remark}
\noindent\emph{Proof.} Most of the results stated in Lemma \ref{fundamental est of g,z,mu} have been collected and proven in \cite[Appendix B]{CaraciCenaSchlein2021} and \cite[Appendix B]{caraciCenaSchlein2022excitation}. Although the potential $u$ here varies as parameters changing, the fact that
\begin{equation}\label{2d crucial}
  \big\vert\Vert u\Vert_{L^\infty}\cdot\ln\mathfrak{a}_u\big\vert\leq C
\end{equation}
for some universal constant $C$ (see (\ref{L^infty u}) and (\ref{est of a_u})) and the fact that $u$ is supported on $\mathcal{B}_1$ rather than a ball with radius tending to infinity, ensure constants shown in the statement of Lemma \ref{fundamental est of g,z,mu} are parameter-free. What left for us is to prove (\ref{est of grad z_h}) and (\ref{est of D^kz_h inside B_1}) for $k=2,\,3,\,4$. Estimate (\ref{est of grad z_h}) just follows from the idea of the proof of \cite[Lemma 7]{CaraciCenaSchlein2021}, and we shall notice that
\begin{equation*}
  \begin{aligned}
  J_1^\prime&=J_0-\frac{1}{r}J_1\\
  Y_1^\prime&=Y_0-\frac{1}{r}Y_1
  \end{aligned}
\end{equation*}
where $J$ and $Y$ are Bessel functions of the first and the second kind (See for example \cite[(8.47)]{tableofintegrals}). As for (\ref{est of D^kz_h inside B_1}), we apply the standard elliptic regularity estimate on equation (\ref{asymptotic energy pde on the ball 2D}) inside of the ball $\mathcal{B}_1$ to get
\begin{equation}\label{2d regular}
  \Vert \nabla_x g_h\Vert_{H^2(\mathcal{B}_1)}\leq C\left(
  \Vert\nabla_x g_h\Vert_{L^\infty(\partial\mathcal{B}_1)}+\Vert
  \nabla_x(\mu_hg_h-\frac{1}{2}ug_h)\Vert_{L^2(\mathcal{B}_1)}\right).
\end{equation}
Using (\ref{L^infty u}), (\ref{est of mu_h}) and (\ref{est of grad z_h}) we can easily deduce
\begin{equation}\label{shit1}
  \Vert\nabla_x g_h\Vert_{L^\infty(\partial\mathcal{B}_1)},\,
  \Vert\mu_h\nabla_x g_h\Vert_{L^2(\mathcal{B}_1)},\,
  \Vert u\nabla_x g_h\Vert_{L^2(\mathcal{B}_1)}\leq \frac{C}{m}.
\end{equation}
On the other hand, according to the definition of $u$ (\ref{define u}) and estimate (\ref{est of grad eta}), we can bound
\begin{equation}\label{shit2}
  \Vert \nabla_x u\Vert_{L^2(\mathcal{B}_1)}\leq C\textit{l}\left(
  \frac{a}{d\textit{l}}\right)^{\frac{3}{2}}.
\end{equation}
Furthermore, by \cite[Appendix B]{CaraciCenaSchlein2021} we have
\begin{equation}\label{shit3}
  \Vert g_h\Vert_{L^\infty(\mathcal{B}_1)}\leq\frac{C\vert\ln\mathfrak{a}_u\vert}{m}.
\end{equation}
Together with the expression of $\mathfrak{a}_u$ (\ref{est of a_u}), we reach
\begin{equation}\label{shit4}
  \Vert g_h\nabla_x u\Vert_{L^2(\mathcal{B}_1)}\leq \frac{C}{m}.
\end{equation}
The estimates above together give (\ref{est of D^kz_h inside B_1}) except for $k=4$. But the $k=4$ case is similar since
\begin{equation}\label{2d regular 2}
  \Vert \nabla_x g_h\Vert_{H^3(\mathcal{B}_1)}\leq C\left(
  \Vert\nabla_x g_h\Vert_{L^\infty(\partial\mathcal{B}_1)}+\Vert
  \nabla_x(\mu_hg_h-\frac{1}{2}ug_h)\Vert_{H^1(\mathcal{B}_1)}\right),
\end{equation}
and we just need additionally using the definition of u (\ref{define u}) and estimates derived in Section \ref{3 3D  scattering equation}.
\begin{equation}\label{shit5}
   \Vert D^2_x u\Vert_{L^2(\mathcal{B}_1)}\leq C\textit{l}\left(
  \frac{a}{d\textit{l}}\right)^{\frac{1}{2}}.
\end{equation}
\begin{flushright}
  {$\Box$}
\end{flushright}

\par Let $p=(\bar{p},p_3)\in 2\pi\mathbb{Z}^3$, where $\bar{p}=(p_1,p_2)\in2\pi\mathbb{Z}^2$, then we define
\begin{equation}\label{define xi_p}
  \xi_p=\left\{\begin{aligned}
&-\widetilde{z}_{h,\bar{p}},\quad&\text{if $p_3=0$},\\
&0,\quad&\text{if $p_3\neq0$}.
\end{aligned}\right.
\end{equation}
Since $z_h$ is real-valued and radially symmetric (in terms of the 2D $x$ variable) we have $\xi_p=\xi_{-p}=\overline{\xi_p}$. We can rewrite (\ref{discrete asymptotic energy pde on the torus 2D}) into
\begin{equation}\label{eqn of xi_p}
  \vert\bar{p}\vert^2\xi_{(\bar{p},0)}+\sum_{\bar{q}\in2\pi\mathbb{Z}^2}
  \xi_{(\bar{q},0)}W_{(\bar{p}-\bar{q},0)}+W_{(\bar{p},0)}=
  \frac{\mu_h}{(d\textit{l})^2}\left(\xi_p+\widehat{\chi^{\mathrm{2D}}_{h}}\left(
  \frac{\bar{p}}{2\pi}\right)\right).
\end{equation}
Let $\xi\in L^2(\Lambda_d)$ be the function with Fourier coefficients $\xi_p$, and $\xi_\perp$ be its projection on $L^2_\perp(\Lambda_d)$. Then by the definition of $\xi_p$ and $\phi_p^{(d)}$, we know that
\begin{equation*}
   \xi(\mathbf{x})=-\frac{1}{\sqrt{d}}\widetilde{z}_h(x).
\end{equation*}
Similar to what we have done to $\eta(\mathbf{x})$, using Lemma \ref{fundamental est of g,z,mu} we can bound
\begin{equation}\label{est of xi}
  \Vert\xi_\perp\Vert^2_2\leq\Vert\xi\Vert^2_2\leq C\Big(d\textit{l}+\frac{h}{m}\Big)^2,
\end{equation}
and
\begin{equation}\label{est of grad xi}
  \Vert\nabla_{\mathbf{x}}\xi_\perp\Vert^2_2=\Vert\nabla_{\mathbf{x}}\xi\Vert^2_2
  \leq \frac{C}{m^2}\ln\Big(1+\frac{h}{d\textit{l}}\Big).
\end{equation}
Moreover
\begin{equation}\label{est of D^kxi}
  \Vert D_{\mathbf{x}}^2\xi_\perp\Vert^2_2
  =\Vert D_{\mathbf{x}}^2\xi\Vert^2_2\leq\frac{C}{(d\textit{l})^2}
  \frac{1}{m^2},\quad
  \Vert D_{\mathbf{x}}^3\xi_\perp\Vert^2_2
  =\Vert D_{\mathbf{x}}^3\xi\Vert^2_2\leq\frac{C}{(d\textit{l})^4}
  \frac{1}{m^2}.
\end{equation}
On the other hand, we can also use (\ref{est of z_h}) to bound
\begin{equation}\label{est of xi_p}
  \vert\xi_p\vert\leq C\Big((d\textit{l})^2+\frac{h^2}{m}\Big).
\end{equation}
Estimate (\ref{est of xi_p}) together with (\ref{est of g_h}) give
\begin{equation}\label{est of xi L^infty}
  \Vert\xi_\perp\Vert_{\infty}\leq\frac{C}{\sqrt{d}},
\end{equation}
since $m\to\infty$ in the Gross-Pitaevskii limit. With (\ref{est of z_h}), (\ref{est of D^kz_h inside B_1}) and (\ref{est of xi_p}), It is also useful to have $L^2$ estimates on the 3D ball $B_{d\textit{l}}$
\begin{equation}\label{est of xi B_1}
  \Vert\xi_\perp\Vert^2_{L^2(B_{d\textit{l}})},\,\Vert\xi\Vert^2
  _{L^2(B_{d\textit{l}})}
  \leq C\textit{l}(d\textit{l})^2,
\end{equation}
and for $1\leq k\leq4$
\begin{equation}\label{est of grad xi B_1}
  \Vert D^k_{\mathbf{x}}\xi_\perp\Vert^2_{L^2(B_{d\textit{l}})}
  =\Vert D^k_{\mathbf{x}}\xi\Vert^2_{L^2(B_{d\textit{l}})}
  \leq\left\{\begin{aligned}
&\frac{C}{m^2}\frac{\textit{l}}{(d\textit{l})^{2(k-1)}}\quad\text{if $1\leq k\leq 3$.}\\
&\frac{C}{m^2}\frac{\textit{l}}{(d\textit{l})^{6}}\frac{d\textit{l}}{a}\quad\text{if $ k=4$.}
\end{aligned}\right.
\end{equation}
By (\ref{est of int ug_h}) and the fact that $g_h=1-z_h$, we can calculate
\begin{equation}\label{useful shit 2d}
  2W_0+2\sum_{p\in2\pi\mathbb{Z}^3}W_p\xi_p
  =\int_{{\mathbb{R}^2}}u_{d\textit{l}}(x)\widetilde{g}_h(x)dx=
  \frac{4\pi}{m}\Big(1+\frac{1}{2m}\Big)+O(m^{-3}).
\end{equation}

\par Parallel to the definition of $W_p$ in (\ref{define W_p}), we let for all $p=(\bar{p},p_3)\in2\pi\mathbb{Z}^3$
\begin{equation}\label{define Y_p}
 Y_p=\left\{ \begin{aligned}
&W_p+\sum_{q\in2\pi\mathbb{Z}^3}\xi_qW_{p-q},\quad p_3\neq0\\
&\frac{\mu_h}{(d\textit{l})^2}\left(\xi_p+\widehat{\chi^{\mathrm{2D}}_{h}}\left(
  \frac{\bar{p}}{2\pi}\right)\right),\quad p_3=0
\end{aligned}\right.
\end{equation}
then (\ref{eqn of xi_p}) can be rewritten as for all $p\in2\pi\mathbb{Z}^3$
\begin{equation}\label{eqn of xi_p rewrt}
  \vert\mathcal{M}_dp\vert^2\xi_p+\sum_{q\in2\pi\mathbb{Z}^3}\xi_qW_{p-q}
+W_p=Y_p.
\end{equation}
Using (\ref{est of mu_h}), (\ref{est of xi_p}) and (\ref{useful shit 2d}), we can bound carefully that
\begin{equation}\label{est of Y_p}
  \vert Y_p\vert\leq \frac{C}{m}.
\end{equation}
Let
\begin{equation}\label{define Y_1 2}
\begin{aligned}
 Y_1(\mathbf{x})&=\sum_{\bar{p}\in2\pi\mathbb{Z}^2}Y_{(\bar{p},0)}\phi_p^{(d)}=
\frac{\mu_h}{(d\textit{l})^2\sqrt{d}}\widetilde{g}_h(x)\chi^{\mathrm{2D}}_h(x)\\
Y_2(\mathbf{x})&=\sum_{\substack{p\in2\pi\mathbb{Z}^3\\
p_3\neq0}}Y_p\phi_p^{(d)}=\Big(W(\mathbf{x})
-\frac{1}{2\sqrt{d}}u_{d\textit{l}}(x)\Big)\widetilde{g}_h(x)\\
Y(\mathbf{x})&=\sum_{p\in2\pi\mathbb{Z}^3}Y_p\phi_p^{(d)}
=Y_1(\mathbf{x})+Y_2(\mathbf{x})
\end{aligned}
\end{equation}
then we can bound
\begin{equation}\label{est of L2L1 Y_1,2}
\begin{aligned}
&\Vert Y_1\Vert_1,\Vert Y_2\Vert_1\leq \frac{C\sqrt{d}}{m}
  &\Vert Y_1\Vert_2\leq \frac{C}{hm},\quad
\Vert Y_2\Vert_2\leq Ca^{\frac{1}{2}}(d\textit{l})^{-\frac{3}{2}}m^{-\frac{1}{2}}.
\end{aligned}
\end{equation}
With all the estimates above, we also derive the following useful estimates following the method used in the proof of Lemma \ref{l1 lemma}.
\begin{lemma}\label{xi_p Y_p lemma}
Assume that $a,d$ and $\frac{a}{d}$ tend to $0$ and $\frac{a}{d\textit{l}}<C$, $\frac{d\textit{l}}{h}<C$ for some universal small constant $C$. Then we have for some universal constant, also denoted by $C$
\begin{align}
  \sum_{p\in2\pi\mathbb{Z}^3}\vert\xi_p\vert&\leq \frac{Ca}{d}\ln(d\textit{l})^{-1},
\label{l1 xi_p}\\
\sum_{p\in2\pi\mathbb{Z}^3\backslash\{0\}}\vert Y_p\vert\cdot\vert\mathcal{M}_dp\vert^{-2}
&\leq C\Big(\frac{1}{m}\ln\frac{1}{h}+\frac{a^{\frac{1}{3}}}{d^{\frac{1}{3}}}
\frac{1}{\textit{l}m^{\frac{2}{3}}}\Big).\label{l1 Y_p}
\end{align}
\end{lemma}

\subsection{Dimensional Coupling Scattering Equation}\label{Dimensional Coupling Scattering Equation Section}
\
\par In the definition of $u_{d\textit{l}}$ (\ref{define u_dl}), it is also intuitive to take the average value of $v_a(\mathbf{x})$ or $v_a(\mathbf{x})\widetilde{f}_\textit{l}(\mathbf{x})$ on the interval $z\in[-\frac{d}{2},\frac{d}{2}]$, rather than $2\sqrt{d}W(\mathbf{x})$. The choice of $2\sqrt{d}W(\mathbf{x})$ is technical such that the intrinsic correlation structure of the 3D to 2D problem will be revealed in our further calculation,  by introducing the difference
\begin{equation}\label{define d(x)}
  \mathfrak{D}(\mathbf{x})=\Big(\frac{1}{2}v_a(\mathbf{x})-
\sqrt{d}W(\mathbf{x})\Big)\xi(\mathbf{x}).
\end{equation}
Notice that since $a\ll d\textit{l}$, we can regard $v_a$ as a function concentrating near the origin, while $2\sqrt{d}W$ flattens $v_a$ from the scale of $a$ to $d\textit{l}$. Hence it is hard to gain a point-wise estimate of $\mathfrak{D}$. But the averaging effects of $v_a$ and $2\sqrt{d}W$ are similar, and can be checked by simply intergrating $v_a$ and $2\sqrt{d}W$ respectively on $\Lambda_d$. To convert the difference $\mathfrak{D}$ which is difficult to evaluate, to some other format with accessible estimates, we need to introduce a dimensional coupling scattering equation. Let
\begin{equation}\label{define k}
  k(\mathbf{x})=\sqrt{d}\eta(\mathbf{x})\xi(\mathbf{x}).
\end{equation}
Using (\ref{eqn of eta_p rewrt}), $k(\mathbf{x})$ satisfies the following equation on $\Lambda_d$
\begin{equation}\label{dimensional coupling scattering equation}
  -\Delta_{\mathbf{x}}k(\mathbf{x})+\frac{1}{2}v_a(\mathbf{x})k(\mathbf{x})
+\mathfrak{D}(\mathbf{x})=q_1(\mathbf{x})+q_2(\mathbf{x})
\end{equation}
where
\begin{equation}\label{define q}
  q_1(\mathbf{x})=-2\sqrt{d}\nabla_{\mathbf{x}}\eta(\mathbf{x})
\cdot\nabla_{\mathbf{x}}\xi(\mathbf{x}),\quad q_2(\mathbf{x})=-\sqrt{d}
\eta(\mathbf{x})\Delta_{\mathbf{x}}\xi(\mathbf{x}).
\end{equation}
We may denote respectively $\mathfrak{D}_p$, $k_p$, $q_{1,p}$ and $q_{2,p}$ the Fourier coefficients of $\mathfrak{D}$, $k$, $q_1$ and $q_2$ on the torus $\Lambda_d$. Notice that we have $k_p=k_{-p}=\overline{k_p}$. Let $q(\mathbf{x})=q_1(\mathbf{x})+q_2(\mathbf{x})$ and $q_p=q_{1,p}+q_{2,p}$, then equation (\ref{dimensional coupling scattering equation}) can be rewritten as
\begin{equation}\label{discrete dimensional coupling scattering equation}
  \vert\mathcal{M}_dp\vert^2k_p+\frac{1}{2\sqrt{d}}
\sum_{q\in2\pi\mathbb{Z}^3}v_{p-q}^{(a,d)}k_q+\mathfrak{D}_p=q_p=q_{1,p}+q_{2,p}.
\end{equation}
From the naive form (\ref{define k}) of $k$, we can intuitively see how different dimensions couple when the space $\Lambda_d$ becomes especially thin. The format of $k$ depicts (up to scalings) a single boson inside a large bosonic system interacting in 3D space, while its movement is strongly limited in one direction. The difference $\mathfrak{D}$ can be interpreted as a compensation for the loss of energy when we replace the original interaction potential $v_a$ with the induced 2D interaction potential $2\sqrt{d}W$. The 2D approximation and this dimensional coupling structure are simply absorbed by the 3D effect and will contribute to the second or lower order energy when $d$ is relatively large. When $d$ enters a especially thin region, Region III for example, their scales will be large enough to compete with the leading order generated by a 3D approximation. The main part, which is a pure 2D approximation analyzed in Section \ref{Induced 2D Scattering Equation}, will modify the classical leading order, while the dimensional coupling structure is the residue characterized by the dimensional coupling scattering equation (\ref{dimensional coupling scattering equation}), and an energy driven by it will become one of the main components of the second order energy.

\par Using estimates of the former two scattering equations, we can collect some useful properties of $k$ and $q_i$ in the next lemma.
\begin{lemma}\label{lemma q}
Let $k$, $q_i$ and $\mathfrak{D}$ be defined above, we have for some certain universal constants $C$
\begin{enumerate}[$(1)$]
  \item We have
        \begin{equation}\label{douche1}
          \vert k(\mathbf{x})\vert\leq \vert\eta(\mathbf{x})\vert.
        \end{equation}
  Hence for some universal constants $C$
        \begin{equation}\label{douche2}
          \Vert k(\mathbf{x})\Vert^2_2\leq Ca^2\textit{l},\quad
\Vert k(\mathbf{x})\Vert_\infty\leq Cd^{-\frac{1}{2}},
\quad \vert k_p\vert\leq
          Cad\textit{l}^2.
        \end{equation}
Moreover, we can  bound
\begin{equation}\label{douche2.5}
  \Vert\nabla_{\mathbf{x}} k\Vert^2_2\leq Cad^{-1}.
\end{equation}
  \item Let $q(\mathbf{x})=q_1(\mathbf{x})+q_2(\mathbf{x})$ and $q_p=q_{1,p}+q_{2,p}$ for $p\in2\pi\mathbb{Z}^3$, then for some universal constants $C$
\begin{equation}\label{douche2.75}
  \Vert q\Vert_1\leq  \frac{C\textit{l}^{\frac{1}{2}}}{m}\sqrt{a},
\quad \Vert q\Vert_2\leq \frac{C}{m}\frac{1}{(d\textit{l})}\sqrt{\frac{a}{d}}.
\end{equation}
Moreover, we have
        \begin{equation}\label{douche3}
          \vert q_p\vert\leq \frac{C\textit{l}^{\frac{1}{2}}}{m}\sqrt{\frac{a}{d}}.
        \end{equation}
        and for $p\neq0$
        \begin{equation}\label{douche4}
          \vert q_p\vert\leq\frac{C\textit{l}^{\frac{1}{2}}}{ma^2}\sqrt{\frac{a}{d}}
          \frac{1}{\vert\mathcal{M}_dp\vert^2}.
        \end{equation}
\end{enumerate}
\end{lemma}
\noindent\emph{Proof.} Estimate (\ref{douche1}) in the first statement of this lemma is obvious since we have $0\leq\sqrt{d}\xi(\mathbf{x})\leq1$, then (\ref{douche2}) follows immediately from (\ref{est of eta and eta_perp}) and (\ref{est of eta_0}). To reach (\ref{douche2.5}), we have
\begin{equation*}
  \Vert\nabla_{\mathbf{x}} k\Vert_2\leq
 \sqrt{d}\Vert\xi\Vert_\infty\Vert\nabla_{\mathbf{x}} \eta\Vert_2
+\sqrt{d}\Vert\eta\Vert_\infty\Vert\nabla_{\mathbf{x}} \xi\Vert_{L^2(B_{d\textit{l}})}
\leq C\big(a^{\frac{1}{2}}d^{-\frac{1}{2}}+m^{-1}\big),
\end{equation*}
where we have used estimates (\ref{est of grad eta}) and (\ref{est of grad xi B_1}). Notice the fact that from (\ref{define m}), we have $\frac{ma}{d}>C$ for some universal constant $C$, hence we conclude (\ref{douche2.5}).

\par For the second statement, we notice that by our choice of Neumann boundary condition (\ref{asymptotic energy pde on the ball}) and the fact that $f_\textit{l}$ is radially symmetric, we have in fact $\nabla_{\mathbf{x}}f_\textit{l}=0$ outside of the 3D ball $B_{\frac{d\textit{l}}{a}}$, which leads to
\begin{equation}\label{douche5}
  \begin{aligned}
  \vert q_{1,p}\vert
  &\leq\frac{1}{\sqrt{d}}\int_{\Lambda_d}\vert q_1(\mathbf{x})\vert d\mathbf{x}
  \leq C\int_{\vert\mathbf{x}\vert\leq d\textit{l}}
  \left\vert\nabla_\mathbf{x}\eta(\mathbf{x})\cdot\nabla_\mathbf{x}\xi(\mathbf{x})
  \right\vert d\mathbf{x}\\
  &\leq C\Vert\nabla_\mathbf{x}\eta\Vert_{L^2(B_{d\textit{l}})}
   \Vert\nabla_\mathbf{x}\xi\Vert_{L^2(B_{d\textit{l}})}
   \leq \frac{C\textit{l}^{\frac{1}{2}}}{m}\sqrt{\frac{a}{d}},
\end{aligned}
\end{equation}
and similarly
\begin{equation}\label{douche6}
  \vert q_{2,p}\vert\leq  C\Vert\eta\Vert_{L^2(B_{d\textit{l}})}
   \Vert \Delta_\mathbf{x}\xi\Vert_{L^2(B_{d\textit{l}})}
   \leq \frac{C}{m}\frac{a}{d},
\end{equation}
where we have used (\ref{est of eta and eta_perp}), (\ref{est of grad eta}) and (\ref{est of grad xi B_1}) in both inequalities. (\ref{douche5}) and (\ref{douche6}) together yield (\ref{douche3}). Notice (\ref{douche5}) and (\ref{douche6}) also include
\begin{equation*}
  \Vert q\Vert_1\leq  \frac{C\textit{l}^{\frac{1}{2}}}{m}\sqrt{a}.
\end{equation*}
\par To prove the $L^2$ bound of $q$ in (\ref{douche2.75}), we first use (\ref{est of grad eta}), (\ref{est of grad z_h}) and the fact that $\xi(\mathbf{x})=-d^{-\frac{1}{2}}z_h\big(x/(d\textit{l})\big)$ to bound
\begin{align}\label{douche100}
  \int_{\Lambda_d}\vert q_1(\mathbf{x})\vert^2d\mathbf{x}
\leq \frac{C}{m^2(d\textit{l})^2}\int_{\Lambda_d}\vert
\nabla_{\mathbf{x}}\eta(\mathbf{x})\vert^2d\mathbf{x}
\leq\frac{Ca}{d}\frac{1}{m^2(d\textit{l})^2}.
\end{align}
On the other hand, we use (\ref{est of grad z_h}) and Sobolev inequality to bound
\begin{align}\label{douche99.5}
  \Vert\Delta_{\mathbf{x}}\xi\Vert_{L^\infty(B_{d\textit{l}})}
=&\frac{1}{\sqrt{d}(d\textit{l})^2}\Vert\Delta_x z_h\Vert_{L^\infty(\mathcal{B}_1)}
\leq \frac{C}{\sqrt{d}(d\textit{l})^2}\Vert\Delta_x z_h\Vert_{H^2(\mathcal{B}_1)}\nonumber\\
\leq& \frac{C}{m\sqrt{d}(d\textit{l})^2}\Big(\frac{d\textit{l}}{a}\Big)^{\frac{1}{2}}.
\end{align}
Therefore from (\ref{est of eta and eta_perp})
\begin{equation}\label{douche99}
   \int_{\Lambda_d}\vert q_2(\mathbf{x})\vert^2d\mathbf{x}
\leq \frac{C}{m^2(d\textit{l})^4}\Big(\frac{d\textit{l}}{a}\Big)
\int_{B_{d\textit{l}}}\vert \eta(\mathbf{x})\vert^2d\mathbf{x}\leq \frac{Ca}{d}\frac{1}{m^2(d\textit{l})^2}.
\end{equation}
(\ref{douche100}) and (\ref{douche99}) together yield (\ref{douche2.75}).
\par To prove (\ref{douche4}), we just need to use additionally the divergence theorem
\begin{equation}\label{douche7}
  \begin{aligned}
&\vert\mathcal{M}_dp\vert^2\vert q_{1,p}\vert=
\left\vert\int_{\Lambda_d}
2(\nabla_\mathbf{x}\eta(\mathbf{x})\cdot\nabla_\mathbf{x}\xi(\mathbf{x}))
\Delta_{\mathbf{x}}e^{-ip^{T}\mathcal{M}_d\mathbf{x}}d\mathbf{x}\right\vert\\
=&2\left\vert\int_{B_{d\textit{l}}}\Delta_{\mathbf{x}}
(\nabla_\mathbf{x}\eta\cdot\nabla_\mathbf{x}\xi)
e^{-ip^{T}\mathcal{M}_d\mathbf{x}}d\mathbf{x}
-\int_{\partial B_{d\textit{l}}}e^{-ip^{T}\mathcal{M}_d\mathbf{x}}
\nabla_\mathbf{x}(\nabla_\mathbf{x}\eta\cdot\nabla_\mathbf{x}\xi)\cdot
\mathbf{n}dS_{\mathbf{x}}\right\vert,
\end{aligned}
\end{equation}
where we have used again that $\nabla_{\mathbf{x}}\eta=0$ when $\vert\mathbf{x}\vert\geq d\textit{l}$. Moreover, with this fact and bounds (\ref{est of grad eta}), (\ref{est of D^2 eta & D^3 eta}), (\ref{est of grad xi B_1}), (\ref{est of w_l and grad w_l}) and (\ref{est of grad z_h}), we can bound the last line of (\ref{douche7}) by
\begin{equation}\label{douche8}
  \vert\mathcal{M}_dp\vert^2\vert q_{1,p}\vert\leq
  \frac{C\textit{l}^{\frac{1}{2}}}{ma^2}\sqrt{\frac{a}{d}}.
\end{equation}
In a similar manner we can bound
\begin{equation}\label{douche9}
 \vert\mathcal{M}_dp\vert^2\vert q_{2,p}\vert\leq
  \frac{C\textit{l}^{\frac{1}{2}}}{ma^2}\sqrt{\frac{a}{d}}\frac{a}{d\textit{l}}.
\end{equation}
\begin{flushright}
  {$\Box$}
\end{flushright}

\par Using the bounds of $q_p$ from the second statement of Lemma \ref{lemma q} together with estimates (\ref{WTH1}) and (\ref{WTH2}) given in the proof of Lemma \ref{l1 lemma}, and most importantly, the equation (\ref{discrete dimensional coupling scattering equation}), we naturally derive the following estimate.
\begin{lemma}\label{q_p lemma}
Assume that $a,d$ and $\frac{a}{d}$ tend to $0$ and $\frac{a}{d\textit{l}}<C$, $\frac{d\textit{l}}{h}<C$ for some universal small constant $C$. Then we have for some universal constant, also denoted by $C$,
\begin{align}
\sum_{p\in2\pi\mathbb{Z}^3\backslash\{0\}}\vert k_p\vert
&\leq C\left(1+\frac{a}{d}\ln(a^{-1})\right),\label{l1 k_p}\\
\sum_{p\in2\pi\mathbb{Z}^3\backslash\{0\}}\vert q_p\vert\cdot\vert\mathcal{M}_dp\vert^{-2}
&\leq C\frac{\textit{l}^{\frac{1}{2}}}{m}
\Big(\textit{l}^{-1}\Big({\frac{a}{d}}\Big)^{\frac{1}{2}}
+\sqrt{\frac{a}{d}}\ln (d\textit{l})^{-1}\Big).
\label{l1 q_p}
\end{align}
\end{lemma}

\section{Excitation Hamiltonians}\label{Excitation Hamiltonians}
\par In this section we lay our strategy of the renormalization illustrated in Figure \ref{layout}). We collect some important properties of renormalized excitation Hamiltonians in propositions in this section. We mainly state the results for Regions I or III. Regions $\mathrm{II}_{\mathrm{I}}$ and $\mathrm{II}_{\mathrm{III}}$ are regarded as intermediate regions, and the corresponding results still apply to these regions without further modifications and specifications. Propositons \ref{quadratic renorm} and \ref{cubic renorm} demonstrate the method of 3D quadratic and cubic renormalizations and describe $\mathcal{G}_N$ and $\mathcal{J}_N$ respectively. They will be proved in details in Sections \ref{2} and \ref{3} successively. Propositions \ref{quasi-2d/3d quadratic renorm} and \ref{dimensional coupling quadratic renorm} process the corresponding quasi-2D and dimensional coupling renormalizaitons and state the results of $\mathcal{M}_N$ and $\mathcal{S}_N$ respectively. We leave their proofs to Sections \ref{5} and \ref{6}. Propositions \ref{Bog renorm} and \ref{Bog renorm III} collect the result of the Bogoliubov transformations, for both Regions I and III, characterize the diagonalized Hamiltonians $\mathcal{Z}_N^I$ and $\mathcal{Z}_N^{III}$ respectively, and hence conclude Theorems \ref{core} and \ref{core III}. We prove them in Sections \ref{4} and \ref{7}.

\par Due to the observation that the the energy of $H_N$ on factorized state $(\varphi_0^{(d)})^{\otimes N}$ is always bigger than the true ground state energy of $H_N$. $(\varphi_0^{(d)})^{\otimes N}$ is not a good approximation to the ground state of $H_N$. The reason that causes such difference is the inter-particle correlation structure. In Region I in the Gross-Pitaevskii regime, the 3D correlation structure of the Hamilton operator is the main driving force that corrects the leading order ground state energy of $H_N$, and contributes to the second order. On the other hand, in Region III in the Gross-Pitaevskii regime, the correlation structure here is way more special and even contains stronger energy than the one in Region I. Here in the region that $d$ decays acutely fast, the main inter-particle correlation structure is not only determined by the 3D effect, but also a quasi-2D effect. 3D and quasi-2D correlation structures together correct the leading order energy to (\ref{leading order}). To compute the ground state energy up to second order, we moreover need to look into a dimensional coupling effect.

\par In order to unearth the energy information inside the formation of the correlation structure, the strategy of renormalization goes as follows. We start by conjugating the Hamilton operator $H_N$ with two unitary operator respectively, the 3D quadratic transformation $e^{B}$ and the 3D cubic transformation $e^{B^\prime}$ with
\begin{align}
  B&=\frac{1}{2}\sum_{p\neq0}\eta_p(a_p^*a_{-p}^*a_0a_0-h.c.),\label{define B}\\
  B^{\prime}&=\sum_{p,q,p+q\neq 0}\eta_p\chi_{\vert\mathcal{M}_dq\vert\leq\kappa}
  (a_{p+q}^*a_{-p}^*a_qa_0-h.c.).\label{define B'}
\end{align}
Here $\kappa$ is a cut-off parameter that may be chosen separately in different regions (even can be infinity in some cases). $\eta_p$ are defined through the 3D scattering equation with the Neumann boundary condition (\ref{asymptotic energy pde on the ball}). We write the excitation Hamilton operator as
\begin{equation}\label{G_N & J_N}
  \mathcal{G}_N=e^{-B}H_Ne^{B},\quad\mathcal{J}_N=e^{-B^\prime}\mathcal{G}_Ne^{B^\prime}.
\end{equation}
In Region I, the above renormalizations actually extract respectively the correct correlation structure hiding in $H_{23}$ and $H_3$ contributing to the first and second order terms of energy, while in Region III, the expectation of $\mathcal{J}_N$ on the factorized state $(\varphi_0^{(d)})^{\otimes N}$ is still of order $N^2\frac{a}{d}> N$, $\mathcal{J}_N$ may in fact become, up to some error terms, a modified Hamiltonian of the form of (\ref{Hamiltonian}) (or equivalently (\ref{split H_N})) whose interaction potential is replaced from $v_a$ to $2\sqrt{d}W$ (See (\ref{rearrange J_N}) below). By (\ref{define W(x)}) and (\ref{L^infty W}) we know $2\sqrt{d}W$ is indeed non-negative, radially symmetric and compactly supported. In particular, a modified non-zero momentum sum of potential operator $H^\prime_4$
\begin{equation}\label{define H'_4}
  H^\prime_4={\sum_{p,q,p+r,q+r\neq0}}W_ra_{p+r}^*a_q^*a_pa_{q+r}
\end{equation}
will emerge in the error estimates of $\mathcal{G}_N$ and $\mathcal{J}_N$ for Region III, such that some part of the higher order energy can be dominated by the modified non-zero momentum sum of potential energy. This kind of potential acts as a transition operator which can be controlled by $H_{21}$ and $H_4$. We will show in Section \ref{7} using the method in \cite[Lemma 2.5]{lieb2005mathematics} that we in fact have
\begin{equation}\label{control H_4' IV}
  H_4^\prime\leq CNH_{21}+CH_4.
\end{equation}
 We describe $\mathcal{G}_N$ and $\mathcal{J}_N$ in the next two propositions. We want to remind readers that the results below is relatively general since we do not require the Gross-Pitaevskii condition in following propositions. In each of Propositons \ref{quadratic renorm} and \ref{cubic renorm}, we will state two results, one of them will be used for the renormalization of Hamiltonian in Regions I and $\mathrm{II}_{\mathrm{I}}$ , the other for Regions III and $\mathrm{II}_{\mathrm{III}}$.
\begin{proposition}\label{quadratic renorm}
Let $v$ be a smooth 3D interaction potential given around (\ref{interaction potential}) (that is non-negative, radially symmetric and compactly supported) with scattering length $\mathfrak{a}_0$. Assume that $a,d$ and $\frac{a}{d}$ tend to $0$ and $N$ tends to infinity. Let $\textit{l}\in(0,\frac{1}{2})$ such that $\frac{d}{a}>\frac{C}{\textit{l}}$ for some universal constant $C$. We assume further that $Na\textit{l}^{\frac{1}{2}}$ tends to $0$ (which can be verified that it holds consistently in all three regions in the Gross-Pitaevskii limit). Then we have
\begin{flushleft}
  $\mathbf{For\; Region\; I}$
\end{flushleft}
\begin{equation}\label{first renorm}
    \mathcal{G}_N=C^B+Q^B\mathcal{N}_+
  +H_{21}+H_4+H_3+H_{23}^{\prime}+\mathcal{E}^B,
\end{equation}
where $C^B$ and $Q^B$ are constants given by
\begin{align}
C^B&=\frac{N(N-1)}{2\sqrt{d}}\Big(v_0^{(a,d)}+\sum_{p\neq0}v_p^{(a,d)}\eta_p\Big)
  +N(N-1)\sum_{p\neq0}W_p\eta_p,\label{first renorm C^B}\\
Q^B&= \frac{N}{\sqrt{d}}\Big(v_0^{(a,d)}-\sum_{p\neq0}v_p^{(a,d)}\eta_p\Big).
\label{first renorm Q^B}
\end{align}
Moreover the renormalized quadratic part $H_{23}^{\prime}$ is defined by
\begin{eqnarray}\label{first renorm H_23'}
  {H}_{23}^{\prime}={\sum_{p\neq0}}W_p(a_p^*a_{-p}^*a_0a_0+h.c.),
\end{eqnarray}
where $W_p$ are defined in (\ref{define W_p}). We also call $H_{23}^\prime$ the correlation remainder since its coefficients $W_p$ emerge from the remainder of (\ref{eqn of eta_p}). Finally the error term $\mathcal{E}^B$ satisfies the bound
\begin{align}
  \pm\mathcal{E}^B\leq & C\Big\{\big(Na^2d^{-2}\textit{l}^{-1}+
N^{2}a^{2}d^{-1}\textit{l}^{\frac{1}{2}}+
N^{\frac{3}{2}}a^{\frac{3}{2}}d^{-\frac{1}{2}}\textit{l}^{\frac{1}{2}}\big)
(\mathcal{N}_++1)\nonumber\\
  &+\big(ad^{-1}+Na^2d^{-2}\textit{l}^{-1}\big)(\mathcal{N}_++1)^2
  +Na^3d^{-1}H_{21}
  +N^{\frac{3}{2}}a^{\frac{3}{2}}d^{-\frac{1}{2}}\textit{l}^{\frac{1}{2}}H_4\Big\}.
\label{first renorm E^B}
\end{align}
\begin{flushleft}
  $\mathbf{For\; Region\; III}$
\end{flushleft}
\begin{align}
    \mathcal{G}_N=&\tilde{C}^B+\tilde{Q}_1^B\mathcal{N}_+
+\tilde{Q}_2^B\mathcal{N}_+(\mathcal{N}_++1)
  +H_{01}+H_{02}+H_{22}\nonumber\\
&+H_{21}+H_4+H_3+H_{23}^{\prime}+\tilde{\mathcal{E}}^B,\label{first renorm III}
\end{align}
where $\tilde{C}^B$, $\tilde{Q}_1^B$ and $\tilde{Q}_2^B$ are constants given by
\begin{align}
\tilde{C}^B&=\frac{N(N-1)}{2\sqrt{d}}\sum_{p\neq0}v_p^{(a,d)}\eta_p
  +N(N-1)\sum_{p\neq0}W_p\eta_p\label{first renorm C^B III}\\
\tilde{Q}^B_1&= -\frac{N}{\sqrt{d}}\sum_{p\neq0}v_p^{(a,d)}\eta_p
  -2N\sum_{p\neq0}W_p\eta_p\label{first renorm Q_1^B III}\\
\tilde{Q}^B_2&= \frac{1}{2\sqrt{d}}\sum_{p\neq0}v_p^{(a,d)}\eta_p
  +\sum_{p\neq0}W_p\eta_p\label{first renorm Q_2^B III}
\end{align}
and the error term $\tilde{\mathcal{E}}^B$ satisfies the bound
\begin{align}
  \pm\tilde{\mathcal{E}}^B\leq&
 C\big(N^{\frac{3}{2}}a^{\frac{3}{2}}d^{-{\frac{1}{2}}}\textit{l}^{\frac{1}{2}}+
N^2a^2d^{-1}\textit{l}^{\frac{1}{2}}\big)(\mathcal{N}_++1)\nonumber\\
&+CN^{\frac{3}{2}}a^{\frac{3}{2}}d^{-{\frac{1}{2}}}\textit{l}^{\frac{1}{2}}
\big(H_4+H_4^\prime\big),\label{first renorm E^B III}
\end{align}
and $H_{23}^\prime$ and $H_4^\prime$ are given in (\ref{first renorm H_23'}) and (\ref{define H'_4}) respectively.
\end{proposition}
\noindent\emph{Proof.} Postponed to Section \ref{2}.

\begin{proposition}\label{cubic renorm}
Under the same configuration of Proposition \ref{quadratic renorm} we have
\begin{flushleft}
  $\mathbf{For\; Region\; I}$
\end{flushleft}
We take $\kappa=\nu d^{-1}$ for some $\nu\geq 1$. Then for some $\alpha>0$ and $0<\gamma<1$ with the further assumptions that $Na^3\kappa^3\textit{l}$ and $N^{\frac{1}{2}}a^{\frac{1}{2}}d^{-\frac{1}{2}}\kappa^{-1}$ tend to 0.
\begin{equation}\label{second renorm}
  \mathcal{J}_N
=C^B+Q^{B^\prime}\mathcal{N}_+
+H_{21}+H_4+H_{23}^{\prime}+\mathcal{E}^{B^\prime},
\end{equation}
where $C^B$ has been defined in (\ref{first renorm C^B}) and $Q^{B^\prime}$ is given by
\begin{equation}\label{second renorm Q^B'}
  Q^{B^\prime}=Q^{B}+\frac{2N}{\sqrt{d}}\sum_{p\neq0}v_p^{(a,d)}\eta_p
=\frac{N}{\sqrt{d}}\Big(v_0^{(a,d)}+\sum_{p\neq0}v_p^{(a,d)}\eta_p\Big).
\end{equation}
Moreover, $H_{23}^{\prime}$ is introduced in (\ref{first renorm H_23'}) and the error term is bounded by
\begin{align}
\pm\mathcal{E}^{B^\prime}
&\leq CNad^{-1}
\Big\{d^\alpha+Na\textit{l}^{\frac{1}{2}}+\kappa^{-2}
+N^{\frac{3}{2}}a^{\frac{3}{2}}d^{-\frac{1}{2}}\textit{l}^{\frac{1}{2}}
+N^{\frac{1}{2}}a^{\frac{1}{2}}d^{-\frac{1}{2}}\kappa^{-1}\nonumber\\
&\quad\quad\quad\quad\quad+(Na^3\kappa^3\textit{l})^\gamma
+a\kappa\big[1+ad^{-1}\ln a^{-1}\big]\Big\}(\mathcal{N}_++1)\nonumber\\
&+CN^{\frac{3}{2}}a^{\frac{3}{2}}d^{-\frac{1}{2}}\textit{l}^{\frac{1}{2}}
(\mathcal{N}_++1)
+CN^{\frac{3}{2}}a^2d^{-2}\textit{l}^{-1}
(\mathcal{N}_++1)^{\frac{3}{2}}\nonumber\\
&+C\Big\{ad^{-1}+Na^2d^{-2}\textit{l}^{-1}
+Na^2d^{-\frac{1}{2}}\kappa^{\frac{3}{2}}\textit{l}^{\frac{1}{2}}
+N(Na^3\kappa^3\textit{l})^{1-\gamma}\nonumber\\
&\quad +ad^{-1}\big[1+ad^{-1}\ln a^{-1}\big]
+Na^2d^{-(2+\alpha)}\big[\textit{l}^{-1}+\ln(d\textit{l})^{-1}\big]\Big\}
(\mathcal{N}_++1)^2\nonumber\\
&+CN^{\frac{3}{2}}a^{\frac{3}{2}}d^{-\frac{1}{2}}\textit{l}^{\frac{1}{2}}
(Na^3\kappa^3\textit{l})^{1-\gamma}(\mathcal{N}_++1)^3
+CNa^3d^{-1}H_{21}\nonumber\\
&+C\big(N^{\frac{3}{2}}a^{\frac{3}{2}}d^{-\frac{1}{2}}\textit{l}^{\frac{1}{2}}
+(Na^3\kappa^3\textit{l})^\gamma\big)H_4\nonumber\\
&+C\big(d^\alpha+Na\textit{l}^{\frac{1}{2}}+\kappa^{-2}
+N^{\frac{1}{2}}a^{\frac{1}{2}}d^{-\frac{1}{2}}\kappa^{-1}\big)\big(H_{21}+H_4\big).
\label{second renorm E^B'}
\end{align}

\begin{flushleft}
  $\mathbf{For\; Region\; III}$
\end{flushleft}
We take $\kappa=\infty$. Assume further $N^{\frac{3}{2}}a^{\frac{7}{6}}d^{-\frac{1}{2}}\textit{l}^{\frac{1}{3}}\to 0$.
\begin{align}
  \mathcal{J}_N
&=\tilde{C}^B+\tilde{Q}^{B}_1\mathcal{N}_+
+\tilde{Q}^{B}_2\mathcal{N}_+(\mathcal{N}_++1)\nonumber\\
&+\frac{1}{\sqrt{d}}\sum_{p,q\neq0}(v^{(a,d)}_p+v^{(a,d)}_{p+q})\eta_p a_q^*a_qa_0^*a_0
+2\sum_{p,q\neq0}(W_p+W_{p+q})\eta_p a_q^*a_qa_0^*a_0\nonumber\\
&+H_{01}+H_{02}+H_{22}
+H_{21}+H_4+H_{23}^{\prime}+H_3^\prime+\tilde{\mathcal{E}}^{B^\prime},
\label{second renorm III}
\end{align}
where $\tilde{C}^B$, $\tilde{Q}_1^{B}$ and $\tilde{Q}_2^{B}$ have been defined in (\ref{first renorm C^B III}), (\ref{first renorm Q_1^B III}) and (\ref{first renorm Q_2^B III}) respectively. $H_3^\prime$ is defined by
\begin{equation}\label{second renorm H_3'}
  H_3^\prime=2\sum_{p,q,p+q\neq0}W_p(a_{p+q}^*a_{-p}^*a_qa_0+h.c.).
\end{equation}
Moreover, the error term is bounded by
\begin{align}
\pm\tilde{\mathcal{E}}^{B^\prime}
&\leq C\big(N^{\frac{3}{2}}a^{\frac{3}{2}}d^{-\frac{1}{2}}\textit{l}^{\frac{1}{2}}
+N^{\frac{5}{2}}a^{\frac{5}{2}}d^{-\frac{3}{2}}\textit{l}^{\frac{1}{2}}
+N^{\frac{1}{2}}a^{\frac{3}{2}}d^{\frac{1}{2}}\textit{l}^{2}
+N^{2}a^{\frac{5}{2}}d^{-1}\textit{l}^{\frac{1}{3}}\big)
(\mathcal{N}_++1)\nonumber\\
&+C\big(Na\textit{l}^{\frac{1}{2}}
+N^{\frac{3}{2}}a^{\frac{7}{6}}d^{-\frac{1}{2}}\textit{l}^{\frac{1}{3}}
+N^{2}a^{\frac{5}{3}}d^{-1}\textit{l}^{\frac{1}{3}}\big)
\big(H_{21}+Nad^{-1}(\mathcal{N}_++1)^2\big)\nonumber\\
&+C\big(Na\textit{l}^{\frac{1}{2}}
+N^{\frac{3}{2}}a^{\frac{7}{6}}d^{-\frac{1}{2}}\textit{l}^{\frac{1}{3}}\big)
H_4
+CN^{\frac{3}{2}}a^{\frac{3}{2}}d^{-\frac{1}{2}}\textit{l}^{\frac{1}{2}}
\big(H_4+H_4^\prime\big).\label{second renorm E^B' III}
\end{align}
\end{proposition}
\noindent\emph{Proof.} Postponed to Section \ref{3}.

\par In Region I, by 3D quadratic and cubic transformation resembling ones in \cite{2018Bogoliubov}, we reach the excitation Hamilton operator $\mathcal{J}_N$ whose expectation on the factorized state is $4\pi\mathfrak{a}_0N(N-1)\frac{a}{d}$, and hence provides the accurate leading order energy in Region I. Moreover, from (\ref{second renorm}) in Propositon \ref{cubic renorm}, the cubic term in $\mathcal{J}_N$ has been eliminated. This allows us to apply the generalized Bogoliubov transformation denoted by $e^{B^{\prime\prime}}$ with
 \begin{equation}\label{define B''}
  B^{\prime\prime}=B(\tau)=\frac{1}{2}\sum_{p\neq0}\tau_p(b_p^*b_{-p}^*-h.c.).
\end{equation}
We thereby write the diagonalized Hamilton operator
\begin{equation}\label{Z^I_N}
  \mathcal{Z}^{I}_N=e^{-B^{\prime\prime}}\mathcal{J}_Ne^{B^{\prime\prime}}.
\end{equation}
The modified creation and annihilation operators $b_p^*$ and $b_p$ for $p\neq0$ are introduced and analyzed in Section \ref{Modified Creation and Annihilation Operators}. The aim of the generalized Bogoliubov transformation is to diagonalize the modified quadratic term $H_{23}^\prime$ in $\mathcal{J}_N$, into an operator $\mathcal{D}$ of the form
\begin{equation*}
  \mathcal{D}=\sum_{p\neq0}\varepsilon_pa_p^*a_p.
\end{equation*}
All the eigenvalues and corresponding eigenfunctions of $\mathcal{D}$ can be explicitly computed due to its elegant diagonal pattern (See Section \ref{proof I} for more details). To explicitly define the operator $B^{\prime\prime}$ so that we can apply the generalized Bogoliubov transformation, we first need some preparations. From Proposition \ref{cubic renorm} and formula (\ref{有用的屎}), we can rewrite
\begin{align}
   Q^{B^\prime}\mathcal{N}_++H_{21}+H_{23}^{\prime}
&=\sum_{p\neq0}\left(F_pa_p^*a_p+\frac{1}{2}G_p(b_p^*b_{-p}^*+h.c.)\right)
+\mathcal{E}^{B^{\prime\prime}}_{res}\nonumber\\
&\eqqcolon\mathcal{Q}^{\prime}+\mathcal{E}^{B^{\prime\prime}}_{res}.
\label{rewrite quadratic}
\end{align}
where $F_p$ and $G_p$ for $p\neq0$ are given by
\begin{equation}\label{define F_p,G_p}
  F_p=\vert\mathcal{M}_dp\vert^2+8\pi\mathfrak{a}_0Nad^{-1},\quad G_p=2NW_p,
\end{equation}
and the error $\mathcal{E}^{B^{\prime\prime}}_{res}$ is bounded by
\begin{equation}\label{E^B''_res}
  \pm\mathcal{E}^{B^{\prime\prime}}_{res}\leq CNa^2d^{-2}\textit{l}^{-1}(\mathcal{N}_++1).
\end{equation}
We can then define coefficients $\tau_p$  by
\begin{equation}\label{define tau_p}
  \tau_p=\frac{1}{4}\ln{\frac{F_p-G_p}{F_p+G_p}}.
\end{equation}
The analysis of $\tau_p$ will be carried out in Section \ref{4}. We first point our that  $\tau_p=\tau_{-p}=\overline{\tau_p}$. Recalling the formula (\ref{action of Bog trans fock}), the action of the Bogoliubov transform $e^{B^{\prime\prime}}$ on a single modified creation or annihilation operator can be calculated explicitly by
\begin{equation}\label{action of Bog trans}
\begin{aligned}
  e^{-B^{\prime\prime}}b_p^*e^{B^{\prime\prime}}&=\cosh\tau_p b_p^*+\sinh\tau_p b_{-p}
  +d_p^*,\\
 e^{-B^{\prime\prime}}b_pe^{B^{\prime\prime}}&=\cosh\tau_p b_p+\sinh\tau_p b_{-p}^*+d_p.
 \end{aligned}
\end{equation}
This formula is the key to the diagonalization of the quadratic operator $\mathcal{Q}^\prime$. The next propostion states the result of Bogoliubov transformation for Region I. Theorem \ref{core} then follows by analyzing the eigenvalues of the diagonalized Hamiltonian $\mathcal{Z}_N^{I}$ using min-max principle. We leave the proof of Theorem \ref{core} to Section \ref{Proof of the Main Theorem for Region I}.
\begin{proposition}\label{Bog renorm}
Under the same configuration of Proposition \ref{cubic renorm} for Region I, assume further the Gross-Pitaevskii condition $Nad^{-1}=1$ for Region I. These assumptions on parameters can now be restated as $N^{-1}$, $a$, $d$ and $N^{-2}\nu^3\textit{l}$ tend to $0$ and $N>C\textit{l}^{-1}$ for $\textit{l}\in(0,\frac{1}{2})$ and $\nu> 1$. We have
\begin{equation}\label{third renorm}
  \mathcal{Z}_N^{I}
=C^{B^{\prime\prime}}+\mathcal{Q}^{\prime\prime}+e^{-B^{\prime\prime}}H_4e^{B^{\prime\prime}}
+\mathcal{E}^{B^{\prime\prime}},
\end{equation}
where
\begin{equation}\label{third renorm Q''}
C^{B^{\prime\prime}}=C^B+\frac{1}{2}\sum_{p\neq0}\left(-F_p+\sqrt{F_p^2-G_p^2}\right),\quad
\mathcal{Q}^{\prime\prime}=\sum_{p\neq0}\sqrt{F_p^2-G_p^2}a_p^*a_p,
\end{equation}
and the error term satisfies the bound
\begin{align}
\pm\mathcal{E}^{B^{\prime\prime}}
&\leq C\Big\{d^{\alpha}+d\big(\textit{l}^{\frac{1}{2}}+\nu^{-1}\big)
+N^{-1}\textit{l}^{-1}+(N^{-2}\nu^3\textit{l})^\gamma\nonumber\\
&\quad\quad\quad+N^{-1}\nu\big(1+N^{-1}\ln a^{-1}\big)\Big\}(\mathcal{N}_++1)
+CN^{-\frac{1}{2}}\textit{l}^{-1}(\mathcal{N}_++1)^{\frac{3}{2}}\nonumber\\
&+C\Big\{N^{-1}\textit{l}^{-1}
+(N^{-2}\nu^3\textit{l})^{\frac{1}{2}}+N(N^{-2}\nu^3\textit{l})^{1-\gamma}
+N^{-1}\big(1+N^{-1}\ln a^{-1}\big)\nonumber\\
&\quad\quad\quad+\big(N^{-1}d^{-\alpha}+N^{-2+\beta}\big)
\big(\textit{l}^{-1}+\ln(d\textit{l})^{-1}\big)\Big\}(\mathcal{N}_++1)^2\nonumber\\
&+CN^{-2+\beta}(\mathcal{N}_++1)^3
+C\Big(d^{\alpha}+d\big(\textit{l}^{\frac{1}{2}}+\nu^{-1}\big)+N^{-\beta}\Big)H_{21}\nonumber\\
&+CN^{-\frac{1}{2}}(\textit{l}^{-1}+\ln(d\textit{l})^{-1})^3(H_{21}+1)\nonumber\\
&+C\Big(d^{\alpha}+d\big(\textit{l}^{\frac{1}{2}}+\nu^{-1}\big)
+(N^{-2}\nu^3\textit{l})^\gamma\Big)
e^{-B^{\prime\prime}}H_4e^{B^{\prime\prime}}\nonumber\\
&+C\Big(d^{\alpha}+d\big(\textit{l}^{\frac{1}{2}}+\nu^{-1}\big)
+N^{-\beta}\Big)
\big(\textit{l}^{-1}+\ln(d\textit{l})^{-1}\big)
\label{third renorm E^B''}
\end{align}
for some $\alpha,\beta>0$ and $0<\gamma<1$.
\end{proposition}
\noindent\emph{Proof.} Postponed to Section \ref{4}.

\par On the other hand in Region III, we can rearrange $\mathcal{J}_N$ stated in (\ref{second renorm III}) in Proposition \ref{cubic renorm} using (\ref{est of eta_0}) to obtain the bound $ \vert v_p^{(a,d)}\eta_0\vert\leq Cd^{\frac{1}{2}}a^2\textit{l}^2$ and using (\ref{eqn of eta_p rewrt}) to merge similar terms
\begin{equation}\label{rearrange J_N}
  \mathcal{J}_N=H_{01}^\prime+H_{02}^{\prime}+H_{22}^\prime+H_{23}^\prime+H_3^\prime
+H_{21}+H_4+\tilde{\mathcal{E}}^{B^\prime}+O(N^2a^2\textit{l}^2),
\end{equation}
where
\begin{align*}
H_{01}^\prime=&\left(W_0+\sum_{p\neq0}W_p\eta_p\right)N(N-1),\quad
H_{02}^\prime=-\left(W_0+\sum_{p\neq0}W_p\eta_p\right)\mathcal{N}_+(\mathcal{N}_+-1),\\
H_{22}^\prime=&2(N-\mathcal{N}_+){\sum_{p\neq0}}
\left(W_p+\sum_{q\neq0}W_{p-q}\eta_q\right)a_p^*a_p,\\
H_{23}^\prime=&{\sum_{p\neq0}}W_p(a_{p}^*a_{-p}^*a_0a_0+h.c.),\quad\quad
H_3^\prime=2{\sum_{p,r,p+r\neq0}}W_r(a_{p+r}^*a_{-r}^*a_pa_0+h.c.)
\end{align*}
Comparing (\ref{rearrange J_N}) with (\ref{split H_N}), we now know $\mathcal{J}_N$ is in some sense a modified Hamiltonian with its interaction potential being subsituted with $2\sqrt{d}W$. Still $\mathcal{J}_N$ is not enough for regaining the correct $N$ leading order in Region III. To this end, we continue to conjugate $\mathcal{J}_N$ with two unitary operators respectively, the quasi-2D quadratic transformation $e^{\tilde{B}}$ and the quasi-2D cubic transformation $e^{\tilde{B}^\prime}$ with
\begin{align}
  \tilde{B}&=\frac{1}{2}\sum_{p\neq0}\xi_p(a_p^*a_{-p}^*a_0a_0-h.c.),\label{define D}\\
  \tilde{B}^{\prime}&=\sum_{p,q,p+q\neq 0}\xi_p(a_{p+q}^*a_{-p}^*a_qa_0-h.c.),\label{define D'}
\end{align}
and we let
\begin{equation}\label{L_N & M_N}
  \mathcal{L}_N=e^{-\tilde{B}}\mathcal{J}_Ne^{\tilde{B}},\quad \mathcal{M}_N=e^{-\tilde{B}^\prime}\mathcal{L}_Ne^{\tilde{B}^{\prime}}.
\end{equation}
$\xi_p$ have been defined through a 2D scattering equation with Neumann condition in Section \ref{Scattering Equations with Neumann condition}. The quadratic and cubic quasi-2D renormalizations extract respectively the correlation structure hiding in $H_{23}^\prime$ and $H_3^\prime$ contributing to the first and second order terms of energy. Through these operations, we effectively correct the energy to $4\pi N^2g$ predicted by (\ref{leading order}). The analysis of $\mathcal{L}_N$ and $\mathcal{M}_N$ resemble the analysis of $\mathcal{G}_N$ and $\mathcal{J}_N$, and we state the result in the next proposition.
\begin{proposition}\label{quasi-2d/3d quadratic renorm}
Under the same configuration of Proposition \ref{cubic renorm} for Region III, That is $N$ tends to infinity, $a$, $d$, $\frac{a}{d}$, $N^{\frac{3}{2}}a^{\frac{7}{6}}d^{-\frac{1}{2}}\textit{l}^{\frac{1}{3}}$ and
$Na\textit{l}^{\frac{1}{2}}$ tend to $0$ and $\frac{d\textit{l}}{a}>C$. Moreover, we demand additionally $\frac{h}{d\textit{l}}>C$, $\frac{Na}{d}>C$, $\frac{ma}{d}>C$ and $N(d\textit{l}+\frac{h}{m})$ and $N^{\frac{3}{2}}a^{\frac{1}{2}}d^{-\frac{1}{2}}(d\textit{l}+\frac{h}{m})^{\frac{2}{3}}$ should tend to $0$. We then have
\begin{equation}\label{fourth/fifth renorm}
  \mathcal{M}_N=N(N-1)\tilde{C}^{\tilde{B}^\prime}+2N\tilde{C}^{\tilde{B}^\prime}\mathcal{N}_+
-3\tilde{C}^{\tilde{B}^\prime}\mathcal{N}_+^2
+H_{21}+H_4+H_{23}^{\prime\prime}+H_3^{\prime\prime}+\tilde{\mathcal{E}}^{\tilde{B}^\prime}
\end{equation}
where
\begin{align}
\tilde{C}^{\tilde{B}^\prime}&=\Big(
W_0+\sum_{p\neq0}W_p\eta_p+\sum_{p\neq0}W_p\xi_p
+\sum_{p\neq0}\widetilde{W}_p\xi_p\Big)\label{fourth/fifth renorm Ctilde ^Btilde'}\\
 H_{23}^{\prime\prime}&=\sum_{p\neq0}\widetilde{W}_p(a_p^*a_{-p}^*a_0a_0+h.c.)
\label{fourth/fifth renorm H_23''}\\
H_{3}^{\prime\prime}&=2\sum_{p,q,p+q\neq0}\widetilde{W}_p(a_{p+q}^*a_{-p}^*a_qa_0+h.c.)
\label{fourth/fifth renorm H_3''}
\end{align}
with for all $p=(\bar{p},p_3)\in2\pi\mathbb{Z}^3$
\begin{equation}\label{fourth/fifth renorm define W tilde}
  \widetilde{W}_p=\left\{\begin{aligned}
&W_p+\frac{1}{2\sqrt{d}}\sum_{q\neq0}v_{p-q}^{(a,d)}\xi_q,\quad p_3\neq0\\
&\frac{\mu_h}{(d\textit{l})^2}\left(\xi_p+\widehat{\chi^{\mathrm{2D}}_{h}}\left(
  \frac{\bar{p}}{2\pi}\right)\right)+\sum_{q\neq0}
\frac{1}{2\sqrt{d}}v_{p-q}^{(a,d)}\xi_q-\sum_{q}W_{p-q}\xi_q,\quad p_3=0
\end{aligned}\right.
\end{equation}
and the error term is bounded by
\begin{align}
 \pm\tilde{\mathcal{E}}^{\tilde{B}^\prime}
\leq& C\Big\{ad^{-1}+N^4a^{\frac{8}{3}}d^{-2}\textit{l}^{\frac{1}{3}}
+N^3a^2d^{-2}(d\textit{l}+hm^{-1})
+N^{\frac{5}{2}}a^{\frac{3}{2}}d^{-\frac{3}{2}}(d\textit{l}+hm^{-1})^{\frac{2}{3}}\nonumber\\
&\quad\quad\quad+N^3m^{-1}(d\textit{l}+hm^{-1})
\Big(\ln\big(1+\frac{h}{d\textit{l}}\big)\Big)^{\frac{1}{2}}\Big\}
(\mathcal{N}_++1)\nonumber\\
&+C\Big\{N^2a^{\frac{5}{3}}d^{-1}\textit{l}^{\frac{1}{3}}
+N^{\frac{7}{2}}a^{\frac{3}{2}}d^{-\frac{3}{2}}(d\textit{l}+hm^{-1})^{\frac{5}{3}}
+N^{2}ad^{-1}(d\textit{l}+hm^{-1})^{\frac{2}{3}}\nonumber\\
&\quad\quad\quad+N^2m^{-1}(d\textit{l}+hm^{-1})
\Big(\ln\big(1+\frac{h}{d\textit{l}}\big)\Big)^{\frac{1}{2}}
+a^2d^{-2}h\ln(d\textit{l})^{-1}\Big\}\nonumber\\
&\times \Big(H_{21}+Nm^{-2}\ln\big(1+\frac{h}{d\textit{l}}\big)
(\mathcal{N}_++1)^2\Big)\nonumber\\
&+C\Big(N^{\frac{3}{2}}a^{\frac{7}{6}}d^{-\frac{1}{2}}\textit{l}^{\frac{1}{3}}
+N^2ad^{-1}(d\textit{l}+hm^{-1})
+N^{\frac{3}{2}}a^{\frac{1}{2}}d^{-\frac{1}{2}}(d\textit{l}+hm^{-1})^{\frac{2}{3}}\Big)
H_4\nonumber\\
&+CN^{\frac{3}{2}}a^{\frac{1}{2}}d^{-\frac{1}{2}}(d\textit{l}+hm^{-1})H_4^\prime.
\label{fourth/fifth renorm Etilde ^Btilde'}
\end{align}

\end{proposition}
\noindent\emph{Proof.} Postponed to Section \ref{5}.

\par The correct leading order $4\pi N^2g$ is now recovered in parts of $N(N-1)\tilde{C}^{\tilde{B}^\prime}$ using (\ref{useful shit 2d}) after a careful choice of parameter $\textit{l}$ and $h$. But the rest of $N(N-1)\tilde{C}^{\tilde{B}^\prime}$ are still of the order $N^2\frac{a}{d}$. Moreover, the cubic term $H_3^{\prime\prime}$ of $\mathcal{M}_N$ can not yet be eliminated like what we have done to $H_3$ in Region I. The reason is that in the region where $d$ decays acutely, simply viewing the system as 3D or 2D are neither a good approximation. Here we discover a 3D-to-2D dimensional coupling correlation structure which in addition contributes to the second order ground state energy. To reveal the energy contribution of this correlation structure, we conjugate $\mathcal{M}_N$ with another two unitary operators, the dimensional coupling quadratic transformation $e^{\mathcal{O}}$ and the dimensional coupling cubic transformation $e^{\mathcal{O}^\prime}$ with
\begin{align}
  \mathcal{O}&=\frac{1}{2}\sum_{p\neq0}k_p(a_p^*a_{-p}^*a_0a_0-h.c.),\label{define O}\\
  \mathcal{O}^{\prime}&=\sum_{p,q,p+q\neq 0}k_p(a_{p+q}^*a_{-p}^*a_qa_0-h.c.),\label{define O'}
\end{align}
and we let
\begin{equation}\label{R_N & S_N}
  \mathcal{R}_N=e^{-\mathcal{O}}\mathcal{M}_Ne^{\mathcal{O}},\quad \mathcal{S}_N=e^{-\mathcal{O}^\prime}\mathcal{R}_Ne^{\mathcal{O}^{\prime}}.
\end{equation}
Again, $k_p$ have been defined in Section \ref{Scattering Equations with Neumann condition} through a dimensional coupling scattering equation. We state the result of these two renormalizations in the next proposition.
\begin{proposition}\label{dimensional coupling quadratic renorm}
Under the same configuration of Proposition \ref{cubic renorm} for Region III, That is $N$ tends to infinity, $a$, $d$, $\frac{a}{d}$, $N^{\frac{3}{2}}a^{\frac{7}{6}}d^{-\frac{1}{2}}\textit{l}^{\frac{1}{3}}$ and
$Na\textit{l}^{\frac{1}{2}}$ tend to $0$ and $\frac{d\textit{l}}{a}>C$. Moreover, we demand additionally $\frac{h}{d\textit{l}}>C$, $\frac{Na}{d}>C$, $\frac{ma}{d}>C$ and $N(d\textit{l}+\frac{h}{m})$ and $N^{\frac{3}{2}}a^{\frac{1}{2}}d^{-\frac{1}{2}}(d\textit{l}+\frac{h}{m})^{\frac{2}{3}}$ should tend to $0$. We then have
\begin{equation}\label{sixth/seventh renorm}
  \mathcal{S}_N=N(N-1)\tilde{C}^{\mathcal{O}^\prime}
+2N\tilde{C}^{\mathcal{O}^\prime}\mathcal{N}_+
-3\tilde{C}^{\mathcal{O}^\prime}\mathcal{N}_+^2
+H_{21}+H_4+H_{23}^{\prime\prime\prime}
+\tilde{\mathcal{E}}^{\mathcal{O}^\prime}
\end{equation}
where
\begin{align}
\tilde{C}^{\mathcal{O}^\prime}&=
\Big(W_0+\sum_{p\neq0}W_p\eta_p+\sum_{p\neq0}\big(W_p+\widetilde{W}_p\big)\xi_p
+\sum_{p\neq0}\big(\widetilde{W}_p+q_p+Y_p\big)k_p\Big)
\label{sixth/seventh renorm Ctilde ^O'}\\
 H_{23}^{\prime\prime\prime}&=\sum_{p\neq0}\big(q_p+Y_p\big)(a_p^*a_{-p}^*a_0a_0+h.c.)
\label{sixth/seventh renorm H_23'''}
\end{align}
Here $q_p$ and $Y_p$ are defined in (\ref{define q}) and (\ref{define Y_p}) respectively. Moreover, the error term is bounded by
\begin{align}
  \pm \tilde{\mathcal{E}}^{\mathcal{O}^\prime}
\leq& C\Big\{ad^{-1}+N^4a^{\frac{8}{3}}d^{-2}\textit{l}^{\frac{1}{3}}
+N^{4}a^{2}d^{-2}(d\textit{l}+hm^{-1})^{\frac{2}{3}}
+N^2a^3d^{-3}h\ln(d\textit{l})^{-1}\nonumber\\
&\quad\quad\quad+N^4ad^{-1}m^{-1}(d\textit{l}+hm^{-1})
\Big(\ln\big(1+\frac{h}{d\textit{l}}\big)\Big)^{\frac{1}{2}}\Big\}
(\mathcal{N}_++1)\nonumber\\
&+CN^{\frac{3}{2}}a^{\frac{3}{2}}d^{-\frac{3}{2}}\textit{l}^{-\frac{1}{2}}m^{-1}
(\mathcal{N}_++1)^{\frac{3}{2}}\nonumber\\
&+CN\vartheta_1^{-1}\Big\{\frac{a\textit{l}\ln (d\textit{l})^{-1}}
{dm^2}+\frac{\ln h^{-1}}{m^2}
+\frac{a^{\frac{1}{3}}}{d^{\frac{1}{3}}m^{\frac{5}{3}}\textit{l}}\Big\}
(\mathcal{N}_++1)^2+C\vartheta_1 H_{21}\nonumber\\
&+C\Big\{N^2a^{\frac{5}{3}}d^{-1}\textit{l}^{\frac{1}{3}}
+N^{\frac{7}{2}}a^{\frac{3}{2}}d^{-\frac{3}{2}}(d\textit{l}+hm^{-1})^{\frac{5}{3}}
+N^{2}ad^{-1}(d\textit{l}+hm^{-1})^{\frac{2}{3}}\nonumber\\
&\quad\quad\quad+N^2m^{-1}(d\textit{l}+hm^{-1})
\Big(\ln\big(1+\frac{h}{d\textit{l}}\big)\Big)^{\frac{1}{2}}
+Nad^{-1}h\Big(1+\frac{a}{d}\ln a^{-1}\Big)\Big\}\nonumber\\
&\times \Big(H_{21}+Nad^{-1}(\mathcal{N}_++1)^2\Big)\nonumber\\
&+C\Big(N^{\frac{3}{2}}a^{\frac{7}{6}}d^{-\frac{1}{2}}\textit{l}^{\frac{1}{3}}
+N^2ad^{-1}(d\textit{l}+hm^{-1})
+N^{\frac{3}{2}}a^{\frac{1}{2}}d^{-\frac{1}{2}}(d\textit{l}+hm^{-1})^{\frac{2}{3}}\Big)
H_4\nonumber\\
&+CN^{\frac{3}{2}}a^{\frac{1}{2}}d^{-\frac{1}{2}}(d\textit{l}+hm^{-1})H_4^\prime.
\label{sixth/seventh renorm Etilde^O'}
\end{align}
for some $\vartheta_1>0$.
\end{proposition}
\noindent\emph{Proof.} Postponed to Section \ref{6}.

\par The first effect of dimensional coupling renormalization is that it compensates the remaining terms in $N(N-1)\tilde{C}^{\tilde{B}^\prime}$ such that they together truly become a second order ground state energy. The detailed analysis is given in Lemma \ref{lemma calculate C^B'''}. Moreover, the cubic term in $\mathcal{S}_N$ has been eliminated, which allows us to apply another generalized Bogoliubov transformation $e^{B^{\prime\prime\prime}}$ with
\begin{equation}\label{define e^B'''}
  B^{\prime\prime\prime}=B(\tilde{\tau})=\frac{1}{2}\sum_{p\neq0}
\tilde{\tau}_p(b_p^*b_{-p}^*-h.c.).
\end{equation}
 We write the diagonalized Hamilton operator
\begin{equation}\label{Z^III_N}
  \mathcal{Z}^{III}_N=e^{-B^{\prime\prime\prime}}\mathcal{S}_Ne^{B^{\prime\prime\prime}}.
\end{equation}
To define coefficients $\tilde{\tau}_p$, similar to (\ref{rewrite quadratic}), we use Proposition \ref{dimensional coupling quadratic renorm} to rewrite
\begin{align}
  {\mathcal{T}}^{\prime}\coloneqq 2N\tilde{C}^{\mathcal{O}^\prime}\mathcal{N}_++H_{21}
+H_{23}^{\prime\prime\prime}
&=\sum_{p\neq0}\left(\tilde{F}_pa_p^*a_p
+\frac{1}{2}\tilde{G}_p(b_p^*b_{-p}^*+h.c.)\right).
\label{rewrite quadratic III}
\end{align}
where $\tilde{F}_p$ and $\tilde{G}_p$ for $p\neq0$ are given by
\begin{equation}\label{define F_p,G_p III}
  \tilde{F}_p=\vert\mathcal{M}_dp\vert^2+2N\tilde{C}^{\mathcal{O}^\prime},\quad \tilde{G}_p=2N(q_p+Y_p),
\end{equation}
We can then define coefficients $\tilde{\tau}_p$  by
\begin{equation}\label{define tau_p III}
  \tilde{\tau}_p=\frac{1}{4}\ln{\frac{\tilde{F}_p-\tilde{G}_p}{\tilde{F}_p+\tilde{G}_p}}.
\end{equation}
Here we also have $\tilde{\tau}_p=\tilde{\tau}_{-p}=\overline{\tilde{\tau}_p}$. We leave the analysis of $\tilde{\tau}_p$ to Section \ref{7}. The next propostion state the result of Bogoliubov transform. We leave the proof of Theorem \ref{core III} to Section \ref{Proof of the Main Theorem for Region III}.
\begin{proposition}\label{Bog renorm III}
Under the same configuration of Proposition \ref{cubic renorm} for Region III, That is $N$ tends to infinity, $a$, $d$, $\frac{a}{d}$, $N^{\frac{3}{2}}a^{\frac{7}{6}}d^{-\frac{1}{2}}\textit{l}^{\frac{1}{3}}$ and
$Na\textit{l}^{\frac{1}{2}}$ tend to $0$ and $\frac{d\textit{l}}{a}>C$. Moreover, we demand additionally $\frac{h}{d\textit{l}}>C$, $\frac{Na}{d}>C$, $\frac{ma}{d}>C$ and $N(d\textit{l}+\frac{h}{m})$ and $N^{\frac{3}{2}}a^{\frac{1}{2}}d^{-\frac{1}{2}}(d\textit{l}+\frac{h}{m})^{\frac{2}{3}}$ should tend to $0$. Assume further that $C^{-1}\leq Nm^{-1}\leq C$ and $N\Big(\frac{h}{m}+\frac{a^2}{d^2\textit{l}}
+\Big(\frac{a}{d}+\frac{1}{h^2m}\Big)(d\textit{l})^2
+\frac{\textit{l}^{\frac{1}{2}}}{m}\sqrt{\frac{a}{d}}\Big)$ tends to $0$. We then have
\begin{equation}\label{eighth renorm Z^III_N}
  \mathcal{Z}^{III}_N
={C}^{B^{\prime\prime\prime}}+
{\mathcal{Q}}^{\prime\prime\prime}
+e^{-B^{\prime\prime\prime}}H_4e^{B^{\prime\prime\prime}}
+{\mathcal{E}}^{B^{\prime\prime\prime}},
\end{equation}
where
\begin{equation}\label{eighth renorm C, Q}
\begin{aligned}
 {C}^{B^{\prime\prime\prime}}&=N(N-1)\tilde{C}^{\mathcal{O}^\prime}
+\frac{1}{2}\sum_{p\neq0}\left(-\tilde{F}_p+\sqrt{\tilde{F}_p^2-\tilde{G}_p^2}\right),\\
{\mathcal{Q}}^{\prime\prime\prime}&=\sum_{p\neq0}\sqrt{\tilde{F}_p^2-\tilde{G}_p^2}a_p^*a_p.
\end{aligned}
\end{equation}
Here the error term is bound by
\begin{align*}
\pm {\mathcal{E}}^{B^{\prime\prime\prime}}
\leq& C\Big\{ad^{-1}+N^4a^{\frac{8}{3}}d^{-2}\textit{l}^{\frac{1}{3}}
+N^{4}a^{2}d^{-2}(d\textit{l}+hm^{-1})^{\frac{2}{3}}
+N^2a^3d^{-3}h\ln(d\textit{l})^{-1}\nonumber\\
&\quad\quad\quad+N^4ad^{-1}m^{-1}(d\textit{l}+hm^{-1})
\Big(\ln\big(1+\frac{h}{d\textit{l}}\big)\Big)^{\frac{1}{2}}+\vartheta_1^2\Big\}
(\mathcal{N}_++1)\nonumber\\
&+CN^{\frac{3}{2}}a^{\frac{3}{2}}d^{-\frac{3}{2}}\textit{l}^{-\frac{1}{2}}m^{-1}
(\mathcal{N}_++1)^{\frac{3}{2}}
+C\Big(\frac{1}{m}+\frac{a^2}{d^2\textit{l}}
+\frac{a}{d}(d\textit{l})^2\Big)(\mathcal{N}_++1)^2\nonumber\\
&+CN\vartheta_1^{-1}\Big\{\frac{a\textit{l}\ln (d\textit{l})^{-1}}
{dm^2}+\frac{\ln h^{-1}}{m^2}
+\frac{a^{\frac{1}{3}}}{d^{\frac{1}{3}}m^{\frac{5}{3}}\textit{l}}\Big\}
(\mathcal{N}_++1)^2\nonumber\\
&+C\vartheta_1 \Big\{H_{21}+N^2\Big(\frac{a\textit{l}
\ln (d\textit{l})^{-1}}
{dm^2}+\frac{\ln h^{-1}}{m^2}
+\frac{a^{\frac{1}{3}}}{d^{\frac{1}{3}}m^{\frac{5}{3}}\textit{l}}\Big)\Big\}\nonumber\\
&+C\Big\{N^2a^{\frac{5}{3}}d^{-1}\textit{l}^{\frac{1}{3}}
+N^{\frac{7}{2}}a^{\frac{3}{2}}d^{-\frac{3}{2}}(d\textit{l}+hm^{-1})
+N^{2}ad^{-1}(d\textit{l}+hm^{-1})^{\frac{2}{3}}\nonumber\\
&\quad\quad\quad+N^2m^{-1}(d\textit{l}+hm^{-1})
\Big(\ln\big(1+\frac{h}{d\textit{l}}\big)\Big)^{\frac{1}{2}}
+Nad^{-1}h\Big(1+\frac{a}{d}\ln a^{-1}\Big)\Big\}\nonumber\\
&\times \Big\{H_{21}+\frac{Na}{d}(\mathcal{N}_++1)^2+N^2\Big(\frac{a\textit{l}
\ln (d\textit{l})^{-1}}
{dm^2}+\frac{\ln h^{-1}}{m^2}
+\frac{a^{\frac{1}{3}}}{d^{\frac{1}{3}}m^{\frac{5}{3}}\textit{l}}\Big)\Big\}\nonumber\\
&+CN^{-\frac{1}{2}}\Big\{\frac{N^2a\textit{l}\ln (d\textit{l})^{-1}}
{dm^2}+\frac{N^2\ln h^{-1}}{m^2}
+\frac{N^2a^{\frac{1}{3}}}{d^{\frac{1}{3}}m^{\frac{5}{3}}\textit{l}}\Big\}^3(H_{21}+1)\nonumber\\
&+C\Big(N^{\frac{3}{2}}a^{\frac{7}{6}}d^{-\frac{1}{2}}\textit{l}^{\frac{1}{3}}
+N^2ad^{-1}(d\textit{l}+hm^{-1})
+N^{\frac{3}{2}}a^{\frac{1}{2}}d^{-\frac{1}{2}}(d\textit{l}+hm^{-1})^{\frac{2}{3}}\Big)
\nonumber\\
&\times e^{-B^{\prime\prime\prime}}H_4e^{B^{\prime\prime\prime}},
\end{align*}
For some $\vartheta_1>0$.
\end{proposition}
\noindent\emph{Proof.} Postponed to Section \ref{7}.

\par With all these preparations, we can prove the two main theorems, Theorems \ref{core} and \ref{core III} in this paper. We conclude Theorem \ref{core} in Section \ref{Proof of the Main Theorem for Region I}, and we complete the proof of Theorem \ref{core III} in Section \ref{Proof of the Main Theorem for Region III}.

\section{Proof of the Main Theorem for Region I}\label{Proof of the Main Theorem for Region I}
\par In this section, we demonstrate how to use Proposition \ref{Bog renorm} to conclude Theorem \ref{core}. We first calculate explicitly the constant $C^{B^{\prime\prime}}$ and the diagonalized operator $\mathcal{Q}^{\prime\prime}$ in Lemma \ref{calculate E_bog}. In particular, we analyze the exact order of the constant $C^{B^{\prime\prime}}$. In Propositon \ref{Optimal BEC}, we give an optimal Bose-Einstein condensation result for Region I using the method of localization, together with the help of Proposition \ref{Bog renorm}. Armed with this inequality, we can officially give the proof of Theorem \ref{core} in Section \ref{proof I}, by comparing the ground state energy of $H_N$ with a diagonalized operator $\mathcal{D}$ shown below. In this section, we mainly concern Region I.The proof for Region $\mathrm{II}_{\mathrm{I}}$ just needs slightly modifications on the proof for Region I and we will provide it in Section \ref{proof II1}.
\subsection{More about the Renormalized Hamiltonian for Region I}
\begin{lemma}\label{calculate E_bog}
Let $v$ be a smooth, radially-symmetric, compactly supported and non-negative function with scattering length $\mathfrak{a}_0$. Let $a,d$ and $\frac{a}{d}$ tend to $0$ with G-P condition i.e. $\frac{Na}{d}=1$. Also let $\textit{l}\in(0,\frac{1}{2})$ such that $\frac{d}{a}>\frac{C}{\textit{l}}$ for some large universal constant $C$.
\begin{enumerate}[$(1)$]
\item $C^{B^{\prime\prime}}$ defined in (\ref{third renorm Q''}) is given by
\begin{equation}\label{calculate C^B''}
\begin{aligned}
  C^{B^{\prime\prime}}=&4\pi(N-1)\mathfrak{a}_0+\mathfrak{e}_d+E_{Bog}^{(d)}\\
&+O(N^{-1}\textit{l}^{-2}+N^{-1}\textit{l}^{-1}\ln(d\textit{l})^{-1}
+d^2\textit{l}+(d\textit{l})^2\ln(d\textit{l})^{-1}),
\end{aligned}
\end{equation}
where
\begin{equation}\label{e_d}
  \mathfrak{e}_d=2\mathfrak{a}_0^2d^2
-\lim_{M\to\infty}\sum_{\substack{p\in\mathbb{Z}^3\backslash\{0\}\\
\vert p_1\vert,\vert p_2\vert,\vert p_3\vert\leq M}}\frac{4\mathfrak{a}_0^2\cos(d\vert\mathcal{M}_dp\vert)}{\vert\mathcal{M}_dp\vert^2},
\end{equation}
and
\begin{equation}\label{E_bod^d}
  E_{Bog}^{(d)}=\frac{1}{2}\sum_{p\in2\pi\mathbb{Z}^3\backslash\{0\}}
e_p^{(d)},
\end{equation}
where
\begin{equation}\label{e_bog^d}
  e_p^{(d)}=-\vert\mathcal{M}_dp\vert^2-8\pi\mathfrak{a}_0+
\sqrt{\vert\mathcal{M}_dp\vert^4+16\pi\mathfrak{a}_0\vert\mathcal{M}_dp\vert^2}
+\frac{(8\pi\mathfrak{a}_0)^2}{2\vert\mathcal{M}_dp\vert^2}.
\end{equation}
Moreover, $\mathfrak{e}_d$ and $E_{bog}^{(d)}$ are exactly of the order $\ln d^{-1}$ and $1$ respectively, and we can write explicitly
\begin{equation}\label{order of e_d}
  \mathfrak{e}_{d}=-8\pi\mathfrak{a}_0^2\ln d^{-1}+O(1).
\end{equation}
\item $\mathcal{Q}^{\prime\prime}$ defined in (\ref{third renorm Q''}) is given by
\begin{equation}\label{calculate Q''}
  \mathcal{Q}^{\prime\prime}=\mathcal{D}+\delta,
\end{equation}
where
\begin{equation}\label{D}
  \mathcal{D}=\sum_{p\in2\pi\mathbb{Z}^3\backslash\{0\}}
\sqrt{\vert\mathcal{M}_dp\vert^4+16\pi\mathfrak{a}_0\vert\mathcal{M}_dp\vert^2}
a_p^*a_p,
\end{equation}
and
\begin{equation}\label{delta}
  \pm\delta\leq C(N^{-1}\textit{l}^{-1}+(d\textit{l})^2)\mathcal{N}_+.
\end{equation}
\end{enumerate}
\end{lemma}
\noindent
\emph{Proof.} The analysis of $C^{B^{\prime\prime}}$ and $\mathcal{Q}^{\prime\prime}$ resemble the ones in \cite{2018Bogoliubov}, but the error estimates here is more subtle due to the extra $d$ dependence in the 3D to 2D problem. We first write $C^{B^{\prime\prime}}$ using the explicit expressions of $F_p$ and $G_p$ defined in (\ref{define F_p,G_p}). Notice we assume additionally that $Nad^{-1}=1$.
\begin{align*}
  C^{B^{\prime\prime}}&=C^B\\
&-\frac{1}{2}\sum_{p\neq0}
\left(\vert\mathcal{M}_dp\vert^2+8\pi\mathfrak{a}_0-
\sqrt{\vert\mathcal{M}_dp\vert^4+16\pi\mathfrak{a}_0\vert\mathcal{M}_dp\vert^2
+64\pi^2\mathfrak{a}_0^2-4N^2W_p^2}\right).
\end{align*}
$C^B$ is given in (\ref{first renorm C^B}). Using (\ref{有用的屎}) we can rewrite it as
\begin{align}\label{122901}
  C^B=4\pi(N-1)\mathfrak{a}_0\left(1+\frac{3}{2}\frac{\mathfrak{a}_0}{N\textit{l}}
\right)+N(N-1)\sum_{p\neq0}W_p\eta_p+O(N^{-1}\textit{l}^{-2}
+d^2\textit{l}^2).
\end{align}
Here we use (\ref{est of eta_0}) to bound $\vert\eta_0\vert$. To evaluate the second term on the right hand side of (\ref{122901}), we use the explicit expression (\ref{define W_p}) to get to
\begin{align}\label{122902}
 N(N-1)\sum_{p\neq0}W_p\eta_p=
N^2\left(\frac{\lambda_\textit{l}}{a^2d}\sum_{p\neq0}
\widehat{\chi_{d\textit{l}}}\left(\frac{\mathcal{M}_dp}{2\pi}\right)
\eta_p+\frac{\lambda_\textit{l}}{a^2}\sum_{p\neq0}\eta_p^2\right)-N\sum_{p\neq0}W_p\eta_p.
\end{align}
We then simplify (\ref{122902}). Using (\ref{sum_pW_peta_p 3dscatt}), we can first bound the last term of (\ref{122902}) by
\begin{align}\label{122903}
-N\sum_{p\neq0}W_p\eta_p=O(N^{-1}\textit{l}^{-1}).
\end{align}
Using (\ref{est of eta and eta_perp}), we can bound the second term of (\ref{122902}) by
\begin{align}\label{122904}
N^2\frac{\lambda_\textit{l}}{a^2}\sum_{p\neq0}\eta_p^2=O(N^{-1}\textit{l}^{-2}).
\end{align}
On the other hand, writing
\begin{equation}\label{define Z_p}
  Z_p=\frac{1}{2\sqrt{d}}\left(v_p^{(a,d)}+\sum_q
v_{p-q}^{(a,d)}\eta_q\right)
\end{equation}
and using (\ref{eqn of eta_p rewrt}), we can rewrite, for $p\neq0$
\begin{equation}\label{122905}
  \eta_p=\frac{1}{\vert\mathcal{M}_dp\vert^2}(W_p-Z_p).
\end{equation}
Plugging (\ref{122905}) into the first term of (\ref{122902}) we have
\begin{align}\label{122906}
  \frac{N^2\lambda_\textit{l}}{a^2d}\sum_{p\neq0}
\widehat{\chi_{d\textit{l}}}\left(\frac{\mathcal{M}_dp}{2\pi}\right)
\eta_p=\frac{N^2\lambda_\textit{l}}{a^2d}\sum_{p\neq0}
\frac{\widehat{\chi_{d\textit{l}}}\left(\frac{\mathcal{M}_dp}{2\pi}\right)}
{\vert\mathcal{M}_dp\vert^2}(W_p-Z_p).
\end{align}
Using again (\ref{define W_p}) together with (\ref{est of lambda_l}) and (\ref{sum W_p M_dp^-2}), we can rewrite the first term on the right hand side of (\ref{122906}) by
\begin{align}\label{123000}
    \frac{N^2\lambda_\textit{l}}{a^2d}\sum_{p\neq0}
\frac{\widehat{\chi_{d\textit{l}}}\left(\frac{\mathcal{M}_dp}{2\pi}\right)}
{\vert\mathcal{M}_dp\vert^2}W_p
=N^2\sum_{p\neq0}\frac{W_p^2}{\vert\mathcal{M}_dp\vert^2}
+O(N^{-1}\textit{l}^{-2}+N^{-1}\textit{l}^{-1}
\ln(d\textit{l})^{-1}).
\end{align}
For the second term on the right hand side of (\ref{122906}) we need some useful estimates. We first notice that
\begin{equation*}
 \sum_p Z_p\phi^{(d)}_p=\frac{1}{2\sqrt{d}}v_a\widetilde{f}_\textit{l},
\end{equation*}
which immediately tells us that for all $p\in2\pi\mathbb{Z}^3$
\begin{equation}\label{est of Z_p}
  \vert Z_p\vert\leq CN^{-1}.
\end{equation}
Using (\ref{有用的屎}), we know that
\begin{equation}\label{est of Z_0-4pia_0N^-1}
  \vert Z_0-4\pi\mathfrak{a}_0N^{-1}\vert\leq CN^{-2}\textit{l}^{-1}.
\end{equation}
Since $v_a$ and $\widetilde{f}_\textit{l}$ are radially symmetric, using Taylor's formula we can bound for all $p\in2\pi\mathbb{Z}^3$
\begin{equation}\label{est of Z_p-Z_0}
  \vert Z_p-Z_0\vert\leq Ca^3d^{-1}\vert\mathcal{M}_dp\vert^2.
\end{equation}
An argument similar to the proof of (\ref{sum W_p M_dp^-2}) derives another useful estimate
\begin{equation}\label{sum chi_dl M_dp^-2}
  \sum_{p\neq0}
\frac{\vert\widehat{\chi_{d\textit{l}}}\left(\frac{\mathcal{M}_dp}{2\pi}\right)\vert}
{\vert\mathcal{M}_dp\vert^2}\leq C(d\textit{l})^3(\textit{l}^{-1}
+\ln(d\textit{l})^{-1}).
\end{equation}
Rewriting
\begin{equation}\label{123001}
  Z_p=4\pi\mathfrak{a}_0N^{-1}+(Z_0-4\pi\mathfrak{a}_0N^{-1})
+(Z_p-Z_0).
\end{equation}
Plugging (\ref{123001}) into the second term on the right hand side of (\ref{122906}), we then evaluate it by (\ref{123002}), (\ref{123003}) and (\ref{123004}) three parts. First we combine (\ref{est of lambda_l}) and (\ref{sum chi_dl M_dp^-2}) to get to
\begin{align}
  -\frac{N^2\lambda_\textit{l}}{a^2d}\sum_{p\neq0}
\frac{\widehat{\chi_{d\textit{l}}}\left(\frac{\mathcal{M}_dp}{2\pi}\right)}
{\vert\mathcal{M}_dp\vert^2}4\pi\mathfrak{a}_0N^{-1}
&=-\frac{12\pi\mathfrak{a}_0^2}{(d\textit{l})^3}\sum_{p\neq0}
\frac{\widehat{\chi_{d\textit{l}}}\left(\frac{\mathcal{M}_dp}{2\pi}\right)}
{\vert\mathcal{M}_dp\vert^2}\nonumber\\
&+O(N^{-1}\textit{l}^{-2}+N^{-1}\textit{l}^{-1}
\ln(d\textit{l})^{-1}).\label{123002}
\end{align}
Secondly, (\ref{est of lambda_l}), (\ref{est of Z_0-4pia_0N^-1}) and (\ref{sum chi_dl M_dp^-2}) together yield
 \begin{align}\label{123003}
   -\frac{N^2\lambda_\textit{l}}{a^2d}\sum_{p\neq0}
\frac{\widehat{\chi_{d\textit{l}}}\left(\frac{\mathcal{M}_dp}{2\pi}\right)}
{\vert\mathcal{M}_dp\vert^2}(Z_0-4\pi\mathfrak{a}_0N^{-1})
=O(N^{-1}\textit{l}^{-2}+N^{-1}\textit{l}^{-1}
\ln(d\textit{l})^{-1}).
 \end{align}
Moreover, splitting high and low momenta at $\epsilon d^{-1}$ for some $\epsilon>1$ to be determined, we combine (\ref{est of lambda_l}), (\ref{moron2}), (\ref{很晦气2}) and (\ref{est of Z_p}) to reach
\begin{align*}
  \left\vert\frac{N^2\lambda_\textit{l}}{a^2d}
\sum_{\vert\mathcal{M}_dp\vert\geq\epsilon d^{-1}}
\frac{\widehat{\chi_{d\textit{l}}}\left(\frac{\mathcal{M}_dp}{2\pi}\right)}
{\vert\mathcal{M}_dp\vert^2}(Z_p-Z_0)\right\vert
\leq C\epsilon^{-1}\textit{l}^{-2},
\end{align*}
and (\ref{est of lambda_l}), (\ref{moron2}), (\ref{很晦气1}) and (\ref{est of Z_p-Z_0}) to get to
\begin{align*}
  \left\vert\frac{N^2\lambda_\textit{l}}{a^2d}
\sum_{0<\vert\mathcal{M}_dp\vert<\epsilon d^{-1}}
\frac{\widehat{\chi_{d\textit{l}}}\left(\frac{\mathcal{M}_dp}{2\pi}\right)}
{\vert\mathcal{M}_dp\vert^2}(Z_p-Z_0)\right\vert
\leq C\frac{a^2}{(d\textit{l})^2}(\epsilon+\ln(\epsilon d^{-1})).
\end{align*}
Taking $\epsilon=N$, we can bound
\begin{align}\label{123004}
  -\frac{N^2\lambda_\textit{l}}{a^2d}
\sum_{p\neq0}
\frac{\widehat{\chi_{d\textit{l}}}\left(\frac{\mathcal{M}_dp}{2\pi}\right)}
{\vert\mathcal{M}_dp\vert^2}(Z_p-Z_0)
=O(N^{-1}\textit{l}^{-2}+N^{-2}\textit{l}^{-2}\ln(N d^{-1})).
\end{align}
Combining (\ref{122901}), (\ref{122902}), (\ref{122903}), (\ref{122904}), (\ref{122906}), (\ref{123000}), (\ref{123001}), (\ref{123002}), (\ref{123003}) and (\ref{123004}) we can write
\begin{align}\label{123005}
C^{B^{\prime\prime}}=&4\pi(N-1)\mathfrak{a}_0\left(1+\frac{3}{2}
\frac{\mathfrak{a}_0}{N\textit{l}}\right)
-\frac{12\pi\mathfrak{a}_0^2}{(d\textit{l})^3}\sum_{p\neq0}
\frac{\widehat{\chi_{d\textit{l}}}\left(\frac{\mathcal{M}_dp}{2\pi}\right)}
{\vert\mathcal{M}_dp\vert^2}+E_{Bog}^{(N,d)}\nonumber\\
&+O(N^{-1}\textit{l}^{-2}+d^2\textit{l}^2+N^{-1}\textit{l}^{-1}
\ln(d\textit{l})^{-1}),
\end{align}
where
\begin{equation}\label{define E_bog^(N,d)}
  E_{Bog}^{(N,d)}=\frac{1}{2}\sum_{p\neq0}e_p^{(N,d)},
\end{equation}
and
\begin{align}
e_p^{(N,d)}=&-
\vert\mathcal{M}_dp\vert^2-8\pi\mathfrak{a}_0+
\sqrt{\vert\mathcal{M}_dp\vert^4+16\pi\mathfrak{a}_0\vert\mathcal{M}_dp\vert^2
+64\pi^2\mathfrak{a}_0^2-4N^2W_p^2}\nonumber\\
&+\frac{2N^2 W_p^2}{\vert\mathcal{M}_dp\vert^2}.\label{define e_bog^(N,d)}
\end{align}

\par To conclude (\ref{calculate C^B''}), we need further simplification of $E_{Bog}^{(N,d)}$. We first estimate $e_p^{(d)}$ and $e_p^{(N,d)}$ given respectively in (\ref{e_bog^d}) and (\ref{define e_bog^(N,d)}) by (\ref{est of e_bog}) and (\ref{est of difference e_bog}). On the one hand, rationalizing the numerator, and using the point-wise bound of $W_p$ (\ref{sum_pW_peta_p 3dscatt}), we can bound for all $p\neq0$ that
\begin{equation}\label{est of e_bog}
  \vert e_p^{(d)}\vert,\vert e_p^{(N,d)}\vert\leq \frac{C}
{\vert\mathcal{M}_dp\vert^4}.
\end{equation}
On the other hand, since $W$ is also radially symmetric, combining (\ref{est of lambda_l}), (\ref{est of int w_l}) and (\ref{define W_p}), we reach an estimate similar to (\ref{est of Z_0-4pia_0N^-1}) and (\ref{est of Z_p-Z_0}) for all $p\in2\pi\mathbb{Z}^3$
\begin{equation}\label{imbecile}
  \vert2NW_p-8\pi\mathfrak{a}_0\vert\leq CN^{-1}\textit{l}^{-1}+C(d\textit{l})^2
\vert\mathcal{M}_dp\vert^2.
\end{equation}
This estimate allows us to rationalize the numerator again to reach
\begin{equation}\label{est of difference e_bog}
  \vert e_p^{(d)}- e_p^{(N,d)}\vert\leq \frac{C}
{N\textit{l}\vert\mathcal{M}_dp\vert^4}+\frac{C(d\textit{l})^2}
{\vert\mathcal{M}_dp\vert^2}.
\end{equation}
Splitting high and low momenta at $\epsilon d^{-1}$ for some $\epsilon>1$ to be determined, we combine (\ref{很晦气2}) and (\ref{est of e_bog}) to get
\begin{align*}
  \frac{1}{2}\left\vert\sum_{\vert\mathcal{M}_dp\vert\geq\epsilon d^{-1}}
(e_p^{(d)}- e_p^{(N,d)})\right\vert\leq C\epsilon^{-1}d^2,
\end{align*}
and combine (\ref{很晦气1}), (\ref{WTH2}) and (\ref{est of difference e_bog}) to get
\begin{align*}
  \frac{1}{2}\left\vert\sum_{0<\vert\mathcal{M}_dp\vert<\epsilon d^{-1}}
(e_p^{(d)}- e_p^{(N,d)})\right\vert\leq CN^{-1}\textit{l}^{-1}
+C(d\textit{l})^2(\epsilon+\ln(\epsilon d^{-1})).
\end{align*}
Taking $\epsilon=\textit{l}^{-1}$, we have
\begin{equation}\label{123006}
  E^{(N,d)}_{Bog}-E^{(d)}_{Bog}=O(N^{-1}\textit{l}^{-1}
+d^2\textit{l}+(d\textit{l})^2\ln(d\textit{l})^{-1}).
\end{equation}
To conclude our proof of (\ref{calculate C^B''}), we move one step further. We denote
\begin{equation}\label{define e_d l}
  \mathfrak{e}_{d,\textit{l}}=6\pi\mathfrak{a}_0^2\textit{l}^{-1}
-\frac{12\pi\mathfrak{a}_0^2}{(d\textit{l})^3}\sum_{p\neq0}
\frac{\widehat{\chi_{d\textit{l}}}\left(\frac{\mathcal{M}_dp}{2\pi}\right)}
{\vert\mathcal{M}_dp\vert^2}.
\end{equation}
 Using the explicit expresssion
\begin{align}
\widehat{(\chi_{d\textit{l}}(\mathbf{x})\vert \mathbf{x}\vert^2)}\left(
\frac{\mathcal{M}_dp}{2\pi}\right)
&=\int_{\mathbb{R}^3}\chi_{d\textit{l}}(\mathbf{x})\vert \mathbf{x}\vert^2
e^{-ip^T\mathcal{M}_d\cdot \mathbf{x}}d\mathbf{x}\nonumber\\
&=4\pi(d\textit{l})^5\left(
-\frac{6\sin\theta}{\theta^5}+\frac{6\cos\theta}{\theta^4}
+\frac{3\sin\theta}{\theta^3}-\frac{\cos\theta}{\theta^2}\right),
\label{fourier chi_dl x^2}
\end{align}
with $\theta=d\textit{l}\vert\mathcal{M}_dp\vert$. (\ref{fourier chi_dl x^2}) together with (\ref{fourier chi_dl}) allow us to argue like \cite[(5.30)]{2018Bogoliubov}:
\begin{align*}
\sum_{p\neq0}
\frac{\widehat{\chi_{d\textit{l}}}\left(\frac{\mathcal{M}_dp}{2\pi}\right)}
{\vert\mathcal{M}_dp\vert^2}
&=4\pi(d\textit{l})^5\sum_{p\neq0}\left(\frac{\sin\theta}{\theta^5}
-\frac{\cos\theta}{\theta^4}\right)\\
&=4\pi(d\textit{l})^5\sum_{p\neq0}\Bigg\{-\frac{1}{6}
\left(-\frac{6\sin\theta}{\theta^5}+\frac{6\cos\theta}{\theta^4}
+\frac{3\sin\theta}{\theta^3}-\frac{\cos\theta}{\theta^2}\right)\\
&\quad\quad\quad
+\frac{1}{2}\left(\frac{\sin\theta}{\theta^3}-\frac{\cos\theta}{\theta^2}\right)
+\frac{1}{3}\frac{\cos\theta}{\theta^2}\Bigg\}\\
&=\lim_{M\to\infty}
\sum_{\substack{p\in2\pi\mathbb{Z}^3\backslash\{0\}\\ \vert p_i\vert\leq2\pi M}}
\Bigg\{-\frac{1}{6}
\widehat{(\chi_{d\textit{l}}(\mathbf{x})\vert \mathbf{x}\vert^2)}\left(
\frac{\mathcal{M}_dp}{2\pi}\right)+\frac{1}{2}
\widehat{\chi_{d\textit{l}}}\left(\frac{\mathcal{M}_dp}{2\pi}\right)\\
&\quad\quad\quad\quad\quad\quad
+\frac{4\pi(d\textit{l})^5}{3}\frac{\cos\theta}{\theta^2}\Bigg\}\\
&=-\frac{8\pi(d\textit{l})^5}{15}+\frac{d(d\textit{l})^2}{2}+
\frac{4\pi(d\textit{l})^3}{3}\lim_{M\to\infty}
\sum_{\substack{p\in2\pi\mathbb{Z}^3\backslash\{0\}\\ \vert p_i\vert\leq2\pi M}}
\frac{\cos\big(d\textit{l}\vert\mathcal{M}_dp\vert\big)}{\vert\mathcal{M}_dp\vert^2}.
\end{align*}

 Therefore we can rewrite
\begin{align}\label{rewrite e_d,l}
\mathfrak{e}_{d,\textit{l}}=I_{d,\textit{l}}+O((d\textit{l})^2),
\end{align}
with
\begin{align}\label{define I_d,l}
  I_{d,\textit{l}}=8\pi^2\mathfrak{a}_0^2(d\textit{l})^2
-\lim_{M\to\infty}16\pi^2\mathfrak{a}_0^2
\sum_{\substack{p\in2\pi\mathbb{Z}^3\backslash\{0\}\\ \vert p_i\vert\leq2\pi M}}
\frac{\cos\left(d\textit{l}\vert\mathcal{M}_dp\vert\right)}
{\vert\mathcal{M}_dp\vert^2}
\end{align}
Using
\begin{equation}\label{fourier chi_dl/x}
  \widehat{\left(\frac{\chi_{d\textit{l}}(\mathbf{x})}
{\vert\mathbf{x}\vert}\right)}\left(\frac{\mathcal{M}_dp}{2\pi}\right)
=4\pi(d\textit{l})^2\left(\frac{1}{\theta^2}-
\frac{\cos\theta}{\theta^2}\right),
\end{equation}
we can argue like \cite[(5.33)]{2018Bogoliubov} to conclude that $I_{d,\textit{l}}$ is independent of $\textit{l}\in(0,\frac{1}{2})$. To be precise, we let
\begin{equation*}
  h(\mathbf{x})=\frac{\chi_{d\textit{l}_1}(\mathbf{x})-\chi_{d\textit{l}_2}(\mathbf{x})}
  {\vert \mathbf{x}\vert}
\end{equation*}
for some $\textit{l}_1,\textit{l}_2\in(0,\frac{1}{2})$, and
\begin{equation*}
  h_p=\int_{\Lambda_d}h(\mathbf{x})\frac{1}{\sqrt{d}}e^{-ip^T\mathcal{M}_d\cdot \mathbf{x}}d\mathbf{x}
\end{equation*}
for $p\in2\pi\mathbb{Z}^3$. We can calculate directly using (\ref{fourier chi_dl/x})
\begin{align*}
 I_{d,\textit{l}_1}-I_{d,\textit{l}_2}
 &=8\pi^2\mathfrak{a}_0^2d^2\big(\textit{l}_1^2-\textit{l}_2^2\big)
 +4\pi\mathfrak{a}_0^2\lim_{M\to\infty}
\sum_{\substack{p\in2\pi\mathbb{Z}^3\backslash\{0\}\\ \vert p_i\vert\leq2\pi M}}
\sqrt{d}h_p\\
&=4\pi^2\mathfrak{a}_0^2\Big(2d^2\big(\textit{l}_1^2-\textit{l}_2^2\big)
-\sqrt{d}h_0+dh(0)\Big)=0.
\end{align*}
Therefore we have
\begin{equation}\label{123007}
  \mathfrak{e}_{d,\textit{l}}=I_{d,(2\pi)^{-1}}+O((d\textit{l})^2)
=\mathfrak{e}_{d}+O((d\textit{l})^2).
\end{equation}
Combining (\ref{123005}), (\ref{123006}), (\ref{define e_d l}), (\ref{rewrite e_d,l}) and (\ref{123007}) we conclude (\ref{calculate C^B''}).
\par To show that $\mathfrak{e}_{d}$ is exactly of order $\ln d^{-1}$, we only need to show that for fixed $\textit{l}$, $\mathfrak{e}_{d,\textit{l}}$ is exactly of order $\ln d^{-1}$. Using (\ref{WTH1}) and (\ref{WTH2}) (taking $\epsilon=\textit{l}^{-1}$ for some fixed $\textit{l}$ and calculating explicitly the constants $C$ in these inequalities), and the fact that
\begin{equation*}
  \left\vert\widehat{\chi_{d\textit{l}}}\left(\frac{\mathcal{M}_dp}{2\pi}\right)
\right\vert\leq \frac{4\pi}{3}(d\textit{l})^3,
\quad \left\vert\widehat{\chi_{d\textit{l}}}\left(\frac{\mathcal{M}_dp}{2\pi}\right)
\right\vert\leq \frac{C(d\textit{l})}{\vert\mathcal{M}_dp\vert^2},
\end{equation*}
we immediately deduce for some fixed $\textit{l}$
\begin{equation*}
  \vert\mathfrak{e}_{d,\textit{l}}\vert\leq8\pi\mathfrak{a}_0^2\ln d^{-1}+O(1),
\end{equation*}
and particularly
\begin{align*}
 \mathfrak{e}_{d,\textit{l}}+\frac{12\pi\mathfrak{a}_0^2}{(d\textit{l})^3}
\sum_{\substack{0<\vert\mathcal{M}_dp\vert<(d\textit{l})^{-1}\\
p_3=0,p_1p_2\neq0}}
\frac{\widehat{\chi_{d\textit{l}}}\left(\frac{\mathcal{M}_dp}{2\pi}\right)}
{\vert\mathcal{M}_dp\vert^2}=O(1).
\end{align*}
Using (\ref{fourier chi_dl}) and Taylor's expansion, we have
\begin{equation*}
  \widehat{\chi_{d\textit{l}}}\left(\frac{\mathcal{M}_dp}{2\pi}\right)
=\frac{4\pi}{3}(d\textit{l})^3+O((d\textit{l})^5\vert\mathcal{M}_dp\vert^2).
\end{equation*}
Hence
\begin{equation*}
  \frac{12\pi\mathfrak{a}_0^2}{(d\textit{l})^3}
\sum_{\substack{0<\vert\mathcal{M}_dp\vert<(d\textit{l})^{-1}\\
p_3=0,p_1p_2\neq0}}
\frac{\widehat{\chi_{d\textit{l}}}\left(\frac{\mathcal{M}_dp}{2\pi}\right)}
{\vert\mathcal{M}_dp\vert^2}=4\mathfrak{a}_0^2
\sum_{\substack{\bar{p}\in\mathbb{Z}^2,p_1p_2\neq0\\
0<\vert\bar{p}\vert< (d\textit{l})^{-1}}}\frac{1}{\vert\bar{p}\vert^2}+O(1).
\end{equation*}
With the fact that
\begin{equation*}
  \sum_{\substack{\bar{p}\in\mathbb{Z}^2,p_1p_2\neq0\\
0<\vert\bar{p}\vert< (d\textit{l})^{-1}}}\frac{1}{\vert\bar{p}\vert^2}
=2\pi\ln(d\textit{l})^{-1}+O(1).
\end{equation*}
We have for some fixed $\textit{l}$
\begin{equation*}
  \mathfrak{e}_{d,\textit{l}}=-8\pi\mathfrak{a}_0^2\ln d^{-1}+O(1).
\end{equation*}
As for $E_{Bog}^{(d)}$, it is obivious that $E_{Bog}^{(d)}\geq C$ since we know that $e^{(d)}_p>0$ for all $p\neq0$ by the rationalizing the numerator . On the other hand,  $E_{Bog}^{(d)}\leq C$ is deduced by combining (\ref{WTH2}) and (\ref{est of e_bog}) (taking $\epsilon=d$). We have finished the proof of the first statement of Lemma \ref{calculate E_bog}.

\par For the proof of (\ref{calculate Q''}), we use again (\ref{imbecile}) and rationalize the numerator to reach
\begin{align*}
  &\left\vert\sqrt{\vert\mathcal{M}_dp\vert^4+16\pi\mathfrak{a}_0\vert\mathcal{M}_dp\vert^2
+64\pi^2\mathfrak{a}_0^2-4N^2W_p^2}
-\sqrt{\vert\mathcal{M}_dp\vert^4+16\pi\mathfrak{a}_0\vert\mathcal{M}_dp\vert^2}\right\vert\\
&\leq \frac{CN\textit{l}}{\vert\mathcal{M}_dp\vert^2}+C(d\textit{l})^2,
\end{align*}
which leads directly to (\ref{calculate Q''}) and (\ref{delta}).
\begin{flushright}
  $\Box$
\end{flushright}

\subsection{Optimal BEC for Region I}\label{subsection Optimal BEC}
\
\par Before we conclude an optimal BEC result, we summarize a Bose-Einstein condensatian result from 3D to 2D in the Gross-Pitaevskii limit proved in \cite[Theorem 1.3]{2005Bosons}. This result will be applied in the proof of our two optimal BEC propositons Proposition \ref{Optimal BEC} in this section and Proposition \ref{Optimal BEC III} in Section \ref{subsection Optimal BEC for Region III}. We can calculate carefully following the idea in it to summarize in the following lemma.
\begin{lemma}[Schnee, Yngvason (2005)]\label{lemma2005}
 For any approximate ground state $\psi_{N,a,d}\in L^2_s(\Lambda_d^N)$ satisfying $\Vert\psi_{N,a,d}\Vert_2=1$ and
\begin{equation}\label{approximate condition}
  \lim_{\text{G-P limit}}\frac{1}{N}\langle H_{N,a,d}\psi_{N,a,d},\psi_{N,a,d}\rangle
=4\pi\mathfrak{a}_0
\end{equation}
Notice from \cite[Theorem 1.1]{2005Bosons} i.e. (\ref{leading order}), the fundamental theorem of the first order ground state energy from 3D to 2D, such family of approximate groud states exists. Then the system of bose gas may exhibit Bose-Einstein condensation phenomenon. Mathematically speaking, there holds
\begin{equation}\label{BEC phenomenon}
 \frac{1}{N} \langle a_0^*a_0\psi_{N,a,d},\psi_{N,a,d}\rangle
=\frac{1}{Nd}\int  \gamma_{N,a,d}(\mathbf{x}_1,\mathbf{x}_1^\prime)
d\mathbf{x}_1d\mathbf{x}_1^\prime\rightarrow 1.
\end{equation}
Here $a_0^*$ and $a_0$ are referred to creation and annihilation operator repectively and will be demonstrated explicitly in Section \ref{Fock Space Formalism}. The one-particle reduced density matrix $\gamma_{N,a,d}$ is defined as
\begin{equation}\label{define density matrix}
  \gamma_{N,a,d}(\mathbf{x}_1,\mathbf{x}_1^\prime)
=N\int_{\Lambda_d^{N-1}}\psi_{N,a,d}(\mathbf{x}_1,\mathbf{x}_2,\dots,\mathbf{x}_N)
\overline{\psi_{N,a,d}}(\mathbf{x}_1^\prime,\mathbf{x}_2,\dots,\mathbf{x}_N)
 d\mathbf{x}_2\dots d\mathbf{x}_N.
\end{equation}
Furthermore, we can estimate the rate of decay as
\begin{equation}\label{rate of decay}
  0\leq 1-\frac{1}{N} \langle a_0^*a_0\psi_{N,a,d},\psi_{N,a,d}\rangle
\leq\left\{\begin{aligned}
&CN^{-\frac{2}{17}}d^{\frac{2}{17}},\quad \text{In Region I}\\
&C\big(N^{-\frac{1}{9}}+N^{-1}\big\vert\ln(Nd^2)\big\vert\big),\quad \text{In Region $\mathrm{II}_{\mathrm{I}}$}\\
&CN^{-s},\quad\text{In Region $\mathrm{II}_{\mathrm{III}}$}\\
&C\left(\frac{a}{d}\right)^{\frac{1}{9}},\quad \text{In Region III}
\end{aligned}
\right.
\end{equation}
for some universal constants $C$ and $s\in (0,1)$ depending only on $t_2$ and may tend to $0$ when $t_2$ tends to $0$ and tend to $\frac{1}{9}$ when $t_2$ tends to $1$.
\end{lemma}

\par Now we let $E_0=4\pi(N-1)\mathfrak{a}_0+\mathfrak{e}_d+E_{Bog}^{(d)}$, $\mathcal{E}=\mathcal{E}^{B^{\prime\prime}}+\delta+(C^{B^{\prime\prime}}-E_0)$ and $\mathcal{U}=e^Be^{B^\prime}e^{B^{\prime\prime}}$, then with Proposition \ref{Bog renorm} and Lemma \ref{calculate E_bog} we can rewrite
\begin{equation}\label{U*H_NU}
 \mathcal{Z}_N^{I}=\mathcal{U}^*H_N\mathcal{U}
=E_0+\mathcal{D}+e^{-B^{\prime\prime}}H_4e^{B^{\prime\prime}}+
\mathcal{E}.
\end{equation}
To prove that the error term $\mathcal{E}$ is actually small we need to prove a result concerning complete Bose-Einstein condensation.
\begin{proposition}\label{Optimal BEC}
Let $N$ tends to infinity while $a,d$ and $\frac{a}{d}$ tend to $0$ with G-P restriction i.e. $\frac{Na}{d}=1$. Assume further that $N^{-1}d^{-2}$ tends to $0$. In other words, we are taking the Gross-Pitaevskii limit in Region I. Then there exists a universal constant $C$ such that
\begin{equation}\label{BEC}
  H_N\geq E_0+C^{-1}\mathcal{N}_+-C.
\end{equation}
\end{proposition}
\noindent
\emph{Proof.} We set in this proof that $\kappa=\nu d^{-1}$ for some fixed, universal but large $\nu>1$. Also we will always fix $\textit{l}\in(0,\frac{1}{2})$ independent of $N$, $a$ and $d$. It is easy to check that in the Gross-Pitaevskii regime in Region I together with our choice of fixed $\textit{l}$ and $\nu$, our assumptions on parameters in Proposition \ref{Bog renorm} are automatically satisfied. We follow the idea in \cite[Proposition 20]{hainzlSchleinTriay2022bogoliubov} to split out the high and low momenta of a test wave function. The method is known as the \textit{localization estimates} in \cite{boccatoBrenCena2020optimal,lewinNamSerfatySolovej2015bogoliubov}. Now we let $f,g:\mathbb{R}\to[0,1]$ be smooth functions such that $f(s)^2+g(s)^2=1$ for all $s\in\mathbb{R}$, and $f(s)=1$ for $s\leq\frac{1}{2}$, $f(s)=0$ for $s\geq1$. For some $M>0$ to be determined, we define
\begin{equation}\label{123101}
  f_M(s)=f(s/M),\quad g_M(s)=g(s/M).
\end{equation}
Then we can calculate directly
\begin{equation}\label{H_N region I}
  H_N=f_M(\mathcal{N}_+)H_Nf_M(\mathcal{N}_+)+
g_M(\mathcal{N}_+)H_Ng_M(\mathcal{N}_+)+\mathcal{E}_M,
\end{equation}
where
\begin{equation}\label{123099}
  \mathcal{E}_M=\frac{1}{2}
\left([f_M(\mathcal{N}_+),[f_M(\mathcal{N}_+),H_N]]
+[g_M(\mathcal{N}_+),[g_M(\mathcal{N}_+),H_N]]\right).
\end{equation}
Note that for any bounded real function $h$ point-wisely defined on $\mathbb{R}$, we can check on Fock space $F_{N,d}$ with (\ref{N_+ on F_N,d}) that
\begin{align*}
  U_N h(\mathcal{N}_+^L)U_N^*=h(U_N\mathcal{N}_+^LU_N^*)
=h(\mathcal{N}_+^F)=\begin{pmatrix}
         h(0) &   &   &   \\
           & h(1) &   &   \\
           &   &\ddots &   \\
           &   &   & h(N)
       \end{pmatrix}.
\end{align*}
We can therefore calculate
\begin{equation}\label{123098}
\begin{aligned}
 [h(\mathcal{N}_+),[h(\mathcal{N}_+),H_N]]=[h(\mathcal{N}_+)-h((\mathcal{N}_+-2)_+)]^2
\frac{1}{2\sqrt{d}}\sum_{p\neq0}v_p^{(a,d)}(a_p^*a_{-p}^*a_0a_0+h.c.)&\\
+[h(\mathcal{N}_+)-h((\mathcal{N}_+-1)_+)]^2
\frac{1}{\sqrt{d}}\sum_{p,r,p+r\neq0}v_r^{(a,d)}(a_{p+r}^*a_{-r}^*a_pa_0+h.c.).&
\end{aligned}
\end{equation}
Taking $h=f_M,g_M$ respectively, we claim that we can, by estimating on Fock space, deduce the bound of $\mathcal{E}_M$
\begin{equation}\label{E_M}
  \pm\mathcal{E}_M\leq CM^{-2}P_X(H_4+N)P_{X}.
\end{equation}
where $P_X$ is the orthogonal projection onto a subspace $X\subset L_s^2(\Lambda_d^N)$ on which there holds $\frac{(M-4)_+}{2}\leq\mathcal{N}_+\leq(M+2)$. Here, $X$ is explicitly given by
\begin{equation}
 X=U_N^*\bigoplus_{n=(M-4)_+/2}^{M+2}L^2_\perp(\Lambda_d)^{\otimes_sn},\label{define X}
\end{equation}
and we can check that $P_X$ commutes with $H_4$ by switching to Fock space. We postpone the proof of claim (\ref{E_M}) to the end of this arguement.

\par Now notice that for all $\psi\in L^2_s(\Lambda_d^N)$, by the definition of $f$ we can verify in the Fock space that for all $n\in\frac{1}{2}\mathbb{N}$ and $n>1$
\begin{equation*}
  \langle\mathcal{N}_+^nf_M(\mathcal{N}_+)\psi,f_M(\mathcal{N}_+)\psi\rangle
\leq M^{n-1}\langle \mathcal{N}_+f_M(\mathcal{N}_+)\psi,f_M(\mathcal{N}_+)\psi\rangle.
\end{equation*}
This simple observation together with Lemmas \ref{control of S_+ conj with e^B}, \ref{control of S_+ conj with e^B'} and \ref{control of S_+ with e^B''} (We apply these lemmas with $t=\pm1$) imply for all $n\in\frac{1}{2}\mathbb{N}$ and $n>1$, that
\begin{align}
  &\langle f_M(\mathcal{N}_+)\mathcal{U}(\mathcal{N}_++1)^n
\mathcal{U}^*f_M(\mathcal{N}_+)\psi,\psi\rangle\nonumber\\
&\leq C(M^{n-1}+1)\langle\mathcal{N}_+f_M(\mathcal{N}_+)\psi,
f_M(\mathcal{N}_+)\psi\rangle
+C\langle f_M(\mathcal{N}_+)\psi,
f_M(\mathcal{N}_+)\psi\rangle\nonumber\\
&\leq C(M^{n-1}+1)\langle f_M(\mathcal{N}_+)\mathcal{U}(\mathcal{N}_++1)
\mathcal{U}^*f_M(\mathcal{N}_+)\psi,\psi\rangle .\label{tempp}
\end{align}
Now we choose $M=N^{\frac{16}{17}}d^{\frac{2}{17}}$. Notice that in the setting of Region I, $M$ tends to infinity. We also denote $\mathcal{N}_+^{\mathcal{U}}=\mathcal{U}^*\mathcal{N}_+\mathcal{U}$ for short. Then using relation (\ref{U*H_NU}) and the fact that $\mathcal{N}_+\leq H_{21}\leq\mathcal{D}$ we obtain
\begin{align}
f_M(\mathcal{N}_+)&H_Nf_M(\mathcal{N}_+)=
\mathcal{U}f_M(\mathcal{N}_+^\mathcal{U})\mathcal{U}^*H_N\mathcal{U}
f_M(\mathcal{N}_+^\mathcal{U})\mathcal{U}^*\nonumber\\
&\geq\mathcal{U}f_M(\mathcal{N}_+^\mathcal{U})
(E_0+H_{21}+e^{-B^{\prime\prime}}H_4e^{B^{\prime\prime}}+\mathcal{E})
f_M(\mathcal{N}_+^\mathcal{U})\mathcal{U}^*.\label{123097}
\end{align}
Notice that in Region I, $Nd^2\gg 1$ (which implies that $\ln N\gtrsim\ln d^{-1}$). We then apply Proposition \ref{Bog renorm} with
\begin{equation*}
  \alpha=\frac{2}{17},\quad \beta=\frac{2}{3},\quad \gamma=\frac{1}{34},
\end{equation*}
together with Lemma \ref{calculate E_bog} and inequality (\ref{tempp}) we obtain
\begin{align}\label{123096}
  \mathcal{U}f_M(\mathcal{N}_+^\mathcal{U})\mathcal{E}
f_M(\mathcal{N}_+^\mathcal{U})\mathcal{U}^*
\geq& -C \mathcal{U}f_M(\mathcal{N}_+^\mathcal{U})
\{d^{\frac{2}{17}}\ln d^{-1}(\mathcal{N}_++1)\nonumber\\
&+d^{\frac{2}{17}}
(H_{21}+e^{-B^{\prime\prime}}H_4e^{B^{\prime\prime}})\}
f_M(\mathcal{N}_+^\mathcal{U})\mathcal{U}^*
\end{align}
Since $\mathcal{N}_+\leq H_{21}$, and (\ref{H_21psi,psi fock}) and (\ref{H_4psi,psi fock}) tell us that $H_{21},H_4\geq0$, we combine (\ref{123097}) and (\ref{123096}) to reach
\begin{equation}\label{mechante1}
  f_M(\mathcal{N}_+)H_Nf_M(\mathcal{N}_+)
\geq f_M(\mathcal{N}_+)^2(E_0+C^{-1}\mathcal{N}_+-C).
\end{equation}

\par Now we turn to the second term on the right hand side of (\ref{H_N region I}). We are going to prove that
\begin{equation}\label{mechante2}
   g_M(\mathcal{N}_+)(H_N-E_0)g_M(\mathcal{N}_+)\geq C^{-1}Ng_M(\mathcal{N}_+)^2
\geq C^{-1}\mathcal{N_+}g_M(\mathcal{N}_+)^2.
\end{equation}
Following the idea of \cite[Proposition 6.1]{boccatoBrenCena2020optimal}, we argue by contradiction. First by the definition of $g_M$, we observe that
\begin{align*}
g_M(\mathcal{N}_+)(H_N-E_0)g_M(\mathcal{N}_+)
\geq\left(\inf_{\psi\in Y,\Vert\psi\Vert_2=1}\frac{1}{N}
\langle(H_N-E_0)\psi,\psi\rangle\right)Ng_M(\mathcal{N}_+)^2.
\end{align*}
Here $Y\subset L_s^2(\Lambda_d^N)$ is explicitly given by
$Y=U_N^*\bigoplus_{n=M/2}^{N}L^2_\perp(\Lambda_d)^{\otimes_sn}$. Then to prove (\ref{mechante2}) it is sufficient to prove
\begin{equation}\label{mechante2.1}
  \inf_{\psi\in Y,\Vert\psi\Vert_2=1}\frac{1}{N}
\langle(H_N-E_0)\psi,\psi\rangle\geq C^{-1}.
\end{equation}
Since we have already known from Lemma \ref{calculate E_bog} and (\ref{leading order more precise}) that
\begin{align*}
  &\inf_{\psi\in Y,\Vert\psi\Vert_2=1}\frac{1}{N}
\langle(H_N-E_0)\psi,\psi\rangle
= \inf_{\psi\in Y,\Vert\psi\Vert_2=1}\frac{1}{N}
\langle H_N\psi,\psi\rangle-4\pi\mathfrak{a}_0+4\pi\mathfrak{a}_0-\frac{E_0}{N}\\
&\geq \inf_{\psi\in L^2_s(\Lambda_d^N),\Vert\psi\Vert_2=1}\frac{1}{N}
\langle H_N\psi,\psi\rangle-4\pi\mathfrak{a}_0+4\pi\mathfrak{a}_0-\frac{E_0}{N}
\stackrel{{\text{G-P limit}}}{\longrightarrow}0,
\end{align*}
then if we assume (\ref{mechante2.1}) is not true, we can find a family of $\{N_j,a_j,d_j\}$ in the Gross-Piaevskii limit in Region I, and $\psi_j\in Y_j$ (the subscript $j$ implies this space Y depends on $j$) with $\Vert\psi_j\Vert_2=1$, such that
\begin{align*}
   \frac{1}{N_j}
\langle H_{N_j}\psi_j,\psi_j\rangle\to4\pi\mathfrak{a}_0.
\end{align*}
That is to say, $\{\psi_j\}$ is a family of approximate ground state wave function. Then by the Bose-Einstein condensation result (\ref{rate of decay}), we know that
\begin{equation}\label{mechante2.2}
  \frac{1}{N_j}\langle\mathcal{N}_+\psi_j,\psi_j\rangle\leq
CN_j^{-\frac{2}{17}}d_j^{\frac{2}{17}}.
\end{equation}
On the other hand, since $\psi_j\in Y_j$, we have
\begin{equation}\label{mechante2.3}
\frac{1}{N_j}\langle\mathcal{N}_+\psi_j,\psi_j\rangle\geq
\frac{M_j}{2N_j}\geq CN_j^{-\frac{1}{17}}d_j^{\frac{2}{17}}
\gg CN_j^{-\frac{2}{17}}d_j^{\frac{2}{17}}
\end{equation}
which contradicts (\ref{mechante2.2}). Hence (\ref{mechante2.1}) holds and thus (\ref{mechante2}).

\par To conclude (\ref{BEC}), we first combine (\ref{E_M}), (\ref{mechante1}) and (\ref{mechante2}) to get to
\begin{equation*}
 H_N\geq E_0+C^{-1}\mathcal{N}_+-C-CM^{-2}H_4
\end{equation*}
with $M=N^{\frac{16}{17}}d^{\frac{2}{17}}$ as chosen before. Conjuagating with $e^B$ and using Lemmas \ref{control of S_+ conj with e^B} and \ref{lemma control of H_4}, we have
\begin{align}
  \mathcal{G}_N&\geq E_0+C^{-1}\mathcal{N}_+-C-CM^{-2}H_4-CM^{-2}N\nonumber\\
&\geq E_0+C^{-1}\mathcal{N}_+-C-CM^{-2}H_4.\label{mechante3.1}
\end{align}
On the other hand, we first fix $\textit{l}\in (0,\frac{1}{2})$ small enough and $\textit{l}$ only depends on the universal constants arising from (\ref{third renorm E^B''}). We then apply Proposition \ref{Bog renorm} with
\begin{equation*}
  \alpha=\frac{\ln\textit{l}}{\ln d},
\quad \beta=-\frac{\ln\textit{l}}{\ln N},
\quad \gamma=-\frac{\ln\textit{l}}{2\ln N},
\end{equation*}
together with Lemma \ref{calculate E_bog} and (\ref{U*H_NU}), and the naive bound
\begin{align*}
  &(\mathcal{N}_++1)^k\leq CN^{k-1}(\mathcal{N}_++1),\\
&0\leq\mathcal{N}_+\leq H_{21}\leq\mathcal{D}
\end{align*}
for any $k\in\frac{1}{2}\mathbb{N}$ and $k>1$, we find that
\begin{equation*}
  \mathcal{Z}_N^{I}\geq E_0+Ce^{-B^{\prime\prime}}H_4e^{B^{\prime\prime}}
-C\ln d^{-1}(\mathcal{N}_++1).
\end{equation*}
Then Lemmas \ref{control of S_+ conj with e^B'}, \ref{lemma e^-B'H_Ne^B'} and \ref{control of S_+ with e^B''} together yield
\begin{equation}\label{mechante3.2}
  \mathcal{G}_N\geq E_0+ CH_4-C\ln d^{-1}(\mathcal{N}_++1).
\end{equation}
Combining (\ref{mechante3.1}) and (\ref{mechante3.2}), we have
\begin{equation*}
  \mathcal{G}_N\geq E_0+C^{-1}\mathcal{N}_+-C.
\end{equation*}
We conclude (\ref{BEC}) using Lemma \ref{control of S_+ conj with e^B}.

\par We are left with the proof of (\ref{E_M}). Letting $h=f_M, g_M$ repectively, due to the choice of $f$ and $g$, we have
\begin{equation}\label{mechante4}
\begin{aligned}
\vert h(s)-h(s-t)\vert&=0,\quad \text{$s<\frac{M}{2}$ or $s>M+t$}\\
 \vert h(s)-h(s-t)\vert&\leq\frac{Ct}{M},\quad\text{$\frac{M}{2}\leq s\leq M+t$}
\end{aligned}
\end{equation}
for all $t>0$. Estimating on the Fock space (for more detailed calculation one can consult the proof of Lemma \ref{control of S_+ conj with e^B}), for any $\psi\in L^2_s(\Lambda_d^N)$ we let
\begin{align*}
  U_N\psi&=(\alpha^{(0)}\dots,\alpha^{(N)}),
\end{align*}
then conjugating with $U_N$ and using relation (\ref{conjugate relation})give
\begin{align*}
  &\frac{1}{2\sqrt{d}}{\sum_{p\neq0}}v_{p}^{(a,d)}
\langle[h(\mathcal{N}_+)-h(\mathcal{N}_+-2)]^2
 a_p^*a_{-p}^*a_0a_0\psi,\psi\rangle\\
=&\frac{1}{2\sqrt{d}}\sum_{n=2}^{N}\sqrt{(N-n)(N-1-n)}
{\sum_{p\neq0}}v_{p}^{(a,d)}
\langle[h(n)-h(n-2)]^2
 a_p^*a_{-p}^*\alpha^{(n-2)},\alpha^{(n)}\rangle\\
=&\frac{1}{2d}\sum_{n=M/2}^{M+2}
[h(n)-h(n-2)]^2\sqrt{n(n-1)}\sqrt{(N-n)(N-1-n)}
\int_{\Lambda_d^n}d\mathbf{x}_1\dots d\mathbf{x}_n\\
&\quad\quad\quad\quad\quad\quad\quad\quad\quad\times
v_a(\mathbf{x}_1-\mathbf{x}_2)
 \alpha^{(n-2)}(\mathbf{x}_3,\dots,\mathbf{x}_{n})
\overline{\alpha^{(n)}}(\mathbf{x}_1,\dots,\mathbf{x}_{n}).
\end{align*}
Using Cauchy-Schwartz inequality together with (\ref{H_4psi,psi fock}) and (\ref{mechante4}) we have
\begin{align*}
&\frac{1}{2\sqrt{d}}\left\vert{\sum_{p\neq0}}v_{p}^{(a,d)}
\langle[h(\mathcal{N}_+)-h(\mathcal{N}_+-2)]^2
 a_p^*a_{-p}^*a_0a_0\psi,\psi\rangle \right\vert\\
&\leq CNM^{-2}d^{-\frac{1}{2}}\Vert v_a\Vert_1^{\frac{1}{2}}
\langle H_4P_X\psi,P_X\psi\rangle^{\frac{1}{2}}
\langle P_X\psi,P_X\psi\rangle^{\frac{1}{2}}
\end{align*}
A similar argument gives
\begin{align*}
&\frac{1}{\sqrt{d}}\left\vert{\sum_{p,r,p+r\neq0}}v_{p}^{(a,d)}
\langle[h(\mathcal{N}_+)-h(\mathcal{N}_+-1)]^2
 a_{p+r}^*a_{-r}^*a_pa_0\psi,\psi\rangle \right\vert\\
&\leq CN^{\frac{1}{2}}M^{-2}d^{-\frac{1}{2}}\Vert v_a\Vert_1^{\frac{1}{2}}
\langle H_4P_X\psi,P_X\psi\rangle^{\frac{1}{2}}
\langle (\mathcal{N}_++1)P_X\psi,P_X\psi\rangle^{\frac{1}{2}}
\end{align*}
Since $\Vert v_a\Vert_1\sim a$, and we demand $\frac{Na}{d}=1$ in the Gross-Pitaevskii limit in Region I, and recall the representation of $\mathcal{E}_M$ in (\ref{123099}) and (\ref{123098}), we reach (\ref{E_M}).
\begin{flushright}
  $\Box$
\end{flushright}

\subsection{Proof of Theorem \ref{core} for Region I}\label{proof I}
\
\par Inspired by \cite{2018Bogoliubov}, we prove Theorem \ref{core} by comparing the eigenvalues of $\mathcal{H}=H_N-E_0$ with $\mathcal{D}$ using min-max principle. If we denote $\{\lambda_j(\mathcal{H})\}$ and $\{\lambda_j(\mathcal{D})\}$ with $j\in\mathbb{N}$ respectively all the eigenvalues (conunting mutiplicity) of $\mathcal{H}$ and $\mathcal{D}$ arranged in the ascending order. We will prove that for any $m\in\mathbb{N}$ such that $\lambda_m(\mathcal{D})\leq\zeta$ with a threshold $1\leq\zeta\leq Cd^{-\frac{1}{2}}$ to be determined, there holds
\begin{equation}\label{threshold estimate}
  \vert\lambda_m(\mathcal{H})-\lambda_m(\mathcal{D})\vert\leq Cd^{\frac{1}{4}}\ln d^{-1}(1+\zeta^3).
\end{equation}
The eigenvalues of the diagonal quadratic operator $\mathcal{D}$ can be thoroughly analyzed. In fact, we denote
\begin{equation*}
  \mathcal{D}=\sum_{p\in2\pi\mathbb{Z}^3}\varepsilon_pa_p^*a_p,
\end{equation*}
with $\varepsilon_p=
\sqrt{\vert\mathcal{M}_dp\vert^4+16\pi\mathfrak{a}_0\vert\mathcal{M}_dp\vert^2}$. Since it is easy to check that
\begin{equation}\label{ONB B}
  \mathfrak{B}=\left\{\prod_{p\in2\pi\mathbb{Z}^3}(a_p^*)^{n_p^{(j)}}1\,:\,j,n_p^{(j)}\in\mathbb{N},
\sum_{p\in2\pi\mathbb{Z}^3}n_p^{(j)}=N\right\}
\end{equation}
is an orthogonal basis of $L^2_s(\Lambda_d^N)$, we can check using Fourier series expansion and formula (\ref{general commutation relations}) that the eigenvalues of $\mathcal{D}$ have the form
\begin{equation}\label{eigen of D}
  \lambda_j(\mathcal{D})=\sum_{p\in2\pi\mathbb{Z}^3}n_p^{(j)}\varepsilon_p
\end{equation}
with $\{n_p^{(j)}\}$ introduced above. The corresponding eigenvector is given by
\begin{equation}\label{eigenvector of D}
  \xi_j=C_j\prod_{p\in2\pi\mathbb{Z}^3}(a_p^*)^{n_p^{(j)}}1
\end{equation}
for some normalization constant $C_j>0$. Since $\varepsilon_0=0$, we know that $\lambda_0(\mathcal{D})=0<\zeta$. Then from (\ref{threshold estimate}), we can simply choose $\zeta$ to be some universal large constant, and we can conclude that since $\lambda_0(\mathcal{H})=E_N-E_0$,
\begin{equation*}
  \vert E_N-E_0\vert\leq Cd^{\frac{1}{4}}\ln d^{-1},
\end{equation*}
which proves Theorem \ref{core} for Region I.
\par The proof of (\ref{threshold estimate}) consists of a lower bound and an upper bound. We first prove the upper bound on $\lambda_m(\mathcal{H})$. Let $Z\subset L^2_s(\Lambda_d^N)$ be the subspace generated by the first $m$ eigenvectors of $\mathcal{D}$ whose form is given in (\ref{eigenvector of D}) and let $P_Z$ be the orthogonal projection onto it. From (\ref{U*H_NU}), we know that
\begin{equation}\label{kuus1}
  \lambda_m(\mathcal{D})\geq P_Z\mathcal{D}P_Z\geq
P_Z(\mathcal{U}^*\mathcal{H}\mathcal{U}-
e^{-B^{\prime\prime}}H_4e^{B^{\prime\prime}}-\mathcal{E})P_Z.
\end{equation}
Notice on the other hand
\begin{equation}\label{kuus1.1}
  P_Z\mathcal{N}_+P_Z\leq P_Z H_{21}P_Z\leq P_Z\mathcal{D}P_Z
\leq \lambda_m(\mathcal{D})\leq\zeta.
\end{equation}
Since the orthgonal projection $P_Z$ can be written explicitly
\begin{equation}\label{define P_Z}
  P_Z=\sum_{j=1}^{m}\tilde{C}_j\prod_{p\in2\pi\mathbb{Z}^3}
(a_p^*)^{n_p^{(j)}}(a_p)^{n_p^{(j)}}
\end{equation}
for some normalization constants $\tilde{C}_j>0$. From (\ref{general commutation relations}) we know that $\mathcal{N}_+$ commutes with $P_Z$. Therefore we can argue by induction that
\begin{equation}\label{kuus1.2}
  P_Z(\mathcal{N}_++1)^{n}P_Z\leq C(1+\zeta^n)
\end{equation}
for any $n\in\frac{1}{2}\mathbb{N}$. Now we consider the limit in the Gross-Pitaevskii regime in Region I (recall that $Nd^2$ is large here), we apply Proposition \ref{Bog renorm} with
\begin{equation*}
  \alpha=1,\quad \beta=1,\quad\gamma=\frac{1}{4}
\end{equation*}
 with $\textit{l}$ and $\nu$ fixed, together with (\ref{kuus1}), (\ref{kuus1.2}) and inequality (\ref{control H_4 with e^B'' 1}) in Lemma \ref{control of S_+ with e^B''} yield
\begin{equation}\label{kuus1.3}
  \lambda_m(\mathcal{D})\geq
P_Z(\mathcal{U}^*\mathcal{H}\mathcal{U}-C
H_4)P_Z-Cd\ln d^{-1}(1+\zeta^3),
\end{equation}
while the expectation of $H_4$ on $Z$ is controllable. From (\ref{eigen of D}), together with the apparent fact that $\varepsilon_p\geq\vert\mathcal{M}_dp\vert^2$ and the requirement that $\lambda_m\leq\zeta$, we know that $n_q^{(m)}=0$ and thus $a_q\xi_m=0$ whenever $\vert\mathcal{M}_dq\vert\geq\zeta^{1/2}$. This implies that $a_qP_Z=0$ whenever $\vert\mathcal{M}_dq\vert\geq\zeta^{1/2}$, and thus leads to the bound for any $\psi\in L^2_s(\Lambda_d^N)$
\begin{align}
\vert\langle H_4P_Z\psi,P_Z\psi\rangle\vert
\leq&\frac{1}{2\sqrt{d}}\sum_{\substack{p,q,r\\
\vert\mathcal{M}_dp\vert,\vert\mathcal{M}_dq\vert,\\\vert\mathcal{M}_dr\vert\leq
C\zeta^{1/2}}}\vert v_r^{(a,d)}
\Vert a_pa_{q+r}P_Z\psi\Vert\Vert a_qa_{p+r}P_Z\psi\Vert\nonumber\\
\leq& Cad^{-1}\zeta\langle (\mathcal{N}_++1)^2P_Z\psi,P_Z\psi\rangle\nonumber\\
\leq&CN^{-1}(1+\zeta^3),\label{bound on H_4 on Z}
\end{align}
where we have additionally assumed $\zeta\leq Cd^{-2}$. Inserting (\ref{bound on H_4 on Z}) into (\ref{kuus1.3}), and applying min-max principle we reach an upper bound on $\lambda_m(\mathcal{H})$
\begin{align}\label{upper bound I}
    \lambda_m(\mathcal{D})
&\geq \sup_{\substack{\psi\in\mathcal{U}Z,\\\Vert\psi\Vert_2=1}}
\langle\mathcal{H}\psi,\psi\rangle-Cd\ln d^{-1}(1+\zeta^3)\nonumber\\
&\geq\inf_{\substack{L\subset L^2_s(\Lambda_d^N)\\
\dim L=m}}\sup_{\substack{\psi\in L\\\Vert\psi\Vert_2=1}}
\langle\mathcal{H}\psi,\psi\rangle-Cd\ln d^{-1}(1+\zeta^3)\nonumber\\
&=\lambda_m(\mathcal{H})-Cd\ln d^{-1}(1+\zeta^3).
\end{align}

\par For the lower bound on $\lambda_m(\mathcal{H})$, we use again the method presented in the proof of Proposition \ref{Optimal BEC}. With the same notations, from (\ref{H_N region I}) we can rewrite
\begin{equation}\label{kuus2.1}
  \mathcal{H}=f_M(\mathcal{N}_+)\mathcal{H}f_M(\mathcal{N}_+)
+g_M(\mathcal{N}_+)\mathcal{H}g_M(\mathcal{N}_+)+\mathcal{E}_M
\end{equation}
for some $M>0$ to be determined and the error term satisfies
\begin{equation}\label{kuus2.2}
  \pm\mathcal{E}_M\leq CM^{-2}P_X(H_4+N)P_X,
\end{equation}
where the subspace $X$ is defined in (\ref{define X}). We now let the space generated by the first $m$ eigenvectors of $\mathcal{H}$ by $V\subset L^2_s(\Lambda_d^N)$ and the orthogonal projection onto it by $P_V$. Notice that this time we can not ensure $P_V$ commutes with $\mathcal{N}_+$ unless $a_0V=0$. One can check this fact by expanding the vectors generating $V$ by the basis $\mathfrak{B}$ introduced in (\ref{ONB B}). From (\ref{kuus2.1}) we immediately obtain
\begin{equation}\label{kuus2.3}
  \lambda_m(\mathcal{H})=\lambda_m(\mathcal{U}^*\mathcal{H}\mathcal{U})
\geq P_V(f_M(\mathcal{N}_+)\mathcal{H}f_M(\mathcal{N}_+)
+g_M(\mathcal{N}_+)\mathcal{H}g_M(\mathcal{N}_+)+\mathcal{E}_M)P_V.
\end{equation}
Now we choose $M=N^{\frac{1}{2}}d^{-\frac{1}{2}}$ in the Gross-Pitaevskii regime in Region I. From Proposition \ref{Optimal BEC} and the definition of $g_M$ we find that
\begin{equation}\label{kuus2.4}
 \begin{aligned}
 P_Vg_M(\mathcal{N}_+)\mathcal{H}g_M(\mathcal{N}_+)P_V&\geq
P_Vg_M(\mathcal{N}_+)(C^{-1}\mathcal{N}_+-C)g_M(\mathcal{N}_+)P_V\\
&\geq P_Vg_M(\mathcal{N}_+)^2P_V(C^{-1}M-C)\geq0.
\end{aligned}
\end{equation}
To bound $P_V\mathcal{E}_MP_V$, we first derive from (\ref{mechante3.2}) together with Lemmas \ref{control of S_+ conj with e^B} and \ref{lemma control of H_4} that
\begin{equation}\label{kuus2.5}
  H_4\leq C(\mathcal{H}+N+\ln d^{-1}(\mathcal{N}_++1)).
\end{equation}
(\ref{kuus2.5}) together with (\ref{BEC}) in Proposition \ref{Optimal BEC} and the upper bound (\ref{upper bound I}) tell us
\begin{equation}\label{kuus2.6}
  P_V H_4P_V\leq C(N+\ln d^{-1}(1+\zeta^3)).
\end{equation}
Combining (\ref{kuus2.6}) with (\ref{kuus2.2}), and noticing that $M=N^{\frac{1}{2}}d^{-\frac{1}{2}}$, we have
\begin{equation}\label{kuus2.7}
  P_V\mathcal{E}_MP_V\geq -Cd(1+\zeta^3).
\end{equation}
To bound $P_Vf_M(\mathcal{N}_+)\mathcal{H}f_M(\mathcal{N}_+)P_V$, we first apply Propostion \ref{Bog renorm} with $\textit{l}$, $\nu$ fixed, as well as
\begin{equation*}
  \alpha=\frac{1}{4},\quad
\beta=\frac{1}{2},\quad
\gamma=\frac{1}{16}.
\end{equation*}
Together with inequality (\ref{tempp}), the choice of $M=N^{\frac{1}{2}}d^{-\frac{1}{2}}$, and the naive fact that $\mathcal{N}_+\leq H_{21}\leq \mathcal{D}$ and $H_4\geq0$ we reach
\begin{align}
 &P_Vf_M(\mathcal{N}_+)\mathcal{H}f_M(\mathcal{N}_+)P_V\nonumber\\
=&P_Vf_M(\mathcal{N}_+)\mathcal{U}
(\mathcal{D}+e^{-B^{\prime\prime}}H_4e^{B^{\prime\prime}}+\mathcal{E})
\mathcal{U}^*f_M(\mathcal{N}_+)P_V\nonumber\\
\geq &(1-Cd^{\frac{1}{4}}\ln d^{-1})
P_Vf_M(\mathcal{N}_+)\mathcal{U}\mathcal{D}
\mathcal{U}^*f_M(\mathcal{N}_+)P_V
-Cd^{\frac{1}{4}}\ln d^{-1}.\label{kuus2.8}
\end{align}
To make use of (\ref{kuus2.8}) and min-max principle, we need to check additionally that the space $f_M(\mathcal{N}_+)V$ is of the same dimension as the space $V$, or in other words, $\dim(f_M(\mathcal{N}_+)V)=m$. Here we directly use a result from \cite{lewinNamSerfatySolovej2015bogoliubov}, and we state it in Lemma \ref{lemma lewinNamSerfatySolovej2015bogoliubov}
\begin{lemma}(\cite[Proposition 6.1 (ii)]{lewinNamSerfatySolovej2015bogoliubov})\label{lemma lewinNamSerfatySolovej2015bogoliubov}
  Let $\mathcal{H}$ be any non-negative operator on $L^2_s(\Lambda_d^N)$ and $V\subset D(\mathcal{H})$ is of finite dimension. $g_M(\mathcal{N}_+)$ and $f_M(\mathcal{N}_+)$ are defined in (\ref{123101}). If $\Vert g_M(\mathcal{N}_+)\vert_V\Vert^2<(\dim V)^{-1}$, then $\dim(f_M(\mathcal{N}_+)V)=\dim V$.
\end{lemma}
Since we can check, in the Fock space that
\begin{align}\label{123102}
 g_M(\mathcal{N}_+)^2\leq CM^{-1}g_M(\mathcal{N}_+)\mathcal{N}_+g_M(\mathcal{N}_+)
\leq CM^{-1}\mathcal{N}_+.
\end{align}
On the other hand, (\ref{BEC}) from Proposition \ref{Optimal BEC} and the upper bound (\ref{upper bound I}) together imply that
\begin{equation}\label{123103}
  P_V\mathcal{N}_+P_V\leq C(1+\zeta)+Cd\ln d^{-1}(1+\zeta^3).
\end{equation}
(\ref{123102}) and (\ref{123103}) together yield
\begin{equation}\label{123104}
  P_Vg_M(\mathcal{N}_+)^2P_V \leq CM^{-1}(1+\zeta+d\ln d^{-1}\zeta^3).
\end{equation}
The right hand side of (\ref{123104}) tends to $0$ in the Gross-Pitaevskii limit in Region I as long as we demand $\zeta\leq Cd^{-\frac{1}{2}}$. Thus Lemma \ref{lemma lewinNamSerfatySolovej2015bogoliubov} guarantees that
\begin{align*}
  \dim(f_M(\mathcal{N}_+)V)=\dim V=m.
\end{align*}
Therefore, we can combine (\ref{kuus2.3}), (\ref{kuus2.4}), (\ref{kuus2.7}) and (\ref{kuus2.8}) and use min-max principle to get
\begin{align}\label{lower bound I}
 \lambda_m(\mathcal{H})
&\geq(1-Cd^{\frac{1}{4}}\ln d^{-1})
\sup_{\substack{\psi\in\mathcal{U}^*f_M(\mathcal{N}_+)V,\\
\Vert\psi\Vert_2=1}}\langle\mathcal{D}\psi,\psi\rangle
-Cd^{\frac{1}{4}}\ln d^{-1}(1+\zeta^3)\nonumber\\
&\geq(1-Cd^{\frac{1}{4}}\ln d^{-1})
\inf_{\substack{L\subset L^2_s(\Lambda_d^N)\\
\dim L=m}}\sup_{\substack{\psi\in L\\\Vert\psi\Vert_2=1}}
\langle\mathcal{D}\psi,\psi\rangle-Cd^{\frac{1}{4}}\ln d^{-1}(1+\zeta^3)\nonumber\\
&\geq \lambda_m(\mathcal{D})-Cd^{\frac{1}{4}}\ln d^{-1}(1+\zeta^3).
\end{align}
(\ref{upper bound I}) and (\ref{lower bound I}) together conclude the claim (\ref{threshold estimate}).
\begin{flushright}
  $\Box$
\end{flushright}

\subsection{Proof of Theorem \ref{core} for Region \texorpdfstring{$\mathrm{II}_{\mathrm{I}}$}{II-I}}
\label{proof II1}
\
\par The arguments carried out in Sections \ref{subsection Optimal BEC} and \ref{proof I} also applies to Region $\mathrm{II}_{\mathrm{I}}$ since the results of Propositons \ref{Bog renorm} and Lemma \ref{calculate C^B''} still hold in the Gross-Pitaevskii regime in Region $\mathrm{II}_{\mathrm{I}}$ as long as we still fix $\textit{l}$ and $\nu$ being universal constants. Here we fix
\begin{equation}\label{choose t1}
  t_1=\frac{1}{72}.
\end{equation}
To prove Theorem \ref{core} for Region $\mathrm{II}_{\mathrm{I}}$, we point out the different choices of parameters in the arguments in Sections \ref{subsection Optimal BEC} and \ref{proof I}. Proceeding as in the proof of Proposition \ref{Optimal BEC}, we choose
\begin{equation*}
  M=N^{\frac{17}{18}}
\end{equation*}
and we apply Propositons \ref{Bog renorm} and Lemma \ref{calculate C^B''} with
\begin{equation*}
  \alpha=-\frac{1}{36}\frac{\ln N}{\ln d},\quad
\beta=\frac{1}{18},\quad
\gamma=\frac{1}{72}.
\end{equation*}
Notice that in Region $\mathrm{II}_{\mathrm{I}}$
\begin{equation*}
  \ln d^{-1}\lesssim N^{t_1},
\end{equation*}
then the optimal BEC (\ref{BEC}) holds in in Region $\mathrm{II}_{\mathrm{I}}$ as well. The rest of Section \ref{subsection Optimal BEC} goes through.

\par As for Section \ref{proof I}, we are also going to prove, for any $m\in\mathbb{N}$ such that $\lambda_m(\mathcal{D})\leq\zeta$ with a threshold
\begin{equation*}
  1\leq\zeta\ll N^{\frac{5}{12}-\frac{t_1}{3}},
\end{equation*}
 there holds
\begin{equation}\label{threshold estimate II1}
  \vert\lambda_m(\mathcal{H})-\lambda_m(\mathcal{D})\vert\leq CN^{-\frac{1}{8}+t_1}(1+\zeta^3).
\end{equation}
With $\textit{l}$ and $\nu$ being universal constants, the upper bound is obtained by applying Propositons \ref{Bog renorm} and Lemma \ref{calculate C^B''} with
\begin{equation*}
  \alpha=-\frac{1}{2}\frac{\ln N}{\ln d},\quad
\beta=1,\quad
\gamma=\frac{1}{4},
\end{equation*}
while we reach the lower bound by choosing
\begin{equation*}
M=N^{\frac{1}{2}+\frac{1}{4}},\quad
  \alpha=-\frac{1}{8}\frac{\ln N}{\ln d},\quad
\beta=\frac{1}{8},\quad
\gamma=\frac{1}{16}.
\end{equation*}
Then the rest of Section \ref{proof I} goes through and concludes Theorem \ref{core} for Region $\mathrm{II}_{\mathrm{I}}$.

\begin{flushright}
  $\Box$
\end{flushright}

\section{Proof of the Main Theorem for Region III}
\par In this section, we establish Theorem \ref{core III} using Proposition \ref{Bog renorm III} to conclude. We first calculate the constant $C^{B^{\prime\prime\prime}}$ and the diagonalized operator $\mathcal{Q}^{\prime\prime\prime}$ in Lemma \ref{lemma calculate C^B'''}. Notice that, different from Lemma \ref{calculate E_bog}, we mainly aim to analyze the order of the constant $C^{B^{\prime\prime\prime}}$, but the format of second order ground state energy approximation is relatively less explicit compared to $C^{B^{\prime\prime}}$ in (\ref{calculate C^B''}) due to the dimensional coupling effect. In Propositon \ref{Optimal BEC III}, we give an optimal Bose-Einstein condensation result for Region III using the method of localization, together with the help of Proposition \ref{Bog renorm III}. Armed with this inequality, we can finally prove Theorem \ref{core III} in Section \ref{proof III}, by comparing the ground state energy of $H_N$ with a diagonalized operator $\tilde{\mathcal{D}}$ shown below. In this section, we mainly concern Region III. The proof for Region $\mathrm{II}_{\mathrm{III}}$ just needs slightly modifications on the proof for Region III and we include it in Section \ref{proof II2}. We remark that results in this section not only hold true for Region III, but also for part of Region $\mathrm{III}^\prime$ (see definition around (\ref{Region I II III})).

\par  Before we set our feet on mathematical proof, we first take a closer look at the Gross-Pitaevskii condition for Regions III and $\mathrm{III}^\prime$. Recall that in Region $\mathrm{III}^\prime$ in the G-P regime, we demand $\frac{d}{a}\lesssim\vert\ln(Nd^2)\vert$ and $Ng=\mathfrak{a}_0$, where $g$ is defined in (\ref{coupling constant g}). These two condition together yield
\begin{equation}\label{G-P III}
  N^{-1}\ll \frac{a}{d}\quad\text{or}\quad \big(1-\mathfrak{a}_0\upsilon\big)\frac{d}{a}=N
\end{equation}
for some universal $\upsilon<0$, and
\begin{equation}\label{G-P III d}
  d=N^{-\frac{1}{2}}e^{-\frac{N}{2}\frac{\upsilon}{\mathfrak{a}_0
\upsilon-1}}.
\end{equation}
This implies in the G-P regime in Regions III and $\mathrm{III}^\prime$, d decays exponentially with respect to $N$. Now recall the definition of $m$ in (\ref{define m}), if we let
\begin{equation}\label{choose l, h III}
  \textit{l}=c\Big(\frac{a}{d}\Big)^\alpha,\quad
h=N^{-\beta}
\end{equation}
for some universal $0\leq\alpha<1$, $\beta\geq 0$ and $0<c<\frac{1}{2}$, we will find
\begin{equation}\label{relation m and N}
  \frac{2m\mathfrak{a}_0}{N}\stackrel{{\text{G-P limit}}}{\longrightarrow}1.
\end{equation}
\label{Proof of the Main Theorem for Region III}
\subsection{More about the Renormalized Hamiltonian for Region III}
\begin{lemma}\label{lemma calculate C^B'''}
Under the same assumptions of Proposition \ref{Bog renorm III}, we have
\begin{enumerate}[$(1)$]
  \item $C^{B^{\prime\prime\prime}}$ given in (\ref{eighth renorm C, Q}) can be written as
\begin{align}\label{calculate C^B'''}
  C^{B^{\prime\prime\prime}}=&\frac{2\pi N(N-1)}{m}
+O\Big(N^2\Big(\frac{a^2}{d^2\textit{l}}
+\Big(\frac{a}{d}+\frac{1}{h^2m}\Big)(d\textit{l})^2
+\frac{\textit{l}^{\frac{1}{2}}}{m}\sqrt{\frac{a}{d}}\Big)\Big)\nonumber\\
&+O\Big(N^2\Big(\frac{a\textit{l}
\ln (d\textit{l})^{-1}}
{dm^2}+\frac{\ln h^{-1}}{m^2}
+\frac{a^{\frac{1}{3}}}{d^{\frac{1}{3}}m^{\frac{5}{3}}\textit{l}}\Big)\Big)
\end{align}
Moreover, if we set $Ng=\mathfrak{a}_0$ and $N\sim\frac{d}{a}$ (i.e. we are taking limit in Rrgion III in the Gross-Pitaevskii regime), and choose
\begin{equation}\label{thm choose l,h6}
 \textit{l}=\frac{1}{4},\quad h=N^{-\frac{13}{2}}.
\end{equation}
then we have
\begin{align}\label{thm E_N III6}
   C^{B^{\prime\prime\prime}}
   =4\pi(N-1)Ng+\mathcal{I}_{N,a,d},
  \end{align}
where the second order term $\mathcal{I}_{N,a,d}$ (or $\mathcal{I}_N$ for short) is given by
\begin{align}\label{thm I_N,a,d III6}
\mathcal{I}_{N}=&(N-1)N\big(\mathcal{C}_N-4\pi g\big)
+\frac{1}{2}\sum_{p\in2\pi\mathbb{Z}^3\backslash\{0\}}\Big\{
-\vert\mathcal{M}_dp\vert^2-2N\mathcal{C}_N\nonumber\\
&+\sqrt{\vert\mathcal{M}_dp\vert^4+4N\mathcal{C}_N\vert\mathcal{M}_dp\vert^2
+4N^2\big(\mathcal{C}_N^2-(q_p+Y_p)\big)}\Big\}.
\end{align}
with
\begin{align}\label{thm C_N III6}
  \mathcal{C}_N=\Big(W_0+\sum_{p\neq0}W_p\eta_p+\sum_{p\neq0}\big(W_p+Y_p+\mathfrak{D}_p\big)\xi_p
+\sum_{p\neq0}\big(2Y_p+\mathfrak{D}_p+q_p\big)k_p\Big).
\end{align}
The coefficients arising in (\ref{thm C_N III6}) are defined around equations (\ref{introduction asymptotic energy pde on the ball}), (\ref{introduction asymptotic energy pde on the ball 2D}) and (\ref{intro dimensional coupling scattering equation}). Moreover, it can be bounded that
\begin{equation}\label{thm bound I_N6}
  \mathcal{I}_{N}=O\Big(N\sqrt{\frac{a}{d}}+\ln N\Big)\ll N.
\end{equation}
Furthermore, the above results still hold true when we improve $\frac{a}{d}\sim N^{-1}$ to
\begin{equation}\label{thm condition III 6}
  N\Big(\frac{a}{d}\Big)^{\frac{19}{18}-r}\to0
\end{equation}
for some $r\in (0,\frac{1}{18})$ (not necessarily fixed).
  \item $\mathcal{Q}^{{\prime\prime\prime}}$ given in (\ref{eighth renorm C, Q}) can be written as
\begin{equation}\label{calculate Q'''}
  \mathcal{Q}^{{\prime\prime\prime}}=\tilde{\mathcal{D}}+\tilde{\delta}
\end{equation}
where
\begin{equation}\label{define Dtilde}
  \tilde{\mathcal{D}}=\sum_{p\neq0}\sqrt{\vert\mathcal{M}_dp\vert^4+\frac{8\pi N}{m}
\vert\mathcal{M}_dp\vert^2}a_p^*a_p
\end{equation}
and $\tilde{\delta}$ is bounded by
\begin{equation}\label{define deltatilde}
  \pm\tilde{\delta}\leq
\Big(h+N\Big(\frac{a^2}{d^2\textit{l}}
+\Big(\frac{a}{d}+\frac{1}{h^2m}\Big)(d\textit{l})^2
+\frac{\textit{l}^{\frac{1}{2}}}{m}\sqrt{\frac{a}{d}}\Big)\Big)\mathcal{N}_+.
\end{equation}
\end{enumerate}
\end{lemma}
\noindent
\emph{Proof.} From the definition of $\tilde{F}_p$ and $\tilde{G}_p$ (\ref{define F_p,G_p III}) and Lemma \ref{lemma tau_p III}, we infer that
\begin{equation*}
  \Big\vert\Big(-\tilde{F}_p+\sqrt{\tilde{F}_p^2-\tilde{G}_p^2}\Big)\Big\vert
\leq \frac{\tilde{G}_p^2}{\tilde{F}_p+\sqrt{\tilde{F}_p^2-\tilde{G}_p^2}}
\leq \frac{CN^2\big(q_p^2+Y_p^2\big)}{\vert\mathcal{M}_dp\vert^2}.
\end{equation*}
Then (\ref{calculate C^B'''}) and (\ref{thm E_N III6}) follow by combining this inequality with Lemmas \ref{xi_p Y_p lemma}, \ref{q_p lemma}, \ref{bound Ctilde^O'} and estimates (\ref{collect1}) and (\ref{collect2}).

\par As for the second part of this lemma, we just need to notice that
\begin{align*}
 &\sqrt{\tilde{F}_p^2-\tilde{G}_p^2}-\sqrt{\vert\mathcal{M}_dp\vert^4+\frac{8\pi N}{m}
\vert\mathcal{M}_dp\vert^2}\\
=&\frac{\Big(4N\tilde{C}^{\mathcal{O}^\prime}-\frac{8\pi N}{m}\Big)\vert\mathcal{M}_dp\vert^2
+\Big(4N^2({\tilde{C}^{\mathcal{O}^\prime}})^2-\tilde{G}_p^2\Big)}
{\sqrt{\tilde{F}_p^2-\tilde{G}_p^2}+\sqrt{\vert\mathcal{M}_dp\vert^4+\frac{8\pi N}{m}
\vert\mathcal{M}_dp\vert^2}}
\end{align*}
This equality with (\ref{est of Ctilde ^O'}) and (\ref{lemma tau_p 1 III}) prove the desired estimates.
\begin{flushright}
  $\Box$
\end{flushright}

\subsection{Optimal BEC for Region III}\label{subsection Optimal BEC for Region III}
\
\par Let $\tilde{E}_0=C^{B^{\prime\prime\prime}}$, $\tilde{\mathcal{E}}=\mathcal{E}^{B^{\prime\prime\prime}}+\tilde{\delta}$, and
\begin{equation}\label{define Utilde}
  \tilde{U}=e^{B}e^{B^\prime}e^{\tilde{B}}e^{\tilde{B}^\prime}e^{\mathcal{O}}
e^{\mathcal{O}^{\prime}}e^{B^{\prime\prime\prime}}.
\end{equation}
Then Proposition \ref{Bog renorm III} and \ref{lemma calculate C^B'''} together lead to
\begin{equation}\label{rewrite Z^III_N}
\mathcal{Z}_N^{{III}}=\tilde{\mathcal{U}}^*H_N\tilde{\mathcal{U}}=
\tilde{E}_0+\tilde{\mathcal{D}}+e^{-B^{\prime\prime\prime}}H_4e^{B^{\prime\prime\prime}}
+\tilde{\mathcal{E}}.
\end{equation}
Now we state the BEC result for Region III parallel to Proposition \ref{Optimal BEC}.
\begin{proposition}\label{Optimal BEC III}
Let $N$ tends to infinity and $a,d$ and $\frac{a}{d}$ tend to $0$ with the G-P restriction for Region $\mathrm{III}^\prime$ i.e. $\frac{d}{a}\lesssim\vert\ln(Nd^2)\vert$ and $Ng=\mathfrak{a}_0$. Let $\textit{l}$ and $h$ be as chosen in (\ref{choose l, h III}) with $\alpha$, $\beta$ and $c$ determined by
\begin{equation}\label{choose alpha beta c}
  \alpha=0,\quad,\beta=\frac{13}{2},\quad c=\frac{1}{4}.
\end{equation}
Assume further that
\begin{equation}\label{condition III}
  N\Big(\frac{a}{d}\Big)^{\frac{19}{18}-r}\to0
\end{equation}
for some $r\in (0,\frac{1}{18})$ (not necessarily fixed). Then there exists a universal constant $C$ such that
\begin{equation}\label{BEC III}
  H_N\geq \tilde{E}_0+C^{-1}\mathcal{N}_+-C.
\end{equation}
\end{proposition}
\noindent
\emph{Proof.} We stick to the notations used in the proof of Proposition \ref{Optimal BEC}. From (\ref{H_N region I}) and (\ref{E_M}) we have
\begin{equation}\label{H_N region III}
   H_N=f_M(\mathcal{N}_+)H_Nf_M(\mathcal{N}_+)+
g_M(\mathcal{N}_+)H_Ng_M(\mathcal{N}_+)+\mathcal{E}_M,
\end{equation}
and
\begin{equation}\label{E_M III}
  \pm\mathcal{E}_M\leq CM^{-2}P_X(H_4+N^2ad^{-1})P_{X}.
\end{equation}
Here $X$ has been defined in (\ref{define X}). Now we choose
\begin{equation}\label{choose M III}
  M=N\Big(\frac{a}{d}\Big)^{\frac{1}{9}-r}.
\end{equation}
It is clear that in Region III, $M$ tends to infinity. We apply Proposition \ref{Bog renorm III} and Lemma \ref{calculate C^B'''} with
\begin{equation}\label{2024010101}
  \vartheta_1=\Big(\frac{a}{d}\Big)^{\frac{1}{18}},\quad
\textit{l}=\frac{1}{4},\quad h=N^{-\frac{13}{2}},
\end{equation}
we then obtain
\begin{align}
f_M(\mathcal{N}_+)&H_Nf_M(\mathcal{N}_+)
\geq\tilde{\mathcal{U}}f_M(\mathcal{N}_+^{\tilde{\mathcal{U}}})
(\tilde{E}_0+H_{21}+e^{-B^{\prime\prime\prime}}H_4e^{B^{\prime\prime\prime}}+\tilde{\mathcal{E}})
f_M(\mathcal{N}_+^{\tilde{\mathcal{U}}})\tilde{\mathcal{U}}^*,\nonumber
\end{align}
with
\begin{align*}
  \tilde{\mathcal{U}}f_M(\mathcal{N}_+^{\tilde{\mathcal{U}}})\tilde{\mathcal{E}}
f_M(\mathcal{N}_+^{\tilde{\mathcal{U}}})\tilde{\mathcal{U}}^*
\geq& -C \tilde{\mathcal{U}}f_M(\mathcal{N}_+^{\tilde{\mathcal{U}}})
\Big\{N\Big(\frac{a}{d}\Big)^{\frac{19}{18}-r}(\mathcal{N}_++1)\\
&+\Big(\frac{a}{d}\Big)^{\frac{1}{18}}H_{21}
+N^{-3}e^{-B^{\prime\prime\prime}}H_4e^{B^{\prime\prime\prime}}\Big\}
f_M(\mathcal{N}_+^{\tilde{\mathcal{U}}})\tilde{\mathcal{U}}^*
\end{align*}
These together yield
\begin{equation}\label{f_M III}
  f_M(\mathcal{N}_+)H_Nf_M(\mathcal{N}_+)
\geq f_M(\mathcal{N}_+)^2(\tilde{E}_0+C^{-1}\mathcal{N}_+-C).
\end{equation}
On the other hand, we can argue by contradiction (see the proof of (\ref{mechante2}) for details) to get
\begin{equation}\label{g_M III}
   g_M(\mathcal{N}_+)(H_N-\tilde{E}_0)g_M(\mathcal{N}_+)\geq C^{-1}Ng_M(\mathcal{N}_+)^2
\geq C^{-1}\mathcal{N_+}g_M(\mathcal{N}_+)^2.
\end{equation}
We then combine (\ref{H_N region III}), (\ref{E_M III}), (\ref{f_M III}) and (\ref{g_M III}) to get, with $M$ chosen in (\ref{choose M III}), that
\begin{equation}\label{temporary1}
  H_N\geq\tilde{E}_0+C^{-1}\mathcal{N}_+-C-CM^{-2}H_4
\end{equation}
and therefore using Lemmas \ref{control of S_+ conj with e^B} and \ref{lemma control of H_4}
\begin{equation}\label{temporary2}
  \mathcal{G}_N\geq \tilde{E}_0+C^{-1}\mathcal{N}_+-C-CM^{-2}H_4.
\end{equation}

\par On the other hand, if we apply Proposition \ref{Bog renorm III} and Lemma \ref{calculate C^B'''} with $\textit{l}$ and $h$ still as in (\ref{2024010101}), while we fix $\vartheta_1$ small but universal, we then find
\begin{equation}\label{temporary3}
  \mathcal{Z}_N^{III}\geq \tilde{E}_0
+Ce^{-B^{\prime\prime\prime}}H_4e^{B^{\prime\prime\prime}}
-C\big(Nad^{-1}+\ln N\big)(\mathcal{N}_++1).
\end{equation}
Conjuagting back using Lemmas \ref{control of S_+ conj with e^B'}, \ref{lemma e^-B'H_Ne^B'}, \ref{control of S_+ conj with e^B tilde}, \ref{lemma control of H_4 with e^B tilde}, \ref{control of S_+ conj with e^B' tilde}, \ref{lemma control of H_N conj with e^Btilde'}, \ref{control of S_+ conj with e^O}, \ref{lemma control of H_4 with e^O}, \ref{control of S_+ conj with e^O'}, \ref{lemma control e^-tO H_21, H_4e^tO} and \ref{lemma control bogoliubov III} stated in the subsequent sections controlling the unitary actions on $(\mathcal{N}_++1)$, $H_{21}$ and $H_4$, we arrive at
\begin{equation}\label{temporary4}
  \mathcal{G}_N\geq \tilde{E}_0+CH_4-C\big(Nad^{-1}+\ln N\big)(\mathcal{N}_++1)
-N^{-\frac{7}{2}}H_{21}-N^2ad^{-1}.
\end{equation}
On the subspace where $H_N\lesssim N^2$ (therefore $H_{21}\leq H_N\lesssim N^2$), we combine (\ref{temporary2}) with (\ref{temporary4}) to get
\begin{equation*}
  \mathcal{G}_N\geq \tilde{E}_0+C^{-1}\mathcal{N}_+-C.
\end{equation*}
Using Lemma \ref{control of S_+ conj with e^B} we have proved (\ref{BEC III}) on the subspace where $H_N\lesssim N^2$, while (\ref{BEC III}) holds true trivially on the subspace where $H_N\gtrsim N^2$. Hence we conclude the proof of Proposition \ref{Optimal BEC III}.

\begin{flushright}
  $\Box$
\end{flushright}

\subsection{Proof of Theorem \ref{core III} for Region III}\label{proof III}
\
\par Similar to Section \ref{proof I}, we are going to prove Theorem \ref{core III} for Region III and part of Region $\mathrm{III}^\prime$ by comparing the eigenvalues of $\tilde{\mathcal{H}}=H_N-\tilde{E}_0$ with $\tilde{\mathcal{D}}$, while the eigenvalues of $\tilde{\mathcal{D}}$ can be explicitly calculated as in (\ref{eigen of D}). We choose as stated in the proof of Proposition \ref{Optimal BEC III} that
\begin{equation*}
 \textit{l}=\frac{1}{4},\quad h=N^{-\frac{13}{2}}
\end{equation*}
and we set
\begin{equation*}
  N\Big(\frac{a}{d}\Big)^{\frac{19}{18}-r}\to0
\end{equation*}
for some $r\in (0,\frac{1}{18})$ (not necessarily fixed). We will show that for any $j\in\mathbb{N}$ such that $\lambda_j(\tilde{\mathcal{D}})\leq\zeta$ with a threshold $1\leq\zeta\ll (\frac{d}{a})^{\frac{5}{12}}$, there holds
\begin{equation}\label{threshold estimate III}
  \vert\lambda_j(\tilde{\mathcal{H}})-\lambda_j(\tilde{\mathcal{D}})\vert\leq C\Big(N\Big(\frac{a}{d}\Big)^{\frac{9}{8}}
+\Big(\frac{a}{d}\Big)^{\frac{1}{8}}\ln N\Big)(1+\zeta^3).
\end{equation}
Then Theorem \ref{core III} follows by choosing $\zeta$ being a universal constant.

\par We follow Section \ref{proof I} to prove (\ref{threshold estimate III}). Let $Z\subset L^2_s(\Lambda_d^N)$ be the subspace generated by the first $j$ eigenvectors of $\tilde{\mathcal{D}}$ and let $P_Z$ be the orthogonal projection onto it. We apply Proposition \ref{Bog renorm III} and Lemma \ref{calculate C^B'''} with $\textit{l}$ and $h$ chosen above and
\begin{equation*}
  \vartheta_1=\Big(\frac{a}{d}\Big)^{\frac{1}{2}}.
\end{equation*}
Then we have
\begin{align}\label{eir1}
  \lambda_j(\tilde{\mathcal{D}})&\geq P_Z\tilde{\mathcal{D}}P_Z\geq
P_Z(\tilde{\mathcal{U}}^*\tilde{\mathcal{H}}\tilde{\mathcal{U}}-
e^{-B^{\prime\prime\prime}}H_4e^{B^{\prime\prime\prime}}-\tilde{\mathcal{E}})P_Z\nonumber\\
&\geq P_Z\tilde{\mathcal{U}}^*\tilde{\mathcal{H}}\tilde{\mathcal{U}}P_Z
-\Big(N\Big(\frac{a}{d}\Big)^{\frac{3}{2}}
+\Big(\frac{a}{d}\Big)^{\frac{1}{2}}\ln N\Big)(1+\zeta^3).
\end{align}
In the last inequality of (\ref{eir1}) we have used inequality (\ref{control H_4 with e^B''' 1}) in Lemma \ref{lemma control bogoliubov III} and inequality (\ref{bound on H_4 on Z}). Then the min-max principle and (\ref{eir1}) together yield
\begin{equation}\label{eir1.1}
   \lambda_j(\tilde{\mathcal{D}})\geq \lambda_j(\tilde{\mathcal{H}})
-\Big(N\Big(\frac{a}{d}\Big)^{\frac{3}{2}}
+\Big(\frac{a}{d}\Big)^{\frac{1}{2}}\ln N\Big)(1+\zeta^3).
\end{equation}

\par As for the other side of the inequality (\ref{threshold estimate III}), we let the space generated by the first $j$ eigenvectors of $\tilde{\mathcal{H}}$ be $V\subset L^2_s(\Lambda_d^N)$ and the orthogonal projection onto it be $P_V$. Using
\begin{equation}\label{eir2.1}
  \tilde{\mathcal{H}}=f_M(\mathcal{N}_+)\tilde{\mathcal{H}}f_M(\mathcal{N}_+)
+g_M(\mathcal{N}_+)\tilde{\mathcal{H}}g_M(\mathcal{N}_+)+{\mathcal{E}_M}
\end{equation}
with
\begin{equation*}
  M=N\Big(\frac{a}{d}\Big)^{\frac{1}{4}},
\end{equation*}
then
\begin{equation}\label{eir2.3}
  \lambda_m(\tilde{\mathcal{H}})=\lambda_m(\tilde{\mathcal{U}}^*
\tilde{\mathcal{H}}\tilde{\mathcal{U}})
\geq P_V(f_M(\mathcal{N}_+)\tilde{\mathcal{H}}f_M(\mathcal{N}_+)
+g_M(\mathcal{N}_+)\tilde{\mathcal{H}}g_M(\mathcal{N}_+)+\mathcal{E}_M)P_V.
\end{equation}
From Lemma \ref{Optimal BEC III} and the definition of $g_M$ we find that
\begin{equation}\label{eir2.4}
 \begin{aligned}
 P_Vg_M(\mathcal{N}_+)\tilde{\mathcal{H}}g_M(\mathcal{N}_+)P_V&\geq
P_Vg_M(\mathcal{N}_+)(C^{-1}\mathcal{N}_+-C)g_M(\mathcal{N}_+)P_V\\
&\geq P_Vg_M(\mathcal{N}_+)^2P_V(C^{-1}M-C)\geq0.
\end{aligned}
\end{equation}
To bound $P_V\mathcal{E}_MP_V$ in (\ref{eir2.3}), we first derive from (\ref{temporary4}) together with Lemmas \ref{control of S_+ conj with e^B} and \ref{lemma control of H_4} that
\begin{equation}\label{eir2.5}
  H_4\leq C\Big(\tilde{\mathcal{H}}
+N^2ad^{-1}+\big(Nad^{-1}+\ln N\big)(\mathcal{N}_++1)+N^{-\frac{7}{2}}H_{21}\Big).
\end{equation}
Moreover, we notice that
\begin{equation}\label{eir 2.55}
  P_VH_{21}P_V\leq P_VH_NP_V=P_V\tilde{H}P_V+P_V\tilde{E}_0P_V.
\end{equation}
We then combine (\ref{eir2.5}), (\ref{eir 2.55}), the estimate of $\tilde{E}_0$ in (\ref{calculate C^B'''}), (\ref{BEC III}) in Lemma \ref{Optimal BEC III} and the already proved bound (\ref{eir1.1}), together to find
\begin{equation}\label{eir2.6}
  P_V H_4P_V\leq C\Big(N^2ad^{-1}+\big(Nad^{-1}+\ln N\big)(1+\zeta^3)\Big).
\end{equation}
Combining (\ref{eir2.6}) with (\ref{E_M III}) we have
\begin{equation}\label{eir2.7}
  P_V\mathcal{E}_MP_V\geq -C\Big(\frac{a}{d}\Big)^{\frac{1}{2}}(1+\zeta^3).
\end{equation}
To bound $P_Vf_M(\mathcal{N}_+)\tilde{\mathcal{H}}f_M(\mathcal{N}_+)P_V$ in (\ref{eir2.3}), we first apply Propostion \ref{Bog renorm III} and set
\begin{equation*}
 \vartheta_1=\Big(\frac{a}{d}\Big)^{\frac{1}{8}},
\end{equation*}
then
\begin{align}
 &P_Vf_M(\mathcal{N}_+)\tilde{\mathcal{H}}f_M(\mathcal{N}_+)P_V
\geq -C\Big(N\Big(\frac{a}{d}\Big)^{\frac{9}{8}}
+\Big(\frac{a}{d}\Big)^{\frac{1}{8}}\ln N\Big)\nonumber\\
&+\Big(1-C\Big(N\Big(\frac{a}{d}\Big)^{\frac{9}{8}}
+\Big(\frac{a}{d}\Big)^{\frac{1}{8}}\ln N\Big)\Big)
P_Vf_M(\mathcal{N}_+)\mathcal{U}\tilde{\mathcal{D}}
\mathcal{U}^*f_M(\mathcal{N}_+)P_V.\label{eir2.8}
\end{align}
From Lemma \ref{lemma lewinNamSerfatySolovej2015bogoliubov} again, since we can check on the Fock space that
\begin{align*}
 g_M(\mathcal{N}_+)^2\leq CM^{-1}g_M(\mathcal{N}_+)\mathcal{N}_+g_M(\mathcal{N}_+)
\leq CM^{-1}\mathcal{N}_+.
\end{align*}
Using Lemma \ref{Optimal BEC III} and the bound (\ref{eir1.1}) we derive
\begin{equation*}
  P_V\mathcal{N}_+P_V\leq C(1+\zeta)+C\Big(N\Big(\frac{a}{d}\Big)^{\frac{3}{2}}
+\Big(\frac{a}{d}\Big)^{\frac{1}{2}}\ln N\Big)(1+\zeta^3).
\end{equation*}
These observations yield
\begin{equation*}
  P_Vg_M(\mathcal{N}_+)^2P_V \leq CM^{-1}\Big\{1+\zeta+\Big(N\Big(\frac{a}{d}\Big)^{\frac{3}{2}}
+\Big(\frac{a}{d}\Big)^{\frac{1}{2}}\ln N\Big)\zeta^3\Big\}.
\end{equation*}
This tends to $0$ in the Gross-Pitaevskii limit in Region I as long as we have $\zeta\ll (\frac{d}{a})^{\frac{5}{12}}$. Then Lemma \ref{lemma lewinNamSerfatySolovej2015bogoliubov} guarantees that
\begin{align*}
  \dim(f_M(\mathcal{N}_+)V)=\dim V=j.
\end{align*}
Therefore we use the above estimates and min-max principle to get
\begin{align}\label{lower bound III}
 \lambda_j(\tilde{\mathcal{H}})
\geq \lambda_j(\tilde{\mathcal{D}})-C\Big(N\Big(\frac{a}{d}\Big)^{\frac{9}{8}}
+\Big(\frac{a}{d}\Big)^{\frac{1}{8}}\ln N\Big).
\end{align}
Hence we conclude the claim (\ref{threshold estimate III}) using (\ref{eir1.1}) and (\ref{lower bound III}) and thus finish the proof of Theorem \ref{core III} for Region III.

\begin{flushright}
  $\Box$
\end{flushright}

\subsection{Proof of Theorem \ref{core III} for Region \texorpdfstring{$\mathrm{II}_{\mathrm{III}}$}{II-III}}
\label{proof II2}
\
\par As long as we still choose
\begin{equation*}
\textit{l}=\frac{1}{4},\quad h=N^{-\frac{13}{2}},
\end{equation*}
 the results of Propositons \ref{Bog renorm III} and Lemma \ref{calculate C^B'''} still hold in the Gross-Pitaevskii regime in Region $\mathrm{II}_{\mathrm{III}}$, and thus the arguments carried out in Sections \ref{subsection Optimal BEC for Region III} and \ref{proof III} still apply to Region $\mathrm{II}_{\mathrm{III}}$.

 \par Here we fix $t_2\in (0,t_1)$ where $t_1$ has been chosen in (\ref{choose t1}). Notice in Region $\mathrm{II}_{\mathrm{III}}$ that we also have relation similar to (\ref{relation m and N})
\begin{equation}\label{relatin m n II}
  \frac{2\mathfrak{a}_0m}{N}\stackrel{{\text{G-P limit}}}{\longrightarrow}1.
\end{equation}

\par To prove Theorem \ref{core III} for Region $\mathrm{II}_{\mathrm{III}}$, it suffices to go through the arguments in Sections \ref{subsection Optimal BEC for Region III} and \ref{proof III} with some different choices of parameters. Proceeding as in the proof of Proposition \ref{Optimal BEC III}, we choose
\begin{equation*}
  M=N^{(1-\frac{s}{2})}
\end{equation*}
where $s\in(0,1)$ is stated in (\ref{rate of decay}). If we apply Propositons \ref{Bog renorm III} and Lemma \ref{calculate C^B'''} with
\begin{equation*}
 \vartheta_1=N^{-\frac{s}{4}},
\end{equation*}
then the optimal BEC (\ref{BEC III}) holds in in Region $\mathrm{II}_{\mathrm{III}}$ as well.

\par Similar to Section \ref{proof III}, we are also going to prove for any $j\in\mathbb{N}$ such that $\lambda_j(\mathcal{D})\leq\zeta$ with a threshold
\begin{equation*}
  1\leq\zeta\ll N^{\frac{11}{30}}\big(\ln N\big)^{-\frac{1}{3}},
\end{equation*}
 there holds
\begin{equation}\label{threshold estimate II2}
  \vert\lambda_j(\tilde{\mathcal{H}})-\lambda_j(\tilde{\mathcal{D}})\vert\leq CN^{-\frac{1}{5}}\ln N(1+\zeta^3).
\end{equation}
With $\textit{l}$ and $\nu$ being chosen above, the upper bound (\ref{eir1}) in this case is obtained by applying Propositons \ref{Bog renorm III} and Lemma \ref{calculate C^B'''} with
\begin{equation*}
  \vartheta_1=N^{-\frac{1}{2}}
\end{equation*}
while we reach the lower bound (\ref{lower bound III}) in this case by choosing
\begin{equation*}
M=N^{\frac{1}{2}+\frac{1}{10}},\quad
 \vartheta_1=N^{-\frac{1}{5}}
\end{equation*}
Then we can conclude the proof of Theorem \ref{core III} for Region $\mathrm{II}_{\mathrm{III}}$.

\begin{flushright}
  $\Box$
\end{flushright}

\section{3D Quadratic Renormalization for Regions I \& III}\label{2}
\par In this section we analyze the excitation Hamiltonian $\mathcal{G}_N$ and prove Propositon \ref{quadratic renorm}. We adopt the notation
\begin{equation*}
  A={\sum_{p\neq 0}}\eta_pa_p^*a_{-p}^*a_0a_0.
\end{equation*}
 By a direct calculation using the definitions of creation and annihilation operators, it is easy to check $A$ is a linear operator on $L^2_s(\Lambda_{d}^N)$ bounded by $N^2\Vert\eta_{\perp}\Vert_2$ for all $N\in\mathbb{N}$. On the other hand, by (\ref{define B}) we have
\begin{equation*}
  B=\frac{1}{2}(A-A^*),
\end{equation*}
hence $B$ is anti-symmetric and $e^B$ is unitary on $L^2_s(\Lambda_{d}^N)$ for any $N\in\mathbb{N}$. We also recall that
\begin{equation*}
  \eta_\perp={\sum_{p\neq0}}\eta_p\phi_p^{(d)}\in L^2_{\perp}(\Lambda_{d}).
\end{equation*}

\par To prove Proposition \ref{quadratic renorm} we split the Hamiltonian operator $H_N$ using (\ref{split H_N}), and analyze respectively their contributions to the ground state energy after we conjugate them with $e^B$. That is, we rewrite the renormalized Hamiltonian $e^{-B}H_Ne^B$ using (\ref{split H_N}) and Newton-Leibniz law
\begin{align}
  e^{-B}H_Ne^B=&H_{01}+e^{-B}(H_{02}+H_{22}+H_3)e^B+
  e^{-B}(H_{21}+H_4+H_{23})e^B\nonumber\\
  =&H_{01}+e^{-B}(H_{02}+H_{22}+H_3)e^B+H_{21}+H_4\nonumber\\
  &+\int_{0}^{1}e^{-tB}
  [H_{21}+H_4,B]e^{tB}dt+e^{-B}H_{23}e^{B}.\label{asfbjsdf}
\end{align}
For 3D quadratic renormalization, we aim to extract energy generated by the 3D correlation structure hidden in the quadratic term $H_{23}$, which is the main driving force of the leading order ground state energy. Therefore the term $e^{-B}(H_{21}+H_4+H_{23})e^B$ is the main object in this section. To compute it precisely, we let
\begin{equation}\label{define Gamma}
  \Gamma=[H_{21}+H_4,B]+H_{23}-{H}_{23}^{\prime},
\end{equation}
where $H_{23}^\prime$ is defined in (\ref{first renorm H_23'}).
\begin{eqnarray}\label{first renorm H_23' dsgvfsdaf}
  {H}_{23}^{\prime}={\sum_{p\neq0}}W_p(a_p^*a_{-p}^*a_0a_0+h.c.),
\end{eqnarray}
Plugging (\ref{define Gamma}) into (\ref{asfbjsdf}) we obtain
\begin{align}
&e^{-B}H_Ne^B=H_{01}+e^{-B}(H_{02}+H_{22}+H_3)e^B+H_{21}+H_4\nonumber\\
&+\int_{0}^{1}e^{-tB}(\Gamma+{H}_{23}^{\prime}-H_{23})e^{tB}dt
+e^{-B}H_{23}e^{B}\nonumber\\
=&H_{01}+e^{-B}(H_{02}+H_{22}+H_3)e^B+H_{21}+H_4+{H}_{23}^{\prime}
+\int_{0}^{1}e^{-tB}\Gamma e^{tB}dt\nonumber\\
&+\int_{0}^{1}(e^{-B}H_{23}e^{B}-e^{-tB}H_{23}e^{tB})dt
+\int_{0}^{1}(e^{-tB}{H}_{23}^{\prime}e^{tB}-{H}_{23}^{\prime})dt\nonumber\\
=&H_{01}+H_{21}+H_4+{H}_{23}^{\prime}+e^{-B}(H_{02}+H_{22}+H_3)e^B
+\int_{0}^{1}e^{-tB}\Gamma e^{tB}dt\nonumber\\
&+\int_{0}^{1}\int_{t}^{1}e^{-sB}[H_{23},B]e^{sB}dsdt
+\int_{0}^{1}\int_{0}^{t}e^{-sB}[{H}_{23}^{\prime},B]e^{sB}dsdt.\label{split e^-BH_Ne^B}
\end{align}

Then the proof of Proposition \ref{quadratic renorm} is done by analyzing each terms on the right-hand side of (\ref{split e^-BH_Ne^B}). We reiterate that we state the results for Regions I or III. As for Regions $\mathrm{II}_{\mathrm{I}}$ and $\mathrm{II}_{\mathrm{III}}$, they are regarded as intermediate regions, and corresponding results can still be applied to these regions without further specifications. In the following lemmas, we bound $e^{-B}H_{02}e^B$ in Lemma \ref{lemma e^-BH_02e^B}, $e^{-B}H_{22}e^B$ in Lemma \ref{lemma e^-BH_22e^B}, $e^{-B}H_{3}e^B$ in Lemma \ref{lemma e^-BH_3e^B}. These three terms stay unchanged up to small errors after conjugating with $e^B$, or can be rewritten in the form of polynomials of $\mathcal{N}_+$. The term containing the difference $\Gamma$ is bounded in Lemma \ref{lemma Gamma}. This term is a negligible error term as we will prove. The contribution of the commutator $[H_{23},B]$ is calculated in Lemma \ref{lemma [H_23,B]}, and the contribution of $[{H}_{23}^{\prime},B]$ is calculated in Lemma \ref{lemma [H_23',B]}. As aforementioned, Lemmas \ref{lemma [H_23,B]} and \ref{lemma [H_23',B]} present the major contributions of the quadratic 3D correlation structure to the first and second order ground state energy, in the form of polynomials of $\mathcal{N}_+$.

\par For our analysis, it is useful to control the action of $e^B$ on the number of excited particles operator $\mathcal{N}_+$. We state the results in Lemma \ref{control of S_+ conj with e^B}. Moreover, although not used in this section, it is also important to have a bound that controls the growth of $H_{21}$ and $H_4$ with respect to the action of $e^B$. We present them in Lemma \ref{lemma control of H_4}.
\begin{lemma}\label{control of S_+ conj with e^B}
Let $\mathcal{N_+}$ be defined on $L_s^2(\Lambda_d^N)$ as stated in (\ref{define of N_+}), then there exist a constant $C_n$ depending only on $n\in\frac{1}{2}\mathbb{N}$ such that: for every $t\in\mathbb{R}$, $N\in\mathbb{N}$, $n\in\frac{1}{2}\mathbb{N}$, $\textit{l}\in(0,\frac{1}{2})$ and $\frac{d}{a}>\frac{C}{\textit{l}}$ for some universal constant $C$. Then we have
\begin{align}
  e^{-tB}(\mathcal{N}_++1)^ne^{tB}&\leq e^{C_nNa\textit{l}^{\frac{1}{2}}\vert t\vert}(\mathcal{N}_++1)^n,\label{e^-tB(N_++1)e^tB}\\
  \pm(e^{-tB}(\mathcal{N}_++1)^ne^{tB}-(\mathcal{N}_++1)^n)&\leq (e^{C_nNa\textit{l}^{\frac{1}{2}}\vert t\vert}-1) (\mathcal{N}_++1)^n.\label{e^-tBN_+e^tB-N_+}
\end{align}
\end{lemma}
\noindent
\emph{Proof.} By a direct calculation we have
\begin{equation*}
  [\mathcal{N}_+,A]
={\sum_{p\neq0}}(\eta_pa_p^*a_{-p}^*a_0a_0+\eta_{-p}a_{-p}^*a_{p}^*a_0a_0)
  =2A,
\end{equation*}
which leads to $[\mathcal{N}_+,B]=A+A^*$.

\par We take up (\ref{e^-tB(N_++1)e^tB}) for $n=1$ first. Let $\psi\in L^2_s(\Lambda_{d}^N)$, and denote
\begin{equation*}
  f(t)=\langle e^{-tB}(\mathcal{N}_++1)e^{tB}\psi,\psi\rangle\geq 0
\end{equation*}
which is a non-negative smooth function for $t\in\mathbb{R}$. Its derivative is given by
\begin{eqnarray}\label{f'(t) eBN_+e-B}
  f^{\prime}(t)=\langle e^{-tB}[(\mathcal{N}_++1),B]e^{tB}\psi,\psi\rangle
  =2\re\langle Ae^{tB}\psi,e^{tB}\psi\rangle.
\end{eqnarray}
Using the unitary operator $U_N$ defined in (\ref{define U_N,d}) we denote
\begin{align*}
  U_{N-2}a_0a_0\psi&=(\alpha^{(0)},\ldots,\alpha^{(N-2)}),\\
  U_N\psi&=(\beta^{(0)},\ldots,\beta^{(N)}).
\end{align*}
We omit the $d$ subscript for succinctness. Calculating directly using definitions we find
\begin{align*}
  &\langle A\psi,\psi\rangle=
  {\sum_{p\neq0}}\eta_p\langle U_Na_p^*a_{-p}^*U_{N-2}^*U_{N-2}a_0a_0\psi,U_N\psi\rangle
  ={\sum_{p\neq0}}\sum_{n=2}^{N}\eta_p\langle a_p^*a_{-p}^*\alpha^{(n-2)},\beta^{(n)}\rangle\\
  =&\sum_{n=2}^{N}\int_{\Lambda_{d}^n}\frac{1}{\sqrt{dn(n-1)}}\sum_{i\neq j}^{n}\eta_\perp(\mathbf{x}_i-\mathbf{x}_j)\alpha^{(n-2)}
(\mathbf{x}_1,\dots,\widehat{\mathbf{x}_i},\widehat{\mathbf{x}_j},\dots,\mathbf{x}_n)
  \overline{\beta^{(n)}}(\mathbf{x}_1,\dots,\mathbf{x}_n)\\
  =&\sum_{n=2}^{N}\sqrt{\frac{n(n-1)}{d}}\int_{\Lambda_{d}^n}
\eta_\perp(\mathbf{x}_1-\mathbf{x}_2)\alpha^{(n-2)}
(\mathbf{x}_3,\dots,\mathbf{x}_n)\overline{\beta^{(n)}}(\mathbf{x}_1,\dots,\mathbf{x}_n).
\end{align*}
The last equality holds since both $\alpha^{(n)}$ and $\beta^{(n)}$ are symmetric functions. Applying Cauchy-Schwartz inequality,
\begin{align*}
  \vert\langle A\psi,\psi\rangle\vert&\leq\sum_{n=2}^{N}\sqrt{\frac{n(n-1)}{d}}
  \left(\int_{\Lambda_{d}^n}\vert\eta_\perp(\mathbf{x}_1-\mathbf{x}_2)\vert^2
  \vert\alpha^{(n-2)}(\mathbf{x}_3,\dots,\mathbf{x}_n)\vert^2\right)^{\frac{1}{2}}
  \left(\int_{\Lambda_{d}^n}\vert\beta^{(n)}\vert^2\right)^{\frac{1}{2}}\nonumber\\
  &=\Vert\eta_\perp\Vert_2\sum_{n=2}^{N}\sqrt{n(n-1)}\Vert\alpha^{(n-2)}\Vert_2
  \Vert\beta^{(n)}\Vert_2\nonumber\\
  &\leq\Vert\eta_\perp\Vert_2\left(\sum_{n=2}^{N}(n-1)
  \Vert\alpha^{(n-2)}\Vert_2^2\right)^{\frac{1}{2}}
  \left(\sum_{n=2}^{N}n\Vert\beta^{(n)}\Vert_2^2\right)^{\frac{1}{2}}\nonumber.
\end{align*}
Switching back to the original $L_s^2(\Lambda_d^N)$ space we get
\begin{align}
  \vert\langle A\psi,\psi\rangle\vert&\leq
  \Vert\eta_\perp\Vert_2\langle (\mathcal{N}_++1)^{\frac{1}{2}}a_0a_0\psi,
(\mathcal{N}_++1)^{\frac{1}{2}}a_0a_0\psi\rangle^{\frac{1}{2}}
  \langle (\mathcal{N}_++1)\psi,\psi\rangle^{\frac{1}{2}}\nonumber\\
  &=\Vert\eta_\perp\Vert_2\langle a_0a_0(\mathcal{N}_++1)^{\frac{1}{2}}\psi,
a_0a_0(\mathcal{N}_++1)^{\frac{1}{2}}\psi\rangle^{\frac{1}{2}}
  \langle (\mathcal{N}_++1)\psi,\psi\rangle^{\frac{1}{2}}\nonumber\\
  &\leq CNa\textit{l}^{\frac{1}{2}}\langle (\mathcal{N}_++1)\psi,\psi\rangle,\label{Abs Apsi,psi}
\end{align}
where we have used the fact that $a_0$ commutes with $\mathcal{N}_+$. In the last inequality above we apply (\ref{norm of a_p and a_p*}) to bound $\Vert a_0\Vert$ and (\ref{est of eta and eta_perp}) to control $\Vert\eta_\perp\Vert_2$ (Notice that (\ref{est of eta and eta_perp}) holds when $\frac{d}{a}>\frac{C}{\textit{l}}$). Combining (\ref{f'(t) eBN_+e-B}) and
(\ref{Abs Apsi,psi}) we  get
\begin{equation}\label{gronwall f' eBN_+e-B}
\vert f^{\prime}(t)\vert\leq CNa\textit{l}^{\frac{1}{2}}\langle (\mathcal{N}_++1)e^{tB}\psi,e^{tB}\psi\rangle
=CNa\textit{l}^{\frac{1}{2}}f(t).
\end{equation}
Since $f^{\prime}$ is real-valued, by Gronwall's inequality we have proved (\ref{e^-tB(N_++1)e^tB}) for $n=1$:
\begin{equation}\label{f(t)<f(0)}
  f(t)\leq e^{CNa\textit{l}^{\frac{1}{2}}\vert t\vert}f(0).
\end{equation}

\par As for (\ref{e^-tBN_+e^tB-N_+}) when $n=1$ we observe that
\begin{equation}\label{int repre e^-tBN_+e^tB-N_+}
  e^{-tB}\mathcal{N}_+e^{tB}-\mathcal{N}_+=\int_{0}^{t}e^{-tB}[\mathcal{N}_+,B]e^{tB}dt.
\end{equation}
Combining (\ref{gronwall f' eBN_+e-B}), (\ref{f(t)<f(0)}) and  (\ref{int repre e^-tBN_+e^tB-N_+}) we reach (\ref{e^-tBN_+e^tB-N_+}) for $n=1$.

\par To prove (\ref{e^-tB(N_++1)e^tB}) for all $n\in\mathbb{N}$, we first observe that
\begin{align}
  [(\mathcal{N}_++1)^n,B]&=\sum_{k=0}^{n-1}(\mathcal{N}_++1)^k[\mathcal{N}_++1,B]
  (\mathcal{N}_++1)^{n-k-1}\nonumber\\
  &=\sum_{k=0}^{n-1}(\mathcal{N}_++1)^k(A+A^*)(\mathcal{N}_++1)^{n-k-1}.\label{[N_++1^n,B]}
\end{align}
We assert that for each non-negative integer $k\leq n-1$ there is an operator inequality
\begin{align}
  \pm[(\mathcal{N}_++1)^k(A+A^*)(\mathcal{N}_++1)^{n-k-1}+&
  (\mathcal{N}_++1)^{n-k-1}(A+A^*)(\mathcal{N}_++1)^k]\nonumber\\
&\leq C_nNa\textit{l}^{\frac{1}{2}}
  (\mathcal{N}_++1)^n.\label{assertion e^-tBN_++1e^tB}
\end{align}
Plugging (\ref{assertion e^-tBN_++1e^tB}) into (\ref{[N_++1^n,B]}) we find
\begin{equation}\label{0103}
  \pm[(\mathcal{N}_++1)^n,B]\leq C_nNa\textit{l}^{\frac{1}{2}}
  (\mathcal{N}_++1)^n.
\end{equation}
A similar argument using Gronwall's inequality proves (\ref{e^-tB(N_++1)e^tB}) for all $n\in\mathbb{N}$.

\par To prove assertion (\ref{assertion e^-tBN_++1e^tB}), we only need to prove for all non-negative integers $k\leq n-1$
\begin{equation}\label{assertion e^-tBN_++1e^tB 2}
  \Vert (\mathcal{N}_++1)^{k-\frac{n}{2}}(A+A^*)(\mathcal{N}_++1)^{\frac{n}{2}-k-1}\Vert\leq C_nNa\textit{l}^{\frac{1}{2}}.
\end{equation}
Switching to Fock space $F_{N,d}$, we get for an arbitrary vector
$(g^{(0)},\dots,g^{(N)})\in F_{N,d}$ such that
\begin{equation*}
  U_N(\mathcal{N}_++1)^{\frac{1}{2}}U_N^*(g^{(0)},\dots,g^{(N)})
  =(\beta^{(0)},\sqrt{2}g^{(1)},\dots,\sqrt{N+1}g^{(N)}),
\end{equation*}
\begin{equation*}
   U_N(\mathcal{N}_+-1)_+^{\frac{1}{2}}U_N^*(g^{(0)},\dots,g^{(N)})
   =(0,0,g^{(3)},\dots,\sqrt{N-1}g^{(N)}),
\end{equation*}
\begin{equation*}
  U_NAU_N^*(g^{(0)},\dots,g^{(N)})={\sum_{p\neq0}}\eta_p
  (0,0,\sqrt{N(N-1)}a_p^*a_{-p}^*g^{(0)},\dots,\sqrt{2\cdot1}a_p^*a_{-p}^*g^{(N-2)}).
\end{equation*}
Then it is easy to deduce
\begin{equation}\label{N_++1^1/2 A 0}
\begin{aligned}
(\mathcal{N}_++1)^{\frac{1}{2}}A&=A(\mathcal{N}_++3)^{\frac{1}{2}},\\
A(\mathcal{N}_++1)^{\frac{1}{2}}&=(\mathcal{N}_+-1)_+^{\frac{1}{2}}A.
\end{aligned}
\end{equation}
By induction, for any $k\in\mathbb{N}$
\begin{equation}\label{N_++1^1/2 A 1}
\begin{aligned}
(\mathcal{N}_++1)^{\frac{k}{2}}A&=A(\mathcal{N}_++3)^{\frac{k}{2}},\\
   A(\mathcal{N}_++1)^{\frac{k}{2}}&=(\mathcal{N}_+-1)_+^{\frac{k}{2}}A.
\end{aligned}
\end{equation}
Let $j=(n-1)/2$, if $k\geq j$ then we use (\ref{N_++1^1/2 A 1}) to get
\begin{align}
  &(\mathcal{N}_++1)^{k-\frac{n}{2}}(A+A^*)(\mathcal{N}_++1)^{\frac{n}{2}-k-1}\nonumber\\
=&(\mathcal{N}_++1)^{-\frac{1}{2}}(\mathcal{N}_++1)^{k-j}(A+A^*)(\mathcal{N}_++1)^{j-k}
  (\mathcal{N}_++1)^{-\frac{1}{2}}\nonumber\\
=&(\mathcal{N}_++1)^{-\frac{1}{2}}A(\mathcal{N}_++1)^{-\frac{1}{2}}
  (\mathcal{N}_++3)^{k-j}(\mathcal{N}_++1)^{j-k}\nonumber\\
  &+(\mathcal{N}_++1)^{-\frac{1}{2}}A^*(\mathcal{N}_++1)^{-\frac{1}{2}}
  (\mathcal{N}_+-1)_+^{k-j}(\mathcal{N}_++1)^{j-k}.\label{一坨大便}
\end{align}
On the one hand, (\ref{Abs   Apsi,psi}) implies that
\begin{align*}
\pm (\mathcal{N}_++1)^{-\frac{1}{2}}(A+A^*)(\mathcal{N}_++1)^{-\frac{1}{2}}&\leq CNa\textit{l}^{\frac{1}{2}},\\
\pm i(\mathcal{N}_++1)^{-\frac{1}{2}}(A-A^*)(\mathcal{N}_++1)^{-\frac{1}{2}}&\leq CNa\textit{l}^{\frac{1}{2}},
\end{align*}
which are equivalent to
\begin{equation}\label{两坨大便}
\begin{aligned}
  \Vert (\mathcal{N}_++1)^{-\frac{1}{2}}A(\mathcal{N}_++1)^{-\frac{1}{2}}\Vert&\leq CNa\textit{l}^{\frac{1}{2}},\\
  \Vert (\mathcal{N}_++1)^{-\frac{1}{2}}A^*(\mathcal{N}_++1)^{-\frac{1}{2}}\Vert&\leq CNa\textit{l}^{\frac{1}{2}}.
\end{aligned}
\end{equation}
On the other hand the spectrum $\sigma(\mathcal{N}_+)=\sigma(U_N\mathcal{N}_+U_N^*)=\{0,1,\dots,N\}$ gives
\begin{equation}\label{三坨大便}
\begin{aligned}
  \Vert(\mathcal{N}_++3)^{k-j}(\mathcal{N}_++1)^{j-k}\Vert&\leq 3^{k-j},\\
  \Vert(\mathcal{N}_+-1)_+^{k-j}(\mathcal{N}_++1)^{j-k}\Vert&\leq 1.
\end{aligned}
\end{equation}
Inserting (\ref{两坨大便}) and (\ref{三坨大便}) into (\ref{一坨大便}) we prove (\ref{assertion e^-tBN_++1e^tB 2}) for $j\leq k\leq n-1$. For the case $0\leq k<j$ we can proceed analogously, and thus we have proved (\ref{assertion e^-tBN_++1e^tB 2}) hence (\ref{assertion e^-tBN_++1e^tB}).

\par What remains for us is to prove (\ref{e^-tB(N_++1)e^tB}) for arbitrary $n\in\frac{1}{2}\mathbb{N}$. Following the above process starting from (\ref{[N_++1^n,B]}), we only need to prove (\ref{e^-tB(N_++1)e^tB}) for $n=\frac{1}{2}$. It follows directly from (\ref{N_++1^1/2 A 0}) that
\begin{equation*}
  [(\mathcal{N}_++1)^{\frac{1}{2}},A]=A((\mathcal{N}_++3)^{\frac{1}{2}}
-(\mathcal{N}_++1)^{\frac{1}{2}}),
\end{equation*}
and it is easy to check
\begin{equation*}
  ((\mathcal{N}_++3)^{\frac{1}{2}}
-(\mathcal{N}_++1)^{\frac{1}{2}})\leq1.
\end{equation*}
Gronwall's inequality we finishes the proof of (\ref{e^-tB(N_++1)e^tB}). To prove (\ref{e^-tBN_+e^tB-N_+}) for arbitrary $n\in\frac{1}{2}\mathbb{N}$, we just need to observe
\begin{equation}\label{no idea}
   e^{-tB}(\mathcal{N}_++1)^ne^{tB}-(\mathcal{N}_++1)^n
=\int_{0}^{t}e^{-tB}[(\mathcal{N}_++1)^n,B]e^{tB}dt.
\end{equation}
Moreover, an argument similar to the proof of (\ref{0103}) yields
\begin{equation}\label{no idea2}
 \pm [(\mathcal{N}_++1)^n,B]\leq C_nNa\textit{l}^{\frac{1}{2}}(\mathcal{N}_++1)^n.
\end{equation}
for arbitrary $n\in\frac{1}{2}\mathbb{N}$. (\ref{no idea}) and (\ref{no idea2}) together with (\ref{e^-tB(N_++1)e^tB}) yield (\ref{e^-tBN_+e^tB-N_+}) for arbitrary $n\in\frac{1}{2}\mathbb{N}$.

\begin{flushright}
  {$\Box$}
\end{flushright}

\par From here on out we will always assume without further specifications that $N$ tends to infinity, $a,d,\frac{a}{d}$ and $Na\textit{l}^{\frac{1}{2}}$ tend to $0$ and $\frac{d}{a}>\frac{C}{\textit{l}}$ for some universal constant $C$ (i.e. under the setting of Proposition \ref{quadratic renorm}). We assume $Na\textit{l}^{\frac{1}{2}}$ tends to $0$ for the technical reason that it can considerably simplify the notations in our error estimates and it can be verified easily that it holds true consistently in all three regions in the Gross-Pitaevskii limit. We state the results regarding terms on the right hand side of (\ref{split e^-BH_Ne^B}) term by term.

\begin{lemma}\label{lemma e^-BH_02e^B}
\
\begin{enumerate}[$(1)$]
  \item $\mathbf{For\; Region\; I}$
\begin{equation}\label{e^-BH_02e^B}
  \pm e^{-B}H_{02}e^B\leq Cad^{-1}(\mathcal{N}_++1)^2.
\end{equation}
  \item $\mathbf{For\; Region\; III}$
\begin{equation}\label{e^-BH_02e^B III}
          e^{-B}H_{02}e^B=H_{02}+\tilde{\mathcal{E}}^B_{02},
        \end{equation}
where
\begin{equation}\label{E^B_02 III}
  \pm\tilde{\mathcal{E}}^B_{02}\leq CN^2a^2d^{-1}\textit{l}^{\frac{1}{2}}(\mathcal{N}_++1).
\end{equation}
\end{enumerate}

\end{lemma}
\noindent\emph{Proof.} By (\ref{v_p^a,d}) and (\ref{e^-tB(N_++1)e^tB}) we have
\begin{align*}
  \pm e^{-B}H_{02}e^B
  &\leq\frac{1}{2\sqrt{d}}\vert v_0^{(a,d)}\vert e^{-B}\mathcal{N}_+(\mathcal{N}_+-1)e^B\\
  &\leq Cad^{-1} e^{-B}(\mathcal{N}_++1)^2e^B\leq Ce^{CNa\textit{l}^{\frac{1}{2}}}ad^{-1}(\mathcal{N}_++1)^2
\end{align*}
which leads to (\ref{e^-BH_02e^B}) since we have assumed $Na\textit{l}^{\frac{1}{2}}$ tends to $0$. On the other hand, since
\begin{equation*}
  \tilde{\mathcal{E}}^B_{02}=e^{-B}H_{02}e^B-H_{02}=-\frac{1}{2\sqrt{d}}v_0^{(a,d)}
[e^{-B}(\mathcal{N}_+-1)\mathcal{N}_+e^B-(\mathcal{N}_+-1)\mathcal{N}_+].
\end{equation*}
We can use (\ref{e^-tBN_+e^tB-N_+}) to reach (\ref{E^B_02 III}).
\begin{flushright}
  $\Box$
\end{flushright}

\begin{lemma}\label{lemma e^-BH_22e^B}
\
\begin{enumerate}[$(1)$]
  \item $\mathbf{For\; Region\; I}$
 \begin{equation}\label{e^-BH_22e^B}
  e^{-B}H_{22}e^B=Nad^{-1}\widehat{v}(0)\mathcal{N}_++\mathcal{E}^{B}_{22},
\end{equation}
where
\begin{eqnarray}\label{E^B_22}
  \pm \mathcal{E}^{B}_{22}\leq CN^2a^2d^{-1}\textit{l}^{\frac{1}{2}}
  (\mathcal{N}_++1)
  +Cad^{-1}(\mathcal{N}_++1)^2+CNa^3d^{-1}H_{21}.
\end{eqnarray}
  \item $\mathbf{For\; Region\; III}$
  \begin{equation}\label{e^-BH_22e^B III}
    e^{-B}H_{22}e^B=H_{22}+\tilde{\mathcal{E}}^{B}_{22},
  \end{equation}
where
\begin{equation}\label{E^B_22 III}
  \pm\tilde{\mathcal{E}}^{B}_{22}\leq CN^2a^2d^{-1}\textit{l}^{\frac{1}{2}}
  (\mathcal{N}_++1)
\end{equation}
\end{enumerate}

\end{lemma}
\noindent\emph{Proof.} We let
\begin{equation}\label{define R}
  R= \frac{1}{\sqrt{d}}{\sum_{p\neq 0}}v_p^{(a,d)}a_p^*a_p,
\end{equation}
then
\begin{equation}\label{rewrite H_22}
  H_{22}=NR-\mathcal{N}_+R.
\end{equation}

\par Let $\psi\in L^2_s(\Lambda_d^N)$, $U_N\psi=(\alpha^{(0)},\dots,\alpha^{(N)})\in F_{N,d}$. Since $a_0\alpha^{(n)}=0$ for all $n$, we can check by Cauchy-Schwartz that
\begin{align}
  \vert\langle R\psi,\psi\rangle\vert
&=\vert\langle U_NRU_N^*U_N\psi,U_N\psi\rangle\vert
  =\left\vert\sum_{n=1}^N\left\langle \frac{1}{\sqrt{d}}{\sum_{p}}v_p^{(a,d)}a_p^*a_p\alpha^{(n)},\alpha^{(n)}\right\rangle\right\vert
  \nonumber\\
  =&d^{-1}\left\vert\sum_{n=1}^N\sum_{i=1}^{n}\int_{\Lambda_{d}^{n+1}}v_a(\mathbf{x}_i-\mathbf{y})
  \alpha^{(n)}(\mathbf{x}_1,\dots,\widehat{\mathbf{x}_i},\mathbf{y},\dots,\mathbf{x}_n)
\overline{\alpha^{(n)}}(\mathbf{x}_1,\dots,\mathbf{x}_n)\right\vert\nonumber\\
  =&d^{-1}\left\vert\sum_{n=1}^N n \int_{\Lambda_{d}^{n+1}}v_a(\mathbf{x}_1-\mathbf{y})
\alpha^{(n)}(\mathbf{y},\mathbf{x}_2,\dots,\mathbf{x}_n)
  \overline{\alpha^{(n)}}(\mathbf{x}_1,\dots,\mathbf{x}_n)\right\vert\nonumber\\
  \leq&d^{-1}\sum_{n=1}^N n \Vert v_a\ast\alpha^{(n)}
(\cdot,\mathbf{x}_2,\dots,\mathbf{x}_n)(\mathbf{x}_1)\Vert_2\Vert \alpha^{(n)}\Vert_2\nonumber\\
  \leq&d^{-1}\sum_{n=1}^N n\Vert v_a\Vert_1\Vert \alpha^{(n)}\Vert_2^2
  \leq Cad^{-1}\langle \mathcal{N}_+\psi,\psi\rangle.\label{010302}
\end{align}
Since $R$ commutes with $\mathcal{N}_+$,
\begin{align}
  \pm e^{-B}\mathcal{N}_+Re^B=& \pm e^{-B}\mathcal{N}_+^{\frac{1}{2}}R\mathcal{N}_+^{\frac{1}{2}}e^B
  \leq Cad^{-1}e^{-B}\mathcal{N}_+^2e^B\nonumber\\
  \leq&Ce^{CNa\textit{l}^{\frac{1}{2}}}ad^{-1}(\mathcal{N}_++1)^2\label{wtf1}
\end{align}
where we have used (\ref{e^-tB(N_++1)e^tB}) in the last inequality.
\par For the action of $e^B$ on the first term of (\ref{rewrite H_22}) we rewrite
\begin{equation*}
  e^{-B}NRe^{B}=NR+N\int_{0}^{1}e^{-tB}[R,B]e^{tB}dt.
\end{equation*}
By a direct calculation we have
\begin{equation*}
  [R,B]=\frac{1}{\sqrt{d}}{\sum_{p\neq 0}}v_p^{(a,d)}\eta_p(a_p^*a_{-p}^*a_0a_0+h.c.).
\end{equation*}
Estimating on Fock space like (\ref{Abs   Apsi,psi}), with the help of (\ref{e^-tB(N_++1)e^tB}) we deduce
\begin{align}
  \pm e^{-tB}[R,B]e^{tB}
  \leq&N\Vert d^{-1}v_a\ast\eta_{\perp}\Vert_2 e^{-tB}(\mathcal{N}_++1)e^{tB}\nonumber\\
  \leq&Ce^{CNa\textit{l}^{\frac{1}{2}}\vert t\vert}Na^2d^{-1}\textit{l}^{\frac{1}{2}}(\mathcal{N}_++1),\label{010303}
\end{align}
which yields directly
\begin{equation}\label{wtf2}
  \pm N\int_{0}^{1}e^{-tB}[R,B]e^{tB}dt\leq Ce^{CNa\textit{l}^{\frac{1}{2}}}N^2a^2d^{-1}\textit{l}^{\frac{1}{2}}(\mathcal{N}_++1).
\end{equation}
Finally by assumptions on $v$, we know that $\widehat{v}$ is a smooth function with bounded derivatives of all orders and $\nabla\widehat{v}(0)=0$. Using Taylor expansion and (\ref{v_p^a,d}) we obtain
\begin{eqnarray*}
  \vert v_p^{(a,d)}-v_0^{(a,d)}\vert
  =ad^{-\frac{1}{2}}\left\vert \widehat{v}\left(\frac{a\mathcal{M}_{d}p}{2\pi}\right)-\widehat{v}(0)\right\vert
  \leq Ca^3d^{-\frac{1}{2}}\vert \mathcal{M}_{d}p\vert^2,
\end{eqnarray*}
that is
\begin{equation}\label{wtf3}
   \pm(NR-Nad^{-1}\widehat{v}(0)\mathcal{N}_+)\leq CNa^3d^{-1}H_{21},
\end{equation}
which holds on the domain of $H_{21}$ since it is not a bounded operator. Then we can define
\begin{eqnarray*}
\mathcal{E}_{22}^B=(NR-Nad^{-1}\widehat{v}(0)\mathcal{N}_+)
+N\int_{0}^{1}e^{-tB}[R,B]e^{tB}dt-e^{-B}\mathcal{N}_+Re^B.
\end{eqnarray*}
Combining estimates (\ref{wtf1}), (\ref{wtf2}) and (\ref{wtf3}) above we can check that $\mathcal{E}_{22}^B$ satisfies (\ref{E^B_22}).
\par On the other hand, we let
\begin{equation*}
  \tilde{\mathcal{E}}_{22}^B=e^{-B}H_{22}e^B-H_{22}=\int_{0}^{1}e^{-tB}
[H_{22},B]e^{tB}dt.
\end{equation*}
With (\ref{rewrite H_22}) and (\ref{wtf2}), we only need to bound
\begin{equation*}
  \int_{0}^{1}e^{-tB}
[\mathcal{N}_+R,B]e^{tB}dt.
\end{equation*}
A calculation gives
\begin{align*}
[\mathcal{N}_+R,B]=&\mathcal{N}_+[R,B]+[\mathcal{N}_+,B]R,
\end{align*}
and we already know from (\ref{010302}) and (\ref{010303}) that
\begin{align*}
\pm R&\leq Cad^{-1}\mathcal{N}_+,\\
\pm [R,B]&\leq CNa^2d^{-1}\textit{l}^{\frac{1}{2}}(\mathcal{N}_++1),
\end{align*}
under which we modify the estimates around (\ref{Abs Apsi,psi}) to get
\begin{equation*}
  \pm [\mathcal{N}_+R,B]\leq CN^2a^2d^{-1}\textit{l}^{\frac{1}{2}}(\mathcal{N}_++1),
\end{equation*}
which together with (\ref{e^-tB(N_++1)e^tB}) yield (\ref{E^B_22 III}).
\begin{flushright}
  $\Box$
\end{flushright}

\begin{lemma}\label{lemma e^-BH_3e^B}
$\mathbf{For\; All\; Regions}$
\begin{equation}\label{e^-BH_3e^B}
  e^{-B}H_{3}e^B=H_3+\mathcal{E}^{B}_{3},
\end{equation}
where
\begin{equation}\label{E^B_3}
\begin{aligned}
  \pm \mathcal{E}^{B}_{3}\leq&
 C(N^{\frac{3}{2}}a^{\frac{3}{2}}d^{-{\frac{1}{2}}}\textit{l}^{\frac{1}{2}}+
N^2a^2d^{-1}\textit{l}^{\frac{1}{2}})(\mathcal{N}_++1)\\
&+CN^{\frac{3}{2}}a^{\frac{3}{2}}d^{-{\frac{1}{2}}}\textit{l}^{\frac{1}{2}}H_4.
\end{aligned}
\end{equation}
\end{lemma}
\noindent\emph{Proof.} We can rewrite
\begin{equation}\label{e^-BH_3e^B 2}
  e^{-B}H_{3}e^B=\frac{1}{\sqrt{d}}{\sum_{p,r,p+r\neq0}}
  v_r^{(a,d)}(e^{-B}a_{p+r}^*a_{-r}^*e^Be^{-B}a_pa_0e^B+h.c.).
\end{equation}
We stress here that the domain of definition of $e^B$ in (\ref{e^-BH_3e^B 2}) may change as the particle number changes.
\par We can expand (\ref{e^-BH_3e^B 2}) for $p,r,p+r\neq0$
\begin{equation*}
  e^{-B}a_{p+r}^*a_{-r}^*e^B=a_{p+r}^*a_{-r}^*+\int_{0}^{1}e^{-tB}[a_{p+r}^*a_{-r}^*,B]e^{tB}dt,
\end{equation*}
\begin{equation*}
  [a_{p+r}^*a_{-r}^*,B]=\eta_{p+r}a_0^*a_0^*a_{-r}^*a_{-(p+r)}+\eta_ra_0^*a_0^*a_{p+r}^*a_r,
\end{equation*}
\begin{equation*}
  e^{-B}a_pa_0e^B=a_pa_0+\int_{0}^{1}e^{-tB}[a_pa_0,B]e^{tB}dt,
\end{equation*}
\begin{equation*}
  [a_pa_0,B]=\eta_pa_{-p}^*a_0a_0a_0-a_p{\sum_{q\neq0}}\eta_qa_0^*a_{q}a_{-q},
\end{equation*}
and define the error
\begin{equation*}
  \mathcal{E}^B_3=e^{-B}H_{3}e^B-H_3
=\sum_{i=1}^{4}\mathcal{E}_{3,i}^B,
\end{equation*}
where
\begin{align*}
  \mathcal{E}_{3,1}^B&=\frac{1}{\sqrt{d}}{\sum_{p,r,p+r\neq0}}v_r^{(a,d)}
  \eta_p\int_{0}^{1}a_{p+r}^*a_{-r}^*e^{-tB}a_{-p}^*a_0a_0a_0e^{tB}dt+h.c.\\
  \mathcal{E}_{3,2}^B&=-\frac{1}{\sqrt{d}}{\sum_{p,r,p+r\neq0}}v_r^{(a,d)}
  \int_{0}^{1}a_{p+r}^*a_{-r}^*e^{-tB}a_p{\sum_{q\neq0}}\eta_qa_0^*a_{q}a_{-q}e^{tB}dt+h.c.\\
  \mathcal{E}_{3,3}^B&=\frac{1}{\sqrt{d}}{\sum_{p,r,p+r\neq0}}v_r^{(a,d)}
  \eta_{p+r}\int_{0}^{1}e^{-tB}a_0^*a_0^*a_{-r}^*a_{-(p+r)}e^{(t-1)B}a_pa_0e^Bdt+h.c.\\
  \mathcal{E}_{3,4}^B&=\frac{1}{\sqrt{d}}{\sum_{p,r,p+r\neq0}}v_r^{(a,d)}
  \eta_r\int_{0}^{1}e^{-tB}a_0^*a_0^*a_{p+r}^*a_re^{(t-1)B}a_pa_0e^Bdt+h.c.
\end{align*}
The estimates of $\mathcal{E}^B_{3,i}$ are all similar, we only do the first one in detail. Let
$\psi\in L^2_s(\Lambda_d^N)$ and
\begin{align*}
 U_N\psi&=(\alpha^{(0)},\ldots,\alpha^{(N)}),\\
 U_{N-2}(e^{-tB}a_{-p}^*a_0a_0a_0e^{tB}\psi)&=(\beta^{(0)}_p(t),\dots,\beta^{(N-2)}_p(t)),\\
 U_{N-3}(a_0a_0a_0e^{tB}\psi)&=(g^{(0)}(t),\dots,g^{(N-3)}(t)).
\end{align*}
Evaluating on Fock space, we have (we always omit the $h.c.$ parts)
\begin{align*}
  \vert\langle\mathcal{E}_{3,1}^B\psi,\psi\rangle\vert
  &=\frac{1}{\sqrt{d}}\left\vert\int_{0}^{1}{\sum_{p,r,p+r\neq0}}v_r^{(a,d)}\eta_p
  \sum_{n=0}^{N-2}\langle a_{p+r}^*a_{-r}^*\beta^{(n)}_p(t),\alpha^{(n+2)}\rangle dt\right\vert\\
  &=\frac{1}{\sqrt{d}}\left\vert\int_{0}^{1}\sum_{n=0}^{N-2}\left\langle {\sum_{p\neq0,r}}v_r^{(a,d)}\eta_p
  a_{p+r}^*a_{-r}^*\beta^{(n)}_p(t),\alpha^{(n+2)}\right\rangle dt\right\vert\\
  &=\frac{1}{\sqrt{d}}\left\vert\int_{0}^{1}\sum_{n=0}^{N-2}\sum_{i\neq j}^{n+2}\frac{1}{\sqrt{(n+1)(n+2)}}
  \int_{\Lambda_d^{(n+2)}}v_a(\mathbf{x}_i-\mathbf{x}_j)
G^{(n)}_{i,j}(t)\overline{\alpha^{(n+2)}}\right\vert,
\end{align*}
where $G^{(n)}_{i,j}$ is given by
\begin{eqnarray*}
  G^{(n)}_{i,j}(t)(\mathbf{x}_1,\dots,\mathbf{x}_{n+2})
  =\sum_{p\neq 0}\eta_p\phi_p^{(d)}(\mathbf{x}_i)\beta^{(n)}_p
(t,\mathbf{x}_1,\dots,\widehat{\mathbf{x}_i},\widehat{\mathbf{x}_j},\dots,\mathbf{x}_{n+2}).
\end{eqnarray*}
By definition, $G^{(n)}_{i,j}$ is symmetric with respect to $(\mathbf{x}_1,\dots,\widehat{\mathbf{x}_i},\widehat{\mathbf{x}_j},\dots,\mathbf{x}_{n+2})$. Using Cauchy-Schwartz we get
\begin{align}
  \vert\langle\mathcal{E}_{3,1}^B\psi,\psi\rangle\vert
   \leq&\frac{1}{\sqrt{d}}\left\vert\int_{0}^{1}\sum_{n=0}^{N-2}{\sqrt{(n+1)(n+2)}}
  \int_{\Lambda_d^{(n+2)}}v_a(\mathbf{x}_1-\mathbf{x}_2)
G^{(n)}_{2,1}(t)\overline{\alpha^{(n+2)}}\right\vert\nonumber\\
  \leq&\frac{1}{\sqrt{d}}\int_{0}^{1}\left(\sum_{n=0}^{N-2}(n+1)(n+2)
\int_{\Lambda_d^{(n+2)}}v_a(\mathbf{x}_1-\mathbf{x}_2)
\vert\alpha^{(n+2)}\vert^2\right)^{\frac{1}{2}}\nonumber\\
  &\times\left(\sum_{n=0}^{N-2}\int_{\Lambda_d^{(n+2)}}v_a(\mathbf{x}_1-\mathbf{x}_2)\vert G^{(n)}_{2,1}(t)\vert^2\right)^{\frac{1}{2}},\label{E_3,1^B}
\end{align}
where we have used the non-negativity of $v$ in the last inequality.
On the one hand we can rewrite the energy of $H_4$ as what we have done in (\ref{H_4psi,psi fock})
\begin{equation}\label{H_4psi,psi}
  \langle H_4\psi,\psi\rangle
  =\frac{1}{2}\sum_{n=2}^{N}n(n-1)\int_{\Lambda_d^n}
v_a(\mathbf{x}_1-\mathbf{x}_2)\vert\alpha^{(n)}\vert^2,
\end{equation}
which can be related to the first factor of (\ref{E_3,1^B}). On the other hand, since both $U_N$ and $e^{-tB}$ are unitary operators, that is,
\begin{align*}
  &\sum_{n=0}^{N-2}\int_{\Lambda_d^{n}}\vert G^{(n)}_{2,1}(t)(\mathbf{x}_1,\dots,\mathbf{x}_{n+2})\vert^2
d\mathbf{x}_3\dots d\mathbf{x}_{n+2}\\
&=\left\Vert\sum_{p\neq0}\eta_p\phi_{p}^{(d)}(\mathbf{x}_2)
U_{N-2}e^{-tB}a_{-p}^*a_0a_0a_0e^{tB}\psi\right\Vert_{F_{N-2,d}}^2\\
&=\left\Vert\sum_{p\neq0}\eta_p\phi_{p}^{(d)}(\mathbf{x}_2)
a_{-p}^*a_0a_0a_0e^{tB}\psi\right\Vert_{L^2(\Lambda_d^{N-2})}^2\\
&=\left\Vert\sum_{p\neq0}\eta_p\phi_{p}^{(d)}(\mathbf{x}_2)U_{N-2}
a_{-p}^*U_{N-3}^*U_{N-3}a_0a_0a_0e^{tB}\psi\right\Vert_{L^2(\Lambda_d^{N-2})}^2.
\end{align*}
Recalling the definition of $g^{(n)}$, we can calculate directly
\begin{align*}
&\sum_{n=0}^{N-2}\int_{\Lambda_d^{n}}\vert G^{(n)}_{2,1}(t)
(\mathbf{x}_1,\dots,\mathbf{x}_{n+2})\vert^2d\mathbf{x}_3\dots d\mathbf{x}_{n+2}\\
&=\sum_{n=0}^{N-3}\int_{\Lambda_d^{n+1}}\left\vert\sum_{p\neq0}\eta_p\phi_p^{(d)}(\mathbf{x}_2)
a_{-p}^*g^{(n)}(t)\right\vert^2d\mathbf{z}_1\dots d\mathbf{z}_{n+1}\\
&=\sum_{n=0}^{N-3}\int_{\Lambda_d^{n+1}}\left\vert\frac{1}{\sqrt{d}\sqrt{n+1}}\sum_{s=1}^{n+1}
\eta_\perp(\mathbf{x}_2-\mathbf{z}_s)g^{(n)}(t)
(t,\mathbf{z}_1,\dots,\widehat{\mathbf{z}_s},\dots,\mathbf{z}_{n+1})\right\vert^2
d\mathbf{z}_1\dots d\mathbf{z}_{n+1}\\
&\leq\sum_{n=0}^{N-3}\frac{n+1}{d}\int_{\Lambda_d^{n+1}}
\vert\eta_\perp(\mathbf{x}_2-\mathbf{z}_1)g^{(n)}
(t,\mathbf{z}_2,\dots,\mathbf{z}_{n+1})\vert^2d\mathbf{z}_1\dots d\mathbf{z}_{n+1}\\
&=d^{-1}\Vert\eta_\perp\Vert^2_2\langle(\mathcal{N}_++1)a_0a_0a_0e^{tB}\psi,
a_0a_0a_0e^{tB}\psi\rangle\\
&\leq Ce^{CNa\textit{l}^{\frac{1}{2}}}N^3d^{-1}\Vert\eta_\perp\Vert^2_2
\langle(\mathcal{N}_++1)\psi,\psi\rangle,
\end{align*}
where we have use (\ref{e^-tB(N_++1)e^tB}) in the last inequality. Then by (\ref{est of eta and eta_perp}), we obtain that
\begin{align}
  &\sum_{n=0}^{N-2}\int_{\Lambda_d^{n+2}}
v_a(\mathbf{x}_1-\mathbf{x}_2)\vert G^{(n)}_{2,1}(t)\vert^2
  =\int_{\Lambda_d^{2}}v_a(\mathbf{x}_1-\mathbf{x}_2)
\sum_{n=0}^{N-2}\int_{\Lambda_d^{n}}\vert G^{(n)}_{2,1}(t)\vert^2\nonumber\\
  &\leq Ce^{CNa\textit{l}^{\frac{1}{2}}}N^3\Vert\eta_\perp\Vert^2_2\Vert v_a\Vert_1\langle(\mathcal{N}_++1)\psi,\psi\rangle
  \leq Ce^{CNa\textit{l}^{\frac{1}{2}}}N^3a^3\textit{l}
\langle(\mathcal{N}_++1)\psi,\psi\rangle.\label{臭狗屎}
\end{align}
Combining (\ref{E_3,1^B}), (\ref{H_4psi,psi}) and (\ref{臭狗屎}) we get
\begin{align*}
  \left\vert\langle\mathcal{E}_{3,1}^B\psi,\psi\rangle\right\vert
  \leq Ce^{CNa\textit{l}^{\frac{1}{2}}}
  N^{\frac{3}{2}}a^{\frac{3}{2}}d^{-\frac{1}{2}}\textit{l}^{\frac{1}{2}}
  \langle H_4\psi,\psi\rangle^{\frac{1}{2}}
\langle(\mathcal{N}_++1)\psi,\psi\rangle^{\frac{1}{2}}.
\end{align*}
By similar computations we can bound the remaining parts with
\begin{align*}
   \left\vert\langle\mathcal{E}_{3,2}^B\psi,\psi\rangle\right\vert
   &\leq Cd^{-\frac{1}{2}}\Vert\eta_\perp\Vert_2\Vert v_a\Vert_1^{\frac{1}{2}}
   \langle(\mathcal{N}_++1)^3a_0^*e^{tB}\psi,
   a_0^*e^{tB}\psi\rangle^{\frac{1}{2}}
\langle H_4\psi,\psi\rangle^{\frac{1}{2}}\\
   &\leq Ce^{CNa\textit{l}^{\frac{1}{2}}}
  N^{\frac{3}{2}}a^{\frac{3}{2}}d^{-\frac{1}{2}}\textit{l}^{\frac{1}{2}}
  \langle H_4\psi,\psi\rangle^{\frac{1}{2}}
\langle(\mathcal{N}_++1)\psi,\psi\rangle^{\frac{1}{2}},
\end{align*}
and for $i=3,4$
\begin{align*}
\left\vert\langle\mathcal{E}_{3,i}^B\psi,\psi\rangle\right\vert
\leq& Cd^{-1}\Vert v_a\ast\eta_\perp\Vert_2
\langle(\mathcal{N}_++1)^2a_0a_0e^{tB}\psi,a_0a_0e^{tB}\psi\rangle^{\frac{1}{2}}\\
&\times\langle(\mathcal{N}_++1)a_0e^{tB}\psi,a_0e^{tB}\psi\rangle^{\frac{1}{2}}\\
\leq& Ce^{CNa\textit{l}^{\frac{1}{2}}}N^2a^2d^{-1}\textit{l}^{\frac{1}{2}}
  \langle(\mathcal{N}_++1)\psi,\psi\rangle.
\end{align*}
With all the estimates above, we reach (\ref{E^B_3}).
\begin{flushright}
  $\Box$
\end{flushright}
\begin{remark}
Notice in the proof of Lemma \ref{lemma e^-BH_3e^B}, we rewrite $e^{-B}H_3e^B$ in the form of (\ref{e^-BH_3e^B 2}) rather than using the more common form
\begin{equation*}
  e^{-B}H_3e^B=H_3+\int_{0}^{1}e^{-tB}[H_3,B]e^{tB}dt
\end{equation*}
then estimating the action of $e^{tB}$ on commutator $[H_3,B]$ instead, like we have done in Lemma \ref{control of S_+ conj with e^B}. This is because it is inevitable that we have to control the action of $e^B$ on $H_4$ if we are going to estimate directly $e^{-tB}[H_3,B]e^{tB}$. But we presumably can not gain a satisfying estimate, and we will show this result in the next lemma. The lemma below is not needed in this section, but interestingly it will be of use in Sections \ref{subsection Optimal BEC} and \ref{subsection Optimal BEC for Region III}, where it will help us prove a result of optimal Bose-Einstein condensation.
\end{remark}

\begin{lemma}\label{lemma control of H_4}
$\mathbf{For\; All\; Regions}$
\begin{align}
  e^{-tB}H_4e^{tB}&\leq C(H_4+N^2ad^{-1}),\label{control of e^-BH_4e^B}\\
e^{-tB}H_{21}e^{tB}&\leq C\big(H_{21}+N^2ad^{-1}(\mathcal{N}_++1)\big).
\label{control of e^-BH_21e^B}
\end{align}
for all $\vert t\vert\leq1$.
\end{lemma}
\noindent
\emph{Proof.} We leave the proof of (\ref{control of e^-BH_21e^B}) to Lemma \ref{lemma control of H_4 with e^B tilde}. We now give a thorough proof of (\ref{control of e^-BH_4e^B}). Calculating directly we have
\begin{align*}
  [H_4,B]=&\frac{1}{\sqrt{d}}{\sum_{p,q,p+r,q+r\neq0}}
  v_r^{(a,d)}\eta_p(a_{p+r}^*a_q^*a_{-p}^*a_{q+r}a_0a_0+h.c.)\\
  &+\frac{1}{2\sqrt{d}}{\sum_{p,q\neq0}}v_{p-q}^{(a,d)}\eta_q(a_p^*a_{-p}^*a_0a_0+h.c.).
\end{align*}
Estimating on the Fock space, for any $\psi\in L^2_s(\Lambda_d^N)$ we can bound the first term by
\begin{align*}
&\frac{1}{\sqrt{d}}\left\vert{\sum_{p,q,p+r,q+r\neq0}}
  v_r^{(a,d)}\eta_p\langle a_{p+r}^*a_q^*a_{-p}^*a_{q+r}a_0a_0
\psi,\psi\rangle\right\vert\\
&\leq Cd^{-\frac{1}{2}}\Vert\eta_\perp\Vert_2\Vert v_a\Vert_1^{\frac{1}{2}}
\langle(\mathcal{N}_++1)^2a_0a_0\psi,a_0a_0\psi\rangle^{\frac{1}{2}}
\langle H_4\psi,\psi\rangle^{\frac{1}{2}}\\
&\leq CN^{\frac{3}{2}}a^{\frac{3}{2}}d^{-\frac{1}{2}}\textit{l}^{\frac{1}{2}}\langle H_4\psi,\psi\rangle^{\frac{1}{2}}
\langle(\mathcal{N}_++1)\psi,\psi\rangle^{\frac{1}{2}},
\end{align*}
and the second term similar to (\ref{Abs Apsi,psi}) by
\begin{align*}
\frac{1}{2\sqrt{d}}\left\vert{\sum_{p,q\neq0}}v_{p-q}^{(a,d)}\eta_q
\langle a_p^*a_{-p}^*a_0a_0\psi,\psi\rangle \right\vert
&\leq C\Vert v_a\Vert_1^{\frac{1}{2}}\Vert\eta_\perp\Vert_\infty
\langle H_4\psi,\psi\rangle^{\frac{1}{2}}
\langle a_0a_0\psi,a_0a_0\psi\rangle^{\frac{1}{2}}\\
&\leq CNa^{\frac{1}{2}}d^{-\frac{1}{2}}
\langle H_4\psi,\psi\rangle^{\frac{1}{2}}
\langle \psi,\psi\rangle^{\frac{1}{2}}.
\end{align*}
Since $\mathcal{N}_+\leq N$ and we always ask that $Na\textit{l}^{\frac{1}{2}}$ is small enough, we have
\begin{align*}
  \pm[H_4,B]\leq  C(H_4+N^2ad^{-1}).
\end{align*}
Then (\ref{control of e^-BH_4e^B}) is derived by Gronwall's inequality.

\begin{flushright}
  $\Box$
\end{flushright}

\begin{lemma}\label{lemma Gamma}
$\mathbf{For\; All\;  Regions}$
\begin{equation}\label{int_0^1 e^-tBGammae^tBdt}
\begin{aligned}
  \pm\int_{0}^{1}e^{-tB}\Gamma e^{tB}dt\leq&
 C(N^{\frac{3}{2}}a^{\frac{3}{2}}d^{-{\frac{1}{2}}}\textit{l}^{\frac{1}{2}}+
N^2a^2d^{-1}\textit{l}^{\frac{1}{2}})(\mathcal{N}_++1)\\
&+CN^{\frac{3}{2}}a^{\frac{3}{2}}d^{-{\frac{1}{2}}}\textit{l}^{\frac{1}{2}}H_4.
\end{aligned}
\end{equation}
\end{lemma}
\noindent
\emph{Proof.} By direct calculations we have
\begin{eqnarray*}
  [H_{21},B]={\sum_{p\neq0}}\vert\mathcal{M}_dp\vert^2\eta_p(a_p^*a_{-p}^*a_0a_0+h.c.),
\end{eqnarray*}
and
\begin{align*}
  [H_4,B]=&\frac{1}{\sqrt{d}}{\sum_{p,q,p+r,q+r\neq0}}
  v_r^{(a,d)}\eta_p(a_{p+r}^*a_q^*a_{-p}^*a_{q+r}a_0a_0+h.c.)\\
  &+\frac{1}{2\sqrt{d}}{\sum_{p,q\neq0}}v_{p-q}^{(a,d)}\eta_q(a_p^*a_{-p}^*a_0a_0+h.c.).
\end{align*}
Using (\ref{eqn of eta_p rewrt}) and (\ref{define Gamma}) we can check that
\begin{align*}
  \Gamma=&\frac{1}{\sqrt{d}}{\sum_{p,q,p+r,q+r\neq0}}v_r^{(a,d)}
  \eta_p(a_{p+r}^*a_q^*a_{-p}^*a_{q+r}a_0a_0+h.c.)\\
&-\frac{1}{2\sqrt{d}}\sum_{p\neq0}v_p^{(a,d)}\eta_0(a_p^*a_{-p}^*a_0a_0+h.c.)\eqqcolon
\Gamma_1+\Gamma_2.
\end{align*}
We can evaluate $e^{-tB}\Gamma_1 e^{tB}$ in the way we have done in Lemma \ref{lemma e^-BH_3e^B} by rewriting
\begin{equation*}
  e^{-tB}\Gamma_1 e^{tB}=\frac{1}{\sqrt{d}}{\sum_{p,q,p+r,q+r\neq0}}v_r^{(a,d)}
  \eta_p(e^{-tB}a_{p+r}^*a_q^*e^{tB}e^{-tB}a_{-p}^*a_{q+r}a_0a_0e^{tB}+h.c.).
\end{equation*}
Expanding the action of $e^{tB}$ using Newton-Leibniz law we get
\begin{equation*}
  \int_{0}^{1}e^{-tB}\Gamma_1 e^{tB}dt=\sum_{i=1}^{7}\mathcal{E}^B_{\Gamma_1,i}+h.c.
\end{equation*}
where the error terms are given respectively by
\begin{align*}
\mathcal{E}^B_{\Gamma_1,1}&=\frac{1}{\sqrt{d}}{\sum_{p,q,p+r,q+r\neq0}}v_r^{(a,d)}
  \eta_p\int_{0}^{1}(a_{p+r}^*a_q^*a_{-p}^*a_{q+r}a_0a_0)dt,\\
\mathcal{E}^B_{\Gamma_1,2}&=\frac{1}{\sqrt{d}}{\sum_{p,q,p+r,q+r\neq0}}v_r^{(a,d)}
  \eta_p\eta_{q+r}\int_{0}^{1}\int_{0}^{t}
(a_{p+r}^*a_q^*e^{-sB}a_{-p}^*a_{-(q+r)}^*a_0a_0a_0a_0e^{sB})dsdt,\\
\mathcal{E}^B_{\Gamma_1,3}&=\frac{1}{\sqrt{d}}{\sum_{p,q,p+r,q+r\neq0}}v_r^{(a,d)}
  \eta_p^2\int_{0}^{1}\int_{0}^{t}
(a_{p+r}^*a_q^*e^{-sB}a_0^*a_0^*a_pa_{q+r}a_0a_0e^{sB})dsdt,\\
\mathcal{E}^B_{\Gamma_1,4}&=\frac{-1}{\sqrt{d}}{\sum_{\substack{p,q,p+r,\\q+r,t\neq0}}}v_r^{(a,d)}
  \eta_p\eta_t\int_{0}^{1}\int_{0}^{t}
(a_{p+r}^*a_q^*e^{-sB}a_{-p}^*a_ta_{-t}a_{q+r}(2a_0^*a_0+1)e^{sB})dsdt,\\
\mathcal{E}^B_{\Gamma_1,5}&=\frac{1}{\sqrt{d}}{\sum_{\substack{p,q,p+r,\\q+r,t\neq0}}}v_r^{(a,d)}
  \eta_p\eta_{p+r}\int_{0}^{1}\int_{0}^{t}
(e^{-sB}a_q^*a_0^*a_0^*a_{-(p+r)}e^{(s-t)B}a_{-p}^*a_{q+r}a_0a_0e^{tB})dsdt,\\
\mathcal{E}^B_{\Gamma_1,6}&=\frac{1}{\sqrt{d}}{\sum_{\substack{p,q,p+r,\\q+r,t\neq0}}}v_r^{(a,d)}
  \eta_p\eta_{q}\int_{0}^{1}\int_{0}^{t}
(e^{-sB}a_{p+r}^*a_0^*a_0^*a_{-q}e^{(s-t)B}a_{-p}^*a_{q+r}a_0a_0e^{tB})dsdt,\\
\mathcal{E}^B_{\Gamma_1,7}&=\frac{1}{\sqrt{d}}\sum_{q,q+r\neq0}v_r^{(a,d)}
  \eta_q\eta_{q+r}\int_{0}^{1}\int_{0}^{t}
(e^{-sB}a_0^*a_0^*e^{(s-t)B}a_{q+r}^*a_{q+r}a_0a_0e^{tB})dsdt.
\end{align*}
Let $\psi\in L^2_s(\Lambda_d^N)$, then we can bound these terms respectively using the methods that have been shown in the proof of Lemma \ref{lemma e^-BH_3e^B}. We recall that we can use (\ref{est of eta and eta_perp}) and (\ref{est of max eta_perp}) to bound the $L^2$ and $L^{\infty}$ norm of $\eta_\perp$, and (\ref{H_4psi,psi}) to reproduce the potential energy on Fock space.
\begin{align*}
\vert\langle\mathcal{E}_{\Gamma_1,1}^B\psi,\psi\rangle\vert&\leq
Cd^{-\frac{1}{2}}\Vert\eta_\perp\Vert_2\Vert v_a\Vert_1^{\frac{1}{2}}
\langle(\mathcal{N}_++1)^2a_0a_0\psi,a_0a_0\psi\rangle^{\frac{1}{2}}
\langle H_4\psi,\psi\rangle^{\frac{1}{2}}\\
&\leq CN^{\frac{3}{2}}a^{\frac{3}{2}}d^{-\frac{1}{2}}\textit{l}^{\frac{1}{2}}\langle H_4\psi,\psi\rangle^{\frac{1}{2}}
\langle(\mathcal{N}_++1)\psi,\psi\rangle^{\frac{1}{2}}.
\end{align*}
Analogously, we have for $i=2,3,4$
\begin{align*}
\vert\langle\mathcal{E}_{\Gamma_1,2}^B\psi,\psi\rangle\vert\leq&
Cd^{-\frac{1}{2}}\Vert\eta_\perp\Vert_2^2\Vert v_a\Vert_1^{\frac{1}{2}}
\int_{0}^{1}\int_{0}^{t}
\langle(\mathcal{N}_++1)^2a_0^4e^{sB}\psi,a_0^4e^{sB}\psi\rangle^{\frac{1}{2}}
\langle H_4\psi,\psi\rangle^{\frac{1}{2}}dsdt\\
\leq&Ce^{CNa\text{l}^{\frac{1}{2}}}N^{\frac{5}{2}}a^{\frac{5}{2}}d^{-\frac{1}{2}}\textit{l}
\langle H_4\psi,\psi\rangle^{\frac{1}{2}}
\langle(\mathcal{N}_++1)\psi,\psi\rangle^{\frac{1}{2}},
\end{align*}
\begin{align*}
\vert\langle\mathcal{E}_{\Gamma_1,3}^B\psi,\psi\rangle\vert\leq&
Cd^{-\frac{1}{2}}\Vert\eta_\perp\Vert_2^2\Vert v_a\Vert_1^{\frac{1}{2}}
\int_{0}^{1}\int_{0}^{t}
\langle(\mathcal{N}_++1)^2a_0^{*2}a_0^2e^{sB}\psi,a_0^{*2}a_0^2e^{sB}\psi\rangle^{\frac{1}{2}}
\langle H_4\psi,\psi\rangle^{\frac{1}{2}}dsdt\\
\leq&Ce^{CNa\text{l}^{\frac{1}{2}}}N^{\frac{5}{2}}a^{\frac{5}{2}}d^{-\frac{1}{2}}\textit{l}
\langle H_4\psi,\psi\rangle^{\frac{1}{2}}
\langle(\mathcal{N}_++1)\psi,\psi\rangle^{\frac{1}{2}},
\end{align*}
and
\begin{align*}
\vert\langle\mathcal{E}_{\Gamma_1,4}^B\psi,\psi\rangle\vert\leq&
Cd^{-\frac{1}{2}}\Vert\eta_\perp\Vert_2^2\Vert v_a\Vert_1^{\frac{1}{2}}
\int_{0}^{1}\int_{0}^{t}
\langle(\mathcal{N}_++1)^4a_0^{*}a_0e^{sB}\psi,a_0^{*}a_0e^{sB}\psi\rangle^{\frac{1}{2}}
\langle H_4\psi,\psi\rangle^{\frac{1}{2}}dsdt\\
\leq&Ce^{CNa\text{l}^{\frac{1}{2}}}N^{\frac{5}{2}}a^{\frac{5}{2}}d^{-\frac{1}{2}}\textit{l}
\langle H_4\psi,\psi\rangle^{\frac{1}{2}}
\langle(\mathcal{N}_++1)\psi,\psi\rangle^{\frac{1}{2}}.
\end{align*}
Moreover, for $i=5,6$
\begin{align*}
\vert\langle\mathcal{E}_{\Gamma_1,i}^B\psi,\psi\rangle\vert\leq&
Cd^{-1}\Vert\eta_\perp\Vert_2^2\Vert v_a\Vert_1
\int_{0}^{1}\int_{0}^{t}dsdt\\
&\times\langle(\mathcal{N}_++1)^2a_0^2e^{sB}\psi,a_0^2e^{sB}\psi\rangle^{\frac{1}{2}}
\langle (\mathcal{N}_++1)^2a_0^2e^{tB}\psi,a_0^2e^{tB}\psi\rangle^{\frac{1}{2}}\\
\leq&Ce^{CNa\text{l}^{\frac{1}{2}}}N^{3}a^{3}d^{-1}\textit{l}
\langle (\mathcal{N}_++1)\psi,\psi\rangle.
\end{align*}
Finally,
\begin{align*}
\vert\langle\mathcal{E}_{\Gamma_1,7}^B\psi,\psi\rangle\vert\leq&
Cd^{-\frac{1}{2}}\Vert\eta_\perp\Vert_2\Vert\eta_\perp\Vert_\infty\Vert v_a\Vert_1
\int_{0}^{1}\int_{0}^{t}dsdt\\
\times\langle(\mathcal{N}_++1)&e^{(t-s)B}a_0^2e^{sB}\psi,
e^{(t-s)B}a_0^2e^{sB}\psi\rangle^{\frac{1}{2}}
\langle (\mathcal{N}_++1)a_0^2e^{tB}\psi,a_0^2e^{tB}\psi\rangle^{\frac{1}{2}}\\
\leq&Ce^{CNa\text{l}^{\frac{1}{2}}}N^{2}a^{2}d^{-1}\textit{l}^{\frac{1}{2}}
\langle (\mathcal{N}_++1)\psi,\psi\rangle.
\end{align*}
\par As for the evaluation of $e^{-tB}\Gamma_2 e^{tB}$, we first rewrite
\begin{equation*}
  \int_{0}^{1}e^{-tB}\Gamma_2 e^{tB}dt=\Gamma_2+
\int_{0}^{1}\int_{0}^{t}e^{-sB}[\Gamma_2,B] e^{sB}dsdt.
\end{equation*}
Estimates similar to those near (\ref{Abs Apsi,psi}) tell us
\begin{align*}
  \vert\langle\Gamma_2\psi,\psi\rangle\vert
\leq& CNd^{-\frac{1}{2}}\vert\eta_0\vert\Vert v_a\Vert_1^{\frac{1}{2}}
\langle H_4\psi,\psi\rangle^{\frac{1}{2}}
\langle\psi,\psi\rangle^{\frac{1}{2}}\\
\leq& CNa^{\frac{3}{2}}d^{\frac{1}{2}}\textit{l}^2\langle H_4\psi,\psi\rangle^{\frac{1}{2}}
\langle\psi,\psi\rangle^{\frac{1}{2}}.
\end{align*}
For the commutator part we first calculate
\begin{align*}
[\Gamma_2,B]=&\frac{1}{4\sqrt{d}}\sum_{p,q\neq0}v_p^{(a,d)}\eta_q\eta_0(-4a_0^*a_0-2)
(a_q^*a_{-q}^*a_{p}a_{-p}+h.c.)\nonumber\\
&+\frac{1}{\sqrt{d}}\sum_{p\neq0}v_p^{(a,d)}\eta_p\eta_0(1+2a^*_pa_p)a_0^*a_0^*a_0a_0.
\end{align*}
Hence
\begin{align*}
\pm\int_{0}^{1}\int_{0}^{t}e^{-sB}[\Gamma_2,B] e^{sB}&dsdt
\leq CN^2a^2\textit{l}^2+CN^2a^3d\textit{l}^4(\mathcal{N}_++1)\\
&+Cad\textit{l}^2[(N^{\frac{3}{2}}a^{\frac{3}{2}}d^{-{\frac{1}{2}}}\textit{l}^{\frac{1}{2}}+
N^2a^2d^{-1}\textit{l}^{\frac{1}{2}})(\mathcal{N}_++1)\\
&+N^{\frac{3}{2}}a^{\frac{3}{2}}d^{-{\frac{1}{2}}}\textit{l}^{\frac{1}{2}}H_4].
\end{align*}
For more details readers can refer to the proof of the next lemma since the only difference between $\Gamma_2$ and $H_{23}$ is that there is a small factor $\eta_0$ attached to $\Gamma_2$. Collecting all the estimates together we reach (\ref{int_0^1 e^-tBGammae^tBdt}).

\begin{flushright}
  $\Box$
\end{flushright}

\begin{lemma}\label{lemma [H_23,B]}
\
\begin{enumerate}[$(1)$]
\item $\mathbf{For\; Region\; I}$
\begin{align}
  \int_{0}^{1}\int_{t}^{1}e^{-sB}[H_{23},B]e^{sB}dsdt
   =& \frac{N(N-1)}{2\sqrt{d}}\sum_{p\neq0}v_p^{(a,d)}\eta_p
   -\left(\frac{N}{\sqrt{d}}\sum_{p\neq0}v_p^{(a,d)}\eta_p\right)\mathcal{N}_+\nonumber\\
   &+\mathcal{E}^B_{[H_{23},B]},\label{答辩一号}
\end{align}
where
\begin{align}
  \pm\mathcal{E}^B_{[H_{23},B]}\leq&
 C(N^{\frac{3}{2}}a^{\frac{3}{2}}d^{-{\frac{1}{2}}}\textit{l}^{\frac{1}{2}}+
N^2a^2d^{-1}\textit{l}^{\frac{1}{2}})(\mathcal{N}_++1)\nonumber\\
&+CN^{\frac{3}{2}}a^{\frac{3}{2}}d^{-{\frac{1}{2}}}\textit{l}^{\frac{1}{2}}H_4
+Cad^{-1}(\mathcal{N}_++1)^2.\label{E^B_23,1}
\end{align}
\item $\mathbf{For\; Region\; III}$
\begin{align}
  \int_{0}^{1}\int_{t}^{1}e^{-sB}[H_{23},B]e^{sB}&dsdt
   = \frac{N(N-1)}{2\sqrt{d}}\sum_{p\neq0}v_p^{(a,d)}\eta_p
   -\left(\frac{N}{\sqrt{d}}\sum_{p\neq0}v_p^{(a,d)}\eta_p\right)\mathcal{N}_+\nonumber\\
   &+\left(\frac{1}{2\sqrt{d}}\sum_{p\neq0}v_p^{(a,d)}\eta_p\right)\mathcal{N}_+(\mathcal{N}_++1)
+\tilde{\mathcal{E}}^B_{[H_{23},B]},\label{答辩一号 III}
\end{align}
where
\begin{align}
  \pm\tilde{\mathcal{E}}^B_{[H_{23},B]}\leq&
 C(N^{\frac{3}{2}}a^{\frac{3}{2}}d^{-{\frac{1}{2}}}\textit{l}^{\frac{1}{2}}+
N^2a^2d^{-1}\textit{l}^{\frac{1}{2}})(\mathcal{N}_++1)\nonumber\\
&+CN^{\frac{3}{2}}a^{\frac{3}{2}}d^{-{\frac{1}{2}}}\textit{l}^{\frac{1}{2}}H_4.\label{E^B_23,1 III}
\end{align}
\end{enumerate}
\end{lemma}
\noindent
\emph{Proof.} To prove (\ref{答辩一号}) we first calculate
\begin{align}
[H_{23},B]=&\frac{1}{4\sqrt{d}}\sum_{p,q\neq0}v_p^{(a,d)}\eta_q(-4a_0^*a_0-2)
(a_q^*a_{-q}^*a_{p}a_{-p}+h.c.)\nonumber\\
&+\frac{1}{\sqrt{d}}\sum_{p\neq0}v_p^{(a,d)}\eta_p(1+2a^*_pa_p)a_0^*a_0^*a_0a_0.
\label{[H_23,B]}
\end{align}
Then using (\ref{a_0*a_0*a_0a_0}) i.e. $a_0^*a_0^*a_0a_0=(N-\mathcal{N}_+)(N-1-\mathcal{N}_+)$, we expand the second term of (\ref{[H_23,B]}). Plugging it into the integral, we have (\ref{答辩一号}) if we define
\begin{equation*}
  \mathcal{E}^B_{[H_{23},B]}=\sum_{i=1}^{4}\mathcal{E}^B_{23,i},
\end{equation*}
where
\begin{align*}
\mathcal{E}^B_{23,1}=&-\frac{2N}{\sqrt{d}}\sum_{p\neq0}v_p^{(a,d)}\eta_p
\int_{0}^{1}\int_{t}^{1}(e^{-sB}\mathcal{N}_+e^{sB}-\mathcal{N}_+)dsdt\nonumber\\
\mathcal{E}^B_{23,2}=&\frac{1}{\sqrt{d}}\sum_{p\neq0}v_p^{(a,d)}\eta_p\int_{0}^{1}
\int_{t}^{1}e^{-sB}\mathcal{N}_+(\mathcal{N}_++1)e^{sB}dsdt\nonumber\\
\mathcal{E}^B_{23,3}=&\frac{2}{\sqrt{d}}\sum_{p\neq0}v_p^{(a,d)}\eta_p\int_{0}^{1}
\int_{t}^{1}e^{-sB}a_p^*a_pa_0^*a_0^*a_0a_0e^{sB}dsdt\nonumber\\
\mathcal{E}^B_{23,4}=&-\frac{1}{2\sqrt{d}}\sum_{p,q\neq0}v_p^{(a,d)}\eta_q\int_{0}^{1}
\int_{t}^{1}e^{-sB}(2a_0^*a_0+1)(a_p^*a_{-p}^*a_qa_{-q}+h.c.)e^{sB}dsdt.\label{E^B_23,1 detail}
\end{align*}
Using (\ref{v_p^a,d}), (\ref{est of eta_0}) and (\ref{有用的屎}) together yields
\begin{equation}\label{sum_p'v_p^a,deta_p}
  \left\vert\sum_{p\neq0}v_p^{(a,d)}\eta_p\right\vert\leq\frac{Ca}{\sqrt{d}},
\end{equation}
via the assumptions that $a,d$ and $\frac{a}{d}$ tend to $0$ and $\frac{d}{a}>\frac{C}{\textit{l}}$. Combining (\ref{sum_p'v_p^a,deta_p}) with Lemma \ref{control of S_+ conj with e^B} yields directly
\begin{equation}\label{010401}
  \pm\mathcal{E}^B_{23,1}\leq CNad^{-1}
(e^{CNa\textit{l}^{\frac{1}{2}}}-1)(\mathcal{N}_++1),
\end{equation}
and
\begin{equation}\label{010402}
  \pm\mathcal{E}^B_{23,2}\leq
Ce^{CNa\textit{l}^{\frac{1}{2}}}ad^{-1}(\mathcal{N}_++1)^2.
\end{equation}
Using (\ref{v_p^a,d}) and (\ref{est of eta_0}) to bound $\vert v_p^{(a,d)}\vert\leq Cad^{-\frac{1}{2}}$ and $\vert\eta_p\vert\leq Cad\textit{l}^2$ respectively and the fact that $a_p^*a_pa_0^*a_0^*a_0a_0\geq 0$ for any $p\neq0$ yields
\begin{align}
\pm\mathcal{E}^B_{23,3}
&\leq Ca^2\textit{l}^2\int_{0}^{1}
\int_{t}^{1}e^{-sB}a_0^*a_0^*a_0a_0\mathcal{N}_+e^{sB}dsdt\nonumber\\
&\leq  Ce^{CNa\textit{l}^{\frac{1}{2}}}N^2a^2\textit{l}^2(\mathcal{N}_++1),\label{010403}
\end{align}
where we have again used (\ref{e^-tB(N_++1)e^tB}) in the last inequality. To estimate $\mathcal{E}_{23,4}^B$ we first rewrite
\begin{align*}
  e^{-sB}a_p^*a_{-p}^*e^{sB}&=a_p^*a_{-p}^*+\int_{0}^{s}e^{-\tau B}
  [a_p^*a_{-p}^*,B]e^{\tau B}d\tau\\
  &=a_p^*a_{-p}^*+\eta_p\int_{0}^{s}e^{-\tau B}a_0^*a_0^*
  (a_p^*a_p+a_{-p}^*a_{-p}+1)e^{\tau B}d\tau.
\end{align*}
Then we argue analogously to the proof of Lemma \ref{lemma e^-BH_3e^B}. Let
\begin{equation*}
  \mathcal{E}_{23,4}^B=\sum_{j=1}^{3}\mathcal{E}_{23,4,j}^B+h.c.
\end{equation*}
with
\begin{align*}
  \mathcal{E}_{23,4,1}^B =& \frac{1}{2\sqrt{d}}\sum_{p,q\neq0}
v_p^{(a,d)}\eta_q\int_{0}^{1}\int_{t}^{1}a_p^*a_{-p}^*e^{-sB}
[a_qa_{-q}(2a_0^*a_0+1)]e^{sB}dsdt \\
  \mathcal{E}_{23,4,2}^B =& \frac{1}{\sqrt{d}}\sum_{p,q\neq0}
v_p^{(a,d)}\eta_p\eta_q\int_{0}^{1}\int_{t}^{1}\int_{0}^{s}
e^{-\tau B}a_0^*a_0^*a_p^*a_pe^{(\tau-s)B}
[a_qa_{-q}(2a_0^*a_0+1)]e^{sB}d\tau dsdt\\
  \mathcal{E}_{23,4,3}^B =& \frac{1}{2\sqrt{d}}\sum_{p,q\neq0}
v_p^{(a,d)}\eta_p\eta_q\int_{0}^{1}\int_{t}^{1}\int_{0}^{s}
e^{-\tau B}a_0^*a_0^*e^{(\tau-s)B}
[a_qa_{-q}(2a_0^*a_0+1)]e^{sB}d\tau dsdt
\end{align*}
Estimating on Fock space, we have for any $\psi\in L^2_s(\Lambda_d)$
\begin{align*}
  \vert\langle\mathcal{E}_{23,4,1}^B\psi,\psi\rangle\vert
&\leq C\Vert v_a\Vert_1^{\frac{1}{2}}\Vert\eta_\perp\Vert_2d^{-\frac{1}{2}}
\langle H_4\psi,\psi\rangle^{\frac{1}{2}}\\
\times&\int_{0}^{1}\int_{t}^{1}
\langle(\mathcal{N}_++1)^2(2a_0^*a_0+1)e^{sB}\psi,(2a_0^*a_0+1)e^{sB}\psi\rangle^{\frac{1}{2}}
dsdt\\
&\leq Ce^{Na\textit{l}^{\frac{1}{2}}}N^{\frac{3}{2}}a^{\frac{3}{2}}
d^{-\frac{1}{2}}\textit{l}^{\frac{1}{2}}
\langle H_4\psi,\psi\rangle^{\frac{1}{2}}
\langle(\mathcal{N}_++1)\psi,\psi\rangle^{\frac{1}{2}},
\end{align*}
\begin{align*}
    \vert\langle\mathcal{E}_{23,4,2}^B\psi,\psi\rangle\vert
&\leq C\Vert v_a\ast\eta_\perp\Vert_2\Vert\eta_\perp\Vert_2d^{-1}
\langle (\mathcal{N}_++1)^2a_0a_0\psi,a_0a_0\psi\rangle^{\frac{1}{2}}\\
\times&\int_{0}^{1}\int_{t}^{1}\int_{0}^{s}
\langle(\mathcal{N}_++1)^2(2a_0^*a_0+1)e^{sB}\psi,(2a_0^*a_0+1)e^{sB}\psi\rangle^{\frac{1}{2}}
d\tau dsdt\\
&\leq Ce^{Na\textit{l}^{\frac{1}{2}}}N^{3}a^{3}
d^{-1}\textit{l}
\langle(\mathcal{N}_++1)\psi,\psi\rangle,
\end{align*}
and
\begin{align*}
  \vert\langle\mathcal{E}_{23,4,3}^B\psi,\psi\rangle\vert
&\leq C\left\vert\sum_{p\neq0}v_p^{(a,d)}\eta_p\right
\vert\Vert\eta_\perp\Vert_2d^{-\frac{1}{2}}\int_{0}^{1}\int_{t}^{1}\int_{0}^{s}d\tau dsdt\\
\times&
\langle(\mathcal{N}_++1)(2a_0^*a_0+1)e^{sB}\psi,(2a_0^*a_0+1)e^{sB}\psi\rangle^{\frac{1}{2}}\\
\times&
\langle (\mathcal{N}_++1)e^{(s-\tau)B}a_0a_0e^{tB}\psi,
e^{(s-\tau)B}a_0a_0e^{tB}\psi\rangle^{\frac{1}{2}}\\
&\leq Ce^{Na\textit{l}^{\frac{1}{2}}}N^{2}a^{2}
d^{-1}\textit{l}^{\frac{1}{2}}
\langle(\mathcal{N}_++1)\psi,\psi\rangle.
\end{align*}
These three estimates together give
\begin{align}
  \pm\mathcal{E}^B_{23,4}\leq&
 C(N^{\frac{3}{2}}a^{\frac{3}{2}}d^{-{\frac{1}{2}}}\textit{l}^{\frac{1}{2}}+
N^2a^2d^{-1}\textit{l}^{\frac{1}{2}})(\mathcal{N}_++1)\nonumber\\
&+CN^{\frac{3}{2}}a^{\frac{3}{2}}d^{-{\frac{1}{2}}}\textit{l}^{\frac{1}{2}}H_4.\label{010404}
\end{align}
Collecting (\ref{010401}), (\ref{010402}), (\ref{010403}) and (\ref{010404}), we have proved (\ref{E^B_23,1}).

\par (\ref{E^B_23,1 III}) is achieved by redefining $\mathcal{E}^B_{23,2}$ as
\begin{equation*}
  \tilde{\mathcal{E}}^B_{23,2}=\frac{1}{\sqrt{d}}\sum_{p\neq0}v_p^{(a,d)}\eta_p\int_{0}^{1}
\int_{t}^{1}[e^{-sB}\mathcal{N}_+(\mathcal{N}_++1)e^{sB}-
\mathcal{N}_+(\mathcal{N}_++1)]dsdt.
\end{equation*}
Using (\ref{e^-tBN_+e^tB-N_+}) and (\ref{sum_p'v_p^a,deta_p}) we find
\begin{align*}
  \pm\tilde{\mathcal{E}}^B_{23,2}\leq& ad^{-1}(e^{Na\textit{l}^{\frac{1}{2}}}-1)
(\mathcal{N}_++1)^2\\
\leq &N^2a^2d^{-1}\textit{l}^{\frac{1}{2}}(\mathcal{N}_++1).
\end{align*}

\begin{flushright}
  $\Box$
\end{flushright}

\begin{lemma}\label{lemma [H_23',B]}
\
\begin{enumerate}[$(1)$]
\item $\mathbf{For\; Region\; I}$
\begin{equation}\label{答辩二号}
   \int_{0}^{1}\int_{0}^{t}e^{-sB}[{H}_{23}^{\prime},B]e^{sB}dsdt
   = N(N-1)\sum_{p\neq0}W_p\eta_p+\mathcal{E}^B_{[H_{23}^\prime,B]},
\end{equation}
where $W_p$ is defined in (\ref{define W_p}) and
\begin{equation}\label{E^B_23,2}
  \pm\mathcal{E}^B_{[H_{23}^\prime,B]}\leq C(N^3a^3d^{-1}\textit{l}
+Na^2d^{-2}\textit{l}^{-1})
(\mathcal{N}_++1)+CNa^2d^{-2}\textit{l}^{-1}(\mathcal{N}_++1)^2,
\end{equation}
\item $\mathbf{For\; Region\; III}$
\begin{align}
   \int_{0}^{1}\int_{0}^{t}e^{-sB}[{H}_{23}^{\prime},B]e^{sB}&dsdt
   = N(N-1)\sum_{p\neq0}W_p\eta_p-2N\sum_{p\neq0}W_p\eta_p\mathcal{N}_+\nonumber\\
&+\sum_{p\neq0}W_p\eta_p\mathcal{N}_+(\mathcal{N}_++1)
+\tilde{\mathcal{E}}^B_{[H_{23}^\prime,B]},\label{答辩二号 III}
\end{align}
where
\begin{align}
  \pm\tilde{\mathcal{E}}^B_{[H_{23}^\prime,B]}\leq&
 C(N^{\frac{3}{2}}a^{\frac{3}{2}}d^{-{\frac{1}{2}}}\textit{l}^{\frac{1}{2}}+
N^2a^2d^{-1}\textit{l}^{\frac{1}{2}})(\mathcal{N}_++1)\nonumber\\
&+CN^{\frac{3}{2}}a^{\frac{3}{2}}d^{-{\frac{1}{2}}}\textit{l}^{\frac{1}{2}}H_4^\prime,
\label{E^B_23,2 III}
\end{align}
and $H_4^\prime$ is defined in (\ref{define H'_4}).
\end{enumerate}
\end{lemma}
\noindent
\emph{Proof.} For the proof of Lemma \ref{lemma [H_23',B]} we can argue similarly to Lemma \ref{lemma [H_23,B]} since we notice that
\begin{align*}
[H_{23}^{\prime},B]=&\frac{1}{2}\sum_{p,q\neq0}W_p\eta_q(-4a_0^*a_0-2)
(a_q^*a_{-q}^*a_{p}a_{-p}+h.c.)\\
&+2\sum_{p\neq0}W_p\eta_p(1+2a^*_pa_p)a_0^*a_0^*a_0a_0.
\end{align*}
Again expanding $a_0^*a_0^*a_0a_0$ we can reach (\ref{答辩二号}) by defining
\begin{equation*}
  \mathcal{E}^B_{[H_{23}^\prime,B]}=\sum_{i=1}^{4}\mathcal{E}^B_{23^\prime,i},
\end{equation*}
where
\begin{align*}
\mathcal{E}^B_{23^\prime,1}=&-4N\sum_{p\neq0}W_p\eta_p
\int_{0}^{1}\int_{t}^{1}e^{-sB}\mathcal{N}_+e^{sB}dsdt\nonumber\\
\mathcal{E}^B_{23^\prime,2}=&2\sum_{p\neq0}W_p\eta_p\int_{0}^{1}
\int_{t}^{1}e^{-sB}\mathcal{N}_+(\mathcal{N}_++1)e^{sB}dsdt\nonumber\\
\mathcal{E}^B_{23^\prime,3}=&4\sum_{p\neq0}W_p\eta_p\int_{0}^{1}
\int_{t}^{1}e^{-sB}a_p^*a_pa_0^*a_0^*a_0a_0e^{sB}dsdt\nonumber\\
\mathcal{E}^B_{23^\prime,4}=&-\sum_{p,q\neq0}W_p\eta_q\int_{0}^{1}
\int_{t}^{1}e^{-sB}(2a_0^*a_0+1)(a_p^*a_{-p}^*a_qa_{-q}+h.c.)e^{sB}dsdt.
\end{align*}
We can rewrite $\mathcal{E}^B_{23^\prime,4}$ similarly
\begin{equation*}
  \mathcal{E}_{23^\prime,4}^B=\sum_{j=1}^{3}\mathcal{E}_{23^\prime,4,j}^B+h.c.
\end{equation*}
with
\begin{align*}
  \mathcal{E}_{23^\prime,4,1}^B =& \sum_{p,q\neq0}
W_p\eta_q\int_{0}^{1}\int_{t}^{1}a_p^*a_{-p}^*e^{-sB}
[a_qa_{-q}(2a_0^*a_0+1)]e^{sB}dsdt \\
  \mathcal{E}_{23^\prime,4,2}^B =& 2\sum_{p,q\neq0}
W_p\eta_p\eta_q\int_{0}^{1}\int_{t}^{1}\int_{0}^{s}
e^{-\tau B}a_0^*a_0^*a_p^*a_pe^{(\tau-s)B}
[a_qa_{-q}(2a_0^*a_0+1)]e^{sB}d\tau dsdt\\
  \mathcal{E}_{23^\prime,4,3}^B =& \sum_{p,q\neq0}
W_p\eta_p\eta_q\int_{0}^{1}\int_{t}^{1}\int_{0}^{s}
e^{-\tau B}a_0^*a_0^*e^{(\tau-s)B}
[a_qa_{-q}(2a_0^*a_0+1)]e^{sB}d\tau dsdt
\end{align*}
By definition (\ref{define W(x)}) we have
\begin{equation*}
  W(\mathbf{x})=\frac{\lambda_{\textit{l}}}{a^2\sqrt{d}}
(\chi_{d\textit{l}}(\mathbf{x})-\widetilde{w}_{\textit{l}}(\mathbf{x})).
\end{equation*}
Using Lemma \ref{fundamental est of v,w,lambda}, (\ref{est of eta and eta_perp}) and (\ref{est of eta_0}) we can estimate that
\begin{equation}\label{L1&L2 norm of W}
 \Vert W\Vert_2\leq Cad^{-2}\textit{l}^{-\frac{3}{2}},\quad
  \Vert W\Vert_1\leq Cad^{-\frac{1}{2}},
\end{equation}
and
\begin{equation}\label{sum_pW_peta_p}
  \vert W_p\vert\leq\frac{Ca}{d},\quad\left\vert\sum_{p\neq0}W_p\eta_p\right\vert\leq Ca^2d^{-2}\textit{l}^{-1}.
\end{equation}
Again by an argument similar to the proof of (\ref{E^B_23,1}) we can prove (\ref{E^B_23,2}). Notice that the only different estimate is that for any $\psi\in L^2_s(\Lambda_d)$
\begin{align*}
  \vert\langle\mathcal{E}_{23^\prime,4,1}^B\psi,\psi\rangle\vert
&\leq C\Vert W\Vert_2\Vert\eta_\perp\Vert_2
\langle (\mathcal{N}_++1)^2\psi,\psi\rangle^{\frac{1}{2}}\\
\times&\int_{0}^{1}\int_{t}^{1}
\langle(\mathcal{N}_++1)^2(2a_0^*a_0+1)e^{sB}\psi,(2a_0^*a_0+1)e^{sB}\psi\rangle^{\frac{1}{2}}
dsdt\\
&\leq Ce^{Na\textit{l}^{\frac{1}{2}}}Na^{2}
d^{-2}\textit{l}^{-1}
\langle(\mathcal{N}_++1)^2\psi,\psi\rangle,
\end{align*}
The proof of (\ref{答辩二号 III}) and (\ref{E^B_23,2 III}) resembles the proof of (\ref{答辩一号 III}) and (\ref{E^B_23,1 III}). We only need to notice that this time $H_4^\prime$ plays the role of $H_4$ and we have a formula similar to (\ref{H_4psi,psi fock}) to reproduce its energy by
\begin{equation}\label{H_4'psi,psi}
  \langle H_4^\prime\psi,\psi\rangle
  =\sum_{n=2}^{N}n(n-1)\int_{\Lambda_d^n}\sqrt{d}
W(\mathbf{x}_1-\mathbf{x}_2)\vert\alpha^{(n)}\vert^2
\end{equation}
for $\psi\in L^2_s(\Lambda_{d}^N)$ and $U_{N}\psi=(\alpha^{(0)},\dots,\alpha^{(N)})$.
\begin{flushright}
  {$\Box$}
\end{flushright}


\par With all the preparations above, we can prove Proposition \ref{quadratic renorm}.
\\
\noindent
\emph{Proof of Proposition \ref{quadratic renorm}.}
\begin{flushleft}
  $\mathbf{For\; Region\; I}$
\end{flushleft}

\par We collect all the lemmas above, if a lemma has two statements we choose the first one (i.e. we combine (\ref{e^-BH_02e^B}), (\ref{e^-BH_22e^B}), (\ref{e^-BH_3e^B}), (\ref{int_0^1 e^-tBGammae^tBdt}), (\ref{答辩一号}) and (\ref{答辩二号})), we then reached
\begin{align*}
  e^{-B}H_Ne^B=&\frac{N(N-1)}{2\sqrt{d}}\left(v_0^{(a,d)}+\sum_{p\neq0}v_p^{(a,d)}\eta_p\right)
  +N(N-1)\sum_{p\neq0}W_p\eta_p\\
  &+\frac{N}{\sqrt{d}}\left(v_0^{(a,d)}-\sum_{p\neq0}v_p^{(a,d)}\eta_p\right)\mathcal{N}_+
+H_{21}+H_4+H_3+H_{23}^{\prime}+\mathcal{E}^B,
\end{align*}
where $\mathcal{E}^B$ is bounded by
\begin{align*}
  \pm\mathcal{E}^B\leq & C\Big\{\big(Na^2d^{-2}\textit{l}^{-1}+
N^{2}a^{2}d^{-1}\textit{l}^{\frac{1}{2}}+
N^{\frac{3}{2}}a^{\frac{3}{2}}d^{-\frac{1}{2}}\textit{l}^{\frac{1}{2}}\big)
(\mathcal{N}_++1)\\
  &+\big(ad^{-1}+Na^2d^{-2}\textit{l}^{-1}\big)(\mathcal{N}_++1)^2
  +Na^3d^{-1}H_{21}
  +N^{\frac{3}{2}}a^{\frac{3}{2}}d^{-\frac{1}{2}}\textit{l}^{\frac{1}{2}}H_4\Big\},
\end{align*}
which are (\ref{first renorm}) and (\ref{first renorm E^B}) as claimed.

\begin{flushleft}
  $\mathbf{For\; Region\; III}$
\end{flushleft}
\par We collect all the lemmas above but choose the second statement (if stated), that is to say we use (\ref{e^-BH_02e^B III}), (\ref{e^-BH_22e^B III}), (\ref{e^-BH_3e^B}), (\ref{int_0^1 e^-tBGammae^tBdt}), (\ref{答辩一号 III}) and (\ref{答辩二号 III}) to derive
\begin{align*}
  e^{-B}H_Ne^B=&\frac{N(N-1)}{2\sqrt{d}}\sum_{p\neq0}v_p^{(a,d)}\eta_p
-\frac{N}{\sqrt{d}}\sum_{p\neq0}v_p^{(a,d)}\eta_p\mathcal{N}_+\\
  &  +\frac{1}{2\sqrt{d}}\sum_{p\neq0}v_p^{(a,d)}\eta_p\mathcal{N}_+(\mathcal{N}_++1)
+N(N-1)\sum_{p\neq0}W_p\eta_p\\
&-2N\sum_{p\neq0}W_p\eta_p\mathcal{N}_++\sum_{p\neq0}W_p\eta_p\mathcal{N}_+(\mathcal{N}_++1)\\
&+H_{01}+H_{02}+H_{22}+H_3+H_{23}^{\prime}+H_{21}+H_4+\tilde{\mathcal{E}}^B,
\end{align*}
where
\begin{align*}
  \pm\tilde{\mathcal{E}}^B\leq&
 C\big(N^{\frac{3}{2}}a^{\frac{3}{2}}d^{-{\frac{1}{2}}}\textit{l}^{\frac{1}{2}}+
N^2a^2d^{-1}\textit{l}^{\frac{1}{2}}\big)(\mathcal{N}_++1)\nonumber\\
&+CN^{\frac{3}{2}}a^{\frac{3}{2}}d^{-{\frac{1}{2}}}\textit{l}^{\frac{1}{2}}
\big(H_4+H_4^\prime\big).
\end{align*}
Thus we conclude (\ref{first renorm III}) and (\ref{first renorm E^B III}).
\begin{flushright}
  {$\Box$}
\end{flushright}





\section{3D Cubic Renormalization for Regions I \& III}\label{3}
\par In this section we analyze the excitation Hamiltonian $\mathcal{J}_N$ and prove Propositon \ref{cubic renorm}. We adopt the notation
\begin{equation*}
  A^{\prime}=\sum_{p,q,p+q\neq0}\eta_p\chi_{\vert\mathcal{M}_dq\vert\leq\kappa}
  a_{p+q}^*a_{-p}^*a_qa_0.
\end{equation*}
The cut-off parameter $\kappa$ will be determined later. One can check that $A^{\prime}$ is also a linear operator on $L^2_s(\Lambda_d^N)$ bounded by $N^2\Vert\eta_\perp\Vert_2$. By (\ref{define B'}) we have
\begin{equation*}
  B^\prime=A^\prime-A^{\prime^*}.
\end{equation*}
Due to the presence of the cut-off parameter $\kappa$, we define the notation $P_\kappa$ which is an orthogonal projection given by
\begin{equation}\label{define P_kappa}
  P_\kappa:\bigoplus_{n=0}^NL^2(\Lambda_d^n)\to
\bigoplus_{n=0}^N\left(L^2(\Lambda_d^{(n-1)})\otimes
\bigoplus_{\vert\mathcal{M}_dp\vert\leq\kappa}\textit{span}\{\phi_p^{(d)}\}\right).
\end{equation}
We also denote each of its components by
\begin{equation*}
  P_\kappa:L^2(\Lambda_d^n)\to L^2(\Lambda_d^{(n-1)})\otimes
\bigoplus_{\vert\mathcal{M}_dp\vert\leq\kappa}\textit{span}\{\phi_p^{(d)}\}.
\end{equation*}
Before we sketch the proof of Proposition \ref{cubic renorm}, we reiterate that we state the results for Regions I or III here. As for Regions $\mathrm{II}_{\mathrm{I}}$ and $\mathrm{II}_{\mathrm{III}}$, they are regarded as intermediate regions, and corresponding results still apply to these regions without further specifications.

\par For Region I, we split $\mathcal{G}_N$ using (\ref{first renorm}) and analyze respectively their contributions to the ground state energy after we
conjugate them with $e^{B^\prime}$. Like what we have done in Section \ref{2}, we rewrite
\begin{align*}
e^{-B^\prime}\mathcal{G}_Ne^{B^\prime}=&
C^B+e^{-B^\prime}(Q^B\mathcal{N}_++H_{23}^\prime+\mathcal{E}^B)e^{B^\prime}+
e^{-B^\prime}(H_{21}+H_4+H_3)e^{B^\prime}\\
=&C^B+e^{-B^\prime}(Q^B\mathcal{N}_++H_{23}^\prime+\mathcal{E}^B)e^{B^\prime}+
H_{21}+H_4\\
&+\int_{0}^{1}e^{-tB^\prime}[H_{21}+H_4,B^\prime]e^{tB^\prime}dt+
e^{-B^\prime}H_3e^{B^\prime}.
\end{align*}
In 3D cubic renormalization, we want to extract energy generated by the 3D correlation structure hidden in the cubic term $H_{3}$, which contributes to the second order ground state energy. Therefore the term $e^{-B^\prime}(H_{21}+H_4+H_{3})e^{B^\prime}$ is the most important part in this Section. To compute it precisely, we let, for Region I
\begin{equation}\label{Gamma'}
  \Gamma^\prime=[H_{21}+H_4,B^\prime]+H_3,
\end{equation}
then
\begin{align}
e^{-B^\prime}\mathcal{G}_Ne^{B^\prime}=&
C^B+e^{-B^\prime}(Q^B\mathcal{N}_++H_{23}^\prime+\mathcal{E}^B)e^{B^\prime}+
H_{21}+H_4\nonumber\\
&+\int_{0}^{1}e^{-tB^\prime}(\Gamma^\prime-H_3)e^{tB^\prime}dt+
e^{-B^\prime}H_3e^{B^\prime}\nonumber\\
=&C^B+e^{-B^\prime}(Q^B\mathcal{N}_++H_{23}^\prime+\mathcal{E}^B)e^{B^\prime}+H_{21}+H_4\nonumber\\
&+\int_{0}^{1}e^{-tB^\prime}\Gamma^\prime e^{tB^\prime}dt
+\int_{0}^{1}\int_{t}^{1}e^{-sB^\prime}[H_3,B^\prime]e^{sB^\prime}dsdt.\label{依托答辩}
\end{align}
The proof of Proposition \ref{cubic renorm} for Region I is done by analyzing each terms on the right-hand side of (\ref{依托答辩}). In the following lemmas, we bound $e^{-B^\prime}Q^B\mathcal{N}_+e^{B^\prime}$ in Corollary \ref{corollary e^-B'(C^B+Q^BN_+)e^B'}, $e^{-B^\prime}\mathcal{E}^Be^{B^\prime}$ in Corollary \ref{corollary e^-B'E^Be^B'}, $e^{-B^\prime}H_{23}^\prime e^{B^\prime}$ in Lemma \ref{lemma e^-B'H_23'e^B'}. These three terms stay unchanged up to small errors after conjugating with $e^{B^\prime}$. The term containing the residue $\Gamma^\prime$ is bounded in Lemma \ref{lemma Gamma'}, and is a negligible error term as we will prove. The contribution of the commutator $[H_3,B^\prime]$ is calculated in Lemma \ref{lemma [H_3,B']}. As stated priorly, Lemma \ref{lemma [H_3,B']} presents the effect of the 3D cubic correlation structure to the second order ground state energy, also in the form of polynomials of $\mathcal{N}_+$.
\par On the other hand, for Region III, since we can no longer neglect the 2D effect, or in other words the term containing the residue $\Gamma^\prime$ will contribute to the second order ground state energy in Region III, we let
\begin{equation}\label{Gamma' III}
  \tilde{\Gamma}^\prime=[H_{21}+H_4,B^\prime]+H_3-H_3^\prime.
\end{equation}
Here we define $H^\prime_3$ as
\begin{equation}\label{define H_3'}
  H_3^\prime=2\sum_{p,q,p+q\neq0}W_p(a_{p+q}^*a_{-p}^*a_qa_0+h.c.)
\end{equation}
Using (\ref{first renorm III}), a similar calculation gives
\begin{align}
 e^{-B^\prime}\mathcal{G}_Ne^{B^\prime}=&
\tilde{C}^B+e^{-B^\prime}\big[\tilde{Q}_1^B\mathcal{N}_+
+\tilde{Q}_2^B\mathcal{N}_+(\mathcal{N}_++1)+\tilde{\mathcal{E}}^B\big]e^{B^\prime}
  +H_{01}\nonumber\\
&+e^{-B^\prime}(H_{02}+H_{22}+H_{23}^{\prime})e^{B^\prime}
+H_{21}+H_4+H_3^\prime\nonumber\\
&+\int_{0}^{1}e^{-tB^\prime}\tilde{\Gamma}^\prime e^{tB^\prime}dt
+\int_{0}^{1}\int_{t}^{1}e^{-sB^\prime}[H_3,B^\prime]e^{sB^\prime}dsdt\nonumber\\
&+\int_{0}^{1}\int_{0}^{t}e^{-sB^\prime}[H_3^\prime,B^\prime]e^{sB^\prime}dsdt.
\label{依托答辩 III}
\end{align}
To prove Proposition \ref{cubic renorm} for Region III, we analyze each terms on the right-hand side of (\ref{依托答辩 III}). We bound $e^{-B^\prime}\big[\tilde{Q}_1^B\mathcal{N}_+
+\tilde{Q}_2^B\mathcal{N}_+(\mathcal{N}_++1)\big]e^{B^\prime}$ in Corollary \ref{corollary e^-B'(C^B+Q^BN_+)e^B'}, $e^{-B^\prime}\tilde{\mathcal{E}}^Be^{B^\prime}$ in Corollary \ref{corollary e^-B'E^Be^B'}, $e^{-B^\prime}H_{23}^\prime e^{B^\prime}$ in Lemma \ref{lemma e^-B'H_23'e^B'} and $e^{-B^\prime}(H_{02}+H_{22})e^{B^\prime}$ in Lemma \ref{lemma e^-B'(H_02+H_22)e^B'}. These four terms stay unchanged up to small errors after conjugating with $e^{B^\prime}$. The term containing the new difference $\tilde{\Gamma}^\prime$ is bounded in Lemma \ref{lemma Gamma'}, and is again proved to be a negligible error term. The contribution of the commutator $[H_3,B^\prime]$ is calculated in Lemma \ref{lemma [H_3,B']}, and the contribution of $[H_3^\prime,B^\prime]$ is calculated in Lemma \ref{lemma [H'_3,B']}. Lemmas \ref{lemma [H_3,B']} and Lemma \ref{lemma [H'_3,B']} present the effect of the cubic 3D correlation structure to the second order ground state energy in Region III, also in the form of polynomials of $\mathcal{N}_+$.

\par We control the action of $e^{B^\prime}$ on the number of excited particles operator $\mathcal{N}_+$ in Lemma \ref{control of S_+ conj with e^B'}. It is also useful in the calculations of 3D cubic renormalization to estimate the growth of kinetic operator $H_{21}$ and the non-zero momentum sum of potential operator $H_4$, and the modified potential operator $H_4^\prime$ defined in (\ref{define H'_4}), with respect to the action of $e^{B^\prime}$. To this end, as well as to compute the residues $\Gamma^\prime$ and $\tilde{\Gamma}^\prime$, we first compute the commutators in Lemma \ref{commutator of H_21,H_4with B'}. We then show the a-priori bounds on the growths of $H_{21}$, $H_4$ and $H_4^\prime$ in Lemma \ref{lemma e^-B'H_Ne^B'}. One can compare this result with Lemma \ref{lemma control of H_4} in 3D quadratic renormalization. In Lemma \ref{lemma e^-B'H_Ne^B'}, the decent bounds on the growths significantly simplify our calculations.

\begin{lemma}\label{control of S_+ conj with e^B'}
Let $\mathcal{N_+}$ be defined on $L_s^2(\Lambda_d^N)$ as stated in (\ref{define of N_+}), then there exist a constant $C_n$ depending only on $n\in\frac{1}{2}\mathbb{N}$ such that: for every $t\in\mathbb{R}$, $N\in\mathbb{N}$, $n\in\frac{1}{2}\mathbb{N}$, $\textit{l}\in(0,\frac{1}{2})$ and $\frac{d}{a}>\frac{C}{\textit{l}}$ for some universal constant $C$, we have
\begin{align}
  e^{-tB^\prime}(\mathcal{N}_++1)^ne^{tB^\prime}&\leq e^{C_nNa\textit{l}^{\frac{1}{2}}\vert t\vert}(\mathcal{N}_++1)^n,\label{e^-tB'(N_++1)e^tB'}\\
  \pm(e^{-tB^\prime}(\mathcal{N}_++1)^ne^{tB^\prime}-(\mathcal{N}_++1)^n)&\leq (e^{C_nNa\textit{l}^{\frac{1}{2}}\vert t\vert}-1) (\mathcal{N}_++1)^n.\label{e^-tB'N_+e^tB'-N_+}
\end{align}
\end{lemma}

\noindent
\emph{Proof.} We follow exactly what we have done in the proof of Lemma \ref{control of S_+ conj with e^B}. We first notice that
\begin{equation*}
  [\mathcal{N}_+,A^{\prime}]=A^{\prime}.
\end{equation*}
Let $\psi\in L^2_s(\Lambda_{d}^N)$, a calculation similar to (\ref{Abs Apsi,psi}) gives
\begin{align}
  \vert\langle A^{\prime}\psi,\psi\rangle\vert
&\leq C\Vert\eta_\perp\Vert_2
\langle U_N(\mathcal{N}_++1)U_N^*P_\kappa U_Na_0\psi,P_\kappa U_Na_0\psi\rangle^{\frac{1}{2}}
\langle (\mathcal{N}_++1)^2\psi,\psi\rangle^{\frac{1}{2}}\nonumber\\
&\leq C\Vert\eta_\perp\Vert_2
\langle (\mathcal{N}_++1)a_0\psi,a_0\psi\rangle^{\frac{1}{2}}
\langle (\mathcal{N}_++1)^2\psi,\psi\rangle^{\frac{1}{2}}\nonumber\\
&\leq CN\Vert\eta_\perp\Vert_2
\langle(\mathcal{N}_++1)\psi,\psi\rangle.\label{A'}
\end{align}
Then (\ref{e^-tB'(N_++1)e^tB'}) and (\ref{e^-tB'N_+e^tB'-N_+}) for $n=1$ follow using Gronwall's inequality and (\ref{est of eta and eta_perp}).
\par (\ref{e^-tB'(N_++1)e^tB'}) for arbitrary $n\in\frac{1}{2}\mathbb{N}$ follows by noticing the facts that
\begin{equation*}
  (\mathcal{N}_++1)^{\frac{1}{2}}A^{\prime}=A^{\prime}(\mathcal{N}_++2)^{\frac{1}{2}},
\end{equation*}
\begin{equation*}
  A^{\prime}(\mathcal{N}_++1)^{\frac{1}{2}}=\mathcal{N}_+^{\frac{1}{2}}A^{\prime}.
\end{equation*}
The remaining proof is just a repeat of the proof of Lemma \ref{control of S_+ conj with e^B}.\hfill  {$\Box$}

\par From here on out without further specification we will always assume that $N$ tends to infinity, $a$, $d$, $\frac{a}{d}$ and $Na\textit{l}^{\frac{1}{2}}$ tend to $0$ and $\frac{d}{a}>\frac{C}{\textit{l}}$ for some universal constant $C$.

\begin{lemma}\label{commutator of H_21,H_4with B'}
\begin{align}
[H_{21},B^{\prime}]=&2\sum_{p,q,p+q\neq0}\eta_p\vert\mathcal{M}_dp\vert^2
\chi_{\vert\mathcal{M}_dq\vert\leq\kappa}
(a_{p+q}^*a_{-p}^*a_qa_0+h.c.)+\mathcal{E}_{21}^{B^\prime},\label{[H_21.B']}\\
{[H_4,B^{\prime}]}=&\frac{1}{\sqrt{d}}\sum_{p,q,p+q\neq0}\left(\sum_{r\neq0}v_{p-r}^{(a,d)}\eta_r
\right)\chi_{\vert\mathcal{M}_dq\vert\leq\kappa}
(a_{p+q}^*a_{-p}^*a_qa_0+h.c.)+\mathcal{E}_{4}^{B^{\prime}},\label{[H_4,B']}\\
{[H_4^\prime,B^{\prime}]}=&2\sum_{p,q,p+q\neq0}\left(\sum_{r\neq0}W_{p-r}\eta_r
\right)\chi_{\vert\mathcal{M}_dq\vert\leq\kappa}
(a_{p+q}^*a_{-p}^*a_qa_0+h.c.)+\mathcal{E}_{4^\prime}^{B^{\prime}},\label{[H_4',B']}
\end{align}
where $H_4^\prime$ is given in (\ref{define H'_4}) and
\begin{enumerate}[$(1)$]
  \item  $\mathbf{For\; Region\; I}$ We take $\kappa=\nu d^{-1}$ for some $\nu\geq 1$, then for some $0<\gamma<1$
\begin{align}
\pm\mathcal{E}_{21}^{B^\prime}\leq& CNa\textit{l}^{\frac{1}{2}}H_{21},\label{E_21^B'}\\
\pm\mathcal{E}_{4}^{B^{\prime}}\leq&C(Na^3\kappa^3\textit{l})^\gamma
 H_4+C(Na^3\kappa^3\textit{l})^{1-\gamma}
(\mathcal{N}_++1)^3.\label{E_4^B'}
\end{align}
  \item  $\mathbf{For\; Region\; III}$ We take $\kappa=\infty$, then
\begin{align}
\pm\mathcal{E}_{21}^{B^\prime}\leq& CNa\textit{l}^{\frac{1}{2}}H_{21},\label{E_21^B' III}\\
\pm\mathcal{E}_{4}^{B^{\prime}}\leq&C(Na\textit{l}^{\frac{1}{2}}
+N^{\frac{3}{2}}a^{\frac{7}{6}}d^{-\frac{1}{2}}\textit{l}^{\frac{1}{3}})
H_4+CN^2a^2d^{-1}\textit{l}^{\frac{1}{2}}
(\mathcal{N}_++1)\nonumber\\
&+CN^{\frac{3}{2}}a^{\frac{7}{6}}d^{-\frac{1}{2}}\textit{l}^{\frac{1}{3}}
H_{21},\label{E_4^B' III}\\
\pm\mathcal{E}_{4^\prime}^{B^{\prime}}\leq&C(Na\textit{l}^{\frac{1}{2}}
+N^{\frac{3}{2}}a^{\frac{7}{6}}d^{-\frac{1}{2}}\textit{l}^{\frac{1}{3}})
H_4^\prime+CN^2a^2d^{-1}\textit{l}^{\frac{1}{2}}
(\mathcal{N}_++1)\nonumber\\
&+CN^{\frac{3}{2}}a^{\frac{7}{6}}d^{-\frac{1}{2}}\textit{l}^{\frac{1}{3}}
H_{21},\label{E_4'^B' III}
\end{align}
\end{enumerate}
\end{lemma}
\noindent
\emph{Proof.} A direct calculation gives
\begin{equation*}
  [H_{21},B^\prime]=2\sum_{p,q,p+q\neq0}\eta_p\left(\vert\mathcal{M}_dp\vert^2
+\mathcal{M}_dp\cdot\mathcal{M}_dq\right)\chi_{\vert\mathcal{M}_dq\vert\leq\kappa}
(a_{p+q}^*a_{-p}^*a_qa_0+h.c.).
\end{equation*}
we therefore define
\begin{equation*}
  \mathcal{E}_{21}^{B^\prime}=2\sum_{p,q,p+q\neq0}\eta_p
(\mathcal{M}_dp\cdot\mathcal{M}_dq)\chi_{\vert\mathcal{M}_dq\vert\leq\kappa}
(a_{p+q}^*a_{-p}^*a_qa_0+h.c.).
\end{equation*}
Let $\psi\in L^2_s(\Lambda_d^N)$, we have
\begin{align*}
&\vert\langle \mathcal{E}_{21}^{B^\prime}\psi,\psi\rangle\vert\\
\leq&4\left\vert\sum_{p,q,p+q\neq0}\eta_p
(\mathcal{M}_dp\cdot\mathcal{M}_dq)\chi_{\vert\mathcal{M}_dq\vert\leq\kappa}
\langle(\mathcal{N}_++1)^{\frac{1}{2}}
a_qa_0\psi,(\mathcal{N}_++1)^{-\frac{1}{2}}a_{-p}a_{p+q}\psi\rangle\right\vert\\
\leq&4\sum_{p,q,p+q\neq0}\vert\eta_p\vert
\vert\mathcal{M}_dp\vert\vert\mathcal{M}_dq\vert\chi_{\vert\mathcal{M}_dq\vert\leq\kappa}
\Vert(\mathcal{N}_++1)^{\frac{1}{2}}
a_qa_0\psi\Vert\cdot\Vert(\mathcal{N}_++1)^{-\frac{1}{2}}a_{-p}a_{p+q}\psi\Vert\\
\leq&C\delta\sum_{p,p+q\neq0}\vert\mathcal{M}_dp\vert^2
\langle a_{p+q}^*a_{-p}^*(\mathcal{N}_++1)^{-1}a_{-p}a_{p+q}\psi,\psi\rangle\\
&+\frac{C}{\delta}\sum_{p,q\neq0}\vert\eta_p\vert^2\cdot\vert\mathcal{M}_dq\vert^2
\chi_{\vert\mathcal{M}_dq\vert\leq\kappa}\langle a_q^*(\mathcal{N}_++1)a_qa_0^*a_0\psi,\psi\rangle\\
\leq&C\delta\langle H_{21}\psi,\psi\rangle+\frac{C}{\delta}N^2\Vert\eta_\perp\Vert^2_2
\langle H_{21}\psi,\psi\rangle.
\end{align*}
Then (\ref{E_21^B'}) and (\ref{E_21^B' III}) follow by noticing (\ref{est of eta and eta_perp}) and taking
$\delta=Na\textit{l}^{\frac{1}{2}}$.
\par Calculating directly also gives (\ref{[H_4,B']}) with
$\mathcal{E}_{4}^{B^{\prime}}=\sum_{i=1}^{6}\mathcal{E}_{4,i}^{B^{\prime}}$ where
\begin{align*}
\mathcal{E}_{4,1}^{B^{\prime}}=&-\frac{1}{2\sqrt{d}}\sum_{\substack{p,q,p+r,q+r,\\s,s+p+r\neq0}}
v_r^{(a,d)}\eta_s\chi_{\vert\mathcal{M}_d(p+r)\vert\leq\kappa}
(a_{s+p+r}^*a_{-s}^*a_q^*a_{q+r}a_pa_0+h.c.),\\
\mathcal{E}_{4,2}^{B^{\prime}}=&-\frac{1}{2\sqrt{d}}\sum_{\substack{p,q,p+r,q+r,\\s,s+q\neq0}}
v_r^{(a,d)}\eta_s\chi_{\vert\mathcal{M}_dq\vert\leq\kappa}
(a_{s+q}^*a_{-s}^*a_{p+r}^*a_{q+r}a_pa_0+h.c.),\\
\mathcal{E}_{4,3}^{B^{\prime}}=&\frac{1}{2\sqrt{d}}\sum_{\substack{p,q,p+r,q+r,\\s,q+r-s\neq0}}
v_r^{(a,d)}\eta_s\chi_{\vert\mathcal{M}_d(q+r-s)\vert\leq\kappa}
(a_{p+r}^*a_{-s}^*a_q^*a_{q+r-s}a_pa_0+h.c.),\\
\mathcal{E}_{4,4}^{B^{\prime}}=&\frac{1}{2\sqrt{d}}\sum_{\substack{p,q,p+r,q+r,\\s,p-s\neq0}}
v_r^{(a,d)}\eta_s\chi_{\vert\mathcal{M}_d(p-s)\vert\leq\kappa}
(a_{p+r}^*a_{-s}^*a_q^*a_{p-s}a_{q+r}a_0+h.c.),\\
\mathcal{E}_{4,5}^{B^{\prime}}=&\frac{1}{2\sqrt{d}}\sum_{\substack{p,q,p+r,q+r,\\s,s-q-r\neq0}}
v_r^{(a,d)}\eta_{q+r}\chi_{\vert\mathcal{M}_ds\vert\leq\kappa}
(a_{p+r}^*a_q^*a_{s-q-r}^*a_pa_sa_0+h.c.),\\
\mathcal{E}_{4,6}^{B^{\prime}}=&\frac{1}{2\sqrt{d}}\sum_{\substack{p,q,p+r,q+r,\\s,p-s\neq0}}
v_r^{(a,d)}\eta_{p}\chi_{\vert\mathcal{M}_d(p-s)\vert\leq\kappa}
(a_{p+r}^*a_q^*a_{-s}^*a_{q+r}a_{p-s}a_0+h.c.).
\end{align*}
By a change of variables one can check that $\mathcal{E}_{4,1}^{B^{\prime}}=\mathcal{E}_{4,2}^{B^{\prime}}$,
$\mathcal{E}_{4,3}^{B^{\prime}}=\mathcal{E}_{4,4}^{B^{\prime}}$, and $\mathcal{E}_{4,5}^{B^{\prime}}=\mathcal{E}_{4,6}^{B^{\prime}}$.
We can estimate the error terms directly using the definition of creation and annihilation operators.

\par The details of the techniques involved in the calculation have already been provided in the same way in the proof of Lemma \ref{lemma e^-BH_3e^B}. For Region I we denote
\begin{equation}\label{zeta}
  \zeta^{(\kappa)}(\mathbf{x})=\frac{1}{\sqrt{d}}\sum_{0<\vert\mathcal{M}_dp\vert\leq\kappa}
\phi_p^{(d)}(\mathbf{x}).
\end{equation}
$\zeta^{(\kappa)}$ is a real-valued, even function bounded by
\begin{equation}\label{L2 zeta}
  \Vert\zeta^{(\kappa)}\Vert_2^2=\sum_{0<\vert\mathcal{M}_dp\vert\leq\kappa}\frac{1}{d}
\leq C\kappa^3.
\end{equation}
We stress here that our prior assumption $\kappa d\geq 1$ without which (\ref{L2 zeta}) won't hold. Then we can estimate $\mathcal{E}_{4,i}^{B^{\prime}}$ respectively. For $i=1,2$ we have
\begin{align*}
  \vert\langle\mathcal{E}_{4,i}^{B^{\prime}}\psi,\psi\rangle\vert
\leq&{C}\Vert\zeta^{(\kappa)}\Vert_2\Vert\eta_\perp\Vert_2\Vert v_a\Vert_1^{\frac{1}{2}}
\langle H_4a_0\psi,a_0\psi\rangle^{\frac{1}{2}}
\langle (\mathcal{N}_++1)^3\psi,\psi\rangle^{\frac{1}{2}}\\
\leq&C(Na^3\kappa^3\textit{l})^\gamma
\langle H_4\psi,\psi\rangle+C(Na^3\kappa^3\textit{l})^{1-\gamma}
\langle(\mathcal{N}_++1)^3\psi,\psi\rangle.
\end{align*}
For some $0<\gamma<1$. As for $i=3,4,5,6$ we have
\begin{align*}
  \vert\langle\mathcal{E}_{4,i}^{B^{\prime}}\psi,\psi\rangle\vert
\leq&{C}\Vert\zeta^{(\kappa)}\Vert_2\Vert\eta_\perp\Vert_2\Vert v_a\Vert_1^{\frac{1}{2}}
\langle H_4\psi,\psi\rangle^{\frac{1}{2}}
\langle (\mathcal{N}_++1)^3a_0\psi,a_0\psi\rangle^{\frac{1}{2}}\\
\leq&C(Na^3\kappa^3\textit{l})^\gamma
\langle H_4\psi,\psi\rangle+C(Na^3\kappa^3\textit{l})^{1-\gamma}
\langle(\mathcal{N}_++1)^3\psi,\psi\rangle.
\end{align*}
This concludes the proof of (\ref{E_4^B'}). As for Region III, since we have put $\kappa=\infty$, we can estimate directly for $i=1,2,3,4$
\begin{align*}
  \vert\langle\mathcal{E}_{4,i}^{B^{\prime}}\psi,\psi\rangle\vert
\leq&C\Vert\eta\Vert_2\langle H_4a_0\psi,a_0\psi\rangle^{\frac{1}{2}}\langle H_4\mathcal{N}_+\psi,\psi\rangle^{\frac{1}{2}}\\
&+\Vert\eta\Vert_2\Vert v_a\Vert_1^{\frac{1}{2}}d^{-\frac{1}{2}}\langle H_4a_0\psi,a_0\psi\rangle^{\frac{1}{2}}\langle (\mathcal{N}_++1)^3\psi,\psi\rangle^{\frac{1}{2}}\\
\leq&CNa\textit{l}^{\frac{1}{2}}\langle H_4\psi,\psi\rangle+
N^2a^2d^{-1}\textit{l}^{\frac{1}{2}}\langle(\mathcal{N}_++1)\psi,\psi\rangle.
\end{align*}
The estimates for $i=5,6$ need a new technique. Taking $\mathcal{E}_{4,6}^{B^{\prime}}$ for example, we first let
\begin{align*}
U_{N-1}a_0\psi=&(\beta^{(0)},\dots,\beta^{(N-1)}),\\
U_N\psi=&(\alpha^{(0)},\dots,\alpha^{(N)}).
\end{align*}
Then we can calculate directly
\begin{align*}
  \langle\mathcal{E}_{4,6}^{B^{\prime}}\psi,\psi\rangle=&
\frac{1}{2}\sum_{n=2}^{N-1}\sqrt{n+1}n(n-1)
\int_{\Lambda_d^{n+1}}d\mathbf{x}_1\dots d\mathbf{x}_{n+1}\\
&\times\eta_\perp(\mathbf{x}_1-\mathbf{x}_3)v_a(\mathbf{x}_1-\mathbf{x}_2)
\beta^{(n)}(\mathbf{x}_2,\dots,\mathbf{x}_{n+1})\overline{\alpha^{(n+1)}}
(\mathbf{x}_1,\dots,\mathbf{x}_{n+1}).
\end{align*}
By Cauchy-Schwartz we have
\begin{align*}
\vert\langle\mathcal{E}_{4,6}^{B^{\prime}}\psi,\psi\rangle\vert
\leq&CN\left(\sum_{n=2}^{N-1}n\int_{\Lambda_d^{n}}
\vert\eta_\perp\vert^2\ast v_a(\mathbf{x}_2-\mathbf{x}_3)
\vert\beta^{(n)}\vert^2\right)^{\frac{1}{2}}\\
&\times\left(\sum_{n=2}^{N-1}n(n+1)
\int_{\Lambda_d^{n+1}}v_a(\mathbf{x}_1-\mathbf{x}_2)\vert\alpha^{(n+1)}
\vert^2\right)^{\frac{1}{2}}.
\end{align*}
The second term on the right-hand side of the inequality can be bounded by $\langle H_4\psi,\psi\rangle^{\frac{1}{2}}$ using (\ref{H_4psi,psi fock}). For the first term we use the Sobolev inequality, since $d$ is small enough and $\Lambda_d\subset\mathbb{R}^3$
\begin{equation*}
  \Vert f\Vert_{L^{2q}(\Lambda_d)}\leq Cd^{\frac{1}{2q}-\frac{1}{2}}\Vert f\Vert_{H^1(\Lambda_d)}
\end{equation*}
for some universal constant $C$ and $1\leq 2q\leq 6$. We then have for $1< q\leq3$
\begin{align*}
&\int_{\Lambda_d^{n}}
\vert\eta_\perp\vert^2\ast v_a(\mathbf{x}_2-\mathbf{x}_3)
\vert\beta^{(n)}\vert^2=
\int_{\Lambda_d^{n-1}}\left\Vert\vert\eta_\perp\vert^2\ast v_a(\cdot-\mathbf{x}_3)
\vert\beta^{(n)}(\cdot,\mathbf{x}_3,\dots,\mathbf{x}_{n+1})\vert^2\right\Vert_1\\
&\leq\int_{\Lambda_d^{n-1}}\left\Vert\vert\eta_\perp\vert^2\ast v_a\right\Vert_{q^\prime}
\left\Vert\beta^{(n)}(\cdot,\mathbf{x}_3,\dots,\mathbf{x}_{n+1})\right\Vert_{2q}^2\\
&\leq Cd^{\frac{1}{q}-1}
\int_{\Lambda_d^{n-1}}\left\Vert\eta_\perp\right\Vert_{2q^\prime}^2\Vert v_a\Vert_1
\left\Vert\beta^{(n)}(\cdot,\mathbf{x}_3,\dots,\mathbf{x}_{n+1})\right\Vert_{H^1}^2
\end{align*}
where $q^\prime=\frac{q}{q-1}$ satisfying $2q^\prime\geq 3$. Moreover, since $a_0\beta^{(n)}=0$, we can use Poincar\'{e}'s inequality to bound
\begin{align}
 \int_{\Lambda_d^{n}}
\vert\eta_\perp\vert^2\ast v_a(\mathbf{x}_2-\mathbf{x}_3)
\vert\beta^{(n)}\vert^2
\leq&Cd^{\frac{1}{q}-1}\left\Vert\eta_\perp\right\Vert_{2}^{\frac{2}{q^\prime}}
\Vert\eta_\perp\Vert_{\infty}^{\frac{2}{q}}\Vert v_a\Vert_1\nonumber\\
&\times\int_{\Lambda_d}\vert\nabla_{\mathbf{x}_2}\beta^{(n)}
(\mathbf{x}_2,\dots,\mathbf{x}_{n+1})\vert^2.\label{Poincare}
\end{align}
Using (\ref{H_21psi,psi fock}) we can bound
\begin{align*}
\vert\langle\mathcal{E}_{4,6}^{B^{\prime}}\psi,\psi\rangle\vert
&\leq CNd^{\frac{1}{2q}-\frac{1}{2}}\left\Vert\eta_\perp\right\Vert_{2}^{\frac{1}{q^\prime}}
\Vert\eta_\perp\Vert_{\infty}^{\frac{1}{q}}\Vert v_a\Vert_1^{\frac{1}{2}}
\langle H_{21}a_0\psi,a_0\psi\rangle^{\frac{1}{2}}\langle H_4\psi,\psi\rangle^{\frac{1}{2}}.
\end{align*}
Taking $q=3$, using (\ref{est of eta and eta_perp}) and (\ref{est of max eta_perp}) we have
\begin{align*}
  \vert\langle\mathcal{E}_{4,6}^{B^{\prime}}\psi,\psi\rangle\vert
\leq CN^{\frac{3}{2}}a^{\frac{7}{6}}d^{-\frac{1}{2}}\textit{l}^{\frac{1}{3}}
\langle H_{21}\psi,\psi\rangle^{\frac{1}{2}}\langle H_4\psi,\psi\rangle^{\frac{1}{2}}
\end{align*}
These estimates finish the proof of (\ref{E_4^B' III}). The proof of the (\ref{E_4'^B' III}) is the same, we only need to substitute the potential $v_a$ with $2\sqrt{d}W$, and notice the fact that by (\ref{L1&L2 norm of W}) we have $\Vert v_a\Vert_1\sim\Vert2\sqrt{d}W\Vert_1\leq Ca$.
\begin{flushright}
  {$\Box$}
\end{flushright}

\begin{lemma}\label{lemma e^-B'H_Ne^B'}
There exists a universal constant $C$
such that for any $\vert t\vert\leq1$,
\begin{enumerate}[$(1)$]
  \item $\mathbf{For\; Region\; I}$ We take $\kappa=\nu d^{-1}$ for some $\nu\geq 1$, then for some $0<\gamma<1$ with the further assumption $Na^3\kappa^3\textit{l}\to0$, we have
\begin{align}
  e^{-tB^\prime}H_{21}e^{tB^\prime}&\leq C(H_{21}+H_4)+CNad^{-1}(\mathcal{N}_++1)
+C(Na^3\kappa^3\textit{l})^{1-\gamma}(\mathcal{N}_++1)^3\nonumber\\
&+CNa^2d^{-2}(\textit{l}^{-1}+\ln[(d\textit{l})^{-1}])(\mathcal{N}_++1)^2
\label{e^-B'H_21e^B'}\\
  e^{-tB^\prime}H_{4}e^{tB^\prime}&\leq CH_4+CNad^{-1}(\mathcal{N}_++1)+
  C(Na^3\kappa^3\textit{l})^{1-\gamma}(\mathcal{N}_++1)^3.\label{e^-B'H_4e^B'}
\end{align}
  \item $\mathbf{For\; Region\; III}$ We take $\kappa=\infty$. Assume further that $N^{\frac{3}{2}}a^{\frac{7}{6}}d^{-\frac{1}{2}}\textit{l}^{\frac{1}{3}}\to 0$, we have
\begin{align}
  e^{-tB^\prime}H_{21}e^{tB^\prime}&\leq CH_{21}+CNad^{-1}(\mathcal{N}_++1)^2
\label{e^-B'H_21e^B' III}\\
  e^{-tB^\prime}H_{4}e^{tB^\prime}&\leq CH_4+CNad^{-1}(\mathcal{N}_++1)\nonumber\\
&+CN^{\frac{3}{2}}a^{\frac{7}{6}}d^{-\frac{1}{2}}\textit{l}^{\frac{1}{3}}
  [H_{21}+Nad^{-1}(\mathcal{N}_++1)^2]\label{e^-B'H_4e^B' III}\\
 e^{-tB^\prime}H_{4}^\prime e^{tB^\prime}&\leq CH_4^\prime+CNad^{-1}(\mathcal{N}_++1)\nonumber\\
&+CN^{\frac{3}{2}}a^{\frac{7}{6}}d^{-\frac{1}{2}}\textit{l}^{\frac{1}{3}}
  [H_{21}+Nad^{-1}(\mathcal{N}_++1)^2].\label{e^-B'H_4'e^B' III}
\end{align}
\end{enumerate}
\end{lemma}
\noindent
\emph{Proof.} We first prove the part regarding Region I, and we start by (\ref{e^-B'H_4e^B'}). With Lemma \ref{commutator of H_21,H_4with B'} we have
\begin{equation*}
  [H_4,B^\prime]=\frac{1}{\sqrt{d}}\sum_{p,q,p+q\neq0}\left(
\sum_{r\neq0}v_{p-r}^{(a,d)}\eta_r
\right)\chi_{\vert\mathcal{M}_dq\vert\leq\kappa}(a_{p+q}^*a_{-p}^*a_qa_0+h.c.)+
\mathcal{E}_{4}^{B^{\prime}},
\end{equation*}
where the error term is bounded as in (\ref{E_4^B'}) for Region I. Let $\psi\in L^2_s(\Lambda_d^N)$, we have
\begin{align}
&\left\vert\frac{1}{\sqrt{d}}\sum_{p,q,p+q\neq0}\left(
\sum_{r\neq0}v_{p-r}^{(a,d)}\eta_r
\right)\chi_{\vert\mathcal{M}_dq\vert\leq\kappa}\langle
a_{p+q}^*a_{-p}^*a_qa_0\psi,\psi\rangle\right\vert\nonumber\\
&\leq C\Vert v_a\Vert_1^{\frac{1}{2}}\Vert\eta_\perp\Vert_{\infty}
\langle U_N(\mathcal{N}_++1)U_N^*P_\kappa U_Na_0\psi,P_\kappa U_Na_0\psi\rangle^{\frac{1}{2}}
\langle H_4\psi,\psi\rangle^{\frac{1}{2}}\nonumber\\
&\leq C\Vert v_a\Vert_1^{\frac{1}{2}}\Vert\eta_\perp\Vert_{\infty}
\langle(\mathcal{N}_++1)a_0\psi,a_0\psi\rangle^{\frac{1}{2}}
\langle H_4\psi,\psi\rangle^{\frac{1}{2}}\nonumber\\
&\leq C\langle H_4\psi,\psi\rangle+CNad^{-1}
\langle(\mathcal{N}_++1)\psi,\psi\rangle\label{temp1}
\end{align}
for $\psi\in L^2_s(\Lambda_d^N)$. Hence if we let $f(t)=\langle e^{-tB^\prime}H_{4}e^{tB^\prime}\psi,\psi\rangle$, we deduce with Lemmas \ref{control of S_+ conj with e^B'}, \ref{commutator of H_21,H_4with B'} and the further assumption $Na^3\kappa^3\textit{l}\to0$, that
\begin{align*}
  \vert f^\prime(t)\vert&=\vert\langle e^{-tB^\prime}[H_{4},B^\prime]e^{tB^\prime}\psi,\psi\rangle\vert\\
&\leq f(t)+CNad^{-1}\langle(\mathcal{N}_++1)\psi,\psi\rangle
+C(Na^3\kappa^3\textit{l})^{1-\gamma}\langle(\mathcal{N}_++1)^3\psi,\psi\rangle.
\end{align*}
Since $f(t)$ is real-valued, we obtain (\ref{e^-B'H_4e^B'}) by Gronwall's inequality.

\par Now for (\ref{e^-B'H_21e^B'}), from Lemma \ref{commutator of H_21,H_4with B'} we have
\begin{equation*}
  [H_{21},B^{\prime}]=2\sum_{p,q,p+q\neq0}\eta_p\vert\mathcal{M}_dp\vert^2
\chi_{\vert\mathcal{M}_dq\vert\leq\kappa}
(a_{p+q}^*a_{-p}^*a_qa_0+h.c.)+\mathcal{E}_{21}^{B^\prime},
\end{equation*}
with the error term bounded as (\ref{E_21^B'}) for Region I. For Region I we use (\ref{eqn of eta_p rewrt}) to rewrite
\begin{equation*}
  [H_{21},B^\prime]=\sum_{i=1}^{3}\Xi_i
+\mathcal{E}_{21}^{B^\prime},
\end{equation*}
where
\begin{align*}
\Xi_1&=-\frac{1}{\sqrt{d}}\sum_{p,q,p+q\neq0}\left(
\sum_{r\neq0}v_{p-r}^{(a,d)}\eta_r
\right)\chi_{\vert\mathcal{M}_dq\vert\leq\kappa}(a_{p+q}^*a_{-p}^*a_qa_0+h.c.),\\
\Xi_2&=-\frac{1}{\sqrt{d}}\sum_{p,q,p+q\neq0}v_p^{(a,d)}
\chi_{\vert\mathcal{M}_dq\vert\leq\kappa}(a_{p+q}^*a_{-p}^*a_qa_0+h.c.),\\
\Xi_3&=2\sum_{p,q,p+q\neq0}W_p\chi_{\vert\mathcal{M}_dq\vert\leq\kappa}
(a_{p+q}^*a_{-p}^*a_qa_0+h.c.).
\end{align*}
The first term $\Xi_1$ has been bounded in (\ref{temp1}). $\Xi_2$ can be bounded analogously
\begin{align*}
  \vert\langle\Xi_2\psi,\psi\rangle\vert
&\leq C\Vert v_a\Vert_1^{\frac{1}{2}}d^{-\frac{1}{2}}
\langle U_N(\mathcal{N}_++1)U_N^*P_\kappa U_Na_0\psi,P_\kappa U_Na_0\psi\rangle^{\frac{1}{2}}
\langle H_4\psi,\psi\rangle^{\frac{1}{2}}\\
&\leq C\Vert v_a\Vert_1^{\frac{1}{2}}d^{-\frac{1}{2}}
\langle(\mathcal{N}_++1)a_0\psi,a_0\psi\rangle^{\frac{1}{2}}
\langle H_4\psi,\psi\rangle^{\frac{1}{2}}\\
&\leq C\langle H_4\psi,\psi\rangle+CNad^{-1}
\langle(\mathcal{N}_++1)\psi,\psi\rangle
\end{align*}
To bound $\Xi_3$ we may use Lemma \ref{W_p lemma} and (\ref{sum_pW_peta_p 3dscatt}) to bound $\vert W_p\vert$
\begin{align*}
&\left\vert\sum_{p,q,p+q\neq0}W_p\chi_{\vert\mathcal{M}_dq\vert\leq\kappa}
\langle a_{p+q}^*a_{-p}^*a_qa_0\psi,\psi\rangle\right\vert\\
=&\left\vert\sum_{p,q,p+q\neq0}W_p\chi_{\vert\mathcal{M}_dq\vert\leq\kappa}\langle (\mathcal{N}_++1)^{\frac{1}{2}}a_qa_0\psi,
(\mathcal{N}_++1)^{-\frac{1}{2}}a_{-p}a_{p+q}\psi\rangle\right\vert\\
\leq&\sum_{p,q,p+q\neq0}\vert W_p\vert\chi_{\vert\mathcal{M}_dq\vert\leq\kappa}
\Vert(\mathcal{N}_++1)^{\frac{1}{2}}a_qa_0\psi\Vert
\cdot\Vert(\mathcal{N}_++1)^{-\frac{1}{2}}a_{-p}a_{p+q}\psi\Vert\\
\leq&\left(\sum_{q,p\neq0}\frac{\vert W_p\vert^2}{\vert\mathcal{M}_dp\vert^2}\langle
a_q^*(\mathcal{N}_++1)a_qa_0^*a_0\psi,\psi\rangle\right)^{\frac{1}{2}}\\
&\times\left(\sum_{p,p+q\neq0}\vert\mathcal{M}_dp\vert^2\langle
a_{p+q}^*a_{-p}^*(\mathcal{N}_++1)^{-1}
a_{-p}a_{p+q}\psi,\psi\rangle\right)^{\frac{1}{2}}\\
\leq&C\langle H_{21}\psi,\psi\rangle+
 CNa^2d^{-2}(\textit{l}^{-1}+\ln[(dl)^{-1}])\langle(\mathcal{N}_++1)^{2}\psi,\psi\rangle.
\end{align*}
 Finally we use Lemma \ref{control of S_+ conj with e^B'}, operator inequality (\ref{e^-B'H_4e^B'}) and Gronwall's inequality to reach (\ref{e^-B'H_21e^B'}). This finishes the proof to Region I.

\par For Region III, we first recall from (\ref{H_21psi,psi fock}) that for $U_N\psi=(\alpha^{(0)},\dots,\alpha^{(N)})$
\begin{align}
  \langle H_{21}\psi,\psi\rangle
=\sum_{n=1}^{N}n\int_{\Lambda_d^n}
\vert\nabla_{\mathbf{x}_1}\alpha^{(n)}\vert^2.\label{H_21psi,psi}
\end{align}
We can use (\ref{H_21psi,psi}) to bound directly
\begin{align*}
&\left\vert2\sum_{p,q,p+q\neq0}\eta_p\vert\mathcal{M}_dp\vert^2
\chi_{\vert\mathcal{M}_dq\vert\leq\kappa}\langle
a_{p+q}^*a_{-p}^*a_qa_0\psi,\psi\rangle\right\vert\\
&\leq C\Vert\nabla\eta_\perp\Vert_2\langle(\mathcal{N}_++1)^2a_0\psi,a_0\psi\rangle^{\frac{1}{2}}
\langle H_{21}\psi,\psi\rangle^{\frac{1}{2}}\\
&\leq C\langle H_{21}\psi,\psi\rangle+CNad^{-1}\langle(\mathcal{N}_++1)^2\psi,\psi\rangle.
\end{align*}
We use (\ref{est of grad eta}) in the last inequality. This together with the bound of $\mathcal{E}_{21}^{B^\prime}$ (\ref{E_21^B' III}), Lemma \ref{control of S_+ conj with e^B'} and the Gronwall's inequality give (\ref{e^-B'H_21e^B' III}).

\par As for (\ref{e^-B'H_4e^B' III}), we combine (\ref{temp1}), the bound of $\mathcal{E}_{4}^{B^\prime}$ (\ref{E_4^B' III}), the estimate on $H_{21}$ (\ref{e^-B'H_21e^B' III}) we just proved above, Lemma \ref{control of S_+ conj with e^B'} and the the Gronwall's inequality achieve it. The proof of (\ref{e^-B'H_4'e^B' III}) is same as (\ref{e^-B'H_4e^B' III}) except for the substitution of potential $v_a$ by $2\sqrt{d}W$. This finishes the proof to Region III.

\begin{flushright}
  {$\Box$}
\end{flushright}

\par Using Lemma \ref{control of S_+ conj with e^B'} and Lemma \ref{lemma e^-B'H_Ne^B'}, and the fact that
\begin{equation}\label{est of Q^B}
  \vert Q^B\vert\leq CNad^{-1},\quad\vert\tilde{Q}_1^B\vert
=2N\vert\tilde{Q}_2^B\vert\leq CNad^{-1}.
\end{equation}
 we immediately deduce the following two corollaries which show that the diagonal term $Q^B\mathcal{N}_+$ and the error term $\mathcal{E}^B$ for Region I, and the terms $\tilde{Q}_1^B\mathcal{N}_++\tilde{Q}_2^B\mathcal{N}_+(\mathcal{N}_++1)$ and $\tilde{\mathcal{E}}^B$ for Region III remain unchanged up to small errors. We want to stress that we have set $a$, $d$, $\frac{a}{d}$ and $Na\textit{l}^{\frac{1}{2}}$ tend to $0$ and $\frac{a}{d\textit{l}}<C$ for some universal constant $C$ in the first place.
\begin{corollary}\label{corollary e^-B'(C^B+Q^BN_+)e^B'}
\
\begin{enumerate}[$(1)$]
  \item $\mathbf{For\; Region\; I}$
\begin{equation}\label{e^-B'(C^B+Q^BN_+)e^B'}
  e^{-B^\prime}Q^B\mathcal{N}_+e^{B^\prime}=Q^B\mathcal{N}_++\mathcal{E}_{diag}^{B^\prime},
\end{equation}
where
\begin{equation}\label{E_diag^B'}
  \pm\mathcal{E}_{diag}^{B^\prime}\leq CN^2a^2d^{-1}\textit{l}^{\frac{1}{2}}(\mathcal{N}_++1).
\end{equation}
  \item $\mathbf{For\; Region\; III}$
\begin{equation}\label{e^-B'(C^B+Q^BN_+)e^B' III}
  e^{-B^\prime}[\tilde{Q}_1^B\mathcal{N}_++\tilde{Q}_2^B\mathcal{N}_+(\mathcal{N}_++1)]
e^{B^\prime}=\tilde{Q}_1^B\mathcal{N}_++\tilde{Q}_2^B\mathcal{N}_+(\mathcal{N}_++1)
+\tilde{\mathcal{E}}_{diag}^{B^\prime},
\end{equation}
where
\begin{equation}\label{E_diag^B' III}
  \pm\tilde{\mathcal{E}}_{diag}^{B^\prime}\leq CN^2a^2d^{-1}\textit{l}^{\frac{1}{2}}(\mathcal{N}_++1).
\end{equation}
\end{enumerate}
\end{corollary}

\begin{corollary}\label{corollary e^-B'E^Be^B'}
\
\begin{enumerate}[$(1)$]
  \item $\mathbf{For\; Region\; I}$ We take $\kappa=\nu d^{-1}$ for some $\nu\geq 1$, then for some $0<\gamma<1$ with the further assumption $Na^3\kappa^3\textit{l}\to0$
\begin{align}
  \pm e^{-B^\prime}\mathcal{E}^B e^{B^\prime}
&\leq C\{Na^2d^{-2}\textit{l}^{-1}+N^2a^2d^{-1}\textit{l}^{\frac{1}{2}}
+N^{\frac{3}{2}}a^{\frac{3}{2}}d^{-\frac{1}{2}}\textit{l}^{\frac{1}{2}}\nonumber\\
&+N^{\frac{5}{2}}a^{\frac{5}{2}}d^{-\frac{3}{2}}\textit{l}^{\frac{1}{2}}\}
(\mathcal{N}_++1)
+C(ad^{-1}+Na^2d^{-2}\textit{l}^{-1})
(\mathcal{N}_++1)^2\nonumber\\
&+CN^{\frac{3}{2}}a^{\frac{3}{2}}d^{-\frac{1}{2}}\textit{l}^{\frac{1}{2}}
(Na^3\kappa^3\textit{l})^{1-\gamma}(\mathcal{N}_++1)^3+Na^3d^{-1}H_{21}\nonumber\\
&+CN^{\frac{3}{2}}a^{\frac{3}{2}}d^{-\frac{1}{2}}\textit{l}^{\frac{1}{2}}
H_4.\label{e^-B'E^Be^B'}
\end{align}
  \item $\mathbf{For\; Region\; III}$ We take $\kappa=\infty$. Assume further that $N^{\frac{3}{2}}a^{\frac{7}{6}}d^{-\frac{1}{2}}\textit{l}^{\frac{1}{3}}\to 0$
\begin{align}
  \pm e^{-B^\prime}\tilde{\mathcal{E}}^B e^{B^\prime}
&\leq C\{N^2a^2d^{-1}\textit{l}^{\frac{1}{2}}
+N^{\frac{3}{2}}a^{\frac{3}{2}}d^{-\frac{1}{2}}\textit{l}^{\frac{1}{2}}
+N^{\frac{5}{2}}a^{\frac{5}{2}}d^{-\frac{3}{2}}\textit{l}^{\frac{1}{2}}\}
(\mathcal{N}_++1)\nonumber\\
&+CN^3a^{\frac{8}{3}}d^{-1}\textit{l}^{\frac{5}{6}}[H_{21}
+Nad^{-1}(\mathcal{N}_++1)^2]\nonumber\\
&+CN^{\frac{3}{2}}a^{\frac{3}{2}}d^{-\frac{1}{2}}\textit{l}^{\frac{1}{2}}
(H_4+H_4^\prime).\label{e^-B'E^Be^B' III}
\end{align}
\end{enumerate}
\end{corollary}

\begin{lemma}\label{lemma e^-B'H_23'e^B'}
\
\begin{equation}\label{e^-B'H_23'e^B'}
  e^{-B^\prime}H_{23}^{\prime}e^{B^\prime}=
H_{23}^\prime+\mathcal{E}_{23}^{B^\prime},
\end{equation}
\begin{enumerate}[$(1)$]
  \item $\mathbf{For\; Region\; I}$ We take $\kappa=\nu d^{-1}$ for some $\nu\geq 1$. The error term is bounded by
\begin{equation}\label{E_23^B'}
  \pm\mathcal{E}_{23}^{\prime}\leq CN^{\frac{3}{2}}a^2d^{-2}\textit{l}^{-1}
(\mathcal{N}_++1)^\frac{3}{2}.
\end{equation}
  \item $\mathbf{For\; Region\; III}$ We take $\kappa=\infty$. Assume further that $N^{\frac{3}{2}}a^{\frac{7}{6}}d^{-\frac{1}{2}}\textit{l}^{\frac{1}{3}}\to 0$. The error term is bounded by
\begin{align}
  \pm\mathcal{E}_{23}^{\prime}
&\leq C\{N^2a^2d^{-1}\textit{l}^{\frac{1}{2}}
+N^{\frac{3}{2}}a^{\frac{3}{2}}d^{-\frac{1}{2}}\textit{l}^{\frac{1}{2}}
+N^{\frac{5}{2}}a^{\frac{5}{2}}d^{-\frac{3}{2}}\textit{l}^{\frac{1}{2}}\}
(\mathcal{N}_++1)\nonumber\\
&+CN^3a^{\frac{8}{3}}d^{-1}\textit{l}^{\frac{5}{6}}[H_{21}
+Nad^{-1}(\mathcal{N}_++1)^2]\nonumber\\
&+CN^{\frac{3}{2}}a^{\frac{3}{2}}d^{-\frac{1}{2}}\textit{l}^{\frac{1}{2}}
H_4^\prime.\label{E_23^B' III}
\end{align}
\end{enumerate}

\end{lemma}
\noindent
\emph{Proof.} The error term can be expressed explicitly by
\begin{equation*}
  \mathcal{E}_{23}^{B^\prime}= e^{-B^\prime}H_{23}^{\prime}e^{B^\prime}-H_{23}^\prime
=\int_{0}^{1}e^{-tB^\prime}[H_{23}^\prime,B^\prime]e^{tB^\prime}dt.
\end{equation*}
Calculating directly we have
\begin{equation*}
  [H_{23}^\prime,B^\prime]=\sum_{i=1}^{3}\mathcal{E}_{23,i}^{B^\prime},
\end{equation*}
where
\begin{align*}
\mathcal{E}_{23,1}^{B^\prime}&=
4\sum_{p,q,p+q\neq0}\eta_pW_p\chi_{\vert\mathcal{M}_dq\vert\leq\kappa}
(a_0^*a_0^*a_{p+q}^*a_pa_qa_0+h.c.),\\
\mathcal{E}_{23,2}^{B^\prime}&=
-2\sum_{p,q,p+q\neq0}\eta_pW_q\chi_{\vert\mathcal{M}_dq\vert\leq\kappa}
(a_{p+q}^*a_{-p}^*a_{-q}^*a_0a_0a_0+h.c.),\\
\mathcal{E}_{23,2}^{B^\prime}&=
-2\sum_{p,q,p+q,r\neq0}\eta_pW_r\chi_{\vert\mathcal{M}_dq\vert\leq\kappa}
(a_{p+q}^*a_{-p}^*a_0^*a_{r}a_{-r}a_q+h.c.).
\end{align*}
Let $\psi\in L^2_s(\Lambda_d^N)$, then we can control these terms for Region I. For $\mathcal{E}_{23,1}^{B^\prime}$, we have
\begin{align*}
\vert\langle\mathcal{E}_{23,1}^{B^\prime}\psi,\psi\rangle\vert
&\leq\frac{C}{\sqrt{d}}\Vert\eta_\perp\ast W\Vert_2
\langle(\mathcal{N}_++1)^\frac{3}{2}\psi,\psi\rangle^{\frac{1}{2}}\\
&\times\langle U_N(\mathcal{N}_++1)^\frac{3}{2}U_N^*P_\kappa U_Na_0^*a_0^*a_0\psi,
P_\kappa U_N a_0^*a_0^*a_0\psi\rangle^{\frac{1}{2}}\\
&\leq\frac{C}{\sqrt{d}}\Vert\eta_\perp\Vert_2\Vert W\Vert_1
\langle(\mathcal{N}_++1)^\frac{3}{2}a_0^*a_0^*a_0\psi,a_0^*a_0^*a_0\psi\rangle^{\frac{1}{2}}
\langle(\mathcal{N}_++1)^\frac{3}{2}\psi,\psi\rangle^{\frac{1}{2}}\\
&\leq CN^{\frac{3}{2}}a^2d^{-1}\textit{l}^{\frac{1}{2}}
\langle(\mathcal{N}_++1)^\frac{3}{2}\psi,\psi\rangle,
\end{align*}
where we have used (\ref{est of eta and eta_perp}) and (\ref{L1&L2 norm of W}) in the last inequality. These two estimates can also help us bound
\begin{align*}
\vert\langle\mathcal{E}_{23,2}^{B^\prime}\psi,\psi\rangle\vert
&\leq C\Vert\eta_\perp\Vert_2\Vert P_\kappa W\Vert_2
\langle(\mathcal{N}_++1)^\frac{3}{2}a_0a_0a_0\psi,a_0a_0a_0\psi\rangle^{\frac{1}{2}}
\langle(\mathcal{N}_++1)^\frac{3}{2}\psi,\psi\rangle^{\frac{1}{2}}\\
&\leq CN^{\frac{3}{2}}a^2d^{-2}\textit{l}^{-1}
\langle(\mathcal{N}_++1)^\frac{3}{2}\psi,\psi\rangle,
\end{align*}
and
\begin{align*}
\vert\langle\mathcal{E}_{23,3}^{B^\prime}\psi,\psi\rangle\vert
&\leq C\Vert\eta_\perp\Vert_2\Vert  W\Vert_2
\langle(\mathcal{N}_++1)^\frac{5}{2}\psi,\psi\rangle^{\frac{1}{2}}\\
&\times\langle U_N(\mathcal{N}_++1)^\frac{5}{2}U_N^*P_\kappa U_Na_0^*\psi,P_\kappa U_Na_0^*\psi\rangle^{\frac{1}{2}}\\
&\leq C\Vert\eta_\perp\Vert_2\Vert  W\Vert_2
\langle (\mathcal{N}_++1)^\frac{5}{2}a_0^*\psi,a_0^*\psi\rangle^{\frac{1}{2}}
\langle(\mathcal{N}_++1)^\frac{5}{2}\psi,\psi\rangle^{\frac{1}{2}}\\
&\leq CN^{\frac{3}{2}}a^2d^{-2}\textit{l}^{-1}
\langle(\mathcal{N}_++1)^\frac{3}{2}\psi,\psi\rangle.
\end{align*}
Then (\ref{E_23^B'}) follows by Lemma \ref{control of S_+ conj with e^B'}.

\par For Region III, we can bound directly
\begin{align*}
\vert\langle\mathcal{E}_{23,1}^{B^\prime}\psi,\psi\rangle\vert
&\leq\frac{C}{\sqrt{d}}\Vert\eta_\perp\ast W\Vert_2
\langle(\mathcal{N}_++1)^\frac{3}{2}a_0^*a_0^*a_0\psi,a_0^*a_0^*a_0\psi\rangle^{\frac{1}{2}}
\langle(\mathcal{N}_++1)^\frac{3}{2}\psi,\psi\rangle^{\frac{1}{2}}\\
&\leq CN^{\frac{3}{2}}a^2d^{-1}\textit{l}^{\frac{1}{2}}
\langle(\mathcal{N}_++1)^\frac{3}{2}\psi,\psi\rangle\\
&\leq CN^{2}a^2d^{-1}\textit{l}^{\frac{1}{2}}
\langle(\mathcal{N}_++1)\psi,\psi\rangle.
\end{align*}
To bound the other two, we recall the definition of $H_4^\prime$ and (\ref{H_4'psi,psi})
\begin{align*}
\vert\langle\mathcal{E}_{23,2}^{B^\prime}\psi,\psi\rangle\vert
&\leq Cd^{-\frac{1}{2}}\Vert\eta_\perp\Vert_2\Vert\sqrt{d}W\Vert_1^{\frac{1}{2}}
\langle(\mathcal{N}_++1)a_0^3\psi,a_0^3\psi\rangle^{\frac{1}{2}}
\langle H_4^\prime\psi,\psi\rangle^{\frac{1}{2}}\\
&\leq CN^{\frac{3}{2}}a^{\frac{3}{2}}d^{-\frac{1}{2}}\textit{l}^{-\frac{1}{2}}
\langle(\mathcal{N}_++1)\psi,\psi\rangle^{\frac{1}{2}}
\langle H_4^\prime\psi,\psi\rangle^{\frac{1}{2}},
\end{align*}
and
\begin{align*}
\vert\langle\mathcal{E}_{23,3}^{B^\prime}\psi,\psi\rangle\vert
&\leq Cd^{-\frac{1}{2}}\Vert\eta_\perp\Vert_2\Vert\sqrt{d}W\Vert_1^{\frac{1}{2}}
\langle(\mathcal{N}_++1)^3\psi,\psi\rangle^{\frac{1}{2}}
\langle H_4^\prime a_0^*\psi,a_0^*\psi\rangle^{\frac{1}{2}}\\
&\leq CN^{\frac{3}{2}}a^{\frac{3}{2}}d^{-\frac{1}{2}}\textit{l}^{-\frac{1}{2}}
\langle(\mathcal{N}_++1)\psi,\psi\rangle^{\frac{1}{2}}
\langle H_4^\prime\psi,\psi\rangle^{\frac{1}{2}}.
\end{align*}
Then (\ref{E_23^B' III}) follows by Lemma \ref{control of S_+ conj with e^B'} and Lemma \ref{lemma e^-B'H_Ne^B'}.
\begin{flushright}
  {$\Box$}
\end{flushright}

\par The next lemma considers the action of $e^{B^\prime}$ on $(H_{02}+H_{22})$ in Region III.
\begin{lemma}\label{lemma e^-B'(H_02+H_22)e^B'}
\
$\mathbf{For\; Region\; III}$ We take $\kappa=\infty$, then we have
\begin{equation}\label{e^-B'(H_02+H_22)e^B' III}
  e^{-B^\prime}(H_{02}+H_{22})e^{B^\prime}=(H_{02}+H_{22})+\tilde{\mathcal{E}}^{B^\prime}_{02+22},
\end{equation}
where
\begin{equation}\label{E^B'_02+22 III}
  \pm\tilde{\mathcal{E}}^{B^\prime}_{02+22}\leq CN^2a^2d^{-1}\textit{l}^{\frac{1}{2}}
(\mathcal{N}_++1).
\end{equation}
\end{lemma}
\noindent
\emph{Proof.} Using Lemma \ref{control of S_+ conj with e^B'} and recalling that
\begin{equation*}
  H_{02}=-\frac{1}{2\sqrt{d}}v_0^{(a,d)}(\mathcal{N}_+-1)\mathcal{N}_+,
\end{equation*}
we immediately deduce
\begin{equation}\label{e^-B'H_02e^B' III}
  e^{-B^\prime}H_{02}e^{B^\prime}=H_{02}+\tilde{\mathcal{E}}^{B^\prime}_{02},
\end{equation}
where
\begin{equation}\label{E^B'_02 III}
  \pm\tilde{\mathcal{E}}^{B^\prime}_{02}\leq CNa^2d^{-1}\textit{l}^{\frac{1}{2}}
(\mathcal{N}_++1)^2\leq CN^2a^2d^{-1}\textit{l}^{\frac{1}{2}}
(\mathcal{N}_++1).
\end{equation}

\par As for $H_{22}$, we rewrite it in the form of (\ref{rewrite H_22}) i.e. $H_{22}=(N-\mathcal{N}_+)R$ where $R$ is defined in (\ref{define R}). Let
\begin{equation*}
  \tilde{\mathcal{E}}^{B^\prime}_{22}=
e^{-B^\prime}H_{22}e^{B^\prime}-H_{22}
=N\int_{0}^{1}e^{-tB^\prime}[R,B^\prime]e^{tB^\prime}dt-
\int_{0}^{1}e^{-tB^\prime}[\mathcal{N}_+R,B^\prime]e^{tB^\prime}dt.
\end{equation*}
We claim with proofs postponed that
\begin{align}
  \pm[R,B^\prime]&\leq CNa^2d^{-1}\textit{l}^{\frac{1}{2}}(\mathcal{N}_++1),\label{[R,B']}\\
\pm\mathcal{N}_+[R,B^\prime]&\leq CN^2a^2d^{-1}\textit{l}^{\frac{1}{2}}(\mathcal{N}_++1),
\label{N_+[R,B']}\\
\pm[\mathcal{N}_+,B^\prime]R&\leq CN^2a^2d^{-1}\textit{l}^{\frac{1}{2}}(\mathcal{N}_++1).
\label{[N_+,B']R}
\end{align}
Then it follows directly from Lemma \ref{control of S_+ conj with e^B'} that
\begin{equation*}
  \pm N\int_{0}^{1}e^{-tB^\prime}[R,B^\prime]e^{tB^\prime}dt\leq
CN^2a^2d^{-1}\textit{l}^{\frac{1}{2}}(\mathcal{N}_++1).
\end{equation*}
On the other hand, we notice that
\begin{equation*}
  [\mathcal{N}_+R,B^\prime]=\mathcal{N}_+[R,B^\prime]+[\mathcal{N}_+,B^\prime]R.
\end{equation*}
Thus we can bound using Lemma \ref{control of S_+ conj with e^B'} again
\begin{equation*}
  \pm\int_{0}^{1}e^{-tB^\prime}[\mathcal{N}_+R,B^\prime]e^{tB^\prime}dt
\leq CN^2a^2d^{-1}\textit{l}^{\frac{1}{2}}(\mathcal{N}_++1).
\end{equation*}
Letting $\tilde{\mathcal{E}}^{B^\prime}_{02+22}
=\tilde{\mathcal{E}}^{B^\prime}_{02}+\tilde{\mathcal{E}}^{B^\prime}_{22}$, we reach (\ref{e^-B'(H_02+H_22)e^B' III}) and (\ref{E^B'_02+22 III}).

\par We now prove (\ref{[R,B']})-(\ref{[N_+,B']R}). For (\ref{[R,B']}), we first notice the fact that
$[R,B^\prime]=[R,A^\prime]+[R,A^\prime]^*$ with
\begin{equation*}
  [R,A^\prime]=\frac{1}{\sqrt{d}}\sum_{p,q,p+q\neq0}\eta_p
(v_p^{(a,d)}+v_{p+q}^{(a,d)}-v_q^{(a,d)})a_{p+q}^*a_{-p}^*a_qa_0.
\end{equation*}
The first term can be bounded similar to (\ref{A'}) for $\psi\in L^2_s(\Lambda_d^N)$ by
\begin{align*}
&\frac{1}{\sqrt{d}}\left\vert\sum_{p,q,p+q\neq0}\eta_pv_p^{(a,d)}
\langle a_{p+q}^*a_{-p}^*a_qa_0\psi,\psi\rangle\right\vert\\
&\leq Cd^{-1}\Vert\eta_\perp\ast v_a\Vert_2
\langle(\mathcal{N}_++1)a_0\psi,a_0\psi\rangle^{\frac{1}{2}}
\langle(\mathcal{N}_++1)^2\psi,\psi\rangle^{\frac{1}{2}}\\
&\leq CNa^2d^{-1}\textit{l}^{\frac{1}{2}}\langle(\mathcal{N}_++1)\psi,\psi\rangle.
\end{align*}
The other two can be bounded respetively by
\begin{align*}
&\frac{1}{\sqrt{d}}\left\vert\sum_{p,q,p+q\neq0}\eta_pv_{p+q}^{(a,d)}
\langle a_{p+q}^*a_{-p}^*a_qa_0\psi,\psi\rangle\right\vert\\
&\leq Cd^{-1}\Vert\eta_\perp\Vert_2\Vert v_a\Vert_1
\langle(\mathcal{N}_++1)a_0\psi,a_0\psi\rangle^{\frac{1}{2}}
\langle(\mathcal{N}_++1)^2\psi,\psi\rangle^{\frac{1}{2}}\\
&\leq CNa^2d^{-1}\textit{l}^{\frac{1}{2}}\langle(\mathcal{N}_++1)\psi,\psi\rangle,
\end{align*}
and
\begin{align*}
&\frac{1}{\sqrt{d}}\left\vert\sum_{p,q,p+q\neq0}\eta_pv_{q}^{(a,d)}
\langle a_{p+q}^*a_{-p}^*a_qa_0\psi,\psi\rangle\right\vert\\
&\leq Cd^{-1}\Vert\eta_\perp\Vert_2\Vert v_a\Vert_1
\langle(\mathcal{N}_++1)a_0\psi,a_0\psi\rangle^{\frac{1}{2}}
\langle(\mathcal{N}_++1)^2\psi,\psi\rangle^{\frac{1}{2}}\\
&\leq CNa^2d^{-1}\textit{l}^{\frac{1}{2}}\langle(\mathcal{N}_++1)\psi,\psi\rangle.
\end{align*}
These three estimates together give (\ref{[R,B']}). The proof of (\ref{N_+[R,B']}) is essentially the same as long as we notice that, by $\mathcal{N}_+=n-a_0^*a_0$ when acting on $L^2_s(\Lambda_d^n)$, we have
\begin{align*}
  \mathcal{N}_+[R,A^\prime]
&=\frac{1}{\sqrt{d}}\sum_{p,q,p+q\neq0}\eta_p
(v_p^{(a,d)}+v_{p+q}^{(a,d)}-v_q^{(a,d)})a_{p+q}^*a_{-p}^*a_q(\mathcal{N}_++1)a_0,\\
[R,A^\prime]\mathcal{N}_+
&=\frac{1}{\sqrt{d}}\sum_{p,q,p+q\neq0}\eta_p
(v_p^{(a,d)}+v_{p+q}^{(a,d)}-v_q^{(a,d)})a_{p+q}^*a_{-p}^*a_q\mathcal{N}_+a_0.
\end{align*}
For the proof of (\ref{[N_+,B']R}), since $[\mathcal{N}_+,B^\prime]=A^\prime+A^{\prime*}$, we slightly modify the calculation in (\ref{A'}) and use the operator inequality (\ref{wtf1}) to find
\begin{align*}
  \vert\langle A^{\prime}R\psi,\psi\rangle\vert
&\leq C\Vert\eta_\perp\Vert_2
\langle (\mathcal{N}_++1)a_0R\psi,a_0R\psi\rangle^{\frac{1}{2}}
\langle (\mathcal{N}_++1)^2\psi,\psi\rangle^{\frac{1}{2}}\\
&\leq CN^2a^2d^{-1}\textit{l}^{\frac{1}{2}}
\langle(\mathcal{N}_++1)\psi,\psi\rangle,
\end{align*}
and
\begin{align*}
  \vert\langle A^{\prime}\psi,R\psi\rangle\vert
&\leq C\Vert\eta_\perp\Vert_2
\langle (\mathcal{N}_++1)a_0\psi,a_0\psi\rangle^{\frac{1}{2}}
\langle (\mathcal{N}_++1)^2R\psi,R\psi\rangle^{\frac{1}{2}}\\
&\leq CN^2a^2d^{-1}\textit{l}^{\frac{1}{2}}
\langle(\mathcal{N}_++1)\psi,\psi\rangle.
\end{align*}
\begin{flushright}
  {$\Box$}
\end{flushright}

\begin{lemma}\label{lemma Gamma'}
\
\begin{enumerate}[$(1)$]
  \item $\mathbf{For\; Region\; I}$ We take $\kappa=\nu d^{-1}$ for some $\nu\geq 1$, then for some $\alpha>0$ and $0<\gamma<1$ with the further assumption $Na^3\kappa^3\textit{l}$ and $N^{\frac{1}{2}}a^{\frac{1}{2}}d^{-\frac{1}{2}}\kappa^{-1}$ tend to 0.
\begin{align}
  \pm\int_{0}^{1}&e^{-tB^\prime}\Gamma^\prime e^{tB^\prime}dt
\leq CN^{\frac{1}{2}}a^{\frac{3}{2}}d^{\frac{1}{2}}\textit{l}^2(\mathcal{N}_++1)
+C(Na^3\kappa^3\textit{l})^{1-\gamma}(\mathcal{N}_++1)^3\nonumber\\
&+CNa^2d^{-(2+\alpha)}(\textit{l}^{-1}+\ln[(d\textit{l})^{-1}])(\mathcal{N}_++1)^2\nonumber\\
&+C(d^\alpha+Na\textit{l}^{\frac{1}{2}}+N^{\frac{1}{2}}a^{\frac{1}{2}}
d^{-\frac{1}{2}}\kappa^{-1})(H_{21}+H_4+Nad^{-1}(\mathcal{N}_++1))\nonumber\\
&+C(N^{\frac{1}{2}}a^{\frac{3}{2}}d^{\frac{1}{2}}\textit{l}^2
+N^{\frac{1}{2}}a^{\frac{1}{2}}d^{-\frac{1}{2}}\kappa^{-1}
+(Na^3\kappa^3\textit{l})^{\gamma})(H_4+Nad^{-1}(\mathcal{N}_++1)).\label{est of Gamma'}
\end{align}
  \item $\mathbf{For\; Region\; III}$ We take $\kappa=\infty$. Assume further that $N^{\frac{3}{2}}a^{\frac{7}{6}}d^{-\frac{1}{2}}\textit{l}^{\frac{1}{3}}\to 0$.
\begin{align}
\pm\int_{0}^{1}e^{-tB^\prime}\tilde{\Gamma}^\prime &e^{tB^\prime}dt
\leq C(N^2a^2d^{-1}\textit{l}^{\frac{1}{2}}
+N^{\frac{1}{2}}a^{\frac{3}{2}}d^{\frac{1}{2}}\textit{l}^2)(\mathcal{N}_++1)\nonumber\\
&+C(Na\textit{l}^{\frac{1}{2}}
+N^{\frac{3}{2}}a^{\frac{7}{6}}d^{-\frac{1}{2}}\textit{l}^{\frac{1}{3}})
\{H_{21}+H_4+Nad^{-1}(\mathcal{N}_++1)^2\}.
\label{est of Gamma' III}
\end{align}
\end{enumerate}

\end{lemma}
\noindent
\emph{Proof.} Since we take $\kappa=\nu d^{-1}$ for some $\nu\geq 1$ in Region I, by (\ref{eqn of eta_p rewrt}) and Lemma \ref{commutator of H_21,H_4with B'}, we find, in Region I,
\begin{equation*}
  \Gamma^\prime=\sum_{i=1}^{3}\Gamma^\prime_i
+\mathcal{E}_{21}^{B^\prime}+\mathcal{E}_{4}^{B^\prime},
\end{equation*}
where
\begin{align*}
 \Gamma^\prime_1&=2\sum_{p,q,p+q\neq0}W_p\chi_{\vert\mathcal{M}_dq\vert\leq\kappa}
(a_{p+q}^*a_{-p}^*a_qa_0+h.c.),\\
\Gamma^\prime_2&=-\frac{1}{\sqrt{d}}
\sum_{p,q,p+q\neq0}v_p^{(a,d)}\eta_0\chi_{\vert\mathcal{M}_dq\vert\leq\kappa}
(a_{p+q}^*a_{-p}^*a_qa_0+h.c.),\\
\Gamma^\prime_3&=\frac{1}{\sqrt{d}}
\sum_{p,q,p+q\neq0}v_p^{(a,d)}\chi_{\vert\mathcal{M}_dq\vert>\kappa}
(a_{p+q}^*a_{-p}^*a_qa_0+h.c.).
\end{align*}
Let $\psi\in L^2_s(\Lambda_d^N)$, using (\ref{sum_pW_peta_p 3dscatt}) and (\ref{sum W_p M_dp^-2}) we can bound
\begin{align*}
&\vert\langle\Gamma_1^\prime\psi,\psi\rangle\vert
=2\left\vert\sum_{p,q,p+q\neq0}W_p\chi_{\vert\mathcal{M}_dq\vert\leq\kappa}
\langle a_{p+q}^*a_{-p}^*a_qa_0\psi,\psi\rangle\right\vert\\
=&2\left\vert\sum_{p,q,p+q\neq0}W_p\chi_{\vert\mathcal{M}_dq\vert\leq\kappa}\langle (\mathcal{N}_++1)^{\frac{1}{2}}a_qa_0\psi,
(\mathcal{N}_++1)^{-\frac{1}{2}}a_{-p}a_{p+q}\psi\rangle\right\vert\\
\leq&\sum_{p,q,p+q\neq0}\vert W_p\vert\chi_{\vert\mathcal{M}_dq\vert\leq\kappa}
\Vert(\mathcal{N}_++1)^{\frac{1}{2}}a_qa_0\psi\Vert
\cdot\Vert(\mathcal{N}_++1)^{-\frac{1}{2}}a_{-p}a_{p+q}\psi\Vert\\
\leq&\left(\sum_{p,q\neq0}\frac{\vert W_p\vert^2}{\vert\mathcal{M}_dp\vert^2}\langle
a_q^*(\mathcal{N}_++1)a_qa_0^*a_0\psi,\psi\rangle\right)^{\frac{1}{2}}\\
&\times\left(\sum_{p,p+q\neq0}\vert\mathcal{M}_dp\vert^2\langle
a_{p+q}^*a_{-p}^*(\mathcal{N}_++1)^{-1}
a_{-p}a_{p+q}\psi,\psi\rangle\right)^{\frac{1}{2}}\\
\leq& CN^{\frac{1}{2}}ad^{-1}(\textit{l}^{-1}+\ln[(d\textit{l})^{-1}])^{\frac{1}{2}}
\langle(\mathcal{N}_++1)^2\psi,\psi\rangle^{\frac{1}{2}}
\langle H_{21}\psi,\psi\rangle^{\frac{1}{2}}\\
\leq& CNa^2d^{-(2+\alpha)}(\textit{l}^{-1}+\ln[(d\textit{l})^{-1}])
\langle(\mathcal{N}_++1)^2\psi,\psi\rangle+
Cd^\alpha \langle H_{21}\psi,\psi\rangle
\end{align*}
for some $\alpha>0$. Secondly, we can bound
\begin{align*}
&\vert\langle\Gamma_2^\prime\psi,\psi\rangle\vert
=\left\vert\frac{\eta_0}{\sqrt{d}}\sum_{p,q,p+q\neq0}
v_p^{(a,d)}\chi_{\vert\mathcal{M}_dq\vert\leq\kappa}
\langle a_{p+q}^*a_{-p}^*a_qa_0\psi,\psi\rangle\right\vert\\
\leq& C\vert\eta_0\vert\Vert v_a\Vert_1^{\frac{1}{2}}d^{-\frac{1}{2}}
\langle U_N(\mathcal{N}_++1)U_N^*P_\kappa U_Na_0\psi,P_\kappa U_Na_0\psi\rangle^{\frac{1}{2}}
\langle H_4\psi,\psi\rangle^{\frac{1}{2}}\\
\leq& C\vert\eta_0\vert\Vert v_a\Vert_1^{\frac{1}{2}}d^{-\frac{1}{2}}
\langle(\mathcal{N}_++1)a_0\psi,a_0\psi\rangle^{\frac{1}{2}}
\langle H_4\psi,\psi\rangle^{\frac{1}{2}}\\
\leq& CN^{\frac{1}{2}}a^{\frac{3}{2}}
d^{\frac{1}{2}}\textit{l}^2\langle(\mathcal{N}_++1+H_4)\psi,\psi\rangle,
\end{align*}
where we have used (\ref{est of eta_0}) to bound $\eta_0$.

\par To bound $\Gamma_3^\prime$ we need a new method. Denote
\begin{align*}
U_N\psi&=(\alpha^{(0)},\dots,\alpha^{(N)}),\\
U_{N-1}a_0\psi&=(\beta^{(0)},\dots,\beta^{(N-1)}),
\end{align*}
we have
\begin{align*}
&\vert\langle\Gamma_3^\prime\psi,\psi\rangle\vert
=\frac{2}{\sqrt{d}}\left\vert
\sum_{p,q,p+q\neq0}v_p^{(a,d)}\chi_{\vert\mathcal{M}_dq\vert>\kappa}
\vert\mathcal{M}_dq\vert^{-1}\vert\mathcal{M}_dq\vert
\langle a_{p+q}^*a_{-p}^*a_qa_0\psi,\psi\rangle\right\vert\\
\leq& Cd^{-\frac{1}{2}}\kappa^{-1}\left\vert\sum_{n=1}^{N-1}
\sum_{p,q}v_p^{(a,d)}\chi_{\vert\mathcal{M}_dq\vert>\kappa}
\vert\mathcal{M}_dq\vert\langle a_{p+q}^*a_{-p}^*a_q\beta^{(n)},
\alpha^{(n+1)}\rangle\right\vert\\
\leq& Cd^{-\frac{1}{2}}\kappa^{-1}\sum_{n=1}^{N-1}
\sqrt{n+1}n\int_{\Lambda_d^{n+1}}v_a(\mathbf{x}_1-\mathbf{x}_2)
\vert\alpha^{(n+1)}(\mathbf{x}_1,\dots,\mathbf{x}_{n+1})\vert\\
&\quad\quad\quad\quad\quad\quad\quad\quad\quad\quad\quad\quad
\times\vert(1-P_\kappa)\{\vert\nabla_{\mathbf{x}_1}\vert\beta^{(n)}
(\cdot,\mathbf{x}_3,\dots,\mathbf{x}_{n+1})\}(\mathbf{x}_1)\vert\\
\leq&Cd^{-\frac{1}{2}}\kappa^{-1}\left(\sum_{n=1}^{N-1}
(n+1)n\int_{\Lambda_d^{n+1}}v_a(\mathbf{x}_1-\mathbf{x}_2)
\vert\alpha^{(n+1)}\vert^2\right)^{\frac{1}{2}}\\
&\quad\quad\quad\quad\quad\quad\quad
\times\left(\sum_{n=1}^{N-1}
n\int_{\Lambda_d^{n+1}}v_a(\mathbf{x}_1-\mathbf{x}_2)
\vert\nabla_{\mathbf{x}_2}\beta^{(n)}(\mathbf{x}_2,\dots,\mathbf{x}_{n+1})
\vert^2\right)^{\frac{1}{2}}.
\end{align*}
Using (\ref{H_4psi,psi fock}) and (\ref{H_21psi,psi fock}) we can bound
\begin{align*}
\vert\langle\Gamma_3^\prime\psi,\psi\rangle\vert
&\leq Cd^{-\frac{1}{2}}\kappa^{-1}\Vert v_a\Vert_1^{\frac{1}{2}}
\langle H_4\psi,\psi\rangle^{\frac{1}{2}}
\langle H_{21}a_0\psi,a_0\psi\rangle^{\frac{1}{2}}\\
&\leq CN^{\frac{1}{2}}a^{\frac{1}{2}}d^{-\frac{1}{2}}\kappa^{-1}
\langle (H_4+H_{21})\psi,\psi\rangle.
\end{align*}
Combining the above result with (\ref{E_21^B'}) and (\ref{E_4^B'}) we deduce
\begin{align*}
  \pm\Gamma^\prime
&\leq CN^{\frac{1}{2}}a^{\frac{3}{2}}d^{\frac{1}{2}}\textit{l}^2(\mathcal{N}_++1)
+CNa^2d^{-(2+\alpha)}(\textit{l}^{-1}+\ln[(d\textit{l})^{-1}])(\mathcal{N}_++1)^2\\
&+C(Na^3\kappa^3\textit{l})^{1-\gamma}(\mathcal{N}_++1)^3
+C(d^\alpha+Na\textit{l}^{\frac{1}{2}}+N^{\frac{1}{2}}a^{\frac{1}{2}}
d^{-\frac{1}{2}}\kappa^{-1})H_{21}\\
&+C(N^{\frac{1}{2}}a^{\frac{3}{2}}d^{\frac{1}{2}}\textit{l}^2
+N^{\frac{1}{2}}a^{\frac{1}{2}}d^{-\frac{1}{2}}\kappa^{-1}
+(Na^3\kappa^3\textit{l})^{\gamma})H_4,
\end{align*}
which together with Lemma \ref{control of S_+ conj with e^B'} and Lemma \ref{lemma e^-B'H_Ne^B'} gives (\ref{est of Gamma'}).

\par As for Region III, since we take $\kappa=\infty$ and define $\tilde{\Gamma}^\prime$ in (\ref{Gamma' III}), we have
\begin{equation*}
  \tilde{\Gamma}^\prime=\Gamma^\prime_2
+\mathcal{E}_{21}^{B^\prime}+\mathcal{E}_{4}^{B^\prime},
\end{equation*}
where $\Gamma^\prime_2$ has been defined as above. Using (\ref{E_21^B' III}) and (\ref{E_4^B' III}), and the estimate of $\Gamma^\prime_2$ given above (notice that in Region III we demand $\kappa=\infty$), we can bound
\begin{align*}
\pm\tilde{\Gamma}^\prime
&\leq C(N^2a^2d^{-1}\textit{l}^{\frac{1}{2}}
+N^{\frac{1}{2}}a^{\frac{3}{2}}d^{\frac{1}{2}}\textit{l}^2)(\mathcal{N}_++1)\\
&+C(Na\textit{l}^{\frac{1}{2}}
+N^{\frac{3}{2}}a^{\frac{7}{6}}d^{-\frac{1}{2}}\textit{l}^{\frac{1}{3}})
(H_{21}+H_4).
\end{align*}
Then Lemma \ref{control of S_+ conj with e^B'} and Lemma \ref{lemma e^-B'H_Ne^B'} give (\ref{est of Gamma' III}).
\begin{flushright}
  {$\Box$}
\end{flushright}

\begin{lemma}\label{lemma [H_3,B']}
\
\begin{enumerate}[$(1)$]
  \item $\mathbf{For\; Region\; I}$ We take $\kappa=\nu d^{-1}$ for some $\nu\geq 1$, then for some $0<\gamma<1$ with the further assumption $Na^3\kappa^3\textit{l}\to0$.
\begin{equation}\label{e^-B'(H_21+H_4+H_3)e^B'}
  \int_{0}^{1}\int_{t}^{1}e^{-sB^\prime}[H_3,B^\prime]e^{sB^\prime}dsdt=
\frac{2N}{\sqrt{d}}\sum_{p\neq0}v_p^{(a,d)}\eta_p\mathcal{N}_+
+\mathcal{E}_{[H_3,B^\prime]}^{B^\prime}.
\end{equation}
The error term satisfies the bound
\begin{align}
  \pm\mathcal{E}_{[H_3,B^\prime]}^{B^\prime}
&\leq
C\{N^{\frac{3}{2}}a^{\frac{3}{2}}d^{-\frac{1}{2}}\textit{l}^{\frac{1}{2}}
+N^2a^2d^{-1}\textit{l}^{\frac{1}{2}}
+N^{\frac{5}{2}}a^{\frac{5}{2}}d^{-\frac{3}{2}}\textit{l}^{\frac{1}{2}}
+Nad^{-1}\kappa^{-2}\nonumber\\
&+Nad^{-1}(Na^3\kappa^3)^{\gamma}+Na^2d^{-1}\kappa[1+ad^{-1}\ln(a^{-1})]
\}(\mathcal{N}_++1)\nonumber\\
&+C\{Na^2d^{-\frac{1}{2}}\kappa^{\frac{3}{2}}\textit{l}^{\frac{1}{2}}
+N(Na^3\kappa^3\textit{l})^{1-\gamma}
+ad^{-1}[1+ad^{-1}\ln(a^{-1})]\nonumber\\
&+Na^2d^{-2}\kappa^{-2}[\textit{l}^{-1}+\ln(d\textit{l})^{-1}]\}
(\mathcal{N}_++1)^2\nonumber\\
&+CN^{\frac{3}{2}}a^{\frac{3}{2}}d^{-\frac{1}{2}}\textit{l}^{\frac{1}{2}}
(Na^3\kappa^3\textit{l})^{1-\gamma}
(\mathcal{N}_++1)^3+C\kappa^{-2} H_{21}\nonumber\\
&+C(N^{\frac{3}{2}}a^{\frac{3}{2}}d^{-\frac{1}{2}}\textit{l}^{\frac{1}{2}}
+(Na^3\kappa^3\textit{l})^{\gamma}+\kappa^{-2})H_4.\label{E_3^B'}
\end{align}
  \item $\mathbf{For\; Region\; III}$ We take $\kappa=\infty$. Assume further that $N^{\frac{3}{2}}a^{\frac{7}{6}}d^{-\frac{1}{2}}\textit{l}^{\frac{1}{3}}\to 0$.
\begin{align}
\int_{0}^{1}\int_{t}^{1}e^{-sB^\prime}[H_3,B^\prime]e^{sB^\prime}dsdt=
\frac{1}{\sqrt{d}}\sum_{p,q\neq0}(v_p^{(a,d)}+v_{p+q}^{(a,d)})
\eta_pa_q^*a_qa_0^*a_0+\tilde{\mathcal{E}}_{[H_3,B^\prime]}^{B^\prime}.
\label{e^-B'(H_21+H_4+H_3)e^B' III}
\end{align}
The error term satisfies the bound
\begin{align}
  \pm\mathcal{E}_{[H_3,B^\prime]}^{B^\prime}
&\leq C\{N^2a^2d^{-1}\textit{l}^{\frac{1}{2}}
+N^{\frac{3}{2}}a^{\frac{3}{2}}d^{-\frac{1}{2}}\textit{l}^{\frac{1}{2}}
+N^{\frac{5}{2}}a^{\frac{5}{2}}d^{-\frac{3}{2}}\textit{l}^{\frac{1}{2}}
+N^{2}a^{\frac{5}{3}}d^{-1}\textit{l}^{\frac{1}{3}}\}
(\mathcal{N}_++1)\nonumber\\
&+C(N^3a^{\frac{8}{3}}d^{-1}\textit{l}^{\frac{5}{6}}
+N^{2}a^{\frac{5}{3}}d^{-1}\textit{l}^{\frac{1}{3}})[H_{21}
+Nad^{-1}(\mathcal{N}_++1)^2]\nonumber\\
&+CN^{\frac{3}{2}}a^{\frac{3}{2}}d^{-\frac{1}{2}}\textit{l}^{\frac{1}{2}}
H_4.\label{E_3^B' III}
\end{align}
\end{enumerate}
\end{lemma}
\noindent
\emph{Proof.} We first calculate
\begin{equation*}
  [H_3,B^{\prime}]=\Upsilon+\sum_{i=1}^{10}\mathcal{E}_{3,i}^{B^\prime},
\end{equation*}
where
\begin{equation*}
  \Upsilon=\frac{2}{\sqrt{d}}\sum_{p,q,p+q\neq0}(v_{p}^{(a,d)}+v_{p+q}^{(a,d)})
\eta_p\chi_{\vert\mathcal{M}_dq\vert\leq\kappa}a_q^*a_qa_0^*a_0,
\end{equation*}
and
\begin{align*}
\mathcal{E}_{3,1}^{B^\prime}&=
\frac{1}{\sqrt{d}}\sum_{\substack{p,q,p+q,\\s,q-s\neq0}}
v_{p}^{(a,d)}\eta_s\chi_{\vert\mathcal{M}_d(q-s)\vert\leq\kappa}
(a_{p+q}^*a_{-p}^*a_{-s}^*a_{q-s}a_0a_0+h.c.),\\
\mathcal{E}_{3,2}^{B^\prime}&=
\frac{1}{\sqrt{d}}\sum_{\substack{p,q,p+q,\\s,s-q\neq0}}
v_{p}^{(a,d)}\eta_q\chi_{\vert\mathcal{M}_ds\vert\leq\kappa}
(a_{p+q}^*a_{-p}^*a_{s-q}^*a_{s}a_0a_0+h.c.),\\
\mathcal{E}_{3,3}^{B^\prime}&=
-\frac{1}{\sqrt{d}}\sum_{\substack{p,q,p+q,\\s,s+p+q\neq0}}
v_{p}^{(a,d)}\eta_s\chi_{\vert\mathcal{M}_d(p+q)\vert\leq\kappa}
(a_{s+p+q}^*a_{-s}^*a_{-p}^*a_{q}a_0a_0+h.c.),\\
\mathcal{E}_{3,4}^{B^\prime}&=
-\frac{1}{\sqrt{d}}\sum_{\substack{p,q,p+q,\\s,s-p\neq0}}
v_{p}^{(a,d)}\eta_s\chi_{\vert\mathcal{M}_dp\vert\leq\kappa}
(a_{s-p}^*a_{-s}^*a_{p+q}^*a_{q}a_0a_0+h.c.),\\
\mathcal{E}_{3,5}^{B^\prime}&=
\frac{1}{\sqrt{d}}\sum_{\substack{p,q,p+q,\\s,p+q-s\neq0}}
v_{p}^{(a,d)}\eta_s\chi_{\vert\mathcal{M}_d(p+q-s)\vert\leq\kappa}
(a_q^*a_{-s}^*a_{-p}a_{p+q-s}a_0^*a_0+h.c.),\\
\mathcal{E}_{3,6}^{B^\prime}&=
\frac{1}{\sqrt{d}}\sum_{\substack{p,q,p+q,\\s,-p-s\neq0}}
v_{p}^{(a,d)}\eta_s\chi_{\vert\mathcal{M}_d(p+s)\vert\leq\kappa}
(a_q^*a_{-s}^*a_{p+q}a_{-p-s}a_0^*a_0+h.c.),\\
\mathcal{E}_{3,7}^{B^\prime}&=
\frac{1}{\sqrt{d}}\sum_{\substack{p,q,p+q,\\s,s-p-q\neq0}}
v_{p}^{(a,d)}\eta_{p+q}\chi_{\vert\mathcal{M}_ds\vert\leq\kappa}
(a_{q}^*a_{s-p-q}^*a_{-p}a_{s}a_0^*a_0+h.c.),\\
\mathcal{E}_{3,8}^{B^\prime}&=
\frac{1}{\sqrt{d}}\sum_{\substack{p,q,p+q,\\s,s+p\neq0}}
v_{p}^{(a,d)}\eta_p\chi_{\vert\mathcal{M}_ds\vert\leq\kappa}
(a_{q}^*a_{p+s}^*a_{p+q}a_{s}a_0^*a_0+h.c.),\\
\mathcal{E}_{3,9}^{B^\prime}&=
-\frac{1}{\sqrt{d}}\sum_{\substack{p,q,p+q,\\s,s+q\neq0}}
v_{p}^{(a,d)}\eta_s\chi_{\vert\mathcal{M}_dq\vert\leq\kappa}
(a_{s+q}^*a_{-s}^*a_{-p}a_{p+q}a_0a_0^*+h.c.),\\
\mathcal{E}_{3,10}^{B^\prime}&=
-\frac{1}{\sqrt{d}}\sum_{\substack{p,q,p+q,\\s,t,s+t\neq0}}
v_{p}^{(a,d)}\eta_s\chi_{\vert\mathcal{M}_dt\vert\leq\kappa}
(a_{s+t}^*a_{-s}^*a_q^*a_ta_{-p}a_{p+q}+h.c.).
\end{align*}

\par For Region I, we can bound $\mathcal{E}_{3,i}^{B^\prime}$ respectively by
\begin{align*}
\vert\langle\mathcal{E}_{3,1}^{B^\prime}\psi,\psi\rangle\vert
&\leq Cd^{-\frac{1}{2}}\Vert\eta_\perp\Vert_2\Vert v_a\Vert_1^{\frac{1}{2}}
\langle H_4\psi,\psi\rangle^{\frac{1}{2}}\\
&\times\langle U_N(\mathcal{N}_++1)^2U_N^*P_\kappa U_N a_0a_0\psi,
P_\kappa U_N a_0a_0\psi\rangle^{\frac{1}{2}}\\
&+Cd^{-\frac{1}{2}}\Vert P_\kappa\eta_\perp\Vert_2\Vert v_a\Vert_1^{\frac{1}{2}}
\langle(\mathcal{N}_++1)^2a_0a_0\psi,a_0a_0\psi\rangle^{\frac{1}{2}}
\langle H_4\psi,\psi\rangle^{\frac{1}{2}}\\
&\leq Cd^{-\frac{1}{2}}\Vert\eta_\perp\Vert_2\Vert v_a\Vert_1^{\frac{1}{2}}
\langle(\mathcal{N}_++1)^2a_0a_0\psi,a_0a_0\psi\rangle^{\frac{1}{2}}
\langle H_4\psi,\psi\rangle^{\frac{1}{2}}\\
&\leq CN^{\frac{3}{2}}a^{\frac{3}{2}}d^{-\frac{1}{2}}\textit{l}^{\frac{1}{2}}
\langle H_4\psi,\psi\rangle^{\frac{1}{2}}
\langle(\mathcal{N}_++1)\psi,\psi\rangle^{\frac{1}{2}},
\end{align*}
and
\begin{align*}
\vert\langle\mathcal{E}_{3,2}^{B^\prime}\psi,\psi\rangle\vert
&\leq Cd^{-\frac{1}{2}}\Vert\eta_\perp\Vert_2\Vert v_a\Vert_1^{\frac{1}{2}}
\langle H_4\psi,\psi\rangle^{\frac{1}{2}}\\
&\times\langle U_N(\mathcal{N}_++1)^2U_N^*P_\kappa U_N a_0a_0\psi,
P_\kappa U_N a_0a_0\psi\rangle^{\frac{1}{2}}\\
&\leq Cd^{-\frac{1}{2}}\Vert\eta_\perp\Vert_2\Vert v_a\Vert_1^{\frac{1}{2}}
\langle(\mathcal{N}_++1)^2a_0a_0\psi,a_0a_0\psi\rangle^{\frac{1}{2}}
\langle H_4\psi,\psi\rangle^{\frac{1}{2}}\\
&\leq CN^{\frac{3}{2}}a^{\frac{3}{2}}d^{-\frac{1}{2}}\textit{l}^{\frac{1}{2}}
\langle H_4\psi,\psi\rangle^{\frac{1}{2}}
\langle(\mathcal{N}_++1)\psi,\psi\rangle^{\frac{1}{2}}.
\end{align*}
Recalling that $\kappa=\nu d^{-1}$ for some $\nu\geq 1$ in Region I, and the bound ($\ref{L2 zeta}$), we have
\begin{align*}
\vert\langle\mathcal{E}_{3,3}^{B^\prime}\psi,\psi\rangle\vert
&\leq \frac{C}{\sqrt{d}}\Vert\eta_\perp\Vert_2\Vert v_a\Vert_1\Vert\zeta^{\kappa}\Vert_2
\langle(\mathcal{N}_++1)^2a_0a_0\psi,a_0a_0\psi\rangle^{\frac{1}{2}}
\langle (\mathcal{N}_++1)^2\psi,\psi\rangle^{\frac{1}{2}}\\
&\leq CNa^2d^{-\frac{1}{2}}\kappa^{\frac{3}{2}}\textit{l}^{\frac{1}{2}}
\langle (\mathcal{N}_++1)^2\psi,\psi\rangle.
\end{align*}
Notice that
\begin{equation*}
  \Vert P_{\kappa,\neq0}v_a\Vert_2^2=
\sum_{0<\vert\mathcal{M}_dp\vert\leq\kappa}\vert v_p^{(a,d)}\vert^2\leq Ca^2\kappa^3,
\end{equation*}
we deduce
\begin{align*}
\vert\langle\mathcal{E}_{3,4}^{B^\prime}\psi,\psi\rangle\vert
&\leq\frac{C}{\sqrt{d}}\Vert\eta_\perp\Vert_2
\Vert P_{\kappa,\neq0}v_a\Vert_2
\langle(\mathcal{N}_++1)^2a_0a_0\psi,a_0a_0\psi\rangle^{\frac{1}{2}}
\langle (\mathcal{N}_++1)^2\psi,\psi\rangle^{\frac{1}{2}}\\
&\leq CNa^2d^{-\frac{1}{2}}\kappa^{\frac{3}{2}}\textit{l}^{\frac{1}{2}}
\langle (\mathcal{N}_++1)^2\psi,\psi\rangle.
\end{align*}
Moreover
\begin{align*}
\vert\langle\mathcal{E}_{3,5}^{B^\prime}\psi,\psi\rangle\vert
&\leq\frac{C}{\sqrt{d}}\Vert\eta_\perp\Vert_2\Vert v_a\Vert_1\Vert\zeta^{\kappa}\Vert_2
\langle(\mathcal{N}_++1)^2a_0^*a_0\psi,a_0^*a_0\psi\rangle^{\frac{1}{2}}
\langle (\mathcal{N}_++1)^2\psi,\psi\rangle^{\frac{1}{2}}\\
&+\frac{C}{d}\Vert P_\kappa\eta_\perp\Vert_2\Vert v_a\Vert_1
\langle(\mathcal{N}_++1)^2a_0^*a_0\psi,a_0^*a_0\psi\rangle^{\frac{1}{2}}
\langle (\mathcal{N}_++1)^2\psi,\psi\rangle^{\frac{1}{2}}\\
&\leq CNa^2d^{-\frac{1}{2}}\kappa^{\frac{3}{2}}\textit{l}^{\frac{1}{2}}
\langle (\mathcal{N}_++1)^2\psi,\psi\rangle+CN^2a^2d^{-1}\textit{l}^{\frac{1}{2}}
(\mathcal{N}_++1),
\end{align*}
\begin{align*}
\vert\langle\mathcal{E}_{3,6}^{B^\prime}\psi,\psi\rangle\vert
&\leq \frac{C}{\sqrt{d}}\Vert\eta_\perp\Vert_2\Vert v_a\Vert_1\Vert\zeta^{\kappa}\Vert_2
\langle(\mathcal{N}_++1)^2a_0^*a_0\psi,a_0^*a_0\psi\rangle^{\frac{1}{2}}
\langle (\mathcal{N}_++1)^2\psi,\psi\rangle^{\frac{1}{2}}\\
& +\frac{C}{\sqrt{d}}\Vert P_\kappa\eta_\perp\Vert_2\vert v_0^{(a,d)}\vert
\langle(\mathcal{N}_++1)^2a_0^*a_0\psi,a_0^*a_0\psi\rangle^{\frac{1}{2}}
\langle (\mathcal{N}_++1)^2\psi,\psi\rangle^{\frac{1}{2}}\\
&\leq CNa^2d^{-\frac{1}{2}}\kappa^{\frac{3}{2}}\textit{l}^{\frac{1}{2}}
\langle (\mathcal{N}_++1)^2\psi,\psi\rangle+CN^2a^2d^{-1}\textit{l}^{\frac{1}{2}}
(\mathcal{N}_++1),
\end{align*}
\begin{align*}
\vert\langle\mathcal{E}_{3,7}^{B^\prime}\psi,\psi\rangle\vert
&\leq \frac{C}{\sqrt{d}}\Vert\eta_\perp\Vert_2\Vert v_a\Vert_1\Vert\zeta^{\kappa}\Vert_2
\langle(\mathcal{N}_++1)^2a_0^*a_0\psi,a_0^*a_0\psi\rangle^{\frac{1}{2}}
\langle (\mathcal{N}_++1)^2\psi,\psi\rangle^{\frac{1}{2}}\\
&\leq CNa^2d^{-\frac{1}{2}}\kappa^{\frac{3}{2}}\textit{l}^{\frac{1}{2}}
\langle (\mathcal{N}_++1)^2\psi,\psi\rangle,
\end{align*}
\begin{align*}
\vert\langle\mathcal{E}_{3,8}^{B^\prime}\psi,\psi\rangle\vert
&\leq \frac{C}{\sqrt{d}}\Vert\eta_\perp\ast v_a\Vert_2\Vert\zeta^{\kappa}\Vert_2
\langle(\mathcal{N}_++1)^2a_0^*a_0\psi,a_0^*a_0\psi\rangle^{\frac{1}{2}}
\langle (\mathcal{N}_++1)^2\psi,\psi\rangle^{\frac{1}{2}}\\
&\leq CNa^2d^{-\frac{1}{2}}\kappa^{\frac{3}{2}}\textit{l}^{\frac{1}{2}}
\langle (\mathcal{N}_++1)^2\psi,\psi\rangle,
\end{align*}
\begin{align*}
\vert\langle\mathcal{E}_{3,9}^{B^\prime}\psi,\psi\rangle\vert
&\leq C\Vert\eta_\perp\Vert_2\Vert v_a\Vert_1^{\frac{1}{2}}\Vert\zeta^{\kappa}\Vert_2
\langle H_4a_0a_0^*\psi,a_0a_0^*\psi\rangle^{\frac{1}{2}}
\langle (\mathcal{N}_++1)^2\psi,\psi\rangle^{\frac{1}{2}}\\
&\leq C(Na^3\kappa^3\textit{l})^\gamma\langle H_4\psi,\psi\rangle+
CN(Na^3\kappa^3\textit{l})^{1-\gamma}
\langle (\mathcal{N}_++1)^2\psi,\psi\rangle,
\end{align*}
and
\begin{align*}
\vert\langle\mathcal{E}_{3,10}^{B^\prime}\psi,\psi\rangle\vert
&\leq C\Vert\eta_\perp\Vert_2\Vert v_a\Vert_1^{\frac{1}{2}}\Vert\zeta^{\kappa}\Vert_2
\langle(\mathcal{N}_++1)^4\psi,\psi\rangle^{\frac{1}{2}}
\langle H_4\psi,\psi\rangle^{\frac{1}{2}}\\
&\leq C(Na^3\kappa^3\textit{l})^\gamma\langle H_4\psi,\psi\rangle+
CN(Na^3\kappa^3\textit{l})^{1-\gamma}
\langle (\mathcal{N}_++1)^2\psi,\psi\rangle,
\end{align*}
for any $0\leq\gamma\leq1$.
\par For the main term $\Upsilon$, we simplify it further in Region I. First we observe that
\begin{equation*}
   \Upsilon=\frac{2}{\sqrt{d}}\sum_{p,q,p+q\neq0}(v_{p}^{(a,d)}+v_{p+q}^{(a,d)})
\eta_p\chi_{\vert\mathcal{M}_dq\vert\leq\kappa}a_q^*a_q(N-\mathcal{N}_+).
\end{equation*}
Noticing the fact that for $q\neq0$
\begin{equation*}
  \langle a_q^*a_q\mathcal{N}_+\psi,\psi\rangle=
\langle a_q^*(\mathcal{N}_++1)a_q\psi,\psi\rangle\geq0.
\end{equation*}
Hence with (\ref{v_p^a,d}) and Lemma \ref{l1 lemma} we have
\begin{align*}
&\left\vert\frac{2}{\sqrt{d}}\sum_{p,q,p+q\neq0}(v_{p}^{(a,d)}+v_{p+q}^{(a,d)})
\eta_p\chi_{\vert\mathcal{M}_dq\vert\leq\kappa}
\langle a_q^*a_q\mathcal{N}_+\psi,\psi\rangle\right\vert\\
&\leq\frac{Ca}{d}\sum_{p\neq0}\vert\eta_p\vert\sum_{q\neq0}\langle a_q^*a_q\mathcal{N}_+\psi,\psi\rangle\\
&\leq Cad^{-1}(1+ad^{-1}\ln(a^{-1}))
\langle \mathcal{N}_+^2\psi,\psi\rangle.
\end{align*}
Moreover, since the potential $v$ is compactly supported we have
\begin{equation*}
  \vert v_{p+q}^{(a,d)}-v_{p}^{(a,d)}\vert
=\frac{a}{\sqrt{d}}\left\vert
\widehat{v}(\frac{a\mathcal{M}_d(p+q)}{2\pi})-\widehat{v}(\frac{a\mathcal{M}_dp}{2\pi})
\right\vert\leq\frac{Ca^2}{\sqrt{d}}\vert\mathcal{M}_dq\vert.
\end{equation*}
Hence
\begin{align*}
&\left\vert\frac{2N}{\sqrt{d}}\sum_{p,q,p+q\neq0}(v_{p+q}^{(a,d)}-v_{p}^{(a,d)})
\eta_p\chi_{\vert\mathcal{M}_dq\vert\leq\kappa}
\langle a_q^*a_q\psi,\psi\rangle\right\vert\\
&\leq \frac{CNa^2}{d}\sum_{p\neq0}\vert\eta_p\vert
\sum_{\vert\mathcal{M}_dq\vert\leq\kappa}
\vert\mathcal{M}_dq\vert\langle a_q^*a_q\psi,\psi\rangle\\\
&\leq CNa^2d^{-1}\kappa(1+ad^{-1}\ln(a^{-1}))\langle \mathcal{N}_+\psi,\psi\rangle.
\end{align*}
Finally we observe that with (\ref{v_p^a,d}) and (\ref{est of eta_0})
\begin{equation*}
 \left\vert\frac{4N}{\sqrt{d}}\sum_{p\neq0}v_p^{(a,d)}\eta_p
\chi_{\vert\mathcal{M}_dp\vert\leq\kappa}\langle a_p^*a_p\psi,\psi\rangle\right\vert
\leq CNa^2\textit{l}^2\langle \mathcal{N}_+\psi,\psi\rangle,
\end{equation*}
and
\begin{equation*}
  \mathcal{N}_+-\sum_{\vert\mathcal{M}_dq\vert\leq\kappa}a_q^*a_q
=\sum_{\vert\mathcal{M}_dq\vert>\kappa}a_q^*a_q\leq\kappa^{-2}H_{21}.
\end{equation*}
With estimates above we arrive at
\begin{align*}
\Upsilon=\frac{4N}{\sqrt{d}}\sum_{p\neq0}v_p^{(a,d)}\eta_p\mathcal{N}_+
+\mathcal{E}_{3,0}^{B^\prime},
\end{align*}
where
\begin{align*}
  \pm\mathcal{E}_{3,0}^{B^\prime}&\leq
Cad^{-1}(1+ad^{-1}\ln(a^{-1}))(\mathcal{N}_++1)\\
&+CNa^2d^{-1}\kappa(1+ad^{-1}\ln(a^{-1}))(\mathcal{N}_++1)^2
+\kappa^{-2}H_{21}.
\end{align*}
Let
\begin{equation*}
  \mathcal{E}_{[H_3,B^\prime]}^{B^\prime}=\int_{0}^{1}\int_{t}^{1}e^{-sB}\sum_{i=0}^{10}
\mathcal{E}_{3,i}^{B^\prime}e^{sB}dsdt.
\end{equation*}
Then using Lemma \ref{control of S_+ conj with e^B'} and Lemma \ref{lemma e^-B'H_Ne^B'} we deduce (\ref{e^-B'(H_21+H_4+H_3)e^B'}) and (\ref{E_3^B'}).
\par For Region III, the bounds of $\mathcal{E}_{3,1}^{B^\prime}$ and $\mathcal{E}_{3,2}^{B^\prime}$ are the same as in Region I, while for the rest, we bound
\begin{align*}
\vert\langle\mathcal{E}_{3,3}^{B^\prime}\psi,\psi\rangle\vert
&\leq Cd^{-\frac{1}{2}}\Vert\eta_\perp\Vert_2\Vert v_a\Vert_1^{\frac{1}{2}}
\langle(\mathcal{N}_++1)^2a_0a_0\psi,a_0a_0\psi\rangle^{\frac{1}{2}}
\langle H_4\psi,\psi\rangle^{\frac{1}{2}}\\
&+Cd^{-1}\Vert\eta_\perp\Vert_2\Vert v_a\Vert_1
\langle(\mathcal{N}_++1)^2a_0a_0\psi,a_0a_0\psi\rangle^{\frac{1}{2}}
\langle (\mathcal{N}_++1)^2\psi,\psi\rangle^{\frac{1}{2}}\\
&\leq C(N^{\frac{3}{2}}a^{\frac{3}{2}}d^{-\frac{1}{2}}\textit{l}^{\frac{1}{2}}
+N^{2}a^{2}d^{-1}\textit{l}^{\frac{1}{2}})
\langle(\mathcal{N}_++1)\psi,\psi\rangle\\
&+CN^{\frac{3}{2}}a^{\frac{3}{2}}d^{-\frac{1}{2}}\textit{l}^{\frac{1}{2}}
\langle H_4\psi,\psi\rangle,
\end{align*}
\begin{align*}
\vert\langle\mathcal{E}_{3,4}^{B^\prime}\psi,\psi\rangle\vert
&\leq Cd^{-\frac{1}{2}}\Vert\eta_\perp\Vert_2\Vert v_a\Vert_1^{\frac{1}{2}}
\langle(\mathcal{N}_++1)^2a_0a_0\psi,a_0a_0\psi\rangle^{\frac{1}{2}}
\langle H_4\psi,\psi\rangle^{\frac{1}{2}}\\
&+Cd^{-\frac{1}{2}}\Vert\eta_\perp\Vert_2\vert v_0^{(a,d)}\vert
\langle(\mathcal{N}_++1)^2a_0a_0\psi,a_0a_0\psi\rangle^{\frac{1}{2}}
\langle (\mathcal{N}_++1)^2\psi,\psi\rangle^{\frac{1}{2}}\\
&\leq C(N^{\frac{3}{2}}a^{\frac{3}{2}}d^{-\frac{1}{2}}\textit{l}^{\frac{1}{2}}
+N^{2}a^{2}d^{-1}\textit{l}^{\frac{1}{2}})
\langle(\mathcal{N}_++1)\psi,\psi\rangle\\
&+CN^{\frac{3}{2}}a^{\frac{3}{2}}d^{-\frac{1}{2}}\textit{l}^{\frac{1}{2}}
\langle H_4\psi,\psi\rangle,
\end{align*}
\begin{align*}
\vert\langle\mathcal{E}_{3,5}^{B^\prime}\psi,\psi\rangle\vert
&\leq Cd^{-\frac{1}{2}}\Vert\eta_\perp\Vert_2\Vert v_a\Vert_1^{\frac{1}{2}}
\langle(\mathcal{N}_++1)^2\psi,\psi\rangle^{\frac{1}{2}}
\langle H_4a_0^*a_0\psi,a_0^*a_0\psi\rangle^{\frac{1}{2}}\\
&+Cd^{-1}\Vert\eta_\perp\Vert_2\Vert v_a\Vert_1
\langle(\mathcal{N}_++1)^2a_0^*a_0\psi,a_0^*a_0\psi\rangle^{\frac{1}{2}}
\langle (\mathcal{N}_++1)^2\psi,\psi\rangle^{\frac{1}{2}}\\
&\leq C(N^{\frac{3}{2}}a^{\frac{3}{2}}d^{-\frac{1}{2}}\textit{l}^{\frac{1}{2}}
+N^{2}a^{2}d^{-1}\textit{l}^{\frac{1}{2}})
\langle(\mathcal{N}_++1)\psi,\psi\rangle\\
&+CN^{\frac{3}{2}}a^{\frac{3}{2}}d^{-\frac{1}{2}}\textit{l}^{\frac{1}{2}}
\langle H_4\psi,\psi\rangle,
\end{align*}
\begin{align*}
\vert\langle\mathcal{E}_{3,6}^{B^\prime}\psi,\psi\rangle\vert
&\leq Cd^{-\frac{1}{2}}\Vert\eta_\perp\Vert_2\Vert v_a\Vert_1^{\frac{1}{2}}
\langle(\mathcal{N}_++1)^2\psi,\psi\rangle^{\frac{1}{2}}
\langle H_4a_0^*a_0\psi,a_0^*a_0\psi\rangle^{\frac{1}{2}}\\
&+Cd^{-\frac{1}{2}}\Vert\eta_\perp\Vert_2\vert v_0^{(a,d)}\vert
\langle(\mathcal{N}_++1)^2a_0^*a_0\psi,a_0^*a_0\psi\rangle^{\frac{1}{2}}
\langle (\mathcal{N}_++1)^2\psi,\psi\rangle^{\frac{1}{2}}\\
&\leq C(N^{\frac{3}{2}}a^{\frac{3}{2}}d^{-\frac{1}{2}}\textit{l}^{\frac{1}{2}}
+N^{2}a^{2}d^{-1}\textit{l}^{\frac{1}{2}})
\langle(\mathcal{N}_++1)\psi,\psi\rangle\\
&+CN^{\frac{3}{2}}a^{\frac{3}{2}}d^{-\frac{1}{2}}\textit{l}^{\frac{1}{2}}
\langle H_4\psi,\psi\rangle,
\end{align*}
\begin{align*}
\vert\langle\mathcal{E}_{3,9}^{B^\prime}\psi,\psi\rangle\vert
&\leq Cd^{-\frac{1}{2}}\Vert\eta_\perp\Vert_2\Vert v_a\Vert_1^{\frac{1}{2}}
\langle(\mathcal{N}_++1)^2\psi,\psi\rangle^{\frac{1}{2}}
\langle H_4a_0a_0^*\psi,a_0a_0^*\psi\rangle^{\frac{1}{2}}\\
&\leq CN^{\frac{3}{2}}a^{\frac{3}{2}}d^{-\frac{1}{2}}\textit{l}^{\frac{1}{2}}
\langle(\mathcal{N}_++1)\psi,\psi\rangle
+CN^{\frac{3}{2}}a^{\frac{3}{2}}d^{-\frac{1}{2}}\textit{l}^{\frac{1}{2}}
\langle H_4\psi,\psi\rangle,
\end{align*}
\begin{align*}
\vert\langle\mathcal{E}_{3,10}^{B^\prime}\psi,\psi\rangle\vert
&\leq Cd^{-\frac{1}{2}}\Vert\eta_\perp\Vert_2\Vert v_a\Vert_1^{\frac{1}{2}}
\langle(\mathcal{N}_++1)^4\psi,\psi\rangle^{\frac{1}{2}}
\langle H_4\psi,\psi\rangle^{\frac{1}{2}}\\
&\leq CN^{\frac{3}{2}}a^{\frac{3}{2}}d^{-\frac{1}{2}}\textit{l}^{\frac{1}{2}}
\langle(\mathcal{N}_++1)\psi,\psi\rangle
+CN^{\frac{3}{2}}a^{\frac{3}{2}}d^{-\frac{1}{2}}\textit{l}^{\frac{1}{2}}
\langle H_4\psi,\psi\rangle,
\end{align*}
and we use Poincar\'{e}'s inequality and Sobolev inequality (see (\ref{Poincare})) to bound
\begin{align*}
\vert\langle\mathcal{E}_{3,7}^{B^\prime}\psi,\psi\rangle\vert
&\leq Cd^{-\frac{5}{6}}\Vert\eta_\perp\Vert_2^{\frac{2}{3}}
\Vert\eta_\perp\Vert_{\infty}^{\frac{1}{3}}
\Vert v_a\Vert_1
\langle(\mathcal{N}_++1)^3\psi,\psi\rangle^{\frac{1}{2}}
\langle H_{21}a_0^*a_0\psi,a_0^*a_0\psi\rangle^{\frac{1}{2}}\\
&\leq CN^{2}a^{\frac{5}{3}}d^{-1}\textit{l}^{\frac{1}{3}}
\langle(\mathcal{N}_++1)\psi,\psi\rangle
+CN^{2}a^{\frac{5}{3}}d^{-1}\textit{l}^{\frac{1}{3}}
\langle H_{21}\psi,\psi\rangle,
\end{align*}
\begin{align*}
\vert\langle\mathcal{E}_{3,8}^{B^\prime}\psi,\psi\rangle\vert
&\leq Cd^{-\frac{5}{6}}\Vert\eta_\perp\Vert_2^{\frac{2}{3}}
\Vert\eta_\perp\Vert_{\infty}^{\frac{1}{3}}
\Vert v_a\Vert_1
\langle(\mathcal{N}_++1)^3\psi,\psi\rangle^{\frac{1}{2}}
\langle H_{21}a_0^*a_0\psi,a_0^*a_0\psi\rangle^{\frac{1}{2}}\\
&\leq CN^{2}a^{\frac{5}{3}}d^{-1}\textit{l}^{\frac{1}{3}}
\langle(\mathcal{N}_++1)\psi,\psi\rangle
+CN^{2}a^{\frac{5}{3}}d^{-1}\textit{l}^{\frac{1}{3}}
\langle H_{21}\psi,\psi\rangle.
\end{align*}

\par For the simplification of $\Upsilon$ in Region III, we first notice that with (\ref{v_p^a,d}) and (\ref{est of eta_0}) we can let
\begin{equation*}
  \Upsilon=\frac{2}{\sqrt{d}}\sum_{p,q\neq0}(v_p^{(a,d)}+v_{p+q}^{(a,d)})
\eta_pa_q^*a_qa_0^*a_0+\tilde{\mathcal{E}}_{3,0}^{B^\prime},
\end{equation*}
with
\begin{equation*}
  \pm\tilde{\mathcal{E}}_{3,0}^{B^\prime}\coloneqq
\pm\frac{2}{\sqrt{d}}\sum_{p\neq0}(v_p^{(a,d)}+v_0^{(a,d)})
\eta_pa_p^*a_pa_0^*a_0\leq CNa^2\textit{l}^2(\mathcal{N}_++1).
\end{equation*}
Moreover, we notice that by Lemma \ref{control of S_+ conj with e^B'} and the fact that $a_0^*a_0=N-\mathcal{N}_+$ when acting on $L^2_s(\Lambda_d^N)$, we can control for $\vert t\vert<1$, that
\begin{align*}
&\pm\frac{2}{\sqrt{d}}\left(e^{-tB^\prime}\sum_{p,q\neq0}v_p^{(a,d)}
\eta_pa_q^*a_qa_0^*a_0e^{tB^\prime}-
\sum_{p,q\neq0}v_p^{(a,d)}
\eta_pa_q^*a_qa_0^*a_0\right)\\
&\leq CN^2a^2d^{-1}\textit{l}^{\frac{1}{2}}(\mathcal{N}_++1).
\end{align*}
Via an argument similar to Lemma \ref{lemma e^-B'(H_02+H_22)e^B'}, where we have controlled the action of $e^{B^\prime}$ on $H_{22}$, we can bound
\begin{align*}
&\pm\frac{2}{\sqrt{d}}\left(e^{-tB^\prime}\sum_{p,q\neq0}v_{p+q}^{(a,d)}
\eta_pa_q^*a_qa_0^*a_0e^{tB^\prime}-
\sum_{p,q\neq0}v_{p+q}^{(a,d)}
\eta_pa_q^*a_qa_0^*a_0\right)\\
&\leq CN^2a^2d^{-1}\textit{l}^{\frac{1}{2}}(\mathcal{N}_++1).
\end{align*}
We only need to replace $v_a$ in the the estimates of error terms in Lemma \ref{lemma e^-B'(H_02+H_22)e^B'} with $\sqrt{d}\eta_\perp v_a$ to reach this estimate. Letting
\begin{align*}
    \tilde{\mathcal{E}}_{[H_3,B^\prime]}^{B^\prime}
=\int_{0}^{1}\int_{t}^{1}e^{-sB}
\sum_{i=1}^{10}\mathcal{E}_{3,i}^{B^\prime}e^{sB}dsdt
+\int_{0}^{1}\int_{t}^{1}(e^{-sB}
\Upsilon e^{sB}-\Upsilon) dsdt-\tilde{\mathcal{E}}_{3,0}^{B^\prime},
\end{align*}
we reach (\ref{e^-B'(H_21+H_4+H_3)e^B' III}) and (\ref{E_3^B' III}).
\begin{flushright}
  {$\Box$}
\end{flushright}

\begin{lemma}\label{lemma [H'_3,B']}
\
$\mathbf{For\; Region\; III}$ We take $\kappa=\infty$. Assume $N^{\frac{3}{2}}a^{\frac{7}{6}}d^{-\frac{1}{2}}\textit{l}^{\frac{1}{3}}\to 0$.
\begin{align}
\int_{0}^{1}\int_{0}^{t}e^{-sB^\prime}[H_3^\prime,B^\prime]e^{sB^\prime}dsdt=
2\sum_{p,q\neq0}(W_p+W_{p+q})
\eta_pa_q^*a_qa_0^*a_0+\tilde{\mathcal{E}}_{[H_3^\prime,B^\prime]}^{B^\prime}.
\label{e^-B'(H_21+H_4+H_3')e^B' III}
\end{align}
The error term satisfies the bound
\begin{align}
  \pm\mathcal{E}_{[H_3^\prime,B^\prime]}^{B^\prime}
&\leq C\{N^2a^2d^{-1}\textit{l}^{\frac{1}{2}}
+N^{\frac{3}{2}}a^{\frac{3}{2}}d^{-\frac{1}{2}}\textit{l}^{\frac{1}{2}}
+N^{\frac{5}{2}}a^{\frac{5}{2}}d^{-\frac{3}{2}}\textit{l}^{\frac{1}{2}}
+N^{2}a^{\frac{5}{3}}d^{-1}\textit{l}^{\frac{1}{3}}\}
(\mathcal{N}_++1)\nonumber\\
&+C(N^3a^{\frac{8}{3}}d^{-1}\textit{l}^{\frac{5}{6}}
+N^{2}a^{\frac{5}{3}}d^{-1}\textit{l}^{\frac{1}{3}})[H_{21}
+Nad^{-1}(\mathcal{N}_++1)^2]\nonumber\\
&+CN^{\frac{3}{2}}a^{\frac{3}{2}}d^{-\frac{1}{2}}\textit{l}^{\frac{1}{2}}
H_4^\prime.\label{E_3'^B' III}
\end{align}
\end{lemma}
\noindent
\emph{Proof.} The proofs of (\ref{e^-B'(H_21+H_4+H_3')e^B' III}) and (\ref{E_3'^B' III}) are essentially the same as (\ref{e^-B'(H_21+H_4+H_3)e^B' III}) and (\ref{E_3^B' III}), we only need to substitute the potential $v_a$ with $2\sqrt{d}W$.

\begin{flushright}
  {$\Box$}
\end{flushright}

\noindent
\emph{Proof of Proposition \ref{cubic renorm}.}
\begin{flushleft}
  $\mathbf{For\; Region\; I}$
\end{flushleft}
\par We collect all the lemmas for Region I above (i.e. we combine (\ref{e^-B'(C^B+Q^BN_+)e^B'}), (\ref{e^-B'E^Be^B'}), (\ref{e^-B'H_23'e^B'}), (\ref{E_23^B'}), (\ref{est of Gamma'}) and (\ref{e^-B'(H_21+H_4+H_3)e^B'})). We take $\kappa=\nu d^{-1}$ for some $\nu\geq 1$, then for some $\alpha>0$ and $0<\gamma<1$ with the further assumption $Na^3\kappa^3\textit{l}$ and $N^{\frac{1}{2}}a^{\frac{1}{2}}d^{-\frac{1}{2}}\kappa^{-1}$ tend to 0. Using (\ref{first renorm})
\begin{equation*}
  e^{-B^\prime}\mathcal{G}_Ne^{B^\prime}
=C^B+\left(Q^{B}+\frac{2N}{\sqrt{d}}\sum_{p\neq0}v_p^{(a,d)}\eta_p\right)\mathcal{N}_+
+H_{21}+H_4+H_{23}^{\prime}+\mathcal{E}^{B^\prime},
\end{equation*}
and the error term is bounded by
\begin{align*}
\pm\mathcal{E}^{B^\prime}
&\leq CNad^{-1}
\{d^\alpha+Na\textit{l}^{\frac{1}{2}}+\kappa^{-2}
+N^{\frac{3}{2}}a^{\frac{3}{2}}d^{-\frac{1}{2}}\textit{l}^{\frac{1}{2}}
+N^{\frac{1}{2}}a^{\frac{1}{2}}d^{-\frac{1}{2}}\kappa^{-1}\\
&\quad\quad\quad\quad\quad+(Na^3\kappa^3\textit{l})^\gamma
+a\kappa[1+ad^{-1}\ln a^{-1}]\}(\mathcal{N}_++1)\\
&+CN^{\frac{3}{2}}a^{\frac{3}{2}}d^{-\frac{1}{2}}\textit{l}^{\frac{1}{2}}
(\mathcal{N}_++1)
+CN^{\frac{3}{2}}a^2d^{-2}\textit{l}^{-1}
(\mathcal{N}_++1)^{\frac{3}{2}}\\
&+C\{ad^{-1}+Na^2d^{-2}\textit{l}^{-1}
+Na^2d^{-\frac{1}{2}}\kappa^{\frac{3}{2}}\textit{l}^{\frac{1}{2}}
+N(Na^3\kappa^3\textit{l})^{1-\gamma}\\
&\quad +ad^{-1}[1+ad^{-1}\ln a^{-1}]
+Na^2d^{-(2+\alpha)}[\textit{l}^{-1}+\ln(d\textit{l})^{-1}]\}
(\mathcal{N}_++1)^2\\
&+CN^{\frac{3}{2}}a^{\frac{3}{2}}d^{-\frac{1}{2}}\textit{l}^{\frac{1}{2}}
(Na^3\kappa^3\textit{l})^{1-\gamma}(\mathcal{N}_++1)^3
+CNa^3d^{-1}H_{21}\\
&+C(N^{\frac{3}{2}}a^{\frac{3}{2}}d^{-\frac{1}{2}}\textit{l}^{\frac{1}{2}}
+(Na^3\kappa^3\textit{l})^\gamma)H_4\\
&+C(d^\alpha+Na\textit{l}^{\frac{1}{2}}+\kappa^{-2}
+N^{\frac{1}{2}}a^{\frac{1}{2}}d^{-\frac{1}{2}}\kappa^{-1})(H_{21}+H_4).
\end{align*}
Then we reach (\ref{second renorm}) and (\ref{second renorm E^B'}).
\begin{flushleft}
  $\mathbf{For\; Region\; III}$
\end{flushleft}
\par We collect all the lemmas for Region I above (i.e. we combine (\ref{e^-B'(C^B+Q^BN_+)e^B' III}), (\ref{e^-B'E^Be^B' III}), (\ref{e^-B'H_23'e^B'}), (\ref{E_23^B' III}), (\ref{e^-B'(H_02+H_22)e^B' III}), (\ref{est of Gamma' III}), (\ref{e^-B'(H_21+H_4+H_3)e^B' III}) and (\ref{e^-B'(H_21+H_4+H_3')e^B' III})). We take $\kappa=\infty$. Assume further $N^{\frac{3}{2}}a^{\frac{7}{6}}d^{-\frac{1}{2}}\textit{l}^{\frac{1}{3}}\to 0$. Using (\ref{first renorm III})
\begin{align*}
  e^{-B^\prime}\mathcal{G}_Ne^{B^\prime}
&=\tilde{C}^B+\tilde{Q}^{B}_1\mathcal{N}_++\tilde{Q}^{B}_2\mathcal{N}_+(\mathcal{N}_++1)\\
&+\frac{1}{\sqrt{d}}\sum_{p,q\neq0}(v^{(a,d)}_p+v^{(a,d)}_{p+q})\eta_p a_q^*a_qa_0^*a_0
+2\sum_{p,q\neq0}(W_p+W_{p+q})\eta_p a_q^*a_qa_0^*a_0\\
&+H_{01}+H_{02}+H_{22}
+H_{21}+H_4+H_{23}^{\prime}+H_3^\prime+\tilde{\mathcal{E}}^{B^\prime},
\end{align*}
and the error term is bounded by
\begin{align*}
\pm\tilde{\mathcal{E}}^{B^\prime}
&\leq C(N^{\frac{3}{2}}a^{\frac{3}{2}}d^{-\frac{1}{2}}\textit{l}^{\frac{1}{2}}
+N^{\frac{5}{2}}a^{\frac{5}{2}}d^{-\frac{3}{2}}\textit{l}^{\frac{1}{2}}
+N^{\frac{1}{2}}a^{\frac{3}{2}}d^{\frac{1}{2}}\textit{l}^{2}
+N^{2}a^{\frac{5}{2}}d^{-1}\textit{l}^{\frac{1}{3}})
(\mathcal{N}_++1)\\
&+C(Na\textit{l}^{\frac{1}{2}}
+N^{\frac{3}{2}}a^{\frac{7}{6}}d^{-\frac{1}{2}}\textit{l}^{\frac{1}{3}}
+N^{2}a^{\frac{5}{3}}d^{-1}\textit{l}^{\frac{1}{3}})
(H_{21}+Nad^{-1}(\mathcal{N}_++1)^2)\\
&+C(Na\textit{l}^{\frac{1}{2}}
+N^{\frac{3}{2}}a^{\frac{7}{6}}d^{-\frac{1}{2}}\textit{l}^{\frac{1}{3}})
H_4
+CN^{\frac{3}{2}}a^{\frac{3}{2}}d^{-\frac{1}{2}}\textit{l}^{\frac{1}{2}}
(H_4+H_4^\prime).
\end{align*}
Then we reach (\ref{second renorm III}) and (\ref{second renorm E^B' III}).

\begin{flushright}
  {$\Box$}
\end{flushright}







\section{Bogoliubov Transformation for Region I}\label{4}
\par  In this section we analyze the diagonalized Hamiltonian $\mathcal{Z}^{I}_N$ and prove Propositon \ref{Bog renorm}. Once we are done, Theorem \ref{core} for Regions I and $\mathrm{II}_{\mathrm{I}}$ is ready (see Section \ref{Proof of the Main Theorem for Region I}). We adopt the notation
\begin{equation*}
  A^{\prime\prime}={\sum_{p\neq 0}}\tau_pb_p^*b_{-p}^*.
\end{equation*}
Using (\ref{define b_p,b_p^* fock}) it is easy to check $ A^{\prime\prime}$ is a linear operator on $L^2_s(\Lambda_{d}^N)$ bounded by $N\Vert\tau\Vert_2$ for all $N\in\mathbb{N}$. Here we let $\tau\in L^2_\perp(\Lambda_d)$ be the function with Fourier coefficients $\tau_p$ for $p\neq0$ i.e.
\begin{equation*}
  \tau={\sum_{p\neq0}}\tau_p\phi_p^{(d)}.
\end{equation*}
  We prove in Lemma \ref{lemma tau_p} that $\tau$ is actually a $L^2$ function. By (\ref{define B''}) we have
\begin{equation*}
   B^{\prime\prime}=\frac{1}{2}( A^{\prime\prime}- A^{{\prime\prime}^*}).
\end{equation*}
Throughout this section we assume that $N$ tends to infinity, $a$, $d$, $\frac{a}{d}$ and $Na\textit{l}^{\frac{1}{2}}$ tend to $0$ and $\frac{d}{a}>\frac{C}{\textit{l}}$. We also want to remind the readers that in Region I we take $\kappa=\nu d^{-1}$ for some $\nu\geq 1$ with the further assumption that $Na^3\kappa^3\textit{l}$ and $N^{\frac{1}{2}}a^{\frac{1}{2}}d^{-\frac{1}{2}}\kappa^{-1}$ tend to 0. To control the action of $e^{B^{\prime\prime}}$ we must first analyze $\{\tau_p\}$, we collect the results in Lemma \ref{lemma tau_p}.
\begin{lemma}\label{lemma tau_p}
Let $F_p$ and $G_p$ be defined in (\ref{define F_p,G_p}), $\tau_p$ be defined in (\ref{define tau_p}). Then we have the followings
\begin{enumerate}[$(1)$]
\item
\begin{equation}\label{lemma tau_p 1}
\vert G_p-8\pi\mathfrak{a}_0Nad^{-1}\vert\leq CNa\textit{l}\vert\mathcal{M}_dp\vert+CNa^2d^{-2}\textit{l}^{-1}.
\end{equation}
Assume further that $Na^2d^{-2}\textit{l}^{-1}$ tends to $0$, we have
\begin{equation}\label{G_p<F_p}
  \vert G_p\vert+c\vert\mathcal{M}_dp\vert^2\leq F_p
\end{equation}
for any fixed $c\in(0,1)$, which implies $\tau_p$ is well-defined from its definition (\ref{define tau_p}). Moreover
\begin{equation}\label{lemma tau_p 2}
\vert G_p\vert\leq CNad^{-1},\quad
\vert G_p\vert\leq\frac{CNad^{-1}}{(d\textit{l})^2\vert\mathcal{M}_dp\vert^2}.
\end{equation}
\item Under the assumption that $Na^2d^{-2}\textit{l}^{-1}$ tends to $0$, we have
\begin{equation}\label{lemma tau_p 3}
  \vert \tau_p\vert\leq C\frac{\vert G_p\vert}{\vert\mathcal{M}_dp\vert^2}.
\end{equation}
Moreover
\begin{equation}\label{lemma tau_p 4}
  \Vert\tau\Vert_2\leq CNad^{-1},
\quad\sum_{p\neq0}\vert\mathcal{M}_dp\vert^2\tau_p^2\leq
CNad^{-1}\sum_{p\neq0}\vert\tau_p\vert\leq CN^2a^2d^{-2}
(\textit{l}^{-1}+\ln(d\textit{l})^{-1}).
\end{equation}
\end{enumerate}
\end{lemma}
\noindent
\emph{Proof.} We can recall from equation (\ref{define W_p}) that $W_p$ is defined as
\begin{equation}\label{rewrite W_p}
  W_p=\frac{\lambda_{\textit{l}}}{a^2d}\widehat{\chi_{d\textit{l}}}\left(
\frac{\mathcal{M}_dp}{2\pi}\right)
+\frac{\lambda_{\textit{l}}}{a^2}\eta_p
\end{equation}
for all $p\in2\pi\mathbb{Z}^3$. With (\ref{define W(x)}) and (\ref{L1&L2 norm of W 3dscatt}) we can bound
\begin{equation}\label{傻}
  \vert W_p-W_0\vert\leq\frac{1}{\sqrt{d}}\int_{\vert x\vert\leq d\textit{l}}
\vert W(x)\vert\cdot\vert e^{-ip^T\mathcal{M}_dx}-1\vert dx\leq Ca\textit{l}\vert
\mathcal{M}_dp\vert.
\end{equation}
On the other hand, combining (\ref{est of lambda_l}) and (\ref{est of eta_0}) we get
\begin{equation}\label{逼}
  \vert W_0-4\pi\mathfrak{a}_0ad^{-1}\vert\leq Ca^2d^{-2}\textit{l}^{-1}.
\end{equation}
Since $G_p=2NW_p$ we reach (\ref{lemma tau_p 1}) by combining (\ref{傻}) and (\ref{逼}). Inequality (\ref{lemma tau_p 1}), the assumption that $Na^2d^{-2}\textit{l}^{-1}$ tends to $0$, and the fact that $\vert\mathcal{M}_dp\vert\geq1$ for all $p\in2\pi\mathbb{Z}^3\backslash\{0\}$ together yield (\ref{G_p<F_p}).

(\ref{lemma tau_p 2}) comes directly from (\ref{sum_pW_peta_p 3dscatt}) and (\ref{moron2}). (\ref{lemma tau_p 3}) is a direct consequence of (\ref{define tau_p}) and (\ref{G_p<F_p}) since
\begin{align*}
\vert\tau_p\vert=\frac{1}{4}\ln\left(1+\frac{2\vert G_p\vert}{F_p-\vert G_p\vert}\right)
\leq \frac{1}{2}\frac{\vert G_p\vert}{F_p-\vert G_p\vert}.
\end{align*}
Using the fact that (see (\ref{WTH2}) taking $\epsilon=d$)
\begin{equation*}
  \sum_{p\neq0}\frac{1}{\vert\mathcal{M}_dp\vert^4}\leq C,
\end{equation*}
together with (\ref{lemma tau_p 2}) and (\ref{lemma tau_p 3}) to bound
\begin{equation*}
  \Vert\tau\Vert_2\leq\left(\sum_{p\neq0}
\frac{\vert G_p\vert^2}{\vert\mathcal{M}_dp\vert^4}\right)^{\frac{1}{2}}
 \leq CNad^{-1}.
\end{equation*}
On the other hand, combining (\ref{lemma tau_p 3}) and (\ref{sum W_p M_dp^-2}) we get
\begin{align*}
\sum_{p\neq0}\vert\mathcal{M}_dp\vert^2\tau_p^2\leq
CNad^{-1}\sum_{p\neq0}\vert\tau_p\vert\leq CN^2a^2d^{-2}
(\textit{l}^{-1}+\ln(d\textit{l})^{-1}).
\end{align*}
This concludes the proof of (\ref{lemma tau_p 4}).
\begin{flushright}
  {$\Box$}
\end{flushright}

\par With Lemma \ref{lemma tau_p} we can bound the action of $e^{B^{\prime\prime}}$ on each of $\mathcal{N}_+$, $H_{21}$ and $H_4$ in the next lemma.
\begin{lemma}\label{control of S_+ with e^B''}
Apply G-P condition for Region I, i.e. we assume $Nad^{-1}=1$. For all $n\in\frac{1}{2}\mathbb{N}$ and $\vert t\vert\leq 1$ we have
\begin{align}
   e^{-tB^{\prime\prime}}(\mathcal{N}_++1)^ne^{tB^{\prime\prime}}
&\leq C_n(\mathcal{N}_++1)^{n},\label{control N_+ with e^B''}\\
e^{-tB^{\prime\prime}}H_{21}e^{tB^{\prime\prime}}
&\leq C(H_{21}+\textit{l}^{-1}+\ln(d\textit{l})^{-1}),\label{control H_21 with e^B''}\\
e^{-tB^{\prime\prime}}H_{4}e^{tB^{\prime\prime}}
&\leq C(H_4+ad^{-1}(\mathcal{N}_++1)^2
+ad^{-1}(\textit{l}^{-1}+\ln(d\textit{l})^{-1})^2).\label{control H_4 with e^B'' 1}
  \end{align}
\end{lemma}
\noindent
\emph{Proof.}
Using (\ref{b_p^*b_q fock}), we find that
\begin{equation}\label{revised conjugate relation}
  \frac{1}{N}a_p^*a_{-p}^*a_0a_0=b_p^*b_{-p}.
\end{equation}
We follow the proof in Lemma \ref{control of S_+ conj with e^B}, (\ref{control N_+ with e^B''}) can be proved by noticing the fact that $[\mathcal{N}_+,A^{\prime\prime}]=2A^{\prime\prime}$ and using (\ref{lemma tau_p 4}) and (\ref{revised conjugate relation}) to bound
\begin{equation*}
  \pm A^{\prime\prime}\leq C\Vert\tau\Vert_2(\mathcal{N}_++1)\leq C(\mathcal{N}_++1).
\end{equation*}
\par To prove (\ref{control H_21 with e^B''}), we calculate the commutator
\begin{equation*}
  [H_{21},B^{\prime\prime}]=\sum_{p\neq0}\vert\mathcal{M}_dp\vert^2\tau_p(b_p^*b_{-p}^*+h.c.).
\end{equation*}
Using Cauchy-Schwartz and (\ref{b_p^*b_q fock}), we have for all $\psi\in L^2_s(\Lambda_d^N)$
\begin{align*}
  &\left\vert\sum_{p\neq0}\vert\mathcal{M}_dp\vert^2\tau_p
\langle b_p^*b_{-p}^*\psi,\psi\rangle\right\vert
\leq \left\vert\sum_{p\neq0}\vert\mathcal{M}_dp\vert^2\tau_p
 \Vert b_{-p}^*\psi\Vert\cdot\Vert b_p\psi\Vert\right\vert\\
&\leq\left(\sum_{p\neq0}\vert\mathcal{M}_dp\vert^2
\langle b_p^*b_p\psi,\psi\rangle\right)^{\frac{1}{2}}
\left(\sum_{p\neq0}\vert\mathcal{M}_dp\vert^2\tau_p^2
\langle (b_p^*b_p+1)\psi,\psi\rangle\right)^{\frac{1}{2}}\\
&\leq \langle H_{21}\psi,\psi\rangle^{\frac{1}{2}}
\left(\sup\vert\tau_p\vert^2\langle H_{21}\psi,\psi\rangle
+\sum_{p\neq0}\tau_p^2\vert\mathcal{M}_dp\vert^2\Vert\psi\Vert^2\right)^{\frac{1}{2}}.
\end{align*}
Here (\ref{b_p^*b_q fock}) implies $b_p^*b_p\leq 2a_p^*a_p$. With Lemma \ref{lemma tau_p} we get
\begin{equation*}
  \pm[H_{21},B^{\prime\prime}]\leq C(H_{21}+\textit{l}^{-1}+\ln(d\textit{l})^{-1}),
\end{equation*}
which implies (\ref{control H_21 with e^B''}) using (\ref{control N_+ with e^B''}) and Gronwall's inequality.

\par Similarly we compute
\begin{align*}
 [H_4,B^{\prime\prime}]=&\frac{1}{2\sqrt{d}}\sum_{p,q\neq0}
v_{p-q}^{(a,d)}\tau_p(b_p^*b_{-p}^*+h.c.)\\
&+\frac{1}{N\sqrt{d}}\sum_{p,q,p+r,q+r\neq0}v_r^{(a,d)}\tau_p
(a_{p+r}^*a_q^*a_{-p}^*a_{q+r}a_0a_0+h.c.)\eqqcolon \Psi_1+\Psi_2
\end{align*}
The first term can be bounded using (\ref{lemma tau_p 4}), (\ref{revised conjugate relation}) and (\ref{H_4psi,psi fock}) by
\begin{align*}
\vert\langle\Psi_1\psi,\psi\rangle\vert
&\leq C\Vert v_a\Vert_1^{\frac{1}{2}}\Vert\tau\Vert_{\infty}
\langle H_4\psi,\psi\rangle^{\frac{1}{2}}
\langle \psi,\psi\rangle^{\frac{1}{2}}\\
&\leq \langle H_4\psi,\psi\rangle
+Cad^{-1}(\textit{l}^{-1}+\ln(d\textit{l})^{-1})^2
\langle \psi,\psi\rangle.
\end{align*}
On the other hand, with (\ref{lemma tau_p 4}) we can bound the second term in the way that we bound $\mathcal{E}^B_{\Gamma_1,1}$ in Lemma \ref{lemma Gamma},
\begin{align*}
\vert\langle\Psi_2\psi,\psi\rangle\vert
&\leq Cd^{-\frac{1}{2}}\Vert v_a\Vert_1^{\frac{1}{2}}
\Vert\tau\Vert_{2}
\langle H_4\psi,\psi\rangle^{\frac{1}{2}}
\langle (\mathcal{N}_++1)^2\psi,\psi\rangle^{\frac{1}{2}}\\
&\leq \langle H_4\psi,\psi\rangle
+Cad^{-1}\langle (\mathcal{N}_++1)^2\psi,\psi\rangle.
\end{align*}
 We use (\ref{control N_+ with e^B''}) and Gronwall's inequality once again to reach (\ref{control H_4 with e^B'' 1}). 
\begin{flushright}
  {$\Box$}
\end{flushright}
\noindent
\emph{Proof of Proposition \ref{Bog renorm}.}
\par In the following, we apply the G-P condition $Nad^{-1}=1$ for Region I. Recall that we demand $N^{-1}$, $a$, $d$, $\frac{a}{d}$ and $Na\textit{l}^{\frac{1}{2}}$ tend to $0$ and $\frac{d}{a}>\frac{C}{\textit{l}}$. Moreover, we also assume that $Na^3\kappa^3\textit{l}$ and $N^{\frac{1}{2}}a^{\frac{1}{2}}d^{-\frac{1}{2}}\kappa^{-1}$ tend to 0 with $\kappa=\nu d^{-1}$ for some $\nu\geq 1$. Notice that under the G-P condition, we can simplify these assumptions to $N^{-1}$, $a$, $d$ and $N^{-2}\nu^3\textit{l}$ tend to $0$ and $N>C\textit{l}^{-1}$.
\par Using (\ref{second renorm}) and (\ref{rewrite quadratic}), we rewrite
\begin{equation*}
  e^{-B^{\prime\prime}}\mathcal{J}_Ne^{B^{\prime\prime}}
=C^B+e^{-B^{\prime\prime}}\mathcal{Q}^\prime e^{B^{\prime\prime}}
+e^{-B^{\prime\prime}}(H_{4}+\mathcal{E}^{B^\prime}
+\mathcal{E}^{B^{\prime\prime}}_{res})e^{B^{\prime\prime}}.
\end{equation*}
The error $e^{-B^{\prime\prime}}(\mathcal{E}^{B^{\prime\prime}}_{res}+\mathcal{E}^{B^\prime})
e^{B^{\prime\prime}}$ part can be estimated using Lemma \ref{control of S_+ with e^B''}.
To make use of relation (\ref{action of Bog trans}) we let
\begin{equation}\label{split Fa_p^*a_p}
  \sum_{p\neq0}F_pa_p^*a_p=\sum_{0<\vert\mathcal{M}_dp\vert\leq M_1}F_pa_p^*a_p
  +\sum_{\vert\mathcal{M}_dp\vert>M_1}F_pa_p^*a_p
\end{equation}
for some $M_1\geq1$ to be determined. Then using (\ref{b_p^*b_q fock}), we can check that
\begin{equation}\label{dsf}
  \sum_{0<\vert\mathcal{M}_dp\vert\leq M_1}F_pa_p^*a_p=
  \sum_{0<\vert\mathcal{M}_dp\vert\leq M_1}F_pb_p^*b_p+\mathcal{E}_{F,1},
\end{equation}
where
\begin{equation}\label{mathcal E_F}
  \pm\mathcal{E}_{F,1}\leq CM_1^2N^{-1}(\mathcal{N}_++1)^2.
\end{equation}
Using (\ref{action of Bog trans fock}), and we let $\gamma_p=\cosh\tau_p,\nu_p=\sinh\tau_p$, we have
\begin{equation}\label{bog F low}
\begin{aligned}
  \sum_{0<\vert\mathcal{M}_dp\vert\leq M_1}F_pe^{-B^{\prime\prime}}b_p^*b_p
  e^{B^{\prime\prime}}=&\sum_{0<\vert\mathcal{M}_dp\vert\leq M_1}
  F_p(\gamma_p^2+\nu_p^2)b_p^*b_p+F_p\gamma_p\nu_p(b_p^*b_{-p}^*+h.c.)\\
  &+\sum_{0<\vert\mathcal{M}_dp\vert\leq M_1}F_p\nu_p^2+\mathcal{E}_{F,2}
\end{aligned}
\end{equation}
where
\begin{equation}\label{mathcal E_F2}
  \pm\mathcal{E}_{F,2}\leq CM_1^2N^{-1}(\textit{l}^{-1}+
  \ln(d\textit{l})^{-1})(\mathcal{N}_++1)^2.
\end{equation}
For a detailed calculation of (\ref{mathcal E_F2}), one can refer the proof of \cite[Lemma 5.3]{2018Bogoliubov}, and we use the fact that $\Vert\tau\Vert_2\leq C$. Also, similar to (\ref{dsf}), since $\vert\tau_p\vert\leq C$, we have
\begin{equation}\label{dsf2}
  \sum_{0<\vert\mathcal{M}_dp\vert\leq M_1}F_p(\gamma_p^2+\nu_p^2)b_p^*b_p=
  \sum_{0<\vert\mathcal{M}_dp\vert\leq M_1}F_p(\gamma_p^2+\nu_p^2)
  a_p^*a_p+\mathcal{E}_{F,3},
\end{equation}
where
\begin{equation}\label{mathcal E_F3}
  \pm\mathcal{E}_{F,3}\leq CM_1^2N^{-1}(\mathcal{N}_++1)^2.
\end{equation}
\par On the other hand, we have
\begin{equation}\label{cuyigvur}
  \begin{aligned}
  \sum_{\vert\mathcal{M}_dp\vert> M_1}F_pe^{-B^{\prime\prime}}a_p^*a_p
  e^{B^{\prime\prime}}=&\sum_{\vert\mathcal{M}_dp\vert> M_1}F_pa_p^*a_p\\
  &+\sum_{\vert\mathcal{M}_dp\vert> M_1}F_p\int_{0}^{1}e^{-tB^{\prime\prime}}
  \tau_p(b_p^*b_{-p}^*+h.c.)
  e^{tB^{\prime\prime}}dt
\end{aligned}
\end{equation}
while using (\ref{action of Bog trans fock}), and we let $\gamma_p(t)=\cosh (t\tau_p),\nu_p=\sinh(t\tau_p)$ we have
\begin{equation}\label{duyfhgreig}
  \begin{aligned}
  &\sum_{\vert\mathcal{M}_dp\vert> M_1}F_p\int_{0}^{1}e^{-tB^{\prime\prime}}
  \tau_p(b_p^*b_{-p}^*+h.c.)
  e^{tB^{\prime\prime}}dt\\
  =&\sum_{\vert\mathcal{M}_dp\vert> M_1}F_p\tau_p\int_{0}^{1}
  (\gamma_p^2(t)+\nu_p^2(t))(b_p^*b_{-p}^*+h.c.)+2\gamma_p(t)\nu_p(t)
  (2b_p^*b_p+1)dt\\
  &+\mathcal{E}_{F,4},
  \end{aligned}
\end{equation}
where
\begin{equation}\label{mathcal EF4}
\begin{aligned}
  \pm\mathcal{E}_{F,4}\leq& CN^{-1}(\textit{l}^{-1}+
  \ln(d\textit{l})^{-1})(\mathcal{N}_++1)^2\\
  &+CN^{-\frac{1}{2}}(\textit{l}^{-1}+\ln(d\textit{l})^{-1})^3(H_{21}+1)
  \end{aligned}
\end{equation}
For a detailed calculation of (\ref{mathcal EF4}), one can refer the proof of \cite[Lemma 5.3]{2018Bogoliubov}, with slight modification. We use the fact that here $\vert F_p\tau_p\vert\leq C\vert G_p\vert$, and estimates (\ref{sum W_p M_dp^-2}) and (\ref{lemma tau_p 4}). Calculating the integral on the right-hand side of (\ref{duyfhgreig}), also notice the fact that $\vert F_p\tau_p\vert\leq C$ and $\vert\tau_p-\gamma_p\nu_p\vert\leq C\tau_p^3$, we have
\begin{equation}\label{seguyfseruikdg}
  \begin{aligned}
  &\sum_{\vert\mathcal{M}_dp\vert> M_1}F_p\tau_p\int_{0}^{1}
  (\gamma_p^2(t)+\nu_p^2(t))(b_p^*b_{-p}^*+h.c.)+2\gamma_p(t)\nu_p(t)
  (2b_p^*b_p+1)dt\\
  &+\sum_{\vert\mathcal{M}_dp\vert> M_1}F_pa_p^*a_p\\
  =&\sum_{\vert\mathcal{M}_dp\vert>M_1}
  F_p(\gamma_p^2+\nu_p^2)a_p^*a_p+F_p\gamma_p\nu_p(b_p^*b_{-p}^*+h.c.)
  +\sum_{\vert\mathcal{M}_dp\vert> M_1}F_p\nu_p^2+\mathcal{E}_{F,5}
  \end{aligned}
\end{equation}
where
\begin{equation}\label{mathcal EF5}
  \pm\mathcal{E}_{F,5}\leq CM_1^{-6}(\mathcal{N}_++1)+CN^{-1}(\mathcal{N}_++1)^2
  +CM_1^{-4}.
\end{equation}
Similar to the calculation from (\ref{duyfhgreig}) to (\ref{mathcal EF5}), we also have
\begin{equation}\label{dsugyedrggfh}
  \begin{aligned}
  \frac{1}{2}\sum_{p\neq0}G_pe^{-B^{\prime\prime}}(b_p^*b_{-p}^*+h.c.)
  e^{B^{\prime\prime}}=&\sum_{p\neq0}
  2G_p\gamma_p\nu_pa_p^*a_p+\frac{1}{2}G_p(\gamma_p^2+\nu_p^2)(b_p^*b_{-p}^*+h.c.)\\
  &+\sum_{p\neq0}G_p\gamma_p\nu_p+\mathcal{E}_{G}
  \end{aligned}
\end{equation}
where
\begin{equation}\label{mathcal G}
  \begin{aligned}
  \pm\mathcal{E}_{G}\leq& CN^{-1}(\textit{l}^{-1}+
  \ln(d\textit{l})^{-1})(\mathcal{N}_++1)^2\\
  &+CN^{-\frac{1}{2}}(\textit{l}^{-1}+\ln(d\textit{l})^{-1})^3(H_{21}+1).
  \end{aligned}
\end{equation}

\par Collecting all the calculations above, and choosing $M_1^2=d^{-\alpha}$, we have
\begin{equation}\label{bogdsyufgy}
\begin{aligned}
 e^{-B^{\prime\prime}}{\mathcal{Q}}^\prime e^{B^{\prime\prime}}
=&\sum_{p\neq0}\left(\nu_p^2F_p+\gamma_p\nu_pG_p\right)
+\sum_{p\neq0}\left((\gamma_p^2+\nu_p^2)F_p+2\gamma_p\nu_pG_p\right)b_p^*b_p\\
&+\sum_{p\neq0}\left(\gamma_p\nu_pF_p+\frac{1}{2}G_p(\gamma_p^2+\nu_p^2)\right)
(b_p^*b_{-p}^*+h.c.)+\mathcal{E}_{\mathcal{Q}^\prime}\\
=&\frac{1}{2}\sum_{p\neq0}\left(-F_p+\sqrt{F_p^2-G_p^2}\right)+
\sum_{p\neq0}\sqrt{F_p^2-G_p^2}a_p^*a_p\mathcal{E}_{\mathcal{Q}^\prime}.
\end{aligned}
\end{equation}
with
\begin{equation}\label{mathcal E Q'}
  \begin{aligned}
  \pm\mathcal{E}_{\mathcal{Q}^\prime}\leq& Cd^{2\alpha}(\mathcal{N}_++1)+
  CN^{-1}(d^{-\alpha}+\textit{l}^{-1}+
  \ln(d\textit{l})^{-1})(\mathcal{N}_++1)^2\\
  &+CN^{-\frac{1}{2}}(\textit{l}^{-1}+\ln(d\textit{l})^{-1})^3(H_{21}+1).
  \end{aligned}
\end{equation}
 Then using (\ref{second renorm E^B'}), (\ref{E^B''_res}), (\ref{mathcal E Q'}) and Lemma \ref{control of S_+ with e^B''}, we can bound
\begin{align*}
\pm\mathcal{E}^{B^{\prime\prime}}
&\leq C\Big\{d^{\alpha}+d\big(\textit{l}^{\frac{1}{2}}+\nu^{-1}\big)
+N^{-1}\textit{l}^{-1}+(N^{-2}\nu^3\textit{l})^\gamma\nonumber\\
&\quad\quad\quad+N^{-1}\nu\big(1+N^{-1}\ln a^{-1}\big)\Big\}(\mathcal{N}_++1)
+CN^{-\frac{1}{2}}\textit{l}^{-1}(\mathcal{N}_++1)^{\frac{3}{2}}\nonumber\\
&+C\Big\{N^{-1}\textit{l}^{-1}
+(N^{-2}\nu^3\textit{l})^{\frac{1}{2}}+N(N^{-2}\nu^3\textit{l})^{1-\gamma}
+N^{-1}\big(1+N^{-1}\ln a^{-1}\big)\nonumber\\
&\quad\quad\quad+\big(N^{-1}d^{-\alpha}+N^{-2+\beta}\big)
\big(\textit{l}^{-1}+\ln(d\textit{l})^{-1}\big)\Big\}(\mathcal{N}_++1)^2\nonumber\\
&+CN^{-2+\beta}(\mathcal{N}_++1)^3
+C\Big(d^{\alpha}+d\big(\textit{l}^{\frac{1}{2}}+\nu^{-1}\big)+N^{-\beta}\Big)H_{21}\nonumber\\
&+CN^{-\frac{1}{2}}(\textit{l}^{-1}+\ln(d\textit{l})^{-1})^3(H_{21}+1)\nonumber\\
&+C\Big(d^{\alpha}+d\big(\textit{l}^{\frac{1}{2}}+\nu^{-1}\big)
+(N^{-2}\nu^3\textit{l})^\gamma\Big)
e^{-B^{\prime\prime}}H_4e^{B^{\prime\prime}}\nonumber\\
&+C\Big(d^{\alpha}+d\big(\textit{l}^{\frac{1}{2}}+\nu^{-1}\big)
+N^{-\beta}\Big)
\big(\textit{l}^{-1}+\ln(d\textit{l})^{-1}\big)
\end{align*}
for some $\alpha,\beta>0$ and $0<\gamma<1$. Hence we have finished the proof of Proposition \ref{Bog renorm}, and Theorem \ref{core} for Regions I and $\mathrm{II}_{\mathrm{I}}$ follows as in Section \ref{Proof of the Main Theorem for Region I}.
\begin{flushright}
  {$\Box$}
\end{flushright}



\section{Quasi-2D Renormalization for Region III}
\label{5}
\par In this section, we compute the quasi-2D renormalization and prove Proposition \ref{quasi-2d/3d quadratic renorm}. We analyze the excitation Hamiltonian $\mathcal{L}_N$ generated by the quadratic quasi-2D renormalization in Section \ref{quasi-2d quadratic}, and we analyze $\mathcal{M}_N$ generated by the cubic quasi-2D renormalization in Section \ref{quasi-2d cubic}. In the grand scheme, the analysis of $\mathcal{L}_N$ and $\mathcal{M}_N$ are similar to those of $\mathcal{G}_N$ and $\mathcal{J}_N$: for the most of the time, we only need to replace the estimates of $\eta$ with the estimates of $\xi$ instead, and substitute the potential $v_a$ with $2\sqrt{d}W$. But there are some technical changes in detail. The point of this section is to extract the energy contribution of the quasi-2D correlation structure and get ready for the dimensional coupling renormalization in Section \ref{6}.

\subsection{Quasi-2D Quadratic Renormalization}\label{quasi-2d quadratic}
\
\par We adopt the notation
\begin{equation*}
  \tilde{A}={\sum_{p\neq 0}}\xi_pa_p^*a_{-p}^*a_0a_0,
\end{equation*}
and also recall that
\begin{equation*}
  \xi_\perp={\sum_{p\neq0}}\xi_p\phi_p^{(d)}\in L^2_{\perp}(\Lambda_{d}).
\end{equation*}
\par As in Section \ref{2}, we rewrite $e^{-\tilde{B}}\mathcal{J}_Ne^{\tilde{B}}$ using (\ref{rearrange J_N})
\begin{align}
  e^{-\tilde{B}}\mathcal{J}_Ne^{\tilde{B}}
=&H_{01}^\prime+H_{21}+H_4+H_{23}^{\prime\prime}
+e^{-\tilde{B}}(H_{02}^\prime+H_{22}^\prime+H_3^\prime)e^{\tilde{B}}
+\int_{0}^{1}e^{-t\tilde{B}}\Theta e^{t\tilde{B}}dt\nonumber\\
&+\int_{0}^{1}\int_{t}^{1}
e^{-s\tilde{B}}[H_{23}^\prime,\tilde{B}] e^{s\tilde{B}}dsdt
+\int_{0}^{1}\int_{0}^{t}
e^{-s\tilde{B}}[H_{23}^{\prime\prime},\tilde{B}] e^{s\tilde{B}}dsdt\nonumber\\
&+e^{-\tilde{B}}\tilde{\mathcal{E}}^{B^\prime}e^{\tilde{B}}+O(N^2a^2\textit{l}^2).
\label{split e^-B tildeJ_Ne^B tilde}
\end{align}
Here we define
\begin{equation}\label{define Theta}
  \Theta=[H_{21}+H_4,\tilde{B}]+H_{23}^\prime-H_{23}^{\prime\prime}
\end{equation}
with $H_{23}^{\prime\prime}$ defined as
\begin{equation}\label{define H_23''}
  H_{23}^{\prime\prime}=\sum_{p\neq0}\widetilde{W}_p(a_p^*a_{-p}^*a_0a_0+h.c.)
\end{equation}
where
\begin{equation}\label{define W tilde}
  \widetilde{W}_p=\left\{\begin{aligned}
&W_p+\frac{1}{2\sqrt{d}}\sum_{q\neq0}v_{p-q}^{(a,d)}\xi_q,\quad p_3\neq0\\
&\frac{\mu_h}{(d\textit{l})^2}\left(\xi_p+\widehat{\chi^{\mathrm{2D}}_{h}}\left(
  \frac{\bar{p}}{2\pi}\right)\right)+\sum_{q\neq0}
\frac{1}{2\sqrt{d}}v_{p-q}^{(a,d)}\xi_q-\sum_{q}W_{p-q}\xi_q,\quad p_3=0
\end{aligned}\right.
\end{equation}
Here $\bar{p}=(p_1,p_2)\in2\pi\mathbb{Z}^2$ and $p=(\bar{p},p_3)\in2\pi\mathbb{Z}^3$. Since the analysis of (\ref{split e^-B tildeJ_Ne^B tilde}) is rather similar to Section \ref{2}, we state them in the up-coming series of lemmas, while we omit the details of proofs except some new estimates which we come upon.

\par In the following lemmas, we bound $e^{-\tilde{B}}\tilde{\mathcal{E}}^{B^\prime}e^{\tilde{B}}$ in Corollary \ref{lemma e^-B tildeEe^B tilde}, $e^{-\tilde{B}}H_{02}^\prime e^{\tilde{B}}$ in Lemma \ref{lemma e^-B tildeH_02'e^B tilde}, $e^{-\tilde{B}}H_{22}^\prime e^{\tilde{B}}$ in Lemma \ref{lemma e^-B tildeH_22'e^B tilde}, and $e^{-\tilde{B}}H_{3}^\prime e^{\tilde{B}}$ in Lemma \ref{lemma e^-B tildeH_3'e^B tilde}. These four terms stay unchanged up to small errors after conjugating with $e^{\tilde{B}}$, or can be rewritten in the form of polynomials of $\mathcal{N}_+$. The term containing the difference $\Theta$ is bounded in Lemma \ref{lemma Theta}, and is proved to be a negligible error term. The contribution of the commutator $[H_{23}^\prime,\tilde{B}]$ is calculated in Lemma \ref{lemma [H_23',B tilde]}, and the contribution of $[H_{23}^{\prime\prime},\tilde{B}]$ is calculated in Lemma \ref{lemma [H_23'',B tilde]}. Lemmas \ref{lemma [H_23',B tilde]} and \ref{lemma [H_23'',B tilde]} present the major contributions of the quadratic quasi-2D correlation structure to the first and second order ground state energy, in the form of polynomials of $\mathcal{N}_+$. The bounds on the growths of $\mathcal{N}_+$, $H_{21}$, $H_4$ and $H_4^\prime$ are useful tools in our analysis, we state these results in Lemmas \ref{control of S_+ conj with e^B tilde} and \ref{lemma control of H_4 with e^B tilde}.
\begin{lemma}\label{control of S_+ conj with e^B tilde}
Let $\mathcal{N_+}$ be defined on $L_s^2(\Lambda_d^N)$ as stated in (\ref{define of N_+}), then there exist a constant $C_n$ depending only on $n\in\frac{1}{2}\mathbb{N}$ such that: for every $t\in\mathbb{R}$, $N\in\mathbb{N}$, $n\in\frac{1}{2}\mathbb{N}$, $\textit{l}\in(0,\frac{1}{2})$ such that $\frac{d\textit{l}}{a}>C$, and $h\in(0,\frac{1}{2})$ such that $\frac{h}{d\textit{l}}>C$ for some universal constant $C$, and we have
\begin{align}
  e^{-t\tilde{B}}(\mathcal{N}_++1)^ne^{t\tilde{B}}&\leq e^{C_nN(d\textit{l}+\frac{h}{m})\vert t\vert}(\mathcal{N}_++1)^n,\label{e^-tB tilde(N_++1)e^tB tilde}\\
  \pm(e^{-t\tilde{B}}(\mathcal{N}_++1)^ne^{t\tilde{B}}-(\mathcal{N}_++1)^n)&\leq (e^{C_nN(d\textit{l}+\frac{h}{m})\vert t\vert}-1) (\mathcal{N}_++1)^n.\label{e^-tB tilde N_+e^tB tilde-N_+}
\end{align}
\end{lemma}
\
\noindent
\emph{Proof.}
  The proof however, is the very same as Lemma \ref{control of S_+ conj with e^B} except this time we use (\ref{est of xi}) instead of (\ref{est of eta and eta_perp}) to bound the $L^2$ norm of coefficients.
  \begin{flushright}
  {$\Box$}
\end{flushright}

  \par Apart from the action on $\mathcal{N}_+$, the actions on $H_{21}$, $H_4$ and $H_4^\prime$ are also needed. We will state the result in the next lemma. In the following analysis throughout the whole section, we are working under the setting of Proposition \ref{cubic renorm} for Region III, that is we set without further specification that $N$ tends to infinity, $a$, $d$, $\frac{a}{d}$, $N^{\frac{3}{2}}a^{\frac{7}{6}}d^{-\frac{1}{2}}\textit{l}^{\frac{1}{3}}$ and
$Na\textit{l}^{\frac{1}{2}}$ tend to $0$ and $\frac{d\textit{l}}{a}>C$. Moreover, we ask additionally $\frac{h}{d\textit{l}}>C$, $\frac{Na}{d}>C$, $\frac{ma}{d}>C$ and $N(d\textit{l}+\frac{h}{m})$ tends to $0$.

\begin{lemma}\label{lemma control of H_4 with e^B tilde}
\begin{align}
e^{-t\tilde{B}}H_{21}e^{t\tilde{B}}&\leq C(H_{21}+N^2m^{-2}\ln(1+h(d\textit{l})^{-1})(\mathcal{N}_++1)),\label{control of e^-B tildeH_21e^B tilde}\\
  e^{-t\tilde{B}}H_4e^{t\tilde{B}}&\leq C(H_4+N^2ad^{-1}),\label{control of e^-B tildeH_4e^B tilde}\\
 e^{-t\tilde{B}}H_4^\prime e^{t\tilde{B}}&\leq C(H_4^\prime+N^2ad^{-1}).\label{control of e^-B tildeH_4'e^B tilde}
\end{align}
for all $\vert t\vert\leq1$.
\end{lemma}
\noindent
\emph{Proof.} For the proof of (\ref{control of e^-B tildeH_4e^B tilde}), one can consult Lemma \ref{lemma control of H_4}, and we will make use of the bound (\ref{est of xi}) and (\ref{est of xi L^infty}). Then (\ref{control of e^-B tildeH_4'e^B tilde}) follows as long as we replace $v_a$ with $2\sqrt{d}W$ in our calculations.
\par For the proof of (\ref{control of e^-B tildeH_21e^B tilde}), we first calculate
\begin{equation*}
  [H_{21},\tilde{B}]={\sum_{p\neq0}}\vert\mathcal{M}_dp\vert^2\xi_p(a_p^*a_{-p}^*a_0a_0+h.c.).
\end{equation*}
Similar to the proof of Lemma \ref{control of S_+ conj with e^B}, for any $\psi\in L^2_s(\Lambda_d^N)$ and
\begin{align*}
  U_{N-2}a_0a_0\psi&=(\alpha^{(0)},\ldots,\alpha^{(N-2)}),\\
  U_N\psi&=(\beta^{(0)},\ldots,\beta^{(N)}),
\end{align*}
we have
\begin{align*}
\langle [H_{21},\tilde{B}]\psi,\psi\rangle
=&-\sum_{n=2}^{N}\sqrt{\frac{n(n-1)}{d}}\int_{\Lambda_{d}^n}
d\mathbf{x}_1\dots d\mathbf{x}_{n}\\
&\quad\quad\quad\times\Delta_{\mathbf{x}_1}\xi_\perp(\mathbf{x}_1-\mathbf{x}_2)\alpha^{(n-2)}
(\mathbf{x}_3,\dots,\mathbf{x}_n)\overline{\beta^{(n)}}(\mathbf{x}_1,\dots,\mathbf{x}_n)\\
=&\sum_{n=2}^{N}\sqrt{\frac{n(n-1)}{d}}\int_{\Lambda_{d}^n}
d\mathbf{x}_1\dots d\mathbf{x}_{n}\\
&\quad\quad\quad\times\nabla_{\mathbf{x}_1}\xi_\perp(\mathbf{x}_1-\mathbf{x}_2)\alpha^{(n-2)}
(\mathbf{x}_3,\dots,\mathbf{x}_n)\nabla_{\mathbf{x}_1}
\overline{\beta^{(n)}}(\mathbf{x}_1,\dots,\mathbf{x}_n).
\end{align*}
Then using (\ref{H_21psi,psi fock}) and Cauchy-Shwartz inequality, we have
\begin{align*}
  \vert\langle [H_{21},\tilde{B}]\psi,\psi\rangle\vert
\leq CN\Vert\nabla_{\mathbf{x}}\xi_\perp\Vert_2
\langle H_{21}\psi,\psi\rangle^{\frac{1}{2}}
\langle (\mathcal{N}_++1)\psi,\psi\rangle^{\frac{1}{2}},
\end{align*}
and (\ref{control of e^-B tildeH_21e^B tilde}) follows by (\ref{est of grad xi}) and Gronwall's inequality.
\begin{flushright}
  {$\Box$}
\end{flushright}

First, as a direct consequence of Lemma \ref{control of S_+ conj with e^B tilde} and Lemma \ref{lemma control of H_4 with e^B tilde}, as well as our aforementioned assumptions on the parameters (which lead to the fact $m^{-2}\ln(1+h(d\textit{l})^{-1})\ll m^{-1}\ll ad^{-1}$), we have the following corollary.
\begin{corollary}\label{lemma e^-B tildeEe^B tilde}
\begin{align}
    e^{-\tilde{B}}\tilde{\mathcal{E}}^{B^{\prime}} e^{\tilde{B}}
&\leq CN^4a^{\frac{8}{3}}d^{-2}\textit{l}^{\frac{1}{3}}(\mathcal{N}_++1)
+CN^2a^{\frac{5}{3}}d^{-1}\textit{l}^{\frac{1}{3}}H_{21}\nonumber\\
&+CN^{\frac{3}{2}}a^{\frac{7}{6}}d^{-\frac{1}{2}}\textit{l}^{\frac{1}{3}}H_4
+CN^{\frac{3}{2}}a^{\frac{3}{2}}d^{-\frac{1}{2}}\textit{l}^{\frac{1}{2}}H_4^\prime.
\label{e^-B tildeEe^B tilde}
\end{align}
\end{corollary}

\par Following the steps carried out in Section \ref{2}, we arrive at
\begin{lemma}\label{lemma e^-B tildeH_02'e^B tilde}
\begin{equation}\label{e^-B tildeH_02'e^B tilde III}
          e^{-\tilde{B}}H_{02}^\prime e^{\tilde{B}}
=-\Big(W_0+\sum_{p\neq0}W_p\eta_p\Big)\mathcal{N}_+^2
+\tilde{\mathcal{E}}^{\tilde{B}}_{02},
\end{equation}
where
\begin{equation}\label{E^B tilde_02 III}
  \pm\tilde{\mathcal{E}}^{\tilde{B}}_{02}\leq C\Big\{N^2ad^{-1}
(d\textit{l}+hm^{-1})+ad^{-1}\Big\}(\mathcal{N}_++1).
\end{equation}
\end{lemma}
\noindent\emph{Proof.} We first have
\begin{equation}\label{0100702}
 \pm \big(e^{-\tilde{B}}H_{02}^\prime e^{\tilde{B}}
-H_{02}^\prime\big)\leq CN^2ad^{-1}
(d\textit{l}+hm^{-1})(\mathcal{N}_++1).
\end{equation}
For the proof of (\ref{0100702}) one can see Lemma \ref{lemma e^-BH_02e^B} for Region III for details. At the same time, we need to use Lemma \ref{control of S_+ conj with e^B tilde} and the bound (\ref{010701}) derived from (\ref{sum_pW_peta_p 3dscatt}) that
\begin{equation}\label{010701}
  \Big\vert W_0+\sum_{p\neq0}W_p\eta_p\Big\vert\leq Cad^{-1}.
\end{equation}
Furthermore, we use again (\ref{010701}) to gain
\begin{equation}\label{010703}
  \pm\Big\{H_{02}^\prime+\Big(W_0+\sum_{p\neq0}W_p\eta_p\Big)\mathcal{N}_+^2
\Big\}\leq Cad^{-1}\mathcal{N}_+.
\end{equation}
We have (\ref{E^B tilde_02 III}) by combining (\ref{0100702}) and (\ref{010703}).
\begin{flushright}
  {$\Box$}
\end{flushright}

\begin{lemma}\label{lemma e^-B tildeH_22'e^B tilde}
\begin{equation}\label{e^-B tildeH_22'e^B tilde III}
          e^{-\tilde{B}}H_{22}^\prime e^{\tilde{B}}
=2\Big(W_0+\sum_{p\neq0}W_{p}\eta_p\Big)(N-\mathcal{N}_+)\mathcal{N}_+
+\tilde{\mathcal{E}}^{\tilde{B}}_{22},
\end{equation}
where
\begin{equation}\label{E^B tilde_22 III}
  \pm\tilde{\mathcal{E}}^{\tilde{B}}_{22}\leq CN^2ad^{-1}
(d\textit{l}+hm^{-1})(\mathcal{N}_++1)+CNad\textit{l}^2H_{21}.
\end{equation}
\end{lemma}
\noindent\emph{Proof.} We first have
\begin{equation}\label{010704}
  \pm\big( e^{-\tilde{B}}H_{22}^\prime e^{\tilde{B}}
-H_{22}^\prime\big)\leq  CN^2ad^{-1}
(d\textit{l}+hm^{-1})(\mathcal{N}_++1).
\end{equation}
We just need to notice that
\begin{equation}\label{010705}
  \sum_{p\in2\pi\mathbb{Z}^3}\Big(
W_p+\sum_{q\neq0}W_{p-q}\eta_q\Big)\phi^{(d)}_p(\mathbf{x})
=W(\mathbf{x})\big(1+\sqrt{d}\eta_\perp(\mathbf{x})\big).
\end{equation}
Using (\ref{L1&L2 norm of W 3dscatt}) and (\ref{est of max eta_perp}), we can bound the $L^1$ norm of (\ref{010705}) by $ad^{-\frac{1}{2}}$. For the rest of the proof of (\ref{010704}), one can see Lemma \ref{lemma e^-BH_22e^B} for Region III for details.
\par Moreover, since (\ref{010705}) is radially symmetric, we have the bound
\begin{align}
  &\Big\vert W_p+\sum_{q\neq0}W_{p-q}\eta_q
-W_0-\sum_{q\neq0}W_{q}\eta_q\Big\vert\nonumber\\
=&\frac{1}{\sqrt{d}}
\Big\vert\int_{\Lambda_d}W(\mathbf{x})\big(1+\sqrt{d}\eta_\perp(\mathbf{x})\big)
\Big(e^{ip^T\mathcal{M}_dp\mathbf{x}}-1\Big)d\mathbf{x}\Big\vert\nonumber\\
\leq&\frac{C}{\sqrt{d}}\vert\mathcal{M}_dp\vert^2\int_{\Lambda_d}
W(\mathbf{x})\vert\mathbf{x}\vert^2d\mathbf{x}
\leq\frac{Ca(d\textit{l})^2}{d}\vert\mathcal{M}_dp\vert^2.\label{010706}
\end{align}
This leads to
\begin{align}\label{010707}
 \pm\Big\{H_{22}^\prime-2(N-\mathcal{N}_+)\mathcal{N}_+
\Big(W_0+\sum_{p\neq0}W_{p}\eta_p\Big)\Big\}\leq Nad\textit{l}^2H_{21}.
\end{align}
Combining (\ref{010704}) and (\ref{010707}), we reach (\ref{E^B tilde_22 III}).
\begin{flushright}
  {$\Box$}
\end{flushright}

\begin{lemma}\label{lemma e^-B tildeH_3'e^B tilde}
\begin{equation}\label{e^-B tildeH_3'e^B tilde III}
          e^{-\tilde{B}}H_{3}^\prime e^{\tilde{B}}
=H_{3}^\prime+\tilde{\mathcal{E}}^{\tilde{B}}_{3},
\end{equation}
where
\begin{equation}\label{E^B tilde_3 III}
  \pm\tilde{\mathcal{E}}^{\tilde{B}}_{3}\leq CN^2ad^{-1}
(d\textit{l}+hm^{-1})(\mathcal{N}_++1+H_4^{\prime}).
\end{equation}
\end{lemma}
\noindent\emph{Proof.} See Lemma \ref{lemma e^-BH_3e^B} for details.
\begin{flushright}
  {$\Box$}
\end{flushright}

\begin{lemma}\label{lemma Theta}
\begin{equation}\label{int_0^1 e^-tB tildeThetae^tB tildedt}
\begin{aligned}
  \pm\int_{0}^{1}e^{-t\tilde{B}}\Theta e^{t\tilde{B}}dt
\leq CN^2ad^{-1}
(d\textit{l}+hm^{-1})(\mathcal{N}_++1)
+CN^{\frac{3}{2}}a^{\frac{1}{2}}d^{-\frac{1}{2}}
(d\textit{l}+hm^{-1})H_4.
\end{aligned}
\end{equation}
\end{lemma}
\noindent\emph{Proof.} See Lemma \ref{lemma Gamma} for details. Using (\ref{eqn of xi_p}) and the definition of $\widetilde{W}_p$ (\ref{define W tilde}), we can calculate
\begin{equation*}
  \Theta=\frac{1}{\sqrt{d}}{\sum_{p,q,p+r,q+r\neq0}}v_r^{(a,d)}
  \xi_p(a_{p+r}^*a_q^*a_{-p}^*a_{q+r}a_0a_0+h.c.).
\end{equation*}
We remark here that $\xi_p=0$ if $p_3=0$ due to its definition (\ref{define xi_p}).
\begin{flushright}
  {$\Box$}
\end{flushright}

\begin{lemma}\label{lemma [H_23',B tilde]}
\begin{align}
   \int_{0}^{1}\int_{t}^{1}e^{-s\tilde{B}}[{H}_{23}^{\prime},\tilde{B}]e^{s\tilde{B}}&dsdt
   = N(N-1)\sum_{p\neq0}W_p\xi_p-2N\sum_{p\neq0}W_p\xi_p\mathcal{N}_+\nonumber\\
&+\sum_{p\neq0}W_p\xi_p\mathcal{N}_+^2
+\tilde{\mathcal{E}}^{\tilde{B}}_{[H_{23}^\prime,\tilde{B}]},\label{[H_23',B tilde]}
\end{align}
where
\begin{align}
  \pm\tilde{\mathcal{E}}^{\tilde{B}}_{[H_{23}^\prime,\tilde{B}]}
\leq& C\big(N^2ad^{-1}
(d\textit{l}+hm^{-1})+ad^{-1}\big)(\mathcal{N}_++1)\nonumber\\
&+CN^{\frac{3}{2}}a^{\frac{1}{2}}d^{-\frac{1}{2}}
(d\textit{l}+hm^{-1})H_4^\prime.
\label{E^B tilde_23,1}
\end{align}
\end{lemma}
\noindent
\emph{Proof.} See Lemmas \ref{lemma [H_23,B]} and \ref{lemma [H_23',B]} for Region III for details. Here we use (\ref{sum_pW_peta_p 3dscatt}), (\ref{est of xi_p}), (\ref{useful shit 2d}) and the assumption $\frac{ma}{d}>C$ to reach the bound
\begin{equation}\label{sum W_pxi_p}
  \left\vert\sum_{p\neq0}W_p\xi_p\right\vert\leq Cad^{-1}.
\end{equation}
\begin{flushright}
  {$\Box$}
\end{flushright}

\begin{lemma}\label{lemma [H_23'',B tilde]}
\begin{align}
   \int_{0}^{1}\int_{0}^{t}e^{-s\tilde{B}}[{H}_{23}^{\prime\prime},\tilde{B}]e^{s\tilde{B}}&dsdt
   = N(N-1)\sum_{p\neq0}\widetilde{W}_p\xi_p
-2N\sum_{p\neq0}\widetilde{W}_p\xi_p\mathcal{N}_+\nonumber\\
&+\sum_{p\neq0}\widetilde{W}_p\xi_p\mathcal{N}_+^2
+\tilde{\mathcal{E}}^{\tilde{B}}_{[H_{23}^{\prime\prime},\tilde{B}]},\label{[H_23'',B tilde]}
\end{align}
where
\begin{align}
  \pm\tilde{\mathcal{E}}^{\tilde{B}}_{[H_{23}^{\prime\prime},\tilde{B}]}
\leq& C\big(N^2ad^{-1}
(d\textit{l}+hm^{-1})+ad^{-1}\big)(\mathcal{N}_++1)\nonumber\\
&+CN^{\frac{3}{2}}a^{\frac{1}{2}}d^{-\frac{1}{2}}
(d\textit{l}+hm^{-1})(H_4+H_4^\prime)\nonumber\\
&+C\frac{N}{m}
\Big(d\textit{l}+\frac{h}{m}\Big)\Big(\ln\big(1+\frac{h}{d\textit{l}}\big)\Big)^{\frac{1}{2}}
(H_{21}+(\mathcal{N}_++1)^3).
\label{E^B tilde_23,2}
\end{align}
\end{lemma}
\noindent
\emph{Proof.} The proof of this result is slightly different from Lemmas \ref{lemma [H_23,B]} and \ref{lemma [H_23',B]}. First we use (\ref{eqn of xi_p}) to divide $\widetilde{W}_p$ into three parts
$\widetilde{W}_p=\widetilde{W}_{p,1}+\widetilde{W}_{p,2}+\widetilde{W}_{p,3}$ where
\begin{align}\label{define W tilde _p,1,2,3}
\widetilde{W}_{p,1}=W_p,\quad
\widetilde{W}_{p,2}=\frac{1}{2\sqrt{d}}\sum_{q\neq0}v_{p-q}^{(a,d)}\xi_q,\quad
\widetilde{W}_{p,3}=\vert\mathcal{M}_dp\vert^2\xi_p.
\end{align}
We therefore let for $i=1,2,3$
\begin{equation}\label{define H_23_123''}
  H_{23,i}^{\prime\prime}=\sum_{p\neq0}\widetilde{W}_{p,i}(a_p^*a_{-p}^*a_0a_0+h.c.).
\end{equation}
We also denote $\widetilde{W}_i=\sum_p\widetilde{W}_{p,i}\phi_p^{(d)}$. Since $\widetilde{W}_{p,1}=W_p$, the calculation of the first part in (\ref{define W tilde _p,1,2,3}) is precisely carried out in Lemma \ref{lemma [H_23',B tilde]}. For the second part in (\ref{define W tilde _p,1,2,3}), the anaylsis is still similar to Lemmas \ref{lemma [H_23,B]} and \ref{lemma [H_23',B]} for Region III, we only need to notice
\begin{equation}\label{sum_d-1/2 v_p-qxiq}
  \frac{1}{2\sqrt{d}}\left\vert\sum_{p,q\neq0}v_{p-q}^{(a,d)}\xi_q\xi_p\right\vert
=\frac{1}{2}\left\vert\int_{\Lambda_d}v_a(\mathbf{x})\xi_{\perp}^2(\mathbf{x})
d\mathbf{x}\right\vert
\leq Cad^{-1},
\end{equation}
and the fact that
\begin{equation}\label{est of v_axi}
  \left\vert\int_{\Lambda_d}v_a(\mathbf{x})\xi_{\perp}(\mathbf{x})
d\mathbf{x}\right\vert\leq Cd^{-\frac{1}{2}}\int_{\Lambda_d}v_a(\mathbf{x})d\mathbf{x}.
\end{equation}
Estimating like Lemmas \ref{lemma [H_23,B]} and \ref{lemma [H_23',B]} for Region III we reach
\begin{align*}
   \int_{0}^{1}\int_{0}^{t}e^{-s\tilde{B}}[{H}_{23,2}^{\prime\prime},\tilde{B}]e^{s\tilde{B}}
&dsdt = N(N-1)\sum_{p\neq0}\widetilde{W}_{p,2}\xi_p
-2N\sum_{p\neq0}\widetilde{W}_{p,2}\xi_p\mathcal{N}_+\nonumber\\
&+\sum_{p\neq0}\widetilde{W}_{p,2}\xi_p\mathcal{N}_+(\mathcal{N}_++1)
+\tilde{\mathcal{E}}^{\tilde{B}}_{[H_{23,2}^{\prime\prime},\tilde{B}]},
\end{align*}
where
\begin{align*}
  \pm\tilde{\mathcal{E}}^{\tilde{B}}_{[H_{23,2}^{\prime\prime},\tilde{B}]}
\leq CN^2ad^{-1}
(d\textit{l}+hm^{-1})(\mathcal{N}_++1)
+CN^{\frac{3}{2}}a^{\frac{1}{2}}d^{-\frac{1}{2}}
(d\textit{l}+hm^{-1})H_4.
\end{align*}
\par For the third part in (\ref{define W tilde _p,1,2,3}), we use
\begin{align*}
[H_{23,3}^{\prime\prime},\tilde{B}]=&\frac{1}{2}\sum_{p,q\neq0}
\widetilde{W}_{p,3}\xi_q(-4a_0^*a_0-2)
(a_q^*a_{-q}^*a_{p}a_{-p}+h.c.)\\
&+2\sum_{p\neq0}\widetilde{W}_{p,3}\xi_p(1+2a^*_pa_p)a_0^*a_0^*a_0a_0.
\end{align*}
Again expanding $a_0^*a_0^*a_0a_0$ can reach
\begin{align*}
   \int_{0}^{1}\int_{0}^{t}e^{-s\tilde{B}}[{H}_{23,3}^{\prime\prime},\tilde{B}]e^{s\tilde{B}}
&dsdt = N(N-1)\sum_{p\neq0}\widetilde{W}_{p,3}\xi_p
-2N\sum_{p\neq0}\widetilde{W}_{p,3}\xi_p\mathcal{N}_+\nonumber\\
&+\sum_{p\neq0}\widetilde{W}_{p,3}\xi_p\mathcal{N}_+(\mathcal{N}_++1)
+\tilde{\mathcal{E}}^{\tilde{B}}_{[H_{23,3}^{\prime\prime},\tilde{B}]},
\end{align*}
by defining
\begin{equation*}
\tilde{\mathcal{E}}^{\tilde{B}}_{[H_{23,3}^{\prime\prime},\tilde{B}]}
=\sum_{i=1}^{4}\tilde{\mathcal{E}}^{\tilde{B}}_{23^{\prime\prime},3,i},
\end{equation*}
where
\begin{align*}
\tilde{\mathcal{E}}^{\tilde{B}}_{23^{\prime\prime},3,1}
=&-4N\sum_{p\neq0}\widetilde{W}_{p,3}\xi_p
\int_{0}^{1}\int_{0}^{t}\Big(e^{-sB}\mathcal{N}_+e^{sB}
-\mathcal{N}_+\Big)dsdt\nonumber\\
\tilde{\mathcal{E}}^{\tilde{B}}_{23^{\prime\prime},3,2}
=&2\sum_{p\neq0}\widetilde{W}_{p,3}\xi_p\int_{0}^{1}
\int_{0}^{t}\Big(e^{-sB}\mathcal{N}_+(\mathcal{N}_++1)e^{sB}
-\mathcal{N}_+(\mathcal{N}_++1)\Big)dsdt\nonumber\\
\tilde{\mathcal{E}}^{\tilde{B}}_{23^{\prime\prime},3,3}
=&4\sum_{p\neq0}\widetilde{W}_{p,3}\xi_p\int_{0}^{1}
\int_{0}^{t}e^{-sB}a_p^*a_pa_0^*a_0^*a_0a_0e^{sB}dsdt\nonumber\\
\tilde{\mathcal{E}}^{\tilde{B}}_{23^{\prime\prime},3,4}
=&-\sum_{p,q\neq0}\widetilde{W}_{p,3}\xi_q\int_{0}^{1}
\int_{0}^{t}e^{-sB}(2a_0^*a_0+1)(a_p^*a_{-p}^*a_qa_{-q}+h.c.)e^{sB}dsdt.
\end{align*}
We rewrite $\tilde{\mathcal{E}}^{\tilde{B}}_{23^{\prime\prime},3,4}$
\begin{equation*}
  \tilde{\mathcal{E}}^{\tilde{B}}_{23^{\prime\prime},3,4}
=\sum_{j=1}^{3}\tilde{\mathcal{E}}^{\tilde{B}}_{23^{\prime\prime},3,4,j}+h.c.
\end{equation*}
with
\begin{align*}
\tilde{\mathcal{E}}^{\tilde{B}}_{23^{\prime\prime},3,4,1}
 =& \sum_{p,q\neq0}
\widetilde{W}_{p,3}\xi_q\int_{0}^{1}\int_{0}^{t}a_p^*a_{-p}^*e^{-sB}
[a_qa_{-q}(2a_0^*a_0+1)]e^{sB}dsdt \\
\tilde{\mathcal{E}}^{\tilde{B}}_{23^{\prime\prime},3,4,2}
=& 2\sum_{p,q\neq0}
\widetilde{W}_{p,3}\xi_p\xi_q\int_{0}^{1}\int_{0}^{t}\int_{0}^{s}
d\tau dsdt\\
&\quad\quad\quad\quad\quad\quad\times
 e^{-\tau B}a_0^*a_0^*a_p^*a_pe^{(\tau-s)B}
[a_qa_{-q}(2a_0^*a_0+1)]e^{sB}\\
\tilde{\mathcal{E}}^{\tilde{B}}_{23^{\prime\prime},3,4,3}
=& \sum_{p,q\neq0}
\widetilde{W}_{p,3}\xi_p\xi_q\int_{0}^{1}\int_{0}^{t}\int_{0}^{s}
e^{-\tau B}a_0^*a_0^*e^{(\tau-s)B}
[a_qa_{-q}(2a_0^*a_0+1)]e^{sB}d\tau dsdt
\end{align*}
Using (\ref{est of grad xi}), the definition of $m$ (\ref{define m}) and our assumptions on the parameters, we have
\begin{equation}\label{est of sum W_p,3 tildexi_p}
  \left\vert\sum_{p\neq0}\widetilde{W}_{p,3}\xi_p\right\vert
=\Vert\nabla_{\mathbf{x}}\xi_{\perp}\Vert_2^2\leq
\frac{C}{m^2}\ln\left(1+\frac{h}{d\textit{l}}\right)\ll\frac{C}{m}\ll\frac{Ca}{d}.
\end{equation}
Hence we can bound using (\ref{e^-tB tilde N_+e^tB tilde-N_+}) for $i=1,2$
\begin{equation*}
  \pm\tilde{\mathcal{E}}^{\tilde{B}}_{23^{\prime\prime},3,i}
\leq  CN^2ad^{-1}
(d\textit{l}+hm^{-1})(\mathcal{N}_++1).
\end{equation*}
From (\ref{est of z_h,p}) we infer that
\begin{equation}\label{est of W_p,3 tilde}
  \vert\widetilde{W}_{p,3}\vert\leq\frac{C}{m}\ll\frac{Ca}{d}.
\end{equation}
This together with (\ref{est of xi_p}) and (\ref{e^-tB tilde(N_++1)e^tB tilde}) yield
\begin{equation*}
  \pm\tilde{\mathcal{E}}^{\tilde{B}}_{23^{\prime\prime},3,3}
\leq CN^2ad^{-1}(d^2\textit{l}^2+h^2m^{-1})(\mathcal{N}_++1).
\end{equation*}
Estimating in Fock space, we have for any $\psi\in L^2_s(\Lambda_d)$ that
\begin{align*}
  \vert\langle\tilde{\mathcal{E}}^{\tilde{B}}_{23^{\prime\prime},3,4,1}\psi,\psi\rangle\vert
&\leq C\Vert \nabla_{\mathbf{x}}\xi\Vert_2\Vert\xi_\perp\Vert_2
\langle H_{21}\psi,\psi\rangle^{\frac{1}{2}}\\
\times&\int_{0}^{1}\int_{0}^{t}
\langle(\mathcal{N}_++1)^3(2a_0^*a_0+1)e^{sB}\psi,(2a_0^*a_0+1)e^{sB}\psi\rangle^{\frac{1}{2}}
dsdt\\
&\leq C\frac{N}{m}
\Big(d\textit{l}+\frac{h}{m}\Big)\Big(\ln\big(1+\frac{h}{d\textit{l}}\big)\Big)^{\frac{1}{2}}
\langle H_{21}\psi,\psi\rangle^{\frac{1}{2}}
\langle(\mathcal{N}_++1)^3\psi,\psi\rangle^{\frac{1}{2}},
\end{align*}
and
\begin{align*}
    \vert\langle\tilde{\mathcal{E}}^{\tilde{B}}_{23^{\prime\prime},3,4,2}\psi,\psi\rangle\vert
&\leq Cd^{-\frac{1}{2}}\Vert\widetilde{W}_3\ast\xi_\perp \Vert_2\Vert\xi_\perp\Vert_2
\langle (\mathcal{N}_++1)^2a_0a_0\psi,a_0a_0\psi\rangle^{\frac{1}{2}}\\
\times&\int_{0}^{1}\int_{0}^{t}\int_{0}^{s}
\langle(\mathcal{N}_++1)^2(2a_0^*a_0+1)e^{sB}\psi,(2a_0^*a_0+1)e^{sB}\psi\rangle^{\frac{1}{2}}
d\tau dsdt\\
&\leq  CN^3ad^{-1}
(d\textit{l}+hm^{-1})^2(\mathcal{N}_++1),
\end{align*}
where we have used (\ref{eqn of xi_p}) to gain the rough bound
\begin{equation}\label{est of widetilde W_3}
  \Vert\widetilde{W}_3\Vert_1\leq Cad^{-\frac{1}{2}}.
\end{equation}
Moreover
\begin{align*}
  \vert\langle\tilde{\mathcal{E}}^{\tilde{B}}_{23^{\prime\prime},3,4,3}\psi,\psi\rangle\vert
&\leq C\left\vert\sum_{p\neq0}\widetilde{W}_{p,3}\xi_p\right
\vert\Vert\xi_\perp\Vert_2\int_{0}^{1}\int_{0}^{t}\int_{0}^{s}d\tau dsdt\\
\times&
\langle(\mathcal{N}_++1)(2a_0^*a_0+1)e^{sB}\psi,(2a_0^*a_0+1)e^{sB}\psi\rangle^{\frac{1}{2}}\\
\times&
\langle (\mathcal{N}_++1)e^{(s-\tau)B}a_0a_0e^{tB}\psi,
e^{(s-\tau)B}a_0a_0e^{tB}\psi\rangle^{\frac{1}{2}}\\
&\leq CN^2ad^{-1}
(d\textit{l}+hm^{-1})(\mathcal{N}_++1).
\end{align*}
We then conclude this lemma by noticing that
\begin{equation}\label{sum W_p tildexi_p}
  \left\vert\sum_{p\neq0}\widetilde{W}_p\xi_p\right\vert\leq Cad^{-1}.
\end{equation}
\begin{flushright}
  {$\Box$}
\end{flushright}

\noindent
\emph{Analysis of $\mathcal{L}_N$.}
\par Summarizing all the estimates above, we have
\begin{align}\label{LN}
  \mathcal{L}_N=e^{-\tilde{B}}\mathcal{J}_Ne^{\tilde{B}}=
\tilde{C}^{\tilde{B}}+\tilde{Q}_1^{\tilde{B}}\mathcal{N}_+
+\tilde{Q}_2^{\tilde{B}}\mathcal{N}_+^2+
H_{21}+H_4+H_{23}^{\prime\prime}+H_3^\prime+\tilde{\mathcal{E}}^{\tilde{B}}
\end{align}
where
\begin{align}
\tilde{C}^{\tilde{B}}&=N(N-1)\Big(
W_0+\sum_{p\neq0}W_p\eta_p+\sum_{p\neq0}W_p\xi_p
+\sum_{p\neq0}\widetilde{W}_p\xi_p\Big)\label{C tilde B tilde}\\
\tilde{Q}_1^{\tilde{B}}&=2N\Big(W_0+\sum_{p\neq0}W_p\eta_p
-\sum_{p\neq0}W_p\xi_p-\sum_{p\neq0}\widetilde{W}_p\xi_p\Big)\label{Q tilde_1 B tilde}\\
\tilde{Q}_2^{\tilde{B}}&=\sum_{p\neq0}W_p\xi_p
+\sum_{p\neq0}\widetilde{W}_p\xi_p-3W_0-3\sum_{p\neq0}W_p\eta_p\label{Q tilde_2 B tilde}
\end{align}
and the error term is bounded by
\begin{align}\label{E tilde B tilde}
 \pm\tilde{\mathcal{E}}^{\tilde{B}}\leq&
C\Big(N^{4}a^{\frac{8}{3}}d^{-2}\textit{l}^{\frac{1}{3}}
+N^2ad^{-1}\big(d\textit{l}+hm^{-1}\big)+ad^{-1}\Big)(\mathcal{N}_++1)\nonumber\\
&+C\frac{N}{m}\Big(d\textit{l}+\frac{h}{m}\Big)
\Big(\ln\big(1+\frac{h}{d\textit{l}}\big)\Big)^{\frac{1}{2}}(\mathcal{N}_++1)^3\nonumber\\
&+C\Big\{N^{2}a^{\frac{5}{3}}d^{-1}\textit{l}^{\frac{1}{3}}
+\frac{N}{m}\Big(d\textit{l}+\frac{h}{m}\Big)
\Big(\ln\big(1+\frac{h}{d\textit{l}}\big)\Big)^{\frac{1}{2}}\Big\}H_{21}\nonumber\\
&+C\Big(N^{\frac{3}{2}}a^{\frac{7}{6}}d^{-\frac{1}{2}}\textit{l}^{\frac{1}{3}}
+N^2ad^{-1}\big(d\textit{l}+hm^{-1}\big)\Big)H_4\nonumber\\
&+CN^{\frac{3}{2}}a^{\frac{1}{2}}d^{-\frac{1}{2}}\big(d\textit{l}+hm^{-1}\big)H_4^\prime.
\end{align}
Moreover, we have the bound
\begin{equation}\label{bound C Q L_N}
  N^2\vert\tilde{Q}_2^{\tilde{B}}\vert\leq C
N\vert\tilde{Q}_1^{\tilde{B}}\vert\leq C\vert\tilde{C}^{\tilde{B}}\vert
\leq CN^2ad^{-1}.
\end{equation}
\begin{flushright}
  {$\Box$}
\end{flushright}
\subsection{Quasi-2D Cubic Renormalization}\label{quasi-2d cubic}
\
\par We adopt the notation
\begin{equation*}
  \tilde{A}^{\prime}=\sum_{p,q,p+q\neq0}\xi_p
  a_{p+q}^*a_{-p}^*a_qa_0,
\end{equation*}
and rewrite $e^{-\tilde{B}^\prime}\mathcal{L}_Ne^{\tilde{B}^\prime}$ using (\ref{LN})
\begin{align}
  e^{-\tilde{B}^\prime}\mathcal{L}_Ne^{\tilde{B}^\prime}
=&\tilde{C}^{\tilde{B}}+H_{21}+H_4+H_{3}^{\prime\prime}
+e^{-\tilde{B}^\prime}\big(\tilde{Q}_1^{\tilde{B}}\mathcal{N}_+
+\tilde{Q}_2^{\tilde{B}}\mathcal{N}_+^2+H_{23}^{\prime\prime}\big)e^{\tilde{B}^\prime}
\nonumber\\
&+\int_{0}^{1}\int_{t}^{1}
e^{-s\tilde{B}^\prime}[H_{3}^\prime,\tilde{B}^\prime] e^{s\tilde{B}^\prime}dsdt
+\int_{0}^{1}\int_{0}^{t}
e^{-s\tilde{B}^\prime}[H_{3}^{\prime\prime},\tilde{B}^\prime]
e^{s\tilde{B}^\prime}dsdt\nonumber\\
&+\int_{0}^{1}e^{-t\tilde{B}^\prime}\Theta^\prime e^{t\tilde{B}^\prime}dt
+e^{-\tilde{B}^\prime}\tilde{\mathcal{E}}^{\tilde{B}}e^{\tilde{B}^\prime}.
\label{split e^-B' tildeL_Ne^B' tilde}
\end{align}
Define
\begin{equation}\label{define Theta'}
  \Theta^\prime=[H_{21}+H_4,\tilde{B}^\prime]+H_{3}^\prime-H_{3}^{\prime\prime}
\end{equation}
with $H_{3}^{\prime\prime}$ defined as
\begin{equation}\label{define H_3''}
  H_{3}^{\prime\prime}=2\sum_{p,q,p+q\neq0}\widetilde{W}_p(a_{p+q}^*a_{-p}^*a_qa_0+h.c.)
\end{equation}
and $\widetilde{W}_p$ defined in (\ref{define W tilde}). Since the analysis of (\ref{split e^-B' tildeL_Ne^B' tilde}) is similar to Section \ref{3}, we state them in the up-coming series of lemmas, while we omit the details of proofs except some new estimates we need.

\par In the following lemmas, we bound $e^{-\tilde{B}^\prime}\tilde{\mathcal{E}}^{\tilde{B}}e^{\tilde{B}^\prime}$ in Corollary \ref{lemma e^Btilde'E tilde^Btildee^-Btilde'}, $e^{-\tilde{B}^\prime}\big(\tilde{Q}_1^{\tilde{B}}\mathcal{N}_+
+\tilde{Q}_2^{\tilde{B}}\mathcal{N}_+^2\big)e^{\tilde{B}^\prime}$ in Corollary \ref{lemma e^-Btilde'Q_1+Q_2e^Btilde'}, and $e^{-\tilde{B}^\prime}H_{23}^{\prime\prime}e^{\tilde{B}^\prime}$ in Lemma \ref{lemma e^-Btilde'H_23''e^Btilde'}. These three terms stay unchanged up to small errors after conjugating with $e^{\tilde{B}^\prime}$. The term containing the difference $\Theta^\prime$ is bounded in Lemma \ref{lemma Theta'}, and is proved to be negligible. The contribution of the commutator $[H_{3}^\prime,\tilde{B}^\prime]$ is calculated in Lemma \ref{[H_3',Btilde']}, and the contribution of $[H_{3}^{\prime\prime},\tilde{B}^\prime]$ is calculated in Lemma \ref{[H_3'',Btilde']}. Lemmas \ref{[H_3',Btilde']} and \ref{[H_3'',Btilde']} present the major contributions of the cubic quasi-2D correlation structure to the second order ground state energy, in the form of polynomials of $\mathcal{N}_+$. Finally, we bound the growths of $\mathcal{N}_+$, $H_{21}$, $H_4$ and $H_4^\prime$ in Lemmas \ref{control of S_+ conj with e^B' tilde} and \ref{lemma control of H_N conj with e^Btilde'}.
\begin{lemma}\label{control of S_+ conj with e^B' tilde}
Let $\mathcal{N_+}$ be defined on $L_s^2(\Lambda_d^N)$ as stated in (\ref{define of N_+}), then there exist a constant $C_n$ depending only on $n\in\frac{1}{2}\mathbb{N}$ such that: for every $t\in\mathbb{R}$, $N\in\mathbb{N}$, $n\in\frac{1}{2}\mathbb{N}$, $\textit{l}\in(0,\frac{1}{2})$ such that $\frac{d\textit{l}}{a}>C$, and $h\in(0,\frac{1}{2})$ such that $\frac{h}{d\textit{l}}>C$ for some universal constant $C$, and we have
\begin{align}
  e^{-t\tilde{B}^\prime}(\mathcal{N}_++1)^ne^{t\tilde{B}^\prime}&\leq e^{C_nN(d\textit{l}+\frac{h}{m})\vert t\vert}(\mathcal{N}_++1)^n,\label{e^-tB' tilde(N_++1)e^tB' tilde}\\
  \pm(e^{-t\tilde{B}^\prime}(\mathcal{N}_++1)^ne^{t\tilde{B}^\prime}-(\mathcal{N}_++1)^n)&\leq (e^{C_nN(d\textit{l}+\frac{h}{m})\vert t\vert}-1) (\mathcal{N}_++1)^n.\label{e^-tB' tilde N_+e^tB' tilde-N_+}
\end{align}
\end{lemma}
\noindent
\emph{Proof.} See the proof of Lemma \ref{control of S_+ conj with e^B'} for details.
\begin{flushright}
  {$\Box$}
\end{flushright}

\par We still work under the same assumptions on parameters as we have made in the analysis of $\mathcal{L}_N$. That is $N$ tends to infinity, $a$, $d$, $\frac{a}{d}$, $N^{\frac{3}{2}}a^{\frac{7}{6}}d^{-\frac{1}{2}}\textit{l}^{\frac{1}{3}}$ and
$Na\textit{l}^{\frac{1}{2}}$ tend to $0$ and $\frac{d\textit{l}}{a}>C$. Moreover, we demand additionally $\frac{h}{d\textit{l}}>C$, $\frac{Na}{d}>C$, $\frac{ma}{d}>C$ and $N(d\textit{l}+\frac{h}{m})$ tends to $0$. The next lemma controls the action of $e^{\tilde{B}^\prime}$ on $H_{21}$, $H_4$ and $H_4^\prime$, with the proof similar to Lemmas \ref{commutator of H_21,H_4with B'} and \ref{lemma e^-B'H_Ne^B'}, and therefore we omit the further details. To ensure the following lemmas hold true, we need to add one more assumption on the parameters, which is $N^{\frac{3}{2}}a^{\frac{1}{2}}d^{-\frac{1}{2}}(d\textit{l}+\frac{h}{m})^{\frac{2}{3}}$ should tend to $0$.

\begin{lemma}\label{lemma control of H_N conj with e^Btilde'}
\begin{align}
  e^{-t\tilde{B}^\prime}H_{21}e^{t\tilde{B}^\prime}
&\leq CH_{21}+C\frac{N}{m^2}\ln\big(1+\frac{h}{d\textit{l}}\big)(\mathcal{N}_++1)^2
\label{e^-Btilde'H_21e^Btilde' III}\\
  e^{-t\tilde{B}^\prime}H_{4}e^{t\tilde{B}^\prime}
&\leq CH_4+CNad^{-1}(\mathcal{N}_++1)\nonumber\\
&+CN^{\frac{3}{2}}a^{\frac{1}{2}}d^{-\frac{1}{2}}(d\textit{l}+\frac{h}{m})^{\frac{2}{3}}
  \Big(H_{21}+\frac{N}{m^2}\ln\big(1+\frac{h}{d\textit{l}}\big)
(\mathcal{N}_++1)^2\Big)\label{e^-Btilde 'H_4e^Btilde' III}\\
 e^{-t\tilde{B}^\prime}H_{4}^\prime e^{t\bar{B}^\prime}&\leq CH_4^\prime+CNad^{-1}(\mathcal{N}_++1)\nonumber\\
&+CN^{\frac{3}{2}}a^{\frac{1}{2}}d^{-\frac{1}{2}}(d\textit{l}+\frac{h}{m})^{\frac{2}{3}}
  \Big(H_{21}+\frac{N}{m^2}\ln\big(1+\frac{h}{d\textit{l}}\big)
(\mathcal{N}_++1)^2\Big).\label{e^-Btilde'H_4'e^Btilde' III}
\end{align}
\end{lemma}
\noindent
\emph{Proof.} See the proofs of Lemmas \ref{commutator of H_21,H_4with B'} and \ref{lemma e^-B'H_Ne^B'} for details.
\begin{flushright}
  {$\Box$}
\end{flushright}

\par As a direct consequence of Lemmas \ref{control of S_+ conj with e^B' tilde} and \ref{lemma control of H_N conj with e^Btilde'} and the estimate (\ref{bound C Q L_N}), we have
\begin{corollary}\label{lemma e^Btilde'E tilde^Btildee^-Btilde'}
\begin{align}\label{e^Btilde'E tilde^Btildee^-Btilde'}
\pm e^{-\tilde{B}^\prime}\tilde{\mathcal{E}}^{\tilde{B}}e^{\tilde{B}^\prime}
&\leq C\Big(N^{4}a^{\frac{8}{3}}d^{-2}\textit{l}^{\frac{1}{3}}
+N^3a^2d^{-2}\big(d\textit{l}+hm^{-1}\big)+ad^{-1}\Big)(\mathcal{N}_++1)\nonumber\\
&+C\frac{N}{m}\Big(d\textit{l}+\frac{h}{m}\Big)
\Big(\ln\big(1+\frac{h}{d\textit{l}}\big)\Big)^{\frac{1}{2}}(\mathcal{N}_++1)^3\nonumber\\
&+C\Big\{N^{2}a^{\frac{5}{3}}d^{-1}\textit{l}^{\frac{1}{3}}
+\frac{N}{m}\Big(d\textit{l}+\frac{h}{m}\Big)
\Big(\ln\big(1+\frac{h}{d\textit{l}}\big)\Big)^{\frac{1}{2}}\nonumber\\
&\quad\quad\quad
+N^{\frac{7}{2}}a^{\frac{3}{2}}d^{-\frac{3}{2}}(d\textit{l}+\frac{h}{m})^{\frac{5}{3}}
\Big\}\Big\{H_{21}+\frac{N}{m^2}
\Big(\ln\big(1+\frac{h}{d\textit{l}}\big)\Big)(\mathcal{N}_++1)^2\Big\}\nonumber\\
&+C\Big(N^{\frac{3}{2}}a^{\frac{7}{6}}d^{-\frac{1}{2}}\textit{l}^{\frac{1}{3}}
+N^2ad^{-1}\big(d\textit{l}+hm^{-1}\big)\Big)H_4\nonumber\\
&+CN^{\frac{3}{2}}a^{\frac{1}{2}}d^{-\frac{1}{2}}\big(d\textit{l}+hm^{-1}\big)H_4^\prime
\end{align}
\end{corollary}

\begin{corollary}\label{lemma e^-Btilde'Q_1+Q_2e^Btilde'}
\begin{equation}\label{e^-Btilde'Q_1+Q_2e^Btilde'}
  e^{-\tilde{B}^\prime}\big(\tilde{Q}_1^{\tilde{B}}\mathcal{N}_+
+\tilde{Q}_2^{\tilde{B}}\mathcal{N}_+^2\big)e^{\tilde{B}^\prime}
=\tilde{Q}_1^{\tilde{B}}\mathcal{N}_+
+\tilde{Q}_2^{\tilde{B}}\mathcal{N}_+^2+\tilde{\mathcal{E}}^{\tilde{B}^\prime}_{diag},
\end{equation}
where
\begin{equation}\label{Etilde^Btilde'_diag}
  \pm\tilde{\mathcal{E}}^{\tilde{B}^\prime}_{diag}
\leq CN^2ad^{-1}\big(d\textit{l}+hm^{-1}\big)(\mathcal{N}_++1).
\end{equation}
\end{corollary}

\par Similar to the analysis of $\mathcal{J}_N$ for Region III, we have the following lemmas.
\begin{lemma}\label{lemma Theta'}
\begin{align}\label{est of Theta'}
  \pm\int_{0}^{1}e^{-t\tilde{B}^\prime}&\Theta^\prime e^{t\tilde{B}^\prime}dt
\leq CN^{\frac{3}{2}}a^{\frac{1}{2}}d^{-\frac{1}{2}}(d\textit{l}+\frac{h}{m})^{\frac{2}{3}}\nonumber\\
&\times  \Big(H_{21}+H_4+Nad^{-1}(\mathcal{N}_++1)
+\frac{N}{m^2}\ln\big(1+\frac{h}{d\textit{l}}\big)
(\mathcal{N}_++1)^2\Big).
\end{align}
\end{lemma}
\noindent
\emph{Proof.} See the proof of Lemma \ref{lemma Gamma'} for Region III for details.
\begin{flushright}
  {$\Box$}
\end{flushright}

\begin{lemma}\label{lemma e^-Btilde'H_23''e^Btilde'}
\begin{equation}\label{e^-Btilde'H_23''e^Btilde'}
  e^{-\tilde{B}^\prime}H_{23}^{\prime\prime} e^{\tilde{B}^\prime}
=H_{23}^{\prime\prime}+\tilde{\mathcal{E}}^{\tilde{B}^\prime}_{23}
\end{equation}
where
\begin{align}\label{Etilde^Btilde'_23}
  \pm\tilde{\mathcal{E}}^{\tilde{B}^\prime}_{23}
&\leq C\Big\{N^{\frac{5}{2}}a^{\frac{3}{2}}d^{-\frac{3}{2}}
(d\textit{l}+\frac{h}{m})+
\frac{N^2}{m}\Big(d\textit{l}+\frac{h}{m}\Big)
\Big(\ln\big(1+\frac{h}{d\textit{l}}\big)\Big)^{\frac{1}{2}}\Big\}
(\mathcal{N}_++1)\nonumber\\
& +C\Big\{N^{3}ad^{-1}
(d\textit{l}+\frac{h}{m})^{\frac{5}{3}}
+\frac{N^2}{m}\Big(d\textit{l}+\frac{h}{m}\Big)
\Big(\ln\big(1+\frac{h}{d\textit{l}}\big)\Big)^{\frac{1}{2}}\Big\}\nonumber\\
&\quad\quad\quad\times\Big\{H_{21}+\frac{N}{m^2}
\Big(\ln\big(1+\frac{h}{d\textit{l}}\big)\Big)(\mathcal{N}_++1)^2\Big\}\nonumber\\
&+N^{\frac{3}{2}}a^{\frac{1}{2}}d^{-\frac{1}{2}}
(d\textit{l}+\frac{h}{m})(H_4+H_4^\prime).
\end{align}
\end{lemma}
\noindent
\emph{Proof.} We use (\ref{define W tilde _p,1,2,3}) to split the estimate into three parts $H_{23}^{\prime\prime}=\sum_{i=1}^{3}H_{23,i}^{\prime\prime}$, and we are going to show
\begin{equation*}
  e^{-\tilde{B}^\prime}H_{23,i}^{\prime\prime} e^{\tilde{B}^\prime}
=H_{23,i}^{\prime\prime}+\tilde{\mathcal{E}}^{\tilde{B}^\prime}_{23,i}.
\end{equation*}
The estimates of the first two parts go in a similar way of the proof of Lemma \ref{lemma e^-B'H_23'e^B'} for Region III as long as we notice that $\Vert\widetilde{W}_i\Vert_1\leq Cad^{-\frac{1}{2}}$ for $i=1,2$. Here, we again use the notation $\widetilde{W}_i=\sum_p\widetilde{W}_{p,i}\phi^{(d)}_p$. The first two error terms can be bounded by
\begin{align*}
 \pm\big(\tilde{\mathcal{E}}^{\tilde{B}^\prime}_{23,1}
+\tilde{\mathcal{E}}^{\tilde{B}^\prime}_{23,2}\big)
&\leq CN^{\frac{5}{2}}a^{\frac{3}{2}}d^{-\frac{3}{2}}
(d\textit{l}+\frac{h}{m})
(\mathcal{N}_++1)
+N^{\frac{3}{2}}a^{\frac{1}{2}}d^{-\frac{1}{2}}
(d\textit{l}+\frac{h}{m})(H_4+H_4^\prime)\\
& +CN^{3}ad^{-1}
(d\textit{l}+\frac{h}{m})^{\frac{5}{3}}
\Big\{H_{21}+\frac{N}{m^2}
\Big(\ln\big(1+\frac{h}{d\textit{l}}\big)\Big)(\mathcal{N}_++1)^2\Big\}.
\end{align*}
 For the analysis of the third part, we write
\begin{equation*}
  \tilde{\mathcal{E}}_{23,3}^{\tilde{B}^\prime}= e^{-\tilde{B}^\prime}H_{23,3}^{\prime\prime}e^{\tilde{B}^\prime}-H_{23,3}^{\prime\prime}
=\int_{0}^{1}e^{-t\tilde{B}^\prime}
[H_{23,3}^{\prime\prime},\tilde{B}^\prime]e^{t\tilde{B}^\prime}dt.
\end{equation*}
Calculating directly gives
\begin{equation*}
  [H_{23,3}^{\prime\prime},\tilde{B}^\prime]
=\sum_{i=1}^{3}\tilde{\mathcal{E}}_{23,3,i}^{\tilde{B}^\prime},
\end{equation*}
where
\begin{align*}
\tilde{\mathcal{E}}_{23,3,1}^{\tilde{B}^\prime}&=
4\sum_{p,q,p+q\neq0}\xi_p\widetilde{W}_{p,3}
(a_0^*a_0^*a_{p+q}^*a_pa_qa_0+h.c.),\\
\tilde{\mathcal{E}}_{23,3,2}^{\tilde{B}^\prime}&=
-2\sum_{p,q,p+q\neq0}\xi_p\widetilde{W}_{q,3}
(a_{p+q}^*a_{-p}^*a_{-q}^*a_0a_0a_0+h.c.),\\
\tilde{\mathcal{E}}_{23,3,3}^{\tilde{B}^\prime}&=
-2\sum_{p,q,p+q,r\neq0}\xi_p\widetilde{W}_{r,3}
(a_{p+q}^*a_{-p}^*a_0^*a_{r}a_{-r}a_q+h.c.).
\end{align*}
Let $\psi\in L^2_s(\Lambda_d^N)$, we can bound directly
\begin{align*}
\vert\langle\tilde{\mathcal{E}}_{23,3,1}^{\tilde{B}^\prime}\psi,\psi\rangle\vert
&\leq\frac{C}{\sqrt{d}}\Vert\xi_\perp\ast \widetilde{W}_3\Vert_2
\langle(\mathcal{N}_++1)^\frac{3}{2}a_0^*a_0^*a_0\psi,a_0^*a_0^*a_0\psi\rangle^{\frac{1}{2}}
\langle(\mathcal{N}_++1)^\frac{3}{2}\psi,\psi\rangle^{\frac{1}{2}}\\
&\leq CN^{2}ad^{-1}(d\textit{l}+hm^{-1})
\langle(\mathcal{N}_++1)\psi,\psi\rangle,
\end{align*}
where we have used (\ref{est of widetilde W_3}) again. To bound the other two, we make use of equation (\ref{H_21psi,psi fock}) and get

\begin{align*}
\vert\langle\tilde{\mathcal{E}}_{23,3,2}^{\tilde{B}^\prime}\psi,\psi\rangle\vert
&\leq C\Vert\xi_\perp\Vert_2\Vert\nabla_{\mathbf{x}}\xi\Vert_2
\langle(\mathcal{N}_++1)^2a_0^3\psi,a_0^3\psi\rangle^{\frac{1}{2}}
\langle H_{21}\psi,\psi\rangle^{\frac{1}{2}}\\
&\leq C\frac{N^2}{m}\Big(d\textit{l}+\frac{h}{m}\Big)
\Big(\ln\big(1+\frac{h}{d\textit{l}}\big)\Big)^{\frac{1}{2}}
\langle(\mathcal{N}_++1)\psi,\psi\rangle^{\frac{1}{2}}
\langle H_{21}\psi,\psi\rangle^{\frac{1}{2}},
\end{align*}
and
\begin{align*}
\vert\langle\tilde{\mathcal{E}}_{23,3,3}^{\tilde{B}^\prime}\psi,\psi\rangle\vert
&\leq C\Vert\xi_\perp\Vert_2\Vert\nabla_{\mathbf{x}}\xi\Vert_2
\langle(\mathcal{N}_++1)^4\psi,\psi\rangle^{\frac{1}{2}}
\langle H_{21} a_0^*\psi,a_0^*\psi\rangle^{\frac{1}{2}}\\
&\leq C\frac{N^2}{m}\Big(d\textit{l}+\frac{h}{m}\Big)
\Big(\ln\big(1+\frac{h}{d\textit{l}}\big)\Big)^{\frac{1}{2}}
\langle(\mathcal{N}_++1)\psi,\psi\rangle^{\frac{1}{2}}
\langle H_{21}\psi,\psi\rangle^{\frac{1}{2}}.
\end{align*}
Hence we conclude this lemma using Lemmas \ref{control of S_+ conj with e^B' tilde} and \ref{lemma control of H_N conj with e^Btilde'}.
\begin{flushright}
  {$\Box$}
\end{flushright}

\begin{lemma}\label{[H_3',Btilde']}
\begin{align}\label{e^-Btilde'(H_21+H_4+H_3')e^Btilde'}
\int_{0}^{1}\int_{t}^{1}e^{-s\tilde{B}^\prime}[H_3^\prime,\tilde{B}^\prime]
e^{s\tilde{B}^\prime}dsdt=4\sum_{p\neq0}W_p
\xi_p\mathcal{N}_+(N-\mathcal{N}_+)
+\tilde{\mathcal{E}}_{[H_3^\prime,\tilde{B}^\prime]}^{\tilde{B}^\prime}.
\end{align}
where
\begin{align}\label{Etilde_[H_3'Btilde']}
 \pm\tilde{\mathcal{E}}_{[H_3^\prime,\tilde{B}^\prime]}^{\tilde{B}^\prime}
&\leq C\Big\{N^{\frac{5}{2}}a^{\frac{3}{2}}d^{-\frac{3}{2}}
(d\textit{l}+\frac{h}{m})+
N^{2}ad^{-1}
(d\textit{l}+\frac{h}{m})^{\frac{2}{3}}\Big\}
(\mathcal{N}_++1)\nonumber\\
& +CN^{2}ad^{-1}
(d\textit{l}+\frac{h}{m})^{\frac{2}{3}}\Big\{H_{21}+\frac{N}{m^2}
\Big(\ln\big(1+\frac{h}{d\textit{l}}\big)\Big)(\mathcal{N}_++1)^2\Big\}\nonumber\\
&+N^{\frac{3}{2}}a^{\frac{1}{2}}d^{-\frac{1}{2}}
(d\textit{l}+\frac{h}{m})H_4^\prime
+CNa^2d^{-2}(d\textit{l})\ln(d\textit{l})^{-1}H_{21}.
\end{align}
\end{lemma}
\noindent
\emph{Proof.} We first have
\begin{align}
\int_{0}^{1}\int_{t}^{1}e^{-s\tilde{B}^\prime}[H_3^\prime,\tilde{B}^\prime]
e^{s\tilde{B}^\prime}dsdt=
2\sum_{p,q\neq0}(W_p+W_{p+q})
\xi_pa_q^*a_qa_0^*a_0+\tilde{\mathcal{E}}_{[H_3^\prime,B^\prime],0}^{\tilde{B}^\prime},
\label{e^-Btilde'(H_21+H_4+H_3')e^Btilde' temp III}
\end{align}
and the error term is bounded by
\begin{align}\label{Etilde_[H_3'Btilde'],0}
\pm\tilde{\mathcal{E}}_{[H_3^\prime,B^\prime],0}^{\tilde{B}^\prime}
&\leq C\Big\{N^{\frac{5}{2}}a^{\frac{3}{2}}d^{-\frac{3}{2}}
(d\textit{l}+\frac{h}{m})+
N^{2}ad^{-1}
(d\textit{l}+\frac{h}{m})^{\frac{2}{3}}\Big\}
(\mathcal{N}_++1)\nonumber\\
& +CN^{2}ad^{-1}
(d\textit{l}+\frac{h}{m})^{\frac{2}{3}}\Big\{H_{21}+\frac{N}{m^2}
\Big(\ln\big(1+\frac{h}{d\textit{l}}\big)\Big)(\mathcal{N}_++1)^2\Big\}\nonumber\\
&+N^{\frac{3}{2}}a^{\frac{1}{2}}d^{-\frac{1}{2}}
(d\textit{l}+\frac{h}{m})H_4^\prime.
\end{align}
See Lemmas \ref{lemma [H_3,B']} and \ref{lemma [H'_3,B']} for Region III for details of (\ref{e^-Btilde'(H_21+H_4+H_3')e^Btilde' temp III}).

\par Furthermore, with the fact that
\begin{equation*}
  \vert W_{p+q}-W_p\vert\leq
C\frac{\vert\mathcal{M}_dq\vert}{\sqrt{d}}
\left\vert\int_{\Lambda_d}W(\mathbf{x})\vert\mathbf{x}\vert d\mathbf{x}\right\vert
\leq \frac{Ca}{d}(d\textit{l})\vert\mathcal{M}_dq\vert,
\end{equation*}
and Lemma \ref{xi_p Y_p lemma}, we have
\begin{align}
  \pm\sum_{p,q\neq0}(W_{p+q}-W_p)\xi_pa_q^*a_qa_0^*a_0
&\leq \frac{CNa}{d}(d\textit{l})\sum_{p\neq0}\vert\xi_p\vert H_{21}\nonumber\\
&\leq CNa^2d^{-2}(d\textit{l})\ln(d\textit{l})^{-1}H_{21}.\label{010799}
\end{align}
Hence we conclude the proof combining (\ref{e^-Btilde'(H_21+H_4+H_3')e^Btilde' temp III}) and (\ref{010799}).
\begin{flushright}
  {$\Box$}
\end{flushright}

\begin{lemma}\label{[H_3'',Btilde']}
\begin{align}\label{e^-Btilde'(H_21+H_4+H_3'')e^Btilde'}
\int_{0}^{1}\int_{0}^{t}e^{-s\tilde{B}^\prime}[H_3^{\prime\prime},\tilde{B}^\prime]
e^{s\tilde{B}^\prime}dsdt=4\sum_{p\neq0}\widetilde{W}_p
\xi_p\mathcal{N}_+(N-\mathcal{N}_+)
+\tilde{\mathcal{E}}_{[H_3^{\prime\prime},\tilde{B}^\prime]}^{\tilde{B}^\prime}.
\end{align}
where
\begin{align}\label{Etilde_[H_3''Btilde']}
 \pm\tilde{\mathcal{E}}_{[H_3^\prime,\tilde{B}^\prime]}^{\tilde{B}^\prime}
&\leq C\Big\{N^{\frac{5}{2}}a^{\frac{3}{2}}d^{-\frac{3}{2}}
(d\textit{l}+\frac{h}{m})+
N^{2}ad^{-1}
(d\textit{l}+\frac{h}{m})^{\frac{2}{3}}\nonumber\\
&\quad\quad\quad+\frac{N^2}{m}\Big(d\textit{l}+\frac{h}{m}\Big)
\Big(\ln\big(1+\frac{h}{d\textit{l}}\big)\Big)^{\frac{1}{2}}\Big\}
(\mathcal{N}_++1)\nonumber\\
& +C\Big\{N^{2}ad^{-1}
(d\textit{l}+\frac{h}{m})^{\frac{2}{3}}
+\frac{N^2}{m}\Big(d\textit{l}+\frac{h}{m}\Big)
\Big(\ln\big(1+\frac{h}{d\textit{l}}\big)\Big)^{\frac{1}{2}}\Big\}\nonumber\\
&\quad\quad\quad\times\Big\{H_{21}+\frac{N}{m^2}
\Big(\ln\big(1+\frac{h}{d\textit{l}}\big)\Big)(\mathcal{N}_++1)^2\Big\}\nonumber\\
&+N^{\frac{3}{2}}a^{\frac{1}{2}}d^{-\frac{1}{2}}
(d\textit{l}+\frac{h}{m})(H_4+H_4^\prime)
+CNa^2d^{-2}h\ln(d\textit{l})^{-1}H_{21}.
\end{align}
\end{lemma}
\noindent
\emph{Proof.} We still divide the estimate into three parts. For the details of the proof however, one can consult the proof of Lemmas \ref{lemma e^-Btilde'H_23''e^Btilde'} and \ref{[H_3',Btilde']}. We replace $W$ by $\widetilde{W}_i$ and use the changed and needed estimates that, for $i=1,2,3$,
\begin{equation}\label{l1 Wtilde}
  \Vert\widetilde{W}_{i}\Vert_1\leq\frac{Ca}{\sqrt{d}},
\end{equation}
and
\begin{align*}
 \vert \widetilde{W}_{p+q,1\,or\,2}-\widetilde{W}_{p,1\,or\,2}\vert&\leq
C\frac{\vert\mathcal{M}_dq\vert}{\sqrt{d}}
\left\vert\int_{\Lambda_d}\widetilde{W}_{1\,or\,2}
(\mathbf{x})\vert\mathbf{x}\vert d\mathbf{x}\right\vert
\leq \frac{Ca}{d}(d\textit{l})\vert\mathcal{M}_dq\vert,\\
  \vert \widetilde{W}_{p+q,3}-\widetilde{W}_{p,3}\vert&\leq
C\frac{\vert\mathcal{M}_dq\vert}{\sqrt{d}}
\left\vert\int_{\Lambda_d}\widetilde{W}_3
(\mathbf{x})\vert\mathbf{x}\vert d\mathbf{x}\right\vert
\leq \frac{Ca}{d}h\vert\mathcal{M}_dq\vert.
\end{align*}
Further details are omitted.
\begin{flushright}
  {$\Box$}
\end{flushright}

\noindent
\emph{Proof of Proposition \ref{quasi-2d/3d quadratic renorm}.}
\par Putting all the estimates above together, we conclude
\begin{align*}
  \mathcal{M}_N=e^{-\tilde{B}^\prime}\mathcal{L}_Ne^{\tilde{B}^\prime}=&
N(N-1)\tilde{C}^{\tilde{B}^\prime}+2N\tilde{C}^{\tilde{B}^\prime}\mathcal{N}_+
-3\tilde{C}^{\tilde{B}^\prime}\mathcal{N}_+^2\nonumber\\
&+H_{21}+H_4+H_{23}^{\prime\prime}+H_3^{\prime\prime}+\tilde{\mathcal{E}}^{\tilde{B}^\prime}
\end{align*}
where
\begin{align*}
\tilde{C}^{\tilde{B}^\prime}=\Big(
W_0+\sum_{p\neq0}W_p\eta_p+\sum_{p\neq0}W_p\xi_p
+\sum_{p\neq0}\widetilde{W}_p\xi_p\Big)
\end{align*}
and the error term is bounded by
\begin{align*}
 \pm\tilde{\mathcal{E}}^{\tilde{B}^\prime}
\leq& C\Big\{ad^{-1}+N^4a^{\frac{8}{3}}d^{-2}\textit{l}^{\frac{1}{3}}
+N^3a^2d^{-2}(d\textit{l}+hm^{-1})
+N^{\frac{5}{2}}a^{\frac{3}{2}}d^{-\frac{3}{2}}(d\textit{l}+hm^{-1})^{\frac{2}{3}}\nonumber\\
&\quad\quad\quad+N^3m^{-1}(d\textit{l}+hm^{-1})
\Big(\ln\big(1+\frac{h}{d\textit{l}}\big)\Big)^{\frac{1}{2}}\Big\}
(\mathcal{N}_++1)\nonumber\\
&+C\Big\{N^2a^{\frac{5}{3}}d^{-1}\textit{l}^{\frac{1}{3}}
+N^{\frac{7}{2}}a^{\frac{3}{2}}d^{-\frac{3}{2}}(d\textit{l}+hm^{-1})^{\frac{5}{3}}
+N^{2}ad^{-1}(d\textit{l}+hm^{-1})^{\frac{2}{3}}\nonumber\\
&\quad\quad\quad+N^2m^{-1}(d\textit{l}+hm^{-1})
\Big(\ln\big(1+\frac{h}{d\textit{l}}\big)\Big)^{\frac{1}{2}}
+Na^2d^{-2}h\ln(d\textit{l})^{-1}\Big\}\nonumber\\
&\times \Big(H_{21}+Nm^{-2}\ln\big(1+\frac{h}{d\textit{l}}\big)
(\mathcal{N}_++1)^2\Big)\nonumber\\
&+C\Big(N^{\frac{3}{2}}a^{\frac{7}{6}}d^{-\frac{1}{2}}\textit{l}^{\frac{1}{3}}
+N^2ad^{-1}(d\textit{l}+hm^{-1})
+N^{\frac{3}{2}}a^{\frac{1}{2}}d^{-\frac{1}{2}}(d\textit{l}+hm^{-1})^{\frac{2}{3}}\Big)
H_4\nonumber\\
&+CN^{\frac{3}{2}}a^{\frac{1}{2}}d^{-\frac{1}{2}}(d\textit{l}+hm^{-1})H_4^\prime.
\end{align*}
Thus we conclude the proof of Proposition \ref{quasi-2d/3d quadratic renorm}.\hfill$\Box$

\section{Dimensional Coupling Renormalization for Region III}\label{6}
\par In this section, we compute the dimensional coupling renormalization and prove Proposition \ref{dimensional coupling quadratic renorm}. We analyze the excitation Hamiltonian $\mathcal{R}_N$ generated by the quadratic dimensional coupling renormalization in Section \ref{dimensional coupling quadratic}, and we analyze $\mathcal{S}_N$ generated by the cubic dimensional coupling renormalization in Section \ref{dimensional coupling cubic}. The direct computations of $\mathcal{R}_N$ and $\mathcal{S}_N$ are still sort of similar to those of $\mathcal{G}_N$ and $\mathcal{J}_N$, however, there are subtle differences in details, which in the end made the problem doable. The key point of this section is to turn the cubic quasi-2D correlation remainder $H_3^{\prime\prime}$, which can not yet be eliminated, to a cubic dimensional coupling correlation remainder $H_{3}^{\prime\prime\prime}$, so that it can be considered as a small error (see Lemma \ref{lemma H_3''' O'}) and we can apply Bogoliubov transformation in Section \ref{7}.

\subsection{Dimensional Coupling Quadratic Renormalization}
\label{dimensional coupling quadratic}
\
\par We apply the notation
\begin{equation*}
  k_\perp={\sum_{p\neq0}}k_p\phi_p^{(d)}\in L^2_{\perp}(\Lambda_{d}).
\end{equation*}
$k_p$ have been defined in Section \ref{Dimensional Coupling Scattering Equation Section}.  Recall the definition of $q_p$ and $\mathfrak{D}_p$ in Section \ref{Dimensional Coupling Scattering Equation Section}, and the definition of $Y_p$ in (\ref{define Y_p}), and the definition of $\widetilde{W}_p$ in (\ref{define W tilde}), we can rewrite
\begin{equation}\label{rewrite Wtilde_p}
  \widetilde{W}_p=Y_p+\mathfrak{D}_p-\frac{1}{2\sqrt{d}}v_p^{(a,d)}\xi_0
\end{equation}
Now we rewrite $e^{-\mathcal{O}}\mathcal{M}_Ne^{\mathcal{O}}$ using (\ref{fourth/fifth renorm})
\begin{align}
  e^{-\mathcal{O}}\mathcal{M}_Ne^{\mathcal{O}}
=&N(N-1)\tilde{C}^{\tilde{B}^\prime}
+e^{-\mathcal{O}}(2N\tilde{C}^{\tilde{B}^\prime}\mathcal{N}_+
-3\tilde{C}^{\tilde{B}^\prime}\mathcal{N}_+^2+H_3^{\prime\prime})e^{\mathcal{O}}\nonumber\\
&+\int_{0}^{1}\int_{t}^{1}
e^{-s\mathcal{O}}[H_{23}^{\prime\prime},\mathcal{O}] e^{s\mathcal{O}}dsdt
+\int_{0}^{1}\int_{0}^{t}
e^{-s\mathcal{O}}[H_{23}^{\prime\prime\prime},\mathcal{O}] e^{s\mathcal{O}}dsdt\nonumber\\
&+\int_{0}^{1}e^{-t\mathcal{O}}\Omega e^{t\mathcal{O}}dt
+e^{-\mathcal{O}}\tilde{\mathcal{E}}^{\tilde{B}^\prime}e^{\mathcal{O}}
+H_{21}+H_4+H_{23}^{\prime\prime\prime},
\label{split e^-OM_Ne^O}
\end{align}
where we have defined
\begin{equation}\label{define Omega}
  \Omega=[H_{21}+H_4,\mathcal{O}]+H_{23}^{\prime\prime}-H_{23}^{\prime\prime\prime}
\end{equation}
and $H_{23}^{\prime\prime\prime}$ by
\begin{equation}\label{define H_23'''}
  H_{23}^{\prime\prime\prime}=
\sum_{p\neq0}\big(q_p+Y_p\big)(a_p^*a_{-p}^*a_0a_0+h.c.).
\end{equation}
Here $q_p$ have been defined in (\ref{define q}), and $Y_p$ have been defined in (\ref{define Y_p}). We state properties of (\ref{split e^-OM_Ne^O}) in the up-coming series of lemmas, while we omit the details of proofs except some new estimates came upon. In the following lemmas, we bound $e^{-\mathcal{O}}(2N\tilde{C}^{\tilde{B}^\prime}\mathcal{N}_+
-3\tilde{C}^{\tilde{B}^\prime}\mathcal{N}_+^2)e^{\mathcal{O}}$ in Corollary \ref{corollary e^-Odiage^O}, $e^{-\mathcal{O}}\tilde{\mathcal{E}}^{\tilde{B}^\prime}e^{\mathcal{O}}$ in Corollary \ref{corollary e^-OEtilde^Btilde'e^O}, and $e^{-\mathcal{O}}H_3^{\prime\prime}e^{\mathcal{O}}$ in Lemma \ref{lemma e^-OH_3''e^O}. These three terms stay unchanged up to small errors after conjugating with $e^{\mathcal{O}}$. The term containing the difference $\Omega$ is bounded in Lemma \ref{lemma int_0^1e^-tOOmegae^tOdt}. and proved to be a negligible error term. The contribution of the commutator $[H_{23}^{\prime\prime},\mathcal{O}]$ is calculated in Lemma \ref{lemma [H_23'',O]}, and the contribution of $[H_{23}^{\prime\prime\prime},\mathcal{O}]$ is calculated in Lemma \ref{lemma [H_23''',O]}. Lemmas \ref{lemma [H_23'',O]} and \ref{lemma [H_23''',O]} present the major contributions of the quadratic dimensional coupling correlation structure to the second order ground state energy, in the form of polynomials of $\mathcal{N}_+$. We bound the growths of $\mathcal{N}_+$, $H_{21}$, $H_4$ and $H_4^\prime$ in Lemmas \ref{control of S_+ conj with e^O} and \ref{lemma control of H_4 with e^O}.

\begin{lemma}\label{control of S_+ conj with e^O}
Let $\mathcal{N_+}$ be defined on $L_s^2(\Lambda_d^N)$ as stated in (\ref{define of N_+}), then there exist a constant $C_n$ depending only on $n\in\frac{1}{2}\mathbb{N}$ such that: for every $t\in\mathbb{R}$, $N\in\mathbb{N}$, $n\in\frac{1}{2}\mathbb{N}$, $\textit{l}\in(0,\frac{1}{2})$ such that $\frac{d\textit{l}}{a}>C$, and $h\in(0,\frac{1}{2})$ such that $\frac{h}{d\textit{l}}>C$ for some universal constant $C$. Then we have
\begin{align}
  e^{-t\mathcal{O}}(\mathcal{N}_++1)^ne^{t\mathcal{O}}&\leq e^{C_nNa\textit{l}^{\frac{1}{2}}\vert t\vert}(\mathcal{N}_++1)^n,\label{e^-tO(N_++1)e^tO}\\
  \pm(e^{-t\mathcal{O}}(\mathcal{N}_++1)^ne^{t\mathcal{O}}-(\mathcal{N}_++1)^n)&\leq (e^{C_nNa\textit{l}^{\frac{1}{2}}\vert t\vert}-1) (\mathcal{N}_++1)^n.\label{e^-tO N_+e^tO-N_+}
\end{align}
\end{lemma}
\noindent
\emph{Proof.} See the proof of Lemma \ref{control of S_+ conj with e^B} for details.
\begin{flushright}
  {$\Box$}
\end{flushright}

\par We always require in this section that $N$ tends to infinity, $a$, $d$, $\frac{a}{d}$, $N^{\frac{3}{2}}a^{\frac{7}{6}}d^{-\frac{1}{2}}\textit{l}^{\frac{1}{3}}$ and $Na\textit{l}^{\frac{1}{2}}$ tend to $0$ and $\frac{d\textit{l}}{a}>C$. Moreover, we ask additionally $\frac{h}{d\textit{l}}>C$, $\frac{Na}{d}>C$, $\frac{ma}{d}>C$ and $N(d\textit{l}+\frac{h}{m})$ and $N^{\frac{3}{2}}a^{\frac{1}{2}}d^{-\frac{1}{2}}(d\textit{l}+\frac{h}{m})^{\frac{2}{3}}$ should tend to $0$. The actions of $e^{\mathcal{O}}$ on $H_{21}$, $H_4$ and $H_4^\prime$ are controlled similarly to Lemma \ref{lemma control of H_4 with e^B tilde}. We state the result while omitting further details in the next lemma. Notice that we make use of Lemma \ref{lemma q} to bound $\Vert k\Vert_2$ and $\Vert \nabla_{\mathbf{x}}k\Vert_2$.
\begin{lemma}\label{lemma control of H_4 with e^O}
\begin{align}
e^{-t\mathcal{O}}H_{21}e^{t\mathcal{O}}&\leq C(H_{21}+N^2ad^{-1}(\mathcal{N}_++1)),\label{control of e^-OH_21e^O}\\
  e^{-t\mathcal{O}}H_4e^{t\mathcal{O}}&\leq C(H_4+N^2ad^{-1}),\label{control of e^-OH_4e^O}\\
 e^{-t\mathcal{O}}H_4^\prime e^{t\mathcal{O}}&\leq C(H_4^\prime+N^2ad^{-1}).\label{control of e^-OH_4'e^O}
\end{align}
for all $\vert t\vert\leq1$.
\end{lemma}
\noindent
\emph{Proof.} See the proofs of Lemmas \ref{lemma control of H_4} and \ref{lemma control of H_4 with e^B tilde} for details.
\begin{flushright}
  {$\Box$}
\end{flushright}

As a direct consequence of Lemmas \ref{control of S_+ conj with e^O} and \ref{lemma control of H_4 with e^O}, and the fact that $\vert\tilde{C}^{\tilde{B}^\prime}\vert\leq Cad^{-1}$, we have
\begin{corollary}\label{corollary e^-Odiage^O}
\begin{equation}\label{e^-Odiage^O}
  e^{-\mathcal{O}}(2N\tilde{C}^{\tilde{B}^\prime}\mathcal{N}_+
-3\tilde{C}^{\tilde{B}^\prime}\mathcal{N}_+^2)e^{\mathcal{O}}
=2N\tilde{C}^{\tilde{B}^\prime}\mathcal{N}_+
-3\tilde{C}^{\tilde{B}^\prime}\mathcal{N}_+^2+\tilde{\mathcal{E}}^{\mathcal{O}}_{diag},
\end{equation}
where
\begin{equation}\label{Etilde^O_diag}
  \pm\tilde{\mathcal{E}}^{\mathcal{O}}_{diag}\leq N^2a^2d^{-1}\textit{l}^{\frac{1}{2}}
(\mathcal{N}_++1).
\end{equation}
\end{corollary}

\begin{corollary}\label{corollary e^-OEtilde^Btilde'e^O}
\begin{align}\label{e^-OEtilde^Btilde'e^O}
 &\pm e^{-\mathcal{O}}\tilde{\mathcal{E}}^{\tilde{B}^\prime}e^{\mathcal{O}}\nonumber\\
\leq& C\Big\{ad^{-1}+N^4a^{\frac{8}{3}}d^{-2}\textit{l}^{\frac{1}{3}}
+N^{4}a^{2}d^{-2}(d\textit{l}+hm^{-1})^{\frac{2}{3}}
+N^2a^3d^{-3}h\ln(d\textit{l})^{-1}\nonumber\\
&\quad\quad\quad+N^4ad^{-1}m^{-1}(d\textit{l}+hm^{-1})
\Big(\ln\big(1+\frac{h}{d\textit{l}}\big)\Big)^{\frac{1}{2}}\Big\}
(\mathcal{N}_++1)\nonumber\\
&+C\Big\{N^2a^{\frac{5}{3}}d^{-1}\textit{l}^{\frac{1}{3}}
+N^{\frac{7}{2}}a^{\frac{3}{2}}d^{-\frac{3}{2}}(d\textit{l}+hm^{-1})^{\frac{5}{3}}
+N^{2}ad^{-1}(d\textit{l}+hm^{-1})^{\frac{2}{3}}\nonumber\\
&\quad\quad\quad+N^2m^{-1}(d\textit{l}+hm^{-1})
\Big(\ln\big(1+\frac{h}{d\textit{l}}\big)\Big)^{\frac{1}{2}}
+Na^2d^{-2}h\ln(d\textit{l})^{-1}\Big\}\nonumber\\
&\times \Big(H_{21}+Nm^{-2}\ln\big(1+\frac{h}{d\textit{l}}\big)
(\mathcal{N}_++1)^2\Big)\nonumber\\
&+C\Big(N^{\frac{3}{2}}a^{\frac{7}{6}}d^{-\frac{1}{2}}\textit{l}^{\frac{1}{3}}
+N^2ad^{-1}(d\textit{l}+hm^{-1})
+N^{\frac{3}{2}}a^{\frac{1}{2}}d^{-\frac{1}{2}}(d\textit{l}+hm^{-1})^{\frac{2}{3}}\Big)
H_4\nonumber\\
&+CN^{\frac{3}{2}}a^{\frac{1}{2}}d^{-\frac{1}{2}}(d\textit{l}+hm^{-1})H_4^\prime.
\end{align}
\end{corollary}

\par The analysis of the rest of (\ref{split e^-OM_Ne^O}) is similar to Section \ref{2}. In this section, we replace the bounds of $\eta$ by $k$, and substitute the estimates of $v_a$ by $2\sqrt{d}\widetilde{W}_i$, $q$ or $Y$, and thus some subtle differences in details arise.

\begin{lemma}\label{lemma e^-OH_3''e^O}
\begin{equation}\label{e^-OH_3''e^O}
  e^{-\mathcal{O}}H_3^{\prime\prime}e^{\mathcal{O}}
=H_3^{\prime\prime}+\tilde{\mathcal{E}}^{\mathcal{O}}_{3},
\end{equation}
where
\begin{align}\label{Etilde^O_3}
  \pm\tilde{\mathcal{E}}^{\mathcal{O}}_{3}\leq& N^2a^2d^{-1}\textit{l}^{\frac{1}{2}}
(\mathcal{N}_++1)+CN^{\frac{3}{2}}a^{\frac{3}{2}}d^{-\frac{1}{2}}\textit{l}^{\frac{1}{2}}
(H_4+H_4^\prime)\nonumber\\
&+CNa\textit{l}^{\frac{1}{2}}\Big(H_{21}+Nm^{-2}\ln\big(1+\frac{h}{d\textit{l}}\big)
(\mathcal{N}_++1)^2\Big).
\end{align}
\end{lemma}
\noindent
\emph{Proof.} Once again, we use (\ref{define W tilde _p,1,2,3}) to split the calculation into three parts. Let
\begin{equation*}
  H_{3,i}^{\prime\prime}=2\sum_{p,q,p+q\neq0}\widetilde{W}_{p,i}(a_{p+q}^*a_{-p}^*a_qa_0+h.c.).
\end{equation*}
We again use the estimates (\ref{l1 Wtilde}) that
\begin{equation*}
  \Vert\widetilde{W}_i\Vert_1\leq Cad^{-\frac{1}{2}}.
\end{equation*}
Now following the calculations given in Lemma \ref{lemma e^-BH_3e^B}, and using Lemma \ref{lemma q} to bound the $\Vert k\Vert_2$, we arrive at
\begin{equation*}
  e^{-\mathcal{O}}\big(H_{3,1}^{\prime\prime}
+H_{3,2}^{\prime\prime}\big)e^{\mathcal{O}}
=H_{3,1}^{\prime\prime}+H_{3,2}^{\prime\prime}+\tilde{\mathcal{E}}^{\mathcal{O}}_{3,1+2},
\end{equation*}
where
\begin{align*}
  \pm\tilde{\mathcal{E}}^{\mathcal{O}}_{3,1+2}
\leq CN^2a^2d^{-1}\textit{l}^{\frac{1}{2}}(\mathcal{N}_++1)
+CN^{\frac{3}{2}}a^{\frac{3}{2}}d^{-\frac{1}{2}}\textit{l}^{\frac{1}{2}}
(H_4+H_4^\prime).
\end{align*}
The estimate to the third part needs a slight modification like Lemmas \ref{lemma [H_23'',B tilde]} and \ref{lemma e^-Btilde'H_23''e^Btilde'}. Following Lemmas \ref{lemma [H_23'',B tilde]} and \ref{lemma e^-Btilde'H_23''e^Btilde'}, we use (\ref{est of grad xi}) to bound $\Vert\nabla_{\mathbf{x}}\xi\Vert_2$ to arrive at
\begin{equation*}
  e^{-\mathcal{O}}H_{3,3}^{\prime\prime}e^{\mathcal{O}}
=H_{3,3}^{\prime\prime}+\tilde{\mathcal{E}}^{\mathcal{O}}_{3,3},
\end{equation*}
where
\begin{align*}
  \pm\tilde{\mathcal{E}}^{\mathcal{O}}_{3,3}
\leq CN^2a^2d^{-1}\textit{l}^{\frac{1}{2}}(\mathcal{N}_++1)
+CNa\textit{l}^{\frac{1}{2}}\Big(H_{21}+Nm^{-2}\ln\big(1+\frac{h}{d\textit{l}}\big)
(\mathcal{N}_++1)^2\Big).
\end{align*}
We then conclude the proof.
\begin{flushright}
  {$\Box$}
\end{flushright}

\begin{lemma}\label{lemma int_0^1e^-tOOmegae^tOdt}
\begin{align}\label{int_0^1e^-tOOmegae^tOdt}
  \pm\int_{0}^{1}e^{-t\mathcal{O}}\Omega e^{t\mathcal{O}}dt
&\leq C\Big(N^{\frac{3}{2}}a^{\frac{3}{2}}d^{-\frac{1}{2}}\textit{l}^{\frac{1}{2}}
+Na^{\frac{1}{2}}d^{-\frac{1}{2}}\big((d\textit{l})^2
+h^2m^{-1}\big)\Big)H_4\nonumber\\
&+CN^{2}a^{2}d^{-1}\textit{l}^{\frac{1}{2}}(\mathcal{N}_++1)
+N^3a^{\frac{3}{2}}d^{-\frac{3}{2}}\big((d\textit{l})^2
+h^2m^{-1}\big).
\end{align}
\end{lemma}
\noindent
\emph{Proof.} Using equation (\ref{discrete dimensional coupling scattering equation}), a calculation similar to Lemma \ref{lemma Gamma} gives
\begin{align*}
  \Omega=&\frac{1}{\sqrt{d}}{\sum_{p,q,p+r,q+r\neq0}}v_r^{(a,d)}
  k_p(a_{p+r}^*a_q^*a_{-p}^*a_{q+r}a_0a_0+h.c.)\\
&-\frac{1}{2\sqrt{d}}\sum_{p\neq0}v_p^{(a,d)}\big(\xi_0+k_0\big)
(a_p^*a_{-p}^*a_0a_0+h.c.)\eqqcolon\Omega_1+\Omega_2.
\end{align*}
The calculation in the proof of Lemma \ref{lemma Gamma} gives
\begin{equation*}
  \pm\int_{0}^{1}e^{-t\mathcal{O}}\Omega_1 e^{t\mathcal{O}}dt\leq
CN^{\frac{3}{2}}a^{\frac{3}{2}}d^{-\frac{1}{2}}\textit{l}^{\frac{1}{2}}H_4
+CN^{2}a^{2}d^{-1}\textit{l}^{\frac{1}{2}}(\mathcal{N}_++1).
\end{equation*}
where we have used (\ref{douche2}) to bound $\Vert k\Vert_2$ and $\Vert k(\mathbf{x})\Vert_\infty$. We can also get
\begin{equation*}
  \pm\Omega_2\leq Na^{\frac{1}{2}}d^{-\frac{1}{2}}\big((d\textit{l})^2
+h^2m^{-1}\big)(H_4+1)
\end{equation*}
by using (\ref{douche2}) and (\ref{est of xi_p}) to bound $\vert k_p\vert$ and $\vert\xi_p\vert$ respectively. Together with Lemma \ref{lemma control of H_4 with e^O} we conclude this Lemma \ref{lemma int_0^1e^-tOOmegae^tOdt}.
\begin{flushright}
  {$\Box$}
\end{flushright}

\begin{lemma}\label{lemma [H_23'',O]}
\begin{align}
   \int_{0}^{1}\int_{t}^{1}e^{-s\mathcal{O}}[{H}_{23}^{\prime\prime},\mathcal{O}]
e^{s\mathcal{O}}&dsdt
   = N(N-1)\sum_{p\neq0}\widetilde{W}_pk_p
-2N\sum_{p\neq0}\widetilde{W}_pk_p\mathcal{N}_+\nonumber\\
&+\sum_{p\neq0}\widetilde{W}_pk_p\mathcal{N}_+^2
+\tilde{\mathcal{E}}^{\mathcal{O}}_{[H_{23}^{\prime\prime},\mathcal{O}]},\label{[H_23'',O]}
\end{align}
where
\begin{align}
  \pm\tilde{\mathcal{E}}^{\mathcal{O}}_{[H_{23}^{\prime\prime},\mathcal{O}]}
\leq& C\big(N^2a^2d^{-1}\textit{l}^{\frac{1}{2}}+ad^{-1}\big)(\mathcal{N}_++1)
+CN^{\frac{3}{2}}a^{\frac{3}{2}}d^{-\frac{1}{2}}\textit{l}^{\frac{1}{2}}
(H_4+H_4^\prime)\nonumber\\
&+C\frac{Na\textit{l}^{\frac{1}{2}}}{m}
\Big(\ln\big(1+\frac{h}{d\textit{l}}\big)\Big)^{\frac{1}{2}}
(H_{21}+(\mathcal{N}_++1)^3).
\label{E^O_23,1}
\end{align}
\end{lemma}
\noindent
\emph{Proof.} See the proof of Lemma \ref{lemma [H_23'',B tilde]} for details. We only need the rough bounds for $i=1,2,3$,
\begin{equation}\label{sum Wtilde_pk_p}
  \left\vert\sum_{p\neq0}\widetilde{W}_{p,i}
k_p\right\vert\leq \Vert k_\perp\Vert_\infty\Vert\widetilde{W}_i\Vert_1\leq Cad^{-1},
\end{equation}
and we use again the estimate (\ref{l1 Wtilde}) that
\begin{equation}\label{Wtilde L1 again}
  \Vert\widetilde{W}_i\Vert_1\leq Cad^{-\frac{1}{2}},\quad
\vert\widetilde{W}_{p,i}\vert\leq Cad^{-1}.
\end{equation}
\begin{flushright}
  {$\Box$}
\end{flushright}

\begin{lemma}\label{lemma [H_23''',O]}
\begin{align}
   \int_{0}^{1}\int_{0}^{t}e^{-s\mathcal{O}}[{H}_{23}^{\prime\prime\prime},\mathcal{O}]
e^{s\mathcal{O}}&dsdt
   = N(N-1)\sum_{p\neq0}\big(q_p+Y_p\big)k_p
-2N\sum_{p\neq0}\big(q_p+Y_p\big)k_p\mathcal{N}_+\nonumber\\
&+\sum_{p\neq0}\big(q_p+Y_p\big)k_p\mathcal{N}_+^2
+\tilde{\mathcal{E}}^{\mathcal{O}}_{[H_{23}^{\prime\prime\prime},\mathcal{O}]},\label{[H_23''',O]}
\end{align}
where
\begin{align}
  \pm\tilde{\mathcal{E}}^{\mathcal{O}}_{[H_{23}^{\prime\prime\prime},\mathcal{O}]}
\leq& C\big(N^2a^2d^{-1}\textit{l}^{\frac{1}{2}}+m^{-1}\big)(\mathcal{N}_++1)
+CNa^{\frac{3}{2}}d^{-\frac{3}{2}}\textit{l}^{-\frac{1}{2}}m^{-1}
(\mathcal{N}_++1)^2\nonumber\\
&+CN^{\frac{3}{2}}a^{\frac{3}{2}}d^{-\frac{1}{2}}\textit{l}^{\frac{1}{2}}H^\prime_4
+C\frac{Na\textit{l}^{\frac{1}{2}}}{m}
\Big(\ln\big(1+\frac{h}{d\textit{l}}\big)\Big)^{\frac{1}{2}}
(H_{21}+(\mathcal{N}_++1)^3).
\label{E^O_23,2}
\end{align}
\end{lemma}
\noindent
\emph{Proof.} We use (\ref{eqn of xi_p rewrt}) to divide $(q_p+Y_p)$ into three parts
\begin{equation}\label{divide q_p+Y_p}
  q_p+Y_p=q_p+\Big(W_p+\sum_q\xi_qW_{p-q}\Big)
+\vert\mathcal{M}_dp\vert^2\xi_p.
\end{equation}
For the calculation of the $q_p$ part, one can see the proof of Lemma \ref{lemma [H_23',B]} for Region I for details. For the calculation of the $W_p+\sum_q\xi_qW_{p-q}$ part, one can see the proof of Lemma \ref{lemma [H_23',B]} for Region III for details. For the calculation of the $\vert\mathcal{M}_dp\vert^2\xi_p=\widetilde{W}_{p,3}$ part, one can see the proof of Lemma \ref{lemma [H_23'',B tilde]} for details. We recall from (\ref{est of L2L1 Y_1,2}), (\ref{douche2.75}) and (\ref{douche3}), which are the estimates that will be useful here:
\begin{equation}\label{collect1}
  \begin{aligned}
&\Vert Y\Vert_1\leq \frac{C\sqrt{d}}{m}
  ,\quad\Vert Y\Vert_2\leq \frac{C}{hm}+ Ca^{\frac{1}{2}}(d\textit{l})^{-\frac{3}{2}}m^{-\frac{1}{2}},
\quad\vert Y_p\vert\leq \frac{C}{m}
\end{aligned}
\end{equation}
and
\begin{equation}\label{collect2}
   \Vert q\Vert_1\leq  \frac{C\textit{l}^{\frac{1}{2}}}{m}\sqrt{a},
\quad \Vert q\Vert_2\leq \frac{C}{m}\frac{1}{(d\textit{l})}\sqrt{\frac{a}{d}},\quad
 \vert q_p\vert\leq \frac{C\textit{l}^{\frac{1}{2}}}{m}\sqrt{\frac{a}{d}}.
\end{equation}
Moreover, we can derive
\begin{equation}\label{collect3}
  \Big\vert\sum_{p\neq0}Y_pk_p\Big\vert\leq\frac{C}{m},\quad
\Big\vert\sum_{p\neq0}q_pk_p\Big\vert\leq
\frac{C\textit{l}^{\frac{1}{2}}}{m}\sqrt{\frac{a}{d}}.
\end{equation}
\begin{flushright}
  {$\Box$}
\end{flushright}

\noindent
\emph{Analysis of $\mathcal{R}_N$.}
\par With all the estimates above, we conclude that
\begin{equation}\label{RN}
  \mathcal{R}_N=e^{-\mathcal{O}}\mathcal{M}_Ne^{\mathcal{O}}
=\tilde{C}^{\mathcal{O}}+\tilde{Q}_1^{\mathcal{O}}\mathcal{N}_+
+\tilde{Q}_2^{\mathcal{O}}\mathcal{N}_+^2+H_{21}+H_4+H_{23}^{\prime\prime\prime}
+H_3^{\prime\prime}+\mathcal{E}^{\mathcal{O}}
\end{equation}
where
\begin{align}
\tilde{C}^{\mathcal{O}}=&
N(N-1)\Big(W_0+\sum_{p\neq0}W_p\eta_p+\sum_{p\neq0}\big(W_p+\widetilde{W}_p\big)\xi_p
+\sum_{p\neq0}\big(\widetilde{W}_p+q_p+Y_p\big)k_p\Big)\nonumber\\
\tilde{Q}_1^{\mathcal{O}}=&2N\Big(W_0+\sum_{p\neq0}W_p\eta_p
+\sum_{p\neq0}\big(W_p+\widetilde{W}_p\big)\xi_p
-\sum_{p\neq0}\big(\widetilde{W}_p+q_p+Y_p\big)k_p\Big)\nonumber\\
\tilde{Q}_2^{\mathcal{O}}=&-3\Big(W_0+\sum_{p\neq0}W_p\eta_p
+\sum_{p\neq0}\big(W_p+\widetilde{W}_p\big)\xi_p
-\frac{1}{3}\sum_{p\neq0}\big(\widetilde{W}_p+q_p+Y_p\big)k_p\Big)\nonumber
\end{align}
and the error term is bounded by
\begin{align}\label{Etilde^O}
  \pm\mathcal{E}^{\mathcal{O}}
\leq& C\Big\{ad^{-1}+N^4a^{\frac{8}{3}}d^{-2}\textit{l}^{\frac{1}{3}}
+N^{4}a^{2}d^{-2}(d\textit{l}+hm^{-1})^{\frac{2}{3}}
+N^2a^3d^{-3}h\ln(d\textit{l})^{-1}\nonumber\\
&\quad\quad\quad+N^4ad^{-1}m^{-1}(d\textit{l}+hm^{-1})
\Big(\ln\big(1+\frac{h}{d\textit{l}}\big)\Big)^{\frac{1}{2}}\Big\}
(\mathcal{N}_++1)\nonumber\\
&+CNa^{\frac{3}{2}}d^{-\frac{3}{2}}\textit{l}^{-\frac{1}{2}}m^{-1}
(\mathcal{N}_++1)^2
+CNa\textit{l}^{\frac{1}{2}}m^{-1}\Big(\ln\big(1+\frac{h}{d\textit{l}}\big)\Big)^{\frac{1}{2}}
(\mathcal{N}_++1)^3\nonumber\\
&+C\Big\{N^2a^{\frac{5}{3}}d^{-1}\textit{l}^{\frac{1}{3}}
+N^{\frac{7}{2}}a^{\frac{3}{2}}d^{-\frac{3}{2}}(d\textit{l}+hm^{-1})^{\frac{5}{3}}
+N^{2}ad^{-1}(d\textit{l}+hm^{-1})^{\frac{2}{3}}\nonumber\\
&\quad\quad\quad+N^2m^{-1}(d\textit{l}+hm^{-1})
\Big(\ln\big(1+\frac{h}{d\textit{l}}\big)\Big)^{\frac{1}{2}}
+Na^2d^{-2}h\ln(d\textit{l})^{-1}\Big\}\nonumber\\
&\times \Big(H_{21}+Nm^{-2}\ln\big(1+\frac{h}{d\textit{l}}\big)
(\mathcal{N}_++1)^2\Big)\nonumber\\
&+C\Big(N^{\frac{3}{2}}a^{\frac{7}{6}}d^{-\frac{1}{2}}\textit{l}^{\frac{1}{3}}
+N^2ad^{-1}(d\textit{l}+hm^{-1})
+N^{\frac{3}{2}}a^{\frac{1}{2}}d^{-\frac{1}{2}}(d\textit{l}+hm^{-1})^{\frac{2}{3}}\Big)
H_4\nonumber\\
&+CN^{\frac{3}{2}}a^{\frac{1}{2}}d^{-\frac{1}{2}}(d\textit{l}+hm^{-1})H_4^\prime.
\end{align}
Moreover, we have the bound
\begin{equation}\label{bound C Q R_N}
  N^2\vert\tilde{Q}_2^{\mathcal{O}}\vert\leq C
N\vert\tilde{Q}_1^{\mathcal{O}}\vert\leq C\vert\tilde{C}^{\mathcal{O}}\vert
\leq CN^2ad^{-1}.
\end{equation}
\begin{flushright}
  {$\Box$}
\end{flushright}

\subsection{Dimensional Coupling Cubic Renormalization}
\label{dimensional coupling cubic}
\
\par  We use (\ref{RN}) to rewrite $e^{-\mathcal{O}^\prime}
\mathcal{R}_Ne^{\mathcal{O}^\prime}$
\begin{align}
 e^{-\mathcal{O}^\prime}\mathcal{R}_Ne^{\mathcal{O}^\prime}
=&\tilde{C}^{\mathcal{O}}+H_{21}+H_4+H_{3}^{\prime\prime\prime}
+e^{-\mathcal{O}^\prime}\big(\tilde{Q}_1^{\mathcal{O}}\mathcal{N}_+
+\tilde{Q}_2^{\mathcal{O}}\mathcal{N}_+^2+H_{23}^{\prime\prime\prime}\big)
e^{\mathcal{O}^\prime}\nonumber\\
&+\int_{0}^{1}\int_{t}^{1}
e^{-s\mathcal{O}^\prime}[H_{3}^{\prime\prime},\mathcal{O}^\prime] e^{s\mathcal{O}^\prime}dsdt
+\int_{0}^{1}\int_{0}^{t}
e^{-s\mathcal{O}^\prime}[H_{3}^{\prime\prime\prime},\mathcal{O}^\prime]
e^{s\mathcal{O}^\prime}dsdt\nonumber\\
&+\int_{0}^{1}e^{-t\mathcal{O}^\prime}\Omega^\prime e^{t\mathcal{O}^\prime}dt
+e^{-\mathcal{O}^\prime}\tilde{\mathcal{E}}^{\mathcal{O}}e^{\mathcal{O}^\prime},
\label{split e^-O'R_Ne^O'}
\end{align}
where we have defined
\begin{equation}\label{define Omega'}
  \Omega^\prime=[H_{21}+H_4,\mathcal{O}^\prime]+H_{3}^{\prime\prime}-H_{3}^{\prime\prime\prime}
\end{equation}
and $H_{3}^{\prime\prime\prime}$ by
\begin{equation}\label{define H_3'''}
  H_{3}^{\prime\prime\prime}=2\sum_{p,q,p+q\neq0}\big(q_p+Y_p\big)(a_{p+q}^*a_{-p}^*a_qa_0+h.c.).
\end{equation}
Here $q_p$ and $Y_p$ are defined in (\ref{define q}) and (\ref{define Y_p}) respectively. The main difference in the proofs of this section is that here we bound the cubic dimensional coupling correlation remainder $H_{3}^{\prime\prime\prime}$ in Lemma \ref{lemma H_3''' O'}, so that we effectively eliminate the cubic term in the excitation Hamiltonian. The left over analysis of (\ref{split e^-OM_Ne^O}) is rather similar to Section \ref{3}, we state them in the up-coming series of lemmas, while we omit the details of proofs except some new estimates came upon. In the following lemmas, we bound $e^{-\mathcal{O}^\prime}\big(\tilde{Q}_1^{\mathcal{O}}\mathcal{N}_+
+\tilde{Q}_2^{\mathcal{O}}\mathcal{N}_+^2\big)
e^{\mathcal{O}^\prime}$ in Corollary \ref{corollary O' diag}, $e^{-\mathcal{O}^\prime}\tilde{\mathcal{E}}^{\mathcal{O}}e^{\mathcal{O}^\prime}$ in Corollary \ref{corollary O' Etilde^O}, and $e^{-\mathcal{O}^\prime}H_{23}^{\prime\prime\prime}
e^{\mathcal{O}^\prime}$ in Lemma \ref{lemma O' H_23'''}. These three terms stay unchanged up to small errors after conjugating with $e^{\mathcal{O}^\prime}$. The term containing the difference $\Omega^\prime$ is bounded in Lemma \ref{lemma Omega' O'}, and is proved to be a negligible error term. The contribution of the commutator $[H_{3}^{\prime\prime},\mathcal{O}^\prime]$ is calculated in Lemma \ref{lemma [H_3'',O']}, and the contribution of $[H_{3}^{\prime\prime\prime},\mathcal{O}^\prime]$ is calculated in Lemma \ref{lemma [H_3''',O']}. Lemmas \ref{lemma [H_3'',O']} and \ref{lemma [H_3''',O']} present the major contributions of the cubic dimensional coupling correlation structure to the second order ground state energy, in the form of polynomials of $\mathcal{N}_+$. We bound the growths of $\mathcal{N}_+$, $H_{21}$, $H_4$ and $H_4^\prime$ in Lemmas \ref{control of S_+ conj with e^O'} and \ref{lemma control e^-tO H_21, H_4e^tO}.
\begin{lemma}\label{control of S_+ conj with e^O'}
Let $\mathcal{N_+}$ be defined on $L_s^2(\Lambda_d^N)$ as stated in (\ref{define of N_+}), then there exist a constant $C_n$ depending only on $n\in\frac{1}{2}\mathbb{N}$ such that: for every $t\in\mathbb{R}$, $N\in\mathbb{N}$, $n\in\frac{1}{2}\mathbb{N}$, $\textit{l}\in(0,\frac{1}{2})$ such that $\frac{d\textit{l}}{a}>C$, and $h\in(0,\frac{1}{2})$ such that $\frac{h}{d\textit{l}}>C$ for some universal constant $C$, and we have
\begin{align}
  e^{-t\mathcal{O}^\prime}(\mathcal{N}_++1)^ne^{t\mathcal{O}^\prime}&\leq e^{C_nNa\textit{l}^{\frac{1}{2}}\vert t\vert}(\mathcal{N}_++1)^n,\label{e^-tO'(N_++1)e^tO'}\\
  \pm(e^{-t\mathcal{O}^\prime}(\mathcal{N}_++1)^ne^{t\mathcal{O}^\prime}-(\mathcal{N}_++1)^n)&\leq (e^{C_nNa\textit{l}^{\frac{1}{2}}\vert t\vert}-1) (\mathcal{N}_++1)^n.\label{e^-tO'N_+e^tO'-N_+}
\end{align}
\end{lemma}
\noindent
\emph{Proof.} See the proof of Lemma \ref{control of S_+ conj with e^B'} for details.
\begin{flushright}
  {$\Box$}
\end{flushright}

\par We want to remind the readers that we are working now under the assumptions that $N$ tends to infinity, $a$, $d$, $\frac{a}{d}$, $N^{\frac{3}{2}}a^{\frac{7}{6}}d^{-\frac{1}{2}}\textit{l}^{\frac{1}{3}}$ and $Na\textit{l}^{\frac{1}{2}}$ tend to $0$ and $\frac{d\textit{l}}{a}>C$. Moreover, we ask additionally $\frac{h}{d\textit{l}}>C$, $\frac{Na}{d}>C$, $\frac{ma}{d}>C$ and $N(d\textit{l}+\frac{h}{m})$ and $N^{\frac{3}{2}}a^{\frac{1}{2}}d^{-\frac{1}{2}}(d\textit{l}+\frac{h}{m})^{\frac{2}{3}}$ should tend to $0$.
\begin{lemma}\label{lemma control e^-tO H_21, H_4e^tO}
\begin{align}
  e^{-t\mathcal{O}^\prime}H_{21}e^{t\mathcal{O}^\prime}&\leq CH_{21}+CNad^{-1}(\mathcal{N}_++1)^2
\label{e^-O'H_21e^O' III}\\
  e^{-t\mathcal{O}^\prime}H_{4}e^{t\mathcal{O}^\prime}&\leq CH_4+CNad^{-1}(\mathcal{N}_++1)\nonumber\\
&+CN^{\frac{3}{2}}a^{\frac{7}{6}}d^{-\frac{1}{2}}\textit{l}^{\frac{1}{3}}
  [H_{21}+Nad^{-1}(\mathcal{N}_++1)^2]\label{e^-O'H_4e^O' III}\\
 e^{-t\mathcal{O}^\prime}H_{4}^\prime e^{t\mathcal{O}^\prime}&\leq CH_4^\prime+CNad^{-1}(\mathcal{N}_++1)\nonumber\\
&+CN^{\frac{3}{2}}a^{\frac{7}{6}}d^{-\frac{1}{2}}\textit{l}^{\frac{1}{3}}
  [H_{21}+Nad^{-1}(\mathcal{N}_++1)^2].\label{e^-O'H_4'e^O' III}
\end{align}
for all $\vert t\vert<1$.
\end{lemma}
\noindent
\emph{Proof.} See the proof of Lemmas \ref{commutator of H_21,H_4with B'} and \ref{lemma e^-B'H_Ne^B'} for details.
\begin{flushright}
  {$\Box$}
\end{flushright}

 As a direct consequence of Lemma \ref{control of S_+ conj with e^O'} and Lemma \ref{lemma control e^-tO H_21, H_4e^tO}, and the estimates (\ref{bound C Q R_N}), we have
\begin{corollary}\label{corollary O' diag}
\begin{equation}\label{O' diag}
  e^{-\mathcal{O}^\prime}\big(\tilde{Q}_1^{\mathcal{O}}\mathcal{N}_+
+\tilde{Q}_2^{\mathcal{O}}\mathcal{N}_+^2\big)
e^{\mathcal{O}^\prime}
=\tilde{Q}_1^{\mathcal{O}}\mathcal{N}_+
+\tilde{Q}_2^{\mathcal{O}}\mathcal{N}_+^2
+\tilde{\mathcal{E}}^{\mathcal{O}^\prime}_{diag},
\end{equation}
where
\begin{equation}\label{Etilde^O'_diag}
  \pm\tilde{\mathcal{E}}^{\mathcal{O}^\prime}_{diag}\leq N^2a^2d^{-1}\textit{l}^{\frac{1}{2}}
(\mathcal{N}_++1).
\end{equation}
\end{corollary}

\begin{corollary}\label{corollary O' Etilde^O}
\begin{align}\label{O' Etilde^O}
  &\pm e^{-\mathcal{O}^\prime}\mathcal{E}^{\mathcal{O}}e^{\mathcal{O}^\prime}\nonumber\\
\leq& C\Big\{ad^{-1}+N^4a^{\frac{8}{3}}d^{-2}\textit{l}^{\frac{1}{3}}
+N^{4}a^{2}d^{-2}(d\textit{l}+hm^{-1})^{\frac{2}{3}}
+N^2a^3d^{-3}h\ln(d\textit{l})^{-1}\nonumber\\
&\quad\quad\quad+N^4ad^{-1}m^{-1}(d\textit{l}+hm^{-1})
\Big(\ln\big(1+\frac{h}{d\textit{l}}\big)\Big)^{\frac{1}{2}}\Big\}
(\mathcal{N}_++1)\nonumber\\
&+CNa^{\frac{3}{2}}d^{-\frac{3}{2}}\textit{l}^{-\frac{1}{2}}m^{-1}
(\mathcal{N}_++1)^2
+CNa\textit{l}^{\frac{1}{2}}m^{-1}\Big(\ln\big(1+\frac{h}{d\textit{l}}\big)\Big)^{\frac{1}{2}}
(\mathcal{N}_++1)^3\nonumber\\
&+C\Big\{N^2a^{\frac{5}{3}}d^{-1}\textit{l}^{\frac{1}{3}}
+N^{\frac{7}{2}}a^{\frac{3}{2}}d^{-\frac{3}{2}}(d\textit{l}+hm^{-1})^{\frac{5}{3}}
+N^{2}ad^{-1}(d\textit{l}+hm^{-1})^{\frac{2}{3}}\nonumber\\
&\quad\quad\quad+N^2m^{-1}(d\textit{l}+hm^{-1})
\Big(\ln\big(1+\frac{h}{d\textit{l}}\big)\Big)^{\frac{1}{2}}
+Na^2d^{-2}h\ln(d\textit{l})^{-1}\Big\}\nonumber\\
&\times \Big(H_{21}+Nad^{-1}(\mathcal{N}_++1)^2\Big)\nonumber\\
&+C\Big(N^{\frac{3}{2}}a^{\frac{7}{6}}d^{-\frac{1}{2}}\textit{l}^{\frac{1}{3}}
+N^2ad^{-1}(d\textit{l}+hm^{-1})
+N^{\frac{3}{2}}a^{\frac{1}{2}}d^{-\frac{1}{2}}(d\textit{l}+hm^{-1})^{\frac{2}{3}}\Big)
H_4\nonumber\\
&+CN^{\frac{3}{2}}a^{\frac{1}{2}}d^{-\frac{1}{2}}(d\textit{l}+hm^{-1})H_4^\prime.
\end{align}
\end{corollary}

\par The analysis of the left over terms in (\ref{split e^-O'R_Ne^O'}) is shown in the following lemmas.
\begin{lemma}\label{lemma O' H_23'''}
\begin{equation}\label{O' H_23'''}
  e^{-\mathcal{O}^\prime}H_{23}^{\prime\prime\prime}
e^{\mathcal{O}^\prime}=H_{23}^{\prime\prime\prime}
+\tilde{\mathcal{E}}^{\mathcal{O}^\prime}_{23},
\end{equation}
where
\begin{align}\label{Etilde^O'_23}
  \pm\tilde{\mathcal{E}}^{\mathcal{O}^\prime}_{diag}&\leq
CN^{\frac{3}{2}}a^{\frac{3}{2}}d^{-\frac{3}{2}}\textit{l}^{-\frac{1}{2}}m^{-1}
(\mathcal{N}_++1)^{\frac{3}{2}}\nonumber\\
&+ C\Big\{N^{\frac{5}{2}}a^{\frac{5}{2}}d^{-\frac{3}{2}}\textit{l}^{\frac{1}{2}}
+\frac{N^2a\textit{l}^{\frac{1}{2}}}{m}
\Big(\ln\big(1+\frac{h}{d\textit{l}}\big)\Big)^{\frac{1}{2}}\Big\}
(\mathcal{N}_++1)\nonumber\\
&+C\Big\{N^3a^{\frac{8}{3}}d^{-1}\textit{l}^{\frac{5}{6}}
+\frac{N^2a\textit{l}^{\frac{1}{2}}}{m}
\Big(\ln\big(1+\frac{h}{d\textit{l}}\big)\Big)^{\frac{1}{2}}\Big\}
\big(H_{21}+Nad^{-1}(\mathcal{N}_++1)^2\big)\nonumber\\
&+CN^{\frac{3}{2}}a^{\frac{3}{2}}d^{-\frac{1}{2}}\textit{l}^{\frac{1}{2}}
H_4^\prime.
\end{align}
\end{lemma}
\noindent
\emph{Proof.} Here we need the estimates in (\ref{collect1}) and (\ref{collect2}). We still use (\ref{divide q_p+Y_p}) to divide the proof into three parts. For the calculation of the $q_p$ part, one can see the proof of Lemma \ref{lemma e^-B'H_23'e^B'} for Region I for details. For the calculation of the $W_p+\sum_q\xi_qW_{p-q}$ part, one can see the proof of Lemma \ref{lemma e^-B'H_23'e^B'} for Region III for details. For the calculation of the $\vert\mathcal{M}_dp\vert^2\xi_p=\widetilde{W}_{p,3}$ part, one can see the proof of Lemma \ref{lemma e^-Btilde'H_23''e^Btilde'} for details.

\begin{flushright}
  {$\Box$}
\end{flushright}

\begin{lemma}\label{lemma Omega' O'}
\begin{align}\label{Omega' O'}
 \int_{0}^{1}e^{-t\mathcal{O}^\prime}\Omega^\prime e^{t\mathcal{O}^\prime}dt
&\leq  CN^2a^2d^{-1}\textit{l}^{\frac{1}{2}}(\mathcal{N}_++1)
+CNa^{\frac{1}{2}}d^{-\frac{1}{2}}\big((d\textit{l})^2+h^2m^{-1}\big)
(H_4+1)\nonumber\\
&+CN^{\frac{3}{2}}a^{\frac{7}{6}}d^{-\frac{1}{2}}\textit{l}^{\frac{1}{3}}
\Big(H_4+H_{21}+Nad^{-1}(\mathcal{N}_++1)^2\Big).
\end{align}
\end{lemma}
\noindent
\emph{Proof.} See the proof of Lemma \ref{lemma Gamma'} for Region III for details. We use here (\ref{rewrite Wtilde_p}) and (\ref{discrete dimensional coupling scattering equation}) to calculate $\Omega^\prime$.

\begin{flushright}
  {$\Box$}
\end{flushright}

\begin{lemma}\label{lemma [H_3'',O']}
\begin{align}\label{[H_3'',O']}
  \int_{0}^{1}\int_{t}^{1}
e^{-s\mathcal{O}^\prime}[H_{3}^{\prime\prime},\mathcal{O}^\prime] e^{s\mathcal{O}^\prime}dsdt
=4\sum_{p\neq0}\widetilde{W}_pk_p\mathcal{N}_+(N-\mathcal{N}_+)
+\tilde{\mathcal{E}}_{[H_3^{\prime\prime},\mathcal{O}^\prime]}^{\mathcal{O}^\prime}
\end{align}
where
\begin{align}\label{Etilde^O'_[H_3'',O']}
 \pm\tilde{\mathcal{E}}_{[H_3^{\prime\prime},\mathcal{O}^\prime]}^{\mathcal{O}^\prime}
&\leq C\Big\{\frac{N^2a\textit{l}^{\frac{1}{2}}}{m}
\Big(\ln\big(1+\frac{h}{d\textit{l}}\big)\Big)^{\frac{1}{2}}
+N^{\frac{5}{2}}a^{\frac{5}{2}}d^{-\frac{3}{2}}\textit{l}^{\frac{1}{2}}
+N^{2}a^{\frac{5}{3}}d^{-1}\textit{l}^{\frac{1}{3}}\Big\}
(\mathcal{N}_++1)\nonumber\\
&+C\Big(\frac{N^2a\textit{l}^{\frac{1}{2}}}{m}
\Big(\ln\big(1+\frac{h}{d\textit{l}}\big)\Big)^{\frac{1}{2}}
+N^{2}a^{\frac{5}{3}}d^{-1}\textit{l}^{\frac{1}{3}}\Big)\big(H_{21}
+Nad^{-1}(\mathcal{N}_++1)^2\big)\nonumber\\
&+CN^{\frac{3}{2}}a^{\frac{3}{2}}d^{-\frac{1}{2}}\textit{l}^{\frac{1}{2}}
(H_4+H_4^\prime)+CNad^{-1}h\Big(1+\frac{a}{d}\ln a^{-1}\Big)H_{21}.
\end{align}
\end{lemma}
\noindent
\emph{Proof.} See the proof of Lemma \ref{[H_3'',Btilde']} for details. Notice that here we have used (\ref{l1 k_p}) to bound $\sum_{p\neq0}\vert k_p\vert$.

\begin{flushright}
  {$\Box$}
\end{flushright}

\begin{lemma}\label{lemma [H_3''',O']}
\begin{align}\label{[H_3''',O']}
  \int_{0}^{1}\int_{0}^{t}
e^{-s\mathcal{O}^\prime}[H_{3}^{\prime\prime\prime},\mathcal{O}^\prime] e^{s\mathcal{O}^\prime}dsdt
=4\sum_{p\neq0}\big(q_p+Y_p\big)k_p\mathcal{N}_+(N-\mathcal{N}_+)
+\tilde{\mathcal{E}}_{[H_3^{\prime\prime\prime},\mathcal{O}^\prime]}^{\mathcal{O}^\prime}
\end{align}
where
\begin{align}\label{Etilde^O'_[H_3''',O']}
 \pm\tilde{\mathcal{E}}_{[H_3^{\prime\prime\prime},\mathcal{O}^\prime]}^{\mathcal{O}^\prime}
&\leq CN^{2}a^{\frac{5}{3}}d^{-1}\textit{l}^{\frac{1}{3}}
\big(H_{21}+Nad^{-1}(\mathcal{N}_++1)^2\big)\nonumber\\
&+CNm^{-1}h\Big(1+\frac{a}{d}\ln a^{-1}\Big)H_{21}
+CNa^{\frac{3}{2}}d^{-\frac{3}{2}}\textit{l}^{-\frac{1}{2}}m^{-1}
(\mathcal{N}_++1)^2\nonumber\\
&+C\big\{N^{\frac{5}{2}}a^{\frac{5}{2}}d^{-\frac{3}{2}}\textit{l}^{\frac{1}{2}}
+N^{2}a^{\frac{5}{3}}d^{-1}\textit{l}^{\frac{1}{3}}\big\}
(\mathcal{N}_++1)
+CN^{\frac{3}{2}}a^{\frac{3}{2}}d^{-\frac{1}{2}}\textit{l}^{\frac{1}{2}}
H_4^\prime.
\end{align}
\end{lemma}
\noindent
\emph{Proof.} We again use (\ref{divide q_p+Y_p}) to divide the proof into three parts. For the calculation of the $q_p$ part, we modify slightly the proof of Lemma \ref{lemma [H_3,B']}, for example we can bound
\begin{equation*}
  \widetilde{\mathcal{E}}_{3^{\prime\prime\prime},1}^{\mathcal{O}^\prime}=
2\sum_{\substack{p,q,p+q,\\s,q-s\neq0}}
q_pk_s
(a_{p+q}^*a_{-p}^*a_{-s}^*a_{q-s}a_0a_0+h.c.)
\end{equation*}
 by
\begin{align*}
\vert\langle\widetilde{\mathcal{E}}_{3^{\prime\prime\prime},1}^{\mathcal{O}^\prime}
\psi,\psi\rangle\vert
&\leq C\Vert k\Vert_2\Vert q\Vert_2
\langle (\mathcal{N}_++1)^2\psi,\psi\rangle^{\frac{1}{2}}
\langle (\mathcal{N}_++1)^2a_0a_0\psi,
a_0a_0\psi\rangle^{\frac{1}{2}}\\
&\leq CNa^{\frac{3}{2}}d^{-\frac{3}{2}}\textit{l}^{-\frac{1}{2}}m^{-1}
\langle (\mathcal{N}_++1)^2\psi,\psi\rangle.
\end{align*}
for all $\psi\in L^2_s(\Lambda_d^N)$. We leave out other redundant calculations. For the calculation of the $W_p+\sum_q\xi_qW_{p-q}$ part, one can see the proof of Lemma \ref{lemma [H'_3,B']} for Region III for details. The calculation of the $\vert\mathcal{M}_dp\vert^2\xi_p=\widetilde{W}_{p,3}$ part is actually same as the $\widetilde{W}_{p,3}$ part of Lemma \ref{lemma [H_3'',O']}, and the error terms have been bounded in (\ref{Etilde^O'_[H_3'',O']}), thus we leave out the result of this part in (\ref{Etilde^O'_[H_3''',O']}).

\par We then combine above analysis with the proof of Lemma \ref{[H_3',Btilde']}. We underline here that we have the estimate
\begin{align*}
  \Big\vert\big(q_{p+q}+Y_{p+q}\big)-\big(q_p+Y_p\big)\Big\vert
\leq C\frac{\vert\mathcal{M}_dq\vert}{\sqrt{d}}
\left\vert\int_{\Lambda_d}\big(q+Y\big)
(\mathbf{x})\vert\mathbf{x}\vert d\mathbf{x}\right\vert
\leq Cm^{-1}h\vert\mathcal{M}_dq\vert.
\end{align*}

\begin{flushright}
  {$\Box$}
\end{flushright}

\par The cubic term can now be considered as an error term. We state the result in the next lemma.
\begin{lemma}\label{lemma H_3''' O'}
\begin{align}\label{H_3''' O'}
\pm H_{3}^{\prime\prime\prime}\leq
 CN\vartheta_1^{-1}\Big\{\frac{a\textit{l}\ln (d\textit{l})^{-1}}
{dm^2}+\frac{\ln h^{-1}}{m^2}
+\frac{a^{\frac{1}{3}}}{d^{\frac{1}{3}}m^{\frac{5}{3}}\textit{l}}\Big\}
(\mathcal{N}_++1)^2+C\vartheta_1 H_{21}
\end{align}
for some $\vartheta_1>0$
\end{lemma}
\noindent
\emph{Proof.} See the bound of $\Gamma_1^\prime$ in the proof of Lemma \ref{lemma Gamma'} for Region I for details. Here we need estimates (\ref{est of Y_p}), (\ref{l1 Y_p}), (\ref{douche3}) and (\ref{l1 q_p}).

\begin{flushright}
  {$\Box$}
\end{flushright}

\noindent
\emph{Proof of Proposition \ref{dimensional coupling quadratic renorm}.}
\par With all the estimates above, we conclude
\begin{align*}
  \mathcal{S}_N=e^{-\mathcal{O}^\prime}\mathcal{R}_Ne^{\mathcal{O}^\prime}
=&N(N-1)\tilde{C}^{\mathcal{O}^\prime}+2N\tilde{C}^{\mathcal{O}^\prime}\mathcal{N}_+
-3\tilde{C}^{\mathcal{O}^\prime}\mathcal{N}_+^2\nonumber\\
&+H_{21}+H_4+H_{23}^{\prime\prime\prime}+\tilde{\mathcal{E}}^{\mathcal{O}^\prime}
\end{align*}
where
\begin{align*}
\tilde{C}^{\mathcal{O}^\prime}=
\Big(W_0+\sum_{p\neq0}W_p\eta_p+\sum_{p\neq0}\big(W_p+\widetilde{W}_p\big)\xi_p
+\sum_{p\neq0}\big(\widetilde{W}_p+q_p+Y_p\big)k_p\Big)
\end{align*}
and the error term is bounded by
\begin{align*}
  \pm \tilde{\mathcal{E}}^{\mathcal{O}^\prime}
\leq& C\Big\{ad^{-1}+N^4a^{\frac{8}{3}}d^{-2}\textit{l}^{\frac{1}{3}}
+N^{4}a^{2}d^{-2}(d\textit{l}+hm^{-1})^{\frac{2}{3}}
+N^2a^3d^{-3}h\ln(d\textit{l})^{-1}\nonumber\\
&\quad\quad\quad+N^4ad^{-1}m^{-1}(d\textit{l}+hm^{-1})
\Big(\ln\big(1+\frac{h}{d\textit{l}}\big)\Big)^{\frac{1}{2}}\Big\}
(\mathcal{N}_++1)\nonumber\\
&+CN^{\frac{3}{2}}a^{\frac{3}{2}}d^{-\frac{3}{2}}\textit{l}^{-\frac{1}{2}}m^{-1}
(\mathcal{N}_++1)^{\frac{3}{2}}\nonumber\\
&+CN\vartheta_1^{-1}\Big\{\frac{a\textit{l}\ln (d\textit{l})^{-1}}
{dm^2}+\frac{\ln h^{-1}}{m^2}
+\frac{a^{\frac{1}{3}}}{d^{\frac{1}{3}}m^{\frac{5}{3}}\textit{l}}\Big\}
(\mathcal{N}_++1)^2+C\vartheta_1 H_{21}\nonumber\\
&+C\Big\{N^2a^{\frac{5}{3}}d^{-1}\textit{l}^{\frac{1}{3}}
+N^{\frac{7}{2}}a^{\frac{3}{2}}d^{-\frac{3}{2}}(d\textit{l}+hm^{-1})^{\frac{5}{3}}
+N^{2}ad^{-1}(d\textit{l}+hm^{-1})^{\frac{2}{3}}\nonumber\\
&\quad\quad\quad+N^2m^{-1}(d\textit{l}+hm^{-1})
\Big(\ln\big(1+\frac{h}{d\textit{l}}\big)\Big)^{\frac{1}{2}}
+Nad^{-1}h\Big(1+\frac{a}{d}\ln a^{-1}\Big)\Big\}\nonumber\\
&\times \Big(H_{21}+Nad^{-1}(\mathcal{N}_++1)^2\Big)\nonumber\\
&+C\Big(N^{\frac{3}{2}}a^{\frac{7}{6}}d^{-\frac{1}{2}}\textit{l}^{\frac{1}{3}}
+N^2ad^{-1}(d\textit{l}+hm^{-1})
+N^{\frac{3}{2}}a^{\frac{1}{2}}d^{-\frac{1}{2}}(d\textit{l}+hm^{-1})^{\frac{2}{3}}\Big)
H_4\nonumber\\
&+CN^{\frac{3}{2}}a^{\frac{1}{2}}d^{-\frac{1}{2}}(d\textit{l}+hm^{-1})H_4^\prime.
\end{align*}
for some $\vartheta_1>0$. Then we conclude the proof of Proposition \ref{dimensional coupling quadratic renorm}, and Region III is now ready for the Bogoliubov transform.

\begin{flushright}
  {$\Box$}
\end{flushright}

\section{Bogoliubov Transformation for Region III}\label{7}

\par In this section, we analyze the diagonalized Hamiltonian $\mathcal{Z}^{III}_N$ and prove Propositon \ref{Bog renorm III}. Recall we always assume $N$ tends to infinity, $a$, $d$, $\frac{a}{d}$, $N^{\frac{3}{2}}a^{\frac{7}{6}}d^{-\frac{1}{2}}\textit{l}^{\frac{1}{3}}$ and
$Na\textit{l}^{\frac{1}{2}}$ tend to $0$ and $\frac{d\textit{l}}{a}>C$. Moreover, we require additionally $\frac{h}{d\textit{l}}>C$, $\frac{Na}{d}>C$, $\frac{ma}{d}>C$ and $N(d\textit{l}+\frac{h}{m})$ and $N^{\frac{3}{2}}a^{\frac{1}{2}}d^{-\frac{1}{2}}(d\textit{l}+\frac{h}{m})^{\frac{2}{3}}$ should tend to $0$. we adopt the notation
\begin{equation*}
  \tilde{\tau}=\sum_{p\neq0}\tilde{\tau}_p\phi_p^{(d)}\in L_\perp^2(\Lambda_d).
\end{equation*}
Before we go on estimating $\tilde{\tau}$, we need first to gain a more subtle bound on the constant $\tilde{C}^{\mathcal{O}^\prime}$ given in (\ref{sixth/seventh renorm Ctilde ^O'}).
\begin{lemma}\label{bound Ctilde^O'}
Constant $\tilde{C}^{\mathcal{O}^\prime}$ given in (\ref{sixth/seventh renorm Ctilde ^O'}) has the form
\begin{equation}\label{est of Ctilde ^O'}
  \tilde{C}^{\mathcal{O}^\prime}=\frac{2\pi}{m}+O\Big(\frac{a^2}{d^2\textit{l}}
+\Big(\frac{a}{d}+\frac{1}{h^2m}\Big)(d\textit{l})^2
+\frac{\textit{l}^{\frac{1}{2}}}{m}\sqrt{\frac{a}{d}}\Big).
\end{equation}
Moreover, if we let
\begin{equation}\label{choose l, h III Bog renorm}
  \textit{l}=c\Big(\frac{a}{d}\Big)^\alpha,\quad
h=N^{-\beta}
\end{equation}
for some universal $0\leq\alpha<1$, $\beta\geq 0$ and $0<c<\frac{1}{2}$, then we have
\begin{align}\label{est of Ctilde^O' 2}
  \tilde{C}^{\mathcal{O}^\prime}=&
\Big(W_0+\sum_{p\neq0}W_p\eta_p+\sum_{p\neq0}\big(W_p+Y_p+\mathfrak{D}_p\big)\xi_p
+\sum_{p\neq0}\big(2Y_p+\mathfrak{D}_p+q_p\big)k_p\Big)\nonumber\\
&+O\Big(\frac{\textit{l}^{\frac{1}{2}}}{m}\sqrt{\frac{a}{d}}\Big)
\end{align}
with
\begin{equation}\label{est of Ctilde^O' 3}
  \big(\tilde{C}^{\mathcal{O}^\prime}-4\pi g\big)=O\Big(\frac{a^2}{d^2\textit{l}}
+\Big(\frac{a}{d}+\frac{1}{h^2m}\Big)(d\textit{l})^2
+\frac{\textit{l}^{\frac{1}{2}}}{m}\sqrt{\frac{a}{d}}+g^2\Big(\ln N+\textit{l}^{-1}\Big)\Big).
\end{equation}
Here $g$ is defined in (\ref{coupling constant g}).
\end{lemma}
\noindent
\emph{Proof.} Recall the definition of $\tilde{C}^{\mathcal{O}^\prime}$
\begin{equation*}
  \tilde{C}^{\mathcal{O}^\prime}=
\Big(W_0+\sum_{p\neq0}W_p\eta_p+\sum_{p\neq0}\big(W_p+\widetilde{W}_p\big)\xi_p
+\sum_{p\neq0}\big(\widetilde{W}_p+q_p+Y_p\big)k_p\Big),
\end{equation*}
and from (\ref{rewrite Wtilde_p}) we have
\begin{equation*}
  \widetilde{W}_p=Y_p+\mathfrak{D}_p-\frac{1}{2\sqrt{d}}v_p^{(a,d)}\xi_0.
\end{equation*}
We prove (\ref{est of Ctilde ^O'}) by dividing $\tilde{C}^{\mathcal{O}^\prime}$ into several parts, and (\ref{est of Ctilde^O' 2}) and (\ref{est of Ctilde^O' 3}) will follow.

\par \ding{172} From (\ref{useful shit 2d}), and estimates (\ref{sum_pW_peta_p 3dscatt}) and (\ref{est of xi_p}), we have
\begin{equation*}
  \tilde{C}^{\mathcal{O}^\prime}_1
\coloneqq W_0+\sum_{p\neq0}W_p\xi_p
=\frac{2\pi}{m}+O\Big(\frac{1}{m^2}
+\frac{a}{d}\big((d\textit{l})^2+h^2m^{-1}\big)\Big).
\end{equation*}

\par \ding{173} From (\ref{sum_pW_peta_p 3dscatt}) we know that
\begin{equation*}
  \tilde{C}^{\mathcal{O}^\prime}_2\coloneqq
\sum_{p\neq0}W_p\eta_p=O\Big(\frac{a^2}{d^2\textit{l}}\Big)
\end{equation*}

\par \ding{174} From (\ref{collect3}) we have
\begin{equation*}
    \tilde{C}^{\mathcal{O}^\prime}_3\coloneqq
\sum_{p\neq0}q_pk_p=O\Big(\frac{\textit{l}^{\frac{1}{2}}}{m}\sqrt{\frac{a}{d}}\Big).
\end{equation*}

\par \ding{175} Using the estimate (\ref{est of xi_p}) and the fact that both $\vert\xi(\mathbf{x})\vert$ and $\vert k(\mathbf{x})\vert$ can be bounded by $Cd^{-\frac{1}{2}}$, we deduce
\begin{equation*}
  \tilde{C}^{\mathcal{O}^\prime}_4\coloneqq
-\frac{1}{2\sqrt{d}}\sum_{p\neq0}v_p^{(a,d)}\xi_0(\xi_p+k_p)
=O\Big(\frac{a}{d}\big((d\textit{l})^2+h^2m^{-1}\big)\Big).
\end{equation*}

\par \ding{176} Write
\begin{equation*}
  \tilde{C}^{\mathcal{O}^\prime}_5\coloneqq
2\sum_{p\neq0}Y_pk_p=2\int_{\Lambda_d}Y(\mathbf{x})k(\mathbf{x})d\mathbf{x}.
\end{equation*}
From (\ref{eqn of xi_p rewrt}) we know that
\begin{equation*}
  Y(\mathbf{x})=-\Delta_{\mathbf{x}}\xi(\mathbf{x})+W(\mathbf{x})\widetilde{g}_h(\mathbf{x}).
\end{equation*}
From the definition of $k$ (\ref{define k}) and estimate (\ref{sum_pW_peta_p 3dscatt}), we have
\begin{equation*}
  \Big\vert\int_{\Lambda_d}W(\mathbf{x})\widetilde{g}_h(\mathbf{x})k(\mathbf{x})
d\mathbf{x}\Big\vert\leq
\Big\vert\int_{\Lambda_d}W(\mathbf{x})\eta(\mathbf{x})
d\mathbf{x}\Big\vert\leq \frac{Ca^2}{d^2\textit{l}}.
\end{equation*}
On the other hand, from (\ref{est of grad xi B_1}) and (\ref{douche2}), we have
\begin{align*}
   &\Big\vert\int_{\Lambda_d}-\Delta_{\mathbf{x}}\xi(\mathbf{x})k(\mathbf{x})
d\mathbf{x}\Big\vert=
 \Big\vert\int_{\Lambda_d}\nabla_{\mathbf{x}}\xi(\mathbf{x})\nabla_{\mathbf{x}}k(\mathbf{x})
d\mathbf{x}\Big\vert\\
&\leq \Vert\nabla_{\mathbf{x}}\xi\Vert_{L^2(B_{d\textit{l}})}
\Vert\nabla_{\mathbf{x}}k\Vert_2\leq \frac{C\textit{l}^{\frac{1}{2}}}{m}
\sqrt{\frac{a}{d}}.
\end{align*}
Therefore
\begin{equation*}
  \tilde{C}^{\mathcal{O}^\prime}_5
=O\Big(\frac{a^2}{d^2\textit{l}}+\frac{\textit{l}^{\frac{1}{2}}}{m}
\sqrt{\frac{a}{d}}\Big).
\end{equation*}

\par \ding{177} Since $\xi_p=0$ when $p_3\neq0$, we can write using (\ref{define Y_p})
\begin{equation*}
  \tilde{C}^{\mathcal{O}^\prime}_6\coloneqq
\sum_{p\neq0}Y_p\xi_p=\sum_{p\neq0}\frac{\mu_h}{(d\textit{l})^2}
\Big(\xi_p+\widehat{\chi^{\mathrm{2D}}_{h}}\left(
  \frac{\bar{p}}{2\pi}\right)\Big)\xi_p.
\end{equation*}
Using (\ref{est of mu_h}) and (\ref{est of xi}), we know that
\begin{equation*}
  \sum_{p\neq0}\frac{\mu_h}{(d\textit{l})^2}
\xi_p^2=\frac{\mu_h}{(d\textit{l})^2}\int_{\Lambda_d}\vert\xi_\perp\vert^2
=O\Big(\frac{1}{h^2m}\big((d\textit{l}^2+h^2m^{-2})\big)\Big)
\end{equation*}
Using (\ref{est of mu_h}) and (\ref{est of xi_p}), we know that
\begin{equation*}
  \sum_{p\neq0}\frac{\mu_h}{(d\textit{l})^2}
\widehat{\chi^{\mathrm{2D}}_{h}}\left(
  \frac{\bar{p}}{2\pi}\right)\xi_p=
-\frac{\mu_h}{(d\textit{l})^2}\Big(\xi_0\widehat{\chi^{\mathrm{2D}}_{h}}(0)+
\int_{\Lambda_{\mathrm{2D}}}\widetilde{z}_h\Big)
=O\Big(\frac{1}{h^2m}\big((d\textit{l}^2+h^2m^{-1})\big)\Big).
\end{equation*}
Therefore
\begin{equation*}
  \tilde{C}^{\mathcal{O}^\prime}_6
=O\Big(\frac{1}{h^2m}\big((d\textit{l}^2+h^2m^{-1})\big)\Big).
\end{equation*}

\par \ding{178} Notice the fact that
\begin{equation*}
  \vert\mathfrak{D}_0\vert\lesssim d^{-\frac{1}{2}}
\Vert\mathfrak{D}\Vert_1\lesssim ad^{-1},
\end{equation*}
 together with (\ref{est of xi_p}) and (\ref{douche2}), we have
\begin{equation*}
  \tilde{C}^{\mathcal{O}^\prime}_7
=\sum_{p\neq0}\mathfrak{D}_p(\xi_p+k_p)
=\sum_{p}\mathfrak{D}_p(\xi_p+k_p)+
O\Big(\frac{a}{d}\big((d\textit{l})^2+h^2m^{-1}\big)\Big).
\end{equation*}
From the definition of $\xi$, $k$ and $\mathfrak{D}$, we can write
\begin{equation*}
  \sum_{p}\mathfrak{D}_p(\xi_p+k_p)=\frac{1}{d}\int_{B_{d\textit{l}}}
\Big(\frac{1}{2}v_a(\mathbf{x})
-\sqrt{d}W(\mathbf{x})\Big)\widetilde{f}_\textit{l}(\mathbf{x})
\widetilde{z}_h^2(x)d\mathbf{x}.
\end{equation*}
Using a Poincar\'{e} type inequality (See for example \cite[Lemma 7.16]{gilbargTrudinger1977elliptic}), we find that inside the 2D ball $\mathcal{B}_{d\textit{l}}$
\begin{align*}
  \vert\widetilde{z}_h^2(x)-(\widetilde{z}_{h}^2)_{avg}\vert&\leq C\int_{\mathcal{B}_{d\textit{l}}}
\vert x-y\vert^{-1}\big\vert\nabla_y(\widetilde{z}_h)^2(y)\big\vert dy\\
&\leq C\int_{\mathcal{B}_{1}}
\big\vert x(d\textit{l})^{-1}-y\vert^{-1}\big\vert
\vert z_h(y)\vert\big\vert\nabla_yz_h(y)\big\vert dy,
\end{align*}
where
\begin{equation*}
  (\widetilde{z}_{h}^2)_{avg}=\frac{1}{\vert\mathcal{B}_{d\textit{l}}\vert}
\int_{\mathcal{B}_{d\textit{l}}}\widetilde{z}_{h}(x)^2dx.
\end{equation*}
Since $x\in\mathcal{B}_{d\textit{l}}$, we have $\vert x/(d\textit{l})\vert\leq 1$. Using (\ref{est of z_h}) and (\ref{est of grad z_h}), we deduce
\begin{align*}
  \vert\widetilde{z}_h^2(x)-(\widetilde{z}_{h}^2)_{avg}\vert
&\leq\frac{C}{m}\int_{\mathcal{B}_2}\vert y\vert^{-1}dy\leq\frac{C}{m},
\end{align*}
and
\begin{equation*}
  0\leq  (\widetilde{z}_{h}^2)_{avg}\leq1.
\end{equation*}
On the other hand, we know from (\ref{est of int vf_l}) that
\begin{equation*}
  \frac{1}{d}\int_{\Lambda_d}\frac{1}{2}v_a(\mathbf{x})\widetilde{f}_\textit{l}(\mathbf{x})
d\mathbf{x}=\frac{4\pi a\mathfrak{a}_0}{d}+O\Big(\frac{a^2}{d^2\textit{l}}\Big),
\end{equation*}
and from (\ref{est of lambda_l}), (\ref{est of int w_l}), (\ref{est of eta and eta_perp}) and (\ref{define W(x)})
\begin{equation*}
   \frac{1}{d}\int_{\Lambda_d}\sqrt{d}W(\mathbf{x})\widetilde{f}_\textit{l}(\mathbf{x})
d\mathbf{x}=\frac{4\pi a\mathfrak{a}_0}{d}+O\Big(\frac{a^2}{d^2\textit{l}}\Big)
\end{equation*}
With all the estimates given above, we arrive at
\begin{equation*}
  \frac{(\widetilde{z}_{h}^2)_{avg}}{d}\int_{B_{d\textit{l}}}
\Big(\frac{1}{2}v_a(\mathbf{x})
-\sqrt{d}W(\mathbf{x})\Big)\widetilde{f}_\textit{l}(\mathbf{x})
d\mathbf{x}=O\Big(\frac{a^2}{d^2\textit{l}}\Big),
\end{equation*}
and
\begin{equation*}
  \frac{1}{d}\int_{B_{d\textit{l}}}
\Big(\frac{1}{2}v_a(\mathbf{x})
-\sqrt{d}W(\mathbf{x})\Big)\widetilde{f}_\textit{l}(\mathbf{x})
\big(\widetilde{z}_h^2(x)-(\widetilde{z}_{h}^2)_{avg}\big)d\mathbf{x}
=O\Big(\frac{a}{md}\Big).
\end{equation*}
Therefore
\begin{equation*}
   \tilde{C}^{\mathcal{O}^\prime}_7=O\Big(\frac{a^2}{d^2\textit{l}}\Big).
\end{equation*}
Hence we have finished the proof of Lemma \ref{bound Ctilde^O'}.

\begin{flushright}
  {$\Box$}
\end{flushright}

\par As a direct consequence of Lemma \ref{bound Ctilde^O'}, we can bound
\begin{corollary}
\begin{equation}\label{dhufgduikjof}
  \pm 3\tilde{C}^{\mathcal{O}^\prime}\mathcal{N}_+^2\leq
C\Big(\frac{1}{m}+\frac{a^2}{d^2\textit{l}}
+\Big(\frac{a}{d}+\frac{1}{h^2m}\Big)(d\textit{l})^2
+\frac{\textit{l}^{\frac{1}{2}}}{m}\sqrt{\frac{a}{d}}\Big)(\mathcal{N}_++1)^2.
\end{equation}
\end{corollary}

\par With the constant $\tilde{C}^{\mathcal{O}^\prime}$ analyzed, we can now go on estimating $\tilde{\tau}$ and the action of Bogoliubov transform. We present the results parallel to Section \ref{4} subsequently.
\begin{lemma}\label{lemma tau_p III}
Let $\tilde{F}_p$ and $\tilde{G}_p$ be defined in (\ref{define F_p,G_p III}), $\tilde{\tau}_p$ be defined in (\ref{define tau_p III}), then there hold the followings
\begin{enumerate}[$(1)$]
\item
\begin{equation}\label{lemma tau_p 1 III}
\vert \tilde{G}_p-4\pi Nm^{-1}\vert\leq
CNm^{-1}h\vert\mathcal{M}_dp\vert+CN\Big(\frac{1}{h^2m}\Big((d\textit{l})^2+\frac{h^2}{m}\Big)
+\frac{\textit{l}^{\frac{1}{2}}}{m}\sqrt{\frac{a}{d}}\Big).
\end{equation}
Assume further that $N\Big(\frac{h}{m}+\frac{a^2}{d^2\textit{l}}
+\Big(\frac{a}{d}+\frac{1}{h^2m}\Big)(d\textit{l})^2
+\frac{\textit{l}^{\frac{1}{2}}}{m}\sqrt{\frac{a}{d}}\Big)$ tends to $0$, we have
\begin{equation}\label{G_p<F_p III}
  \vert \tilde{G}_p\vert+c\vert\mathcal{M}_dp\vert^2\leq \tilde{F}_p
\end{equation}
for any fixed $c\in(0,1)$, which implies $\tilde{\tau}_p$ is well-defined. Moreover
\begin{equation}\label{lemma tau_p 2 III}
\vert \tilde{G}_p\vert\leq CNm^{-1}.
\end{equation}
\item Under the assumption that $N\Big(\frac{h}{m}+\frac{a^2}{d^2\textit{l}}
+\Big(\frac{a}{d}+\frac{1}{h^2m}\Big)(d\textit{l})^2
+\frac{\textit{l}^{\frac{1}{2}}}{m}\sqrt{\frac{a}{d}}\Big)$ tends to $0$, we have
\begin{equation}\label{lemma tau_p 3 III}
  \vert \tilde{\tau}_p\vert\leq C\frac{\vert \tilde{G}_p\vert}{\vert\mathcal{M}_dp\vert^2}
\leq CN\frac{\big(\vert q_p\vert+\vert Y_p\vert\big)}{\vert\mathcal{M}_dp\vert^2} .
\end{equation}
Moreover
\begin{equation}\label{lemma tau_p 4 III}
  \Vert\tilde{\tau}\Vert_2\leq CNm^{-1},
\end{equation}
and
\begin{equation}\label{lemma tau_p 5 III}
\Vert\tilde{\tau}\Vert_\infty\leq \frac{CN}{\sqrt{d}}
\Big(\frac{1}{m}\ln\frac{1}{h}+\frac{a^{\frac{1}{3}}}{d^{\frac{1}{3}}}
\frac{1}{\textit{l}m^{\frac{2}{3}}}
+\frac{\textit{l}^{-\frac{1}{2}}}{m}
\Big({\frac{a}{d}}\Big)^{\frac{1}{6}}
+\frac{\textit{l}^{\frac{1}{2}}}{m}\sqrt{\frac{a}{d}}
\ln (d\textit{l})^{-1}\Big).
\end{equation}
\end{enumerate}
\end{lemma}
\noindent
\emph{Proof.} See the proof of Lemma \ref{lemma tau_p} for details. Here we use Lemma \ref{bound Ctilde^O'} and estimates given in (\ref{collect1}), (\ref{collect2}), Section \ref{Induced 2D Scattering Equation} and Section \ref{Dimensional Coupling Scattering Equation Section}.

\begin{flushright}
  {$\Box$}
\end{flushright}

\begin{lemma}\label{lemma control bogoliubov III}
Assume further that $N\Big(\frac{h}{m}+\frac{a^2}{d^2\textit{l}}
+\Big(\frac{a}{d}+\frac{1}{h^2m}\Big)(d\textit{l})^2
+\frac{\textit{l}^{\frac{1}{2}}}{m}\sqrt{\frac{a}{d}}\Big)$ tends to $0$ and $C^{-1}\leq Nm^{-1}\leq C$, we have for all $n\in\frac{1}{2}\mathbb{N}$ and $\vert t\vert\leq1$
\begin{align}
   &e^{-tB^{\prime\prime\prime}}(\mathcal{N}_++1)^ne^{tB^{\prime\prime\prime}}
\leq C_n(\mathcal{N}_++1)^{n},\label{control N_+ with e^B'''}\\
&e^{-tB^{\prime\prime\prime}}H_{21}e^{tB^{\prime\prime\prime}}
\leq C\Big\{H_{21}+N^2\Big(\frac{a\textit{l}
\ln (d\textit{l})^{-1}}
{dm^2}+\frac{\ln h^{-1}}{m^2}
+\frac{a^{\frac{1}{3}}}{d^{\frac{1}{3}}m^{\frac{5}{3}}\textit{l}}\Big)\Big\},
\label{control H_21 with e^B'''}\\
&e^{-tB^{\prime\prime\prime}}H_{4}e^{tB^{\prime\prime\prime}}
\leq C\Big\{H_4+ad^{-1}(\mathcal{N}_++1)^2\nonumber\\
&\quad
+\frac{N^2a}{d}\Big(\frac{1}{m}\ln\frac{1}{h}+\frac{a^{\frac{1}{3}}}{d^{\frac{1}{3}}}
\frac{1}{\textit{l}m^{\frac{2}{3}}}
+\frac{\textit{l}^{-\frac{1}{2}}}{m}
\Big({\frac{a}{d}}\Big)^{\frac{1}{6}}
+\frac{\textit{l}^{\frac{1}{2}}}{m}\sqrt{\frac{a}{d}}
\ln (d\textit{l})^{-1}\Big)^2\Big\}.
\label{control H_4 with e^B''' 1}
\end{align}
\end{lemma}
\noindent
\emph{Proof.} See the proof of Lemma \ref{control of S_+ with e^B''} for details. Here we use Lemma \ref{lemma tau_p III}, \ref{xi_p Y_p lemma} and \ref{q_p lemma}, and estimates (\ref{collect1}) and (\ref{collect2}).

\begin{flushright}
  {$\Box$}
\end{flushright}

\par We also present that the control of modified non-zero momentum sum of potential operator $H_4^\prime$, which is actually controlled by $H_4$ and $H_{21}$.
\begin{lemma}\label{lemam control H_4'}
As long as $a$, $d$ and $\frac{a}{d}$ tend to $0$, $N$ tends to infinity, and $\frac{a}{d\textit{l}}<C$ for $\textit{l}\in(0,\frac{1}{2})$ and some small but universal constant $C$, then there exists another unversal constant, also denoted as $C$, such that
\begin{equation}\label{control H_4'}
  H_4^\prime\leq CNH_{21}+CH_4.
\end{equation}
\end{lemma}
\noindent
\emph{Proof.} We claim, for any $\psi\in L^2(\Lambda_d)$
\begin{equation}\label{claim control H_4'}
  \int_{\Lambda_d}\vert\nabla_{\mathbf{x}}\psi\vert^2
+\frac{1}{2}v_a(\mathbf{x})\vert\psi\vert^2d\mathbf{x}\geq
C\int_{\Lambda_d}\sqrt{d}W(\mathbf{x})\vert\psi\vert^2d\mathbf{x}.
\end{equation}
With (\ref{claim control H_4'}) holding true, a simple change of variables gives for any $\mathbf{y}\in\Lambda_d$ (Remember we demand $\Lambda_d$ to be a torus)
\begin{equation*}
  \int_{\Lambda_d}\vert\nabla_{\mathbf{x}}\psi\vert^2
+\frac{1}{2}v_a(\mathbf{x}-\mathbf{y})\vert\psi\vert^2d\mathbf{x}\geq
C\int_{\Lambda_d}\sqrt{d}W(\mathbf{x}-\mathbf{y})\vert\psi\vert^2d\mathbf{x}.
\end{equation*}
Then we combine (\ref{H_21psi,psi fock}), (\ref{H_4psi,psi fock}) and (\ref{H_4'psi,psi}) to reach (\ref{control H_4'}).

\par We then tend to proving (\ref{claim control H_4'}). Recall that we have assumed the interaction potential $v$ to be supported on a 3D ball $B_{R_0}$. From \cite[Lemma 2.5]{lieb2005mathematics} (for a more mathematically rigorous proof one can see \cite[Lemma 3.1]{lieb2005mathematics}), we know that
\begin{equation}\label{claim control H_4' 1}
  \int_{B_{d\textit{l}}}\vert\nabla_{\mathbf{x}}\psi\vert^2
+\frac{1}{2}v_a(\mathbf{x})\vert\psi\vert^2d\mathbf{x}\geq
C\int_{B_{d\textit{l}}\backslash B_{aR_0}}\sqrt{d}W(\mathbf{x})\vert\psi\vert^2d\mathbf{x}.
\end{equation}
On the other hand, we claim that
\begin{equation}\label{claim control H_4' 2}
  \int_{B_{aR_0}}\vert\nabla_{\mathbf{x}}\psi\vert^2
+\frac{1}{2}v_a(\mathbf{x})\vert\psi\vert^2d\mathbf{x}\geq
C\int_{B_{aR_0}}\sqrt{d}W(\mathbf{x})\vert\psi\vert^2d\mathbf{x}.
\end{equation}
We argue by contradiction. Since (\ref{claim control H_4' 2}) holds trivially when $\mathfrak{a}_0$, the scattering length of $v$, vanishes (which implies that $W=0$), we then assume without lost of generality that $v$ non-zero potential. Assume there exists a family $\{\psi_j,a_j,d_j,\textit{l}_j\}$ for $j\in\mathbb{N}$, such that
\begin{equation*}
  \int_{B_{R_0}}\vert\nabla_{\mathbf{x}}\psi_j\vert^2
+\frac{1}{2}v(\mathbf{y})\vert\psi_j\vert^2d\mathbf{y}\leq
j^{-1}\frac{a_j^3}{(d_j\textit{l}_j)^3},
\end{equation*}
with
\begin{equation*}
  \int_{B_{R_0}}a_j^2\sqrt{d_j}W(a_j\mathbf{y})\vert\psi_j\vert^2d\mathbf{y}
=\frac{a_j^3}{(d_j\textit{l}_j)^3}
\end{equation*}
From (\ref{est of f_l}) and (\ref{L^infty W}), we have inside the ball $B_{R_0}$
\begin{equation*}
  0\leq C_1\frac{a_j^3}{(d_j\textit{l}_j)^3}
\leq a_j^2\sqrt{d_j}W(a_j\mathbf{y})\leq C_2\frac{a_j^3}{(d_j\textit{l}_j)^3}
\end{equation*}
which implies that
\begin{equation*}
  0\leq C_1\leq \int_{B_{R_0}}\vert\psi_j\vert^2\leq C_2.
\end{equation*}
Then a contradiction would arise if we let $j$ go to infinity. Combing (\ref{claim control H_4' 1}) and (\ref{claim control H_4' 2}) we have reached (\ref{claim control H_4'}).
\begin{flushright}
  {$\Box$}
\end{flushright}



\noindent
\emph{Proof of Proposition \ref{Bog renorm III}.}
\par With all these lemmas proven above and the further assumptions on the parameters that $N\Big(\frac{h}{m}+\frac{a^2}{d^2\textit{l}}
+\Big(\frac{a}{d}+\frac{1}{h^2m}\Big)(d\textit{l})^2
+\frac{\textit{l}^{\frac{1}{2}}}{m}\sqrt{\frac{a}{d}}\Big)$ tends to $0$ and $C^{-1}\leq Nm^{-1}\leq C$, we deduce
\begin{equation*}
  e^{-B^{\prime\prime\prime}}\mathcal{S}_Ne^{B^{\prime\prime\prime}}
=N(N-1)\tilde{C}^{\mathcal{O}^\prime}+
e^{-B^{\prime\prime\prime}}{\mathcal{T}}^\prime e^{B^{\prime\prime\prime}}
+e^{-B^{\prime\prime\prime}}H_4e^{B^{\prime\prime\prime}}
+\tilde{\mathcal{E}}^{B^{\prime\prime\prime}}.
\end{equation*}
Here we take the error term to be
\begin{equation*}
  \tilde{{\mathcal{E}}}^{B^{\prime\prime\prime}}
=e^{-B^{\prime\prime\prime}}\big(\widetilde{\mathcal{T}}
-\widetilde{\mathcal{T}}^\prime+\tilde{\mathcal{E}}^{\mathcal{O}^\prime}
-3\tilde{C}^{\mathcal{O}^\prime}\mathcal{N}_+^2\big)e^{B^{\prime\prime\prime}},
\end{equation*}
which is bound by
\begin{align*}
\pm \tilde{{\mathcal{E}}}^{B^{\prime\prime\prime}}
\leq& C\Big\{ad^{-1}+N^4a^{\frac{8}{3}}d^{-2}\textit{l}^{\frac{1}{3}}
+N^{4}a^{2}d^{-2}(d\textit{l}+hm^{-1})^{\frac{2}{3}}
+N^2a^3d^{-3}h\ln(d\textit{l})^{-1}\nonumber\\
&\quad\quad\quad+N^4ad^{-1}m^{-1}(d\textit{l}+hm^{-1})
\Big(\ln\big(1+\frac{h}{d\textit{l}}\big)\Big)^{\frac{1}{2}}\Big\}
(\mathcal{N}_++1)\nonumber\\
&+CN^{\frac{3}{2}}a^{\frac{3}{2}}d^{-\frac{3}{2}}\textit{l}^{-\frac{1}{2}}m^{-1}
(\mathcal{N}_++1)^{\frac{3}{2}}
+C\Big(\frac{1}{m}+\frac{a^2}{d^2\textit{l}}
+\frac{a}{d}(d\textit{l})^2\Big)(\mathcal{N}_++1)^2\nonumber\\
&+CN\vartheta_1^{-1}\Big\{\frac{a\textit{l}\ln (d\textit{l})^{-1}}
{dm^2}+\frac{\ln h^{-1}}{m^2}
+\frac{a^{\frac{1}{3}}}{d^{\frac{1}{3}}m^{\frac{5}{3}}\textit{l}}\Big\}
(\mathcal{N}_++1)^2\nonumber\\
&+C\vartheta_1 \Big\{H_{21}+N^2\Big(\frac{a\textit{l}
\ln (d\textit{l})^{-1}}
{dm^2}+\frac{\ln h^{-1}}{m^2}
+\frac{a^{\frac{1}{3}}}{d^{\frac{1}{3}}m^{\frac{5}{3}}\textit{l}}\Big)\Big\}\nonumber\\
&+C\Big\{N^2a^{\frac{5}{3}}d^{-1}\textit{l}^{\frac{1}{3}}
+N^{\frac{7}{2}}a^{\frac{3}{2}}d^{-\frac{3}{2}}(d\textit{l}+hm^{-1})
+N^{2}ad^{-1}(d\textit{l}+hm^{-1})^{\frac{2}{3}}\nonumber\\
&\quad\quad\quad+N^2m^{-1}(d\textit{l}+hm^{-1})
\Big(\ln\big(1+\frac{h}{d\textit{l}}\big)\Big)^{\frac{1}{2}}
+Nad^{-1}h\Big(1+\frac{a}{d}\ln a^{-1}\Big)\Big\}\nonumber\\
&\times \Big\{H_{21}+\frac{Na}{d}(\mathcal{N}_++1)^2+N^2\Big(\frac{a\textit{l}
\ln (d\textit{l})^{-1}}
{dm^2}+\frac{\ln h^{-1}}{m^2}
+\frac{a^{\frac{1}{3}}}{d^{\frac{1}{3}}m^{\frac{5}{3}}\textit{l}}\Big)\Big\}\nonumber\\
&+C\Big(N^{\frac{3}{2}}a^{\frac{7}{6}}d^{-\frac{1}{2}}\textit{l}^{\frac{1}{3}}
+N^2ad^{-1}(d\textit{l}+hm^{-1})
+N^{\frac{3}{2}}a^{\frac{1}{2}}d^{-\frac{1}{2}}(d\textit{l}+hm^{-1})^{\frac{2}{3}}\Big)
\nonumber\\
&\times e^{-B^{\prime\prime\prime}}H_4e^{B^{\prime\prime\prime}},
\end{align*}
for some $\vartheta_1>0$. Moreover, we can calculate similar to (\ref{bogdsyufgy}) to reach
\begin{align*}
 e^{-B^{\prime\prime\prime}}{\mathcal{T}}^\prime e^{B^{\prime\prime\prime}}
=&\frac{1}{2}\sum_{p\neq0}\left(-\tilde{F}_p+\sqrt{\tilde{F}_p^2-\tilde{G}_p^2}\right)+
\sum_{p\neq0}\sqrt{\tilde{F}_p^2-\tilde{G}_p^2}a_p^*a_p+\mathcal{E}_{\mathcal{T}^\prime}.
\end{align*}
where
\begin{align*}
  \pm\mathcal{E}_{\mathcal{T}^\prime}\leq& C\vartheta_1^2(\mathcal{N}_++1)+
  CN\vartheta_1^{-1}\Big\{\frac{a\textit{l}\ln (d\textit{l})^{-1}}
{dm^2}+\frac{\ln h^{-1}}{m^2}
+\frac{a^{\frac{1}{3}}}{d^{\frac{1}{3}}m^{\frac{5}{3}}\textit{l}}\Big\}
(\mathcal{N}_++1)^2\\
&+CN^{-\frac{1}{2}}\Big\{\frac{N^2a\textit{l}\ln (d\textit{l})^{-1}}
{dm^2}+\frac{N^2\ln h^{-1}}{m^2}
+\frac{N^2a^{\frac{1}{3}}}{d^{\frac{1}{3}}m^{\frac{5}{3}}\textit{l}}\Big\}^3(H_{21}+1).
\end{align*}
We then let $\mathcal{E}^{B^{\prime\prime\prime}}=\tilde{\mathcal{E}}^{B^{\prime\prime\prime}}
+\mathcal{E}_{\mathcal{T}^\prime}$, and we finish the proof of Proposition \ref{Bog renorm III}.\hfill $\Box$

\section*{Acknowledgments}

X. Chen was supported in part by NSF grant DMS-2005469 and a Simons fellowship numbered 916862, and Z. Zhang was partially supported by NSF of China under Grant 12171010 and 12288101. We appreciate the delightful discussions with Zhifang Xu, Shizhong Zhang, Chushun Tian and Tin-Lun Ho, who have provided us many useful physical backgrounds and helped us enhance the understanding to the physical meanings of this problem. We also appreciate the helpful discussion with Arnaud Triay, as well as his careful reading, checking and kind comments
to this work.

\bigskip

%

\bibliographystyle{abbrv}
\bibliography{ref}

\end{document}